\documentclass[10pt,a4paper]{iopart}\usepackage{iopams}  
\usepackage{xcolor}
\usepackage[normalem]{ulem}

\usepackage{environ}
\NewEnviron{deletedParagraphs}{\color{red}{\sout{\BODY}}}{}
\NewEnviron{addedParagraphs}{\color{blue}{\BODY}}{}

\usepackage{cancel}

\usepackage{soul}

\usepackage{ulem}

\usepackage{graphicx}
\usepackage{mathrsfs} 

\usepackage{epsfig}

\begin{document}
\maketitle




\title[Vacuum Potentials for Proton and Electron. 
Appendix D: QEM Theory of IVBs, Higgs]{Vacuum Potentials for the Two Only Permanent Free Particles, Proton and Electron. 
Appendix D: QEM Theory of IVBs, Higgs}

\author{J.X. Zheng-Johansson}

\vspace{0.5cm}

\noindent
{\bf {\large{Contents}}} 
\\
{\bf{\small Part A \quad Vacuum Potentials for the Two Only Permanent Free Particles, Proton and }}
\\
\qquad\qquad {\bf{\small Electron. Pair Productions }} \dotfill\quad
 1-17
\\
{\bf{\small  Part B \quad A Microscopic Theory of the Neutron}} \dotfill\quad  18-37
\\
{\bf{\small Part C \quad A Quantum Electromagnetic Theory of the Pions, Muons and Their Emitting}}
\\
\qquad\qquad  {\bf{\small Particles (I)}}
\dotfill\quad  38-46
\\
{\bf{\small Part D \quad A Quantum Electromagnetic Theory of the Intermediate Vector Bosons and the }}
\\
\qquad\qquad {\bf{\small Higgs }} 
\dotfill\quad  46-58

\vspace{2cm}

\noindent
{\large{Part A} (Published in  {\it J. Phys.: Conf. Ser.} {\bf 343} 012135, 2012.)}

\title[Vacuum Potentials for Proton and Electron. Pair Productions]{Vacuum Potentials for the Two Only Permanent Free Particles, Proton and Electron. Pair Productions}

\author{J.X. Zheng-Johansson}
\address{Institute of Fundamental Physics Research
}

 
\def\gtsim{\gtrsim} 
\def\ltsim{\lessim} 

\def\psub{{\mbox{\tiny{$+$}} \hspace{-0.12cm}}}
\def\p{{\mbox{\scriptsize{$+$}} \hspace{-0.03cm}}}
\def\pe{\p e}
\def\pq{\p q}


\def\vsub{{\mbox{\tiny{v}}}}
\def\uv{u}
\def\av{u}
\def\sig{\sigma}

\def\pv{{p_{\vsub}}}
\def\vp{\pv}
\def\nv{{n_{\vsub}}}
\def\nvef{{n_{\vsub}^{{\rm ef}}}}
\def\vn{\nv}

\def\rbar{{\bar{r}}}
\def\pbar{{\bar{p}}}
\def\ebar{{\bar{e}}}
\def\abar{{\bar{\a}}}

\def\c{s}
\def\s{i}
\def\qc{q}

\def\Gcal{\mathcal{G}}
\def\Ab{{\bf{A}}}
\def\cb{{\bf c}}
\def\gw{\w}
\def\rr{r}
\def\em{e}
\def\ab{a}
\def\Ocal{\mathcal{O}}
\def\qz{q}
\def\a{\alpha}
\def\abar{{\bar{\a}}}
\def\nee{{\mbox{\scriptsize{-\hspace{-0.3mm}e}}}}
\def\qt{{qt}}
\def\fz{\phi}
\def\ft{\theta}
\def\f{\mathcal{A}}
\def\Asub{{\mbox{\tiny${\f}$}}}

\def\Acal{\mathscr{A}}
\def\Acalw{\Acal'}
\def\A{A}
\def\vphiq{\psi_q}
\def\etam{{\mbox{\small{${\mathcal{Q}}$}}}}
\def\dash{{\mbox{\tiny{$_-$}}}}
\def\bcal{\mathbin{{\etam}\mkern-8.5mu^{_{\mbox{\small{$\dash$}}}}\hspace{-0.04cm} }}
\def\b{\etam}
\def\rhow{\rho'}
\def\engw{\eng'}
\def\engb{\tilde{\eng}_\vphi}
\def\engqa{\eng_q}
\def\engqan{\tilde{\eng}_{qn}}
\def\Aqa{\tilde{A}_q}
\def\enga{\tilde{\eng}}
\def\velqz{\vel_{qz}}
\def\rhoqz{\rho_{q}}
\def\Dqz{D_{q}}
\def\db{{\bf d}}
\def\pb{{\bf p}}

\def\osub{{\mbox{\small{$0$}}}}
\def\N{{\mbox{\tiny{$N$}}}}
\def\Nsm{{\mbox{\scriptsize{$N$}}}}
\def\Ncal{{\cal N}}
\def\od{{\rm od}}
\def\eve{{\rm ev}}
\def\vtheta{\vartheta}
\def\phiz{\fz}
\def\vphiz{\phi}
\def\phix{\varphi}

\def\nh{n}
\def\fa{f}
\def\zrm{\mbox{\rm{z}}}
\def\minus{\mbox{-}}
\def\m{{\mbox{-}}}
\def\Rbqt{{\bf R}}
\def\a{\alpha}
\def\ph{{{\rm ph}}}
\def\ph{{\rm{ph}}}
\def\ub{{\bf u}}
\def\v{{\mbox{\footnotesize{\rm{v}}}}}
\def\vrm{{\rm v}}
\def\vac{\v}
\def\strvac{\supvac}
\def\supvac{\v}
\def\Nsub{{\mbox{\tiny${N}$}}}
\def\Zsub{{\mbox{\tiny${Z}$}}}
\def\kin{{\rm kin}}
\def\Y{Z}
\def\Z{Z}
\def\Pcal{{\mathcal{P}}}
\def\Fcal{{b_m}}
\def\bfrak{b_\v}
\def\ii{n}
\def\Esub{{\mbox{\tiny${E}$}}}
\def\cEphi{\Ccal_{\Esub \vphi}}
\def\cEu{\Ccal}
\def\cuphi{A_q}
\def\Ccal{{\mathcal{C}}}

\def\Bcal{{\mathcal{B}}}
\def\Xsub{{\mbox{\tiny${X}$}}}
\def\Ucal{{\mathcal{U}}}
\def\subempty{\empty}
\def\str{\nu}
\def\strempt{{}}

\def\prim{\prime}
\def\ArticleLabel-lp{16}
\def\Authorname{J.X. Zheng-Johansson}
\def\Cyc{\zrm_\vphi}
\def\vir{{\rm vir}}
\def\AppA{}
\def\AppChDy{}
\def\AppB{}
\def\AppC{}
\def\citeUnif{xx}
\def\citeHDDMunich{5} 
\def\vel{\upsilon}
\def\Eb{{\bf{E}}}
\def\Bb{{\bf{B}}}
\def\obs{{\rm obs}}
\def\ev{\epsilon}
\def\ke{\kappa}
\def\Omegavel{\mathbin{{\mit\Omega}\mkern-13.mu^{_{\mbox{$-$}}}\hspace{-0.08cm}{}_d }}
\def\Wvel{\mathbin{{\mit\Omega}\mkern-13.mu^{_{\mbox{$-$}}}\hspace{-0.08cm}{}_d }}
\def\wvel{\varpi_d}

\def\q{\mathbin{q\mkern-11mu-}}
\def\empty{{\mbox{\tiny${\emptyset}$}}}
\def\nf{n}
\def\Kcal{{\math{K}}}
\def\Xb{{\bf{X}}}
\def\CL{{\mbox{\tiny{{\it C.L}}}}}
\def\imr{{\rm im}}
\def\pl{{\mbox{\tiny{$\|$}}}}
\def\pp{{\mbox{\tiny{$\bot$}}}}
\def\Thm{\vartheta}
\def\TThm{{\mit{\Theta}}}
\def\Thetam{{\mit{\Theta}}}
\def\veli{{\mathscr{V}}}
\def\velib{{\pmb{\veli}}}
\def\Xim{{\mit{\Xi}}}
\def\Xima{\xi}
\def\Vel{W}
\def\Velb{{\bf{\Vel}}}
\def\Pw{{\psitot_p}}
\def\PW{{\Psim_p}}

\def\Mcal{{\mathfrak{M}}}
\def\nablab{{\pmb{\nabla}}}
\def\velb{{\bf{v}}}
\def\rb{{\bf{r}}}
\def\jb{{\bf{j}}}
\def\fb{{\bf{f}}}
\def\Fb{{\bf{F}}}
\def\wrm{{\rm w}}
\def\med{{\med}}
\def\mfrak{{\mathfrak{M}}}
\def\Mfrak{{\mathfrak{M}}}
\def\Mfk{{\mathfrak{M}}}

\def\Jcal{j}
\def\Jcalb{{\bf{j}}}
\def\Jobs{j_{\rm obs}}
\def\jobs{\Jobs}
\def\jr{j_{\rm obs}}
\def\obs{{\rm{obs}} }
\def\qm{{\rm qm}}
\def\Dscr{D_{\qm}}



\def\W{{\mit \Omega}}
\def\g{\gamma{}}

\begin{abstract}
\def\eng{{\varepsilon}}
\def\a{\alpha}
\def\w{\omega}
The two only species of isolatable, smallest, or unit charges  $+e$ and $-e$ present in nature will interact with a polarisable dielectric vacuum  through two uniquely defined vacuum potential energy functions. All of the non-composite  subatomic particles containing one-unit charges, $+e$ or $-e$, in terms of the IED model  are therefore generated by the unite charges of either sign, of zero rest masses, oscillating in either of the two unique vacuum potential fields.
In this paper we give a first principles treatment of the dynamics of a specified charge $q$ in a dielectric vacuum. Based on the solutions for the charge, combined with previous solutions for the radiation fields,  we derive the vacuum potential energy function for the specified charge, which is quadratic and  consists of quantised potential energy levels. 
This therefore gives  rise to sharply defined charge  oscillation frequencies  and accordingly sharply-defined masses of the IED particles. 
By further combining with relevant experimental properties as input information, we determine the IED particles built from the charges $+e$ and $-e$ at their first excited states in the respective vacuum potential wells, togather with their radiation electromagntic waves, to be the proton and the electron, the observationally two only stable (permanently lived) and "free" particles containing one-unit charges. The formation conditions for their antiparticles as produced in pair productions can  be accordingly determined. The characteristics of formation conditions of all of the other more  energetic non-composite subatomic particles can also be recognised. We finally discuss the energy condition for pair production, which requires two successive energy supplies to (1) first disintegrate the  bound pair of vaculeon charges $+e,-e$ composing a vacuuon of the vacuum and (2) impart masses to the disintegrated charges.
 \end{abstract}


\def\utot{\mathscr{U}}
\def\Vscr{\mathscr{V}}
\def\V{\mathscr{V}}
\def\Vvq{V_{\v q}}
\def\Vvqo{V_{\v q0}}

\def\uscr{\mathscr{U}}
\def\uscrw{\uscr'}
\def\ubscr{\pmb{\mathscr{U}}}
        \def\ubscrq{{\pmb{\mathscr{U}}\hspace{-0.14cm}}_q}
         \def\ubscrqsq{{\pmb{\mathscr{U}}\hspace{-0.14cm}}_q{\hspace{-0.05cm}}}

\def\Zcal{{\mathcal{Z}}}
\def\Kcal{{\mathcal{K}}}
\def\kappab{\pmb{\kappa}}
\def\gd{{\mathcal{G}}}
\def\lb{{\bf l}}
\def\vb{{\bf v}}
\def\Rb{{\bf R}}
\def\pd{\partial}
\def\vphi{\varphi}
\def\psitot{{\mathcal{Y}}}
\def\psiR{\widetilde{\psi}}
\def\psiL{\widetilde{\psi}^{{\rm vir}}}
\def\psitotR{\widetilde{\psitot}}
\def\psitotL{\widetilde{\psitot}^{{\rm vir}}}
\def\PhimR{\widetilde{ {\mit \Phi}}}
\def\PsimR{\widetilde{ {\mit \Psi}}}
\def\PsimL{{\widetilde{ {\mit \Psi}}}^{{\rm vir}}}
\def\a{\alpha}
\def\apl{a}
\def\aplF{}
\def\uav{\bar{u}}
\def\D{\Delta}
\def\t{t}
\def\x{x}
\def\z{z}
\def\y{y}

\def\th{\theta}
\def\r{{\mbox{\tiny${R}$}}}
\def\re{{\mbox{\tiny${R}$}}}
\def\Fmed{F_{{\rm a.med}}}
\def\med{{\rm med}}
\def\Lw{L_{\varphi}}

\def\Efb{{\bf E}}
\def\Bfb{{\bf B}}
\def\Ac{ \varphi}
\def\Xsub{{\mbox{\tiny${X}$}}}
\def\Xssub{{\mbox{\tiny${X}$}}}
\def\Tssub{{\mbox{\tiny${T}$}}}
\def\Kb{{\bf{K}}}
\def\kb{{\bf{k}}}
\def\Ksub{{\mbox{\tiny${K}$}}}
\def\W{{\mit \Omega}}
\def\Wd{\W_d{}}
\def\Nu{{\cal V}}
\def\Nud{\Nu_d{}}
\def\Eng{{\cal E}}
\def\eng{{\varepsilon}}
\def\Acuni{\Ac_{{\Ksub}^\dagsup}^{\dagsup}}
\def\unduni{\Ac_{{\Ksub}^\dagger}^{\dagsup}}
\def\Acauni{\Ac_{{\Ksub}^\ddagsup}^{\ddagsup}}
\def\Acunim{{\Ac_{{\Ksub}^\dagsup}^{\dagsup *}}}
\def\undunim{{\Ac_{{\Ksub}^\dagsup}^{\dagsup}}^*}
\def\Acaunim{{\Ac_{{\Ksub}^\ddagsup}^{\ddagsup *}}}
\def\pd{\partial}
\def\Ad{ {\mit \psi}}
\def\psim{ {\mit \psi}}
\def\Kd{K_d{}}
\def\Lam{{\mit \Lambda}}
\def\lam{\lambda}
\def\dagsup{{\mbox{\tiny${\dagger}$}}}
\def\ddagsup{{\mbox{\tiny${\ddagger}$}}}
\def\psimKdK{\psim_{\Ksub,\Kdsub}}
\def\w{\omega{}}
\def\wdlow{\omega_d }
\def\g{\gamma{}} 
\def\Phim{{\mit \Phi}}
\def\Psim{{\mit \Psi}}     
\def\Psima{{\mit\Psi}}

\def\arm{{\rm a}}
\def\brm{{\rm b}}
\def\crm{{\rm c}}
\def\drm{{\rm d}}
\def\erm{{\rm e}}
\def\frm{{\rm f}}
\def\grm{{\rm g}}
\def\hrm{{\rm h}}
\def\lf{\left}
\def\rt{\right}
\def\Kdsub{{\mbox{\tiny${K_d}$}}}
\def\psimkd{\psim_{\kdsub}}
\def\psimKd{\psim_{\Kdsub}}
\def\hquad{ \ \ } 
\def\Taum{{\mit \Gamma}}

\setcounter{equation}{0}
\setcounter{section}{0}
\setcounter{figure}{0}
\section{Introduction}\label{sec-intr}
Up to the present several hundreds of  isolatable subatomic particles along with their antiparticles have been discovered, of these the very energetic and (or)  short lived ones existing only in high energy accelerators and cosmic ray radiation\cite{ParticleDataGroup2010}. Of these observational particles,  the proton (E Rutherford, 1919\cite{ParticleDataGroup2010}) and the electron (J J Thomson, 1897\cite{ParticleDataGroup2010})  are the only two particle species  containing one-unit charges which are stable, or permanently lived, and  "free" (i.e. available for building the usual materials with no need of "extraction") in the vacuum;  they are the building constituents of all atoms. While conceding such "privileged" status only to  these two particular opposite charged particles, nature differentiates the two by  unequal masses, with  the proton being about 1836 times heavier  than the electron. Nature differentiates their opposite charged antiparticles,  antiproton  and   positron, with a similar  mass asymmetry, and nevertheless appears to admit both with a similar permanent lifetime expectation. Although, if pair productions are the only sources of their creations in the real physical world,  the antiprotons would  appear to be prominently missing, and the positrons appear to be similarly missing, or "hidden" in the vacuum. The fundamental  reason for  this selective, asymmetric preference of nature for our  physical world is up to the present not explained.

This selective, asymmetric characteristic of the particle system  has been one essential  constraint imposed from the beginning upon the construction of an {\it internally electrodynamic} (IED) {\it particle model} and  {\it vacuuonic vacuum structure},  which the author carried out  in recent work \cite{Unif1}-\cite{Unif1sho} based on overall relevant experimental observations as input information. According to the construction, briefly, a single-charged matter particle like the electron, proton, etc., is composed  of (i) a point-like  charge (as source) of a  zero rest mass but of an oscillation of  characteristic  frequency  and (ii) the  electromagnetic waves 
generated by the  oscillating charge. And the vacuum is filled of electrically neutral but polarisable building entities,  vacuuons (to be detailed in Sec. \ref{Sec-Vnps}), separated at a mean distance  $b_\v$; this vacuum is an electrically polarisable dielectric medium. Representations of the IED particle based  mainly on solutions for the electromagnetic wave component  have been the subjects of previous investigations  \cite{Unif1}-\cite{Unif1qmprob}, which have yielded  predictions  of a  range of the long established basic properties and relations of particles under corresponding  conditions.

In this paper, in terms of  first principles solutions for the charge to be obtained first (Sec. \ref{Sec-VacPot}) and for the electromagnetic wave component of an IED particle obtained previously \cite{Unif1}-\cite{Unif1mass} we formally derive in Sec. \ref{Sec-VacPot} the vacuum potential energy function for a specified  charge $q$. Further combining with relevant experimental properties for particles as input information, we parameterise  the vacuum potential energy functions for the  two unit charges  $+e,-e$, and determine accordingly the dynamical states of the two only stable, "free" particles formed therein out  of the two respective charges,  the proton and electron, and their antiparticles; and we elucidate the characteristics of the remaining,  more energetic subatomic particles containing one-unit charges (Sec. \ref{sec-pot-para}).  Finally, we determine the vacuuonic potential energy functions and elucidate  the energy condition for pair production (Sec. \ref{Sec-Vnps}).
 
\section{Vacuum potential energy functions}\label{Sec-VacPot}

The vacuum is according to \cite{Unif1,Unif1Vac,Unif1Vacdiel} a substantial medium constituted of electrically neutral but polarisable 
 vacuuons that are densely packed relative to one another. This vacuum will be represented in  a three-dimensional flat euclidean space ($\Rb^3$), spanned by three Cartesian coordinates $X,Y,Z$ fixed in the vacuum. 
In this vacuum, in an interstice formed by vacuuons centred at position $\Rb_i$ there presents a charge $q$. 
The charge has been in the past time driven into motion by an applied  force  $\Fb_{app.0}
$, and thus endowed with  a total mechanical energy $\eng_{q}$. From time $t=0$ the force $\Fb_{app.0} $ has ceased action. So the charge will hereafter tend to  spontaneously move about.  
The charge is for the present assumed to be prevented from radiating and thus will maintain  the energy  $\eng_q$ through the course. 
It can be readily extended  to the general radiating case later.

The charge $q$ (together with its radiation field) is to eventually form a simple matter particle, like an electron and proton, etc in terms of the IED model. The charge will serve as the generating source charge of the matter particle; and 
its spontaneous motion  be the internal motion of the resulting matter particle. 
With the matter particle so formed, the internal energy will not be incorrectly twice counted only if  the charge has a zero rest mass. The charge will instead  have  dynamical mass ($\Mcal_{q}$) as a result of the spontaneous  motion of the charge, which pertains to the internal process of the matter particle.

To furnish a realistic model of matter particle,  the charge  needs further be \cite{Unif1} point like relative to its radiation waves, and yet be 
an extensive spinning liquid-like entity, or whirlpool, extending 
across the interstice region ($-\frac{b_\v}{2}, \frac{b_\v}{2}$) of vacuuons \cite{Unif1,Unif1sho}; $b_\v \approx 1 \times 10^{-18}$ m by a crude estimate based on experiment\cite{bv}. 
This extensive feature of the charge is necessary so as to conform to the overall basic experimental properties of charge, including  spin \cite{Unif1} and   the quantisation of energy (see below). The dynamics of $q$ at the scale $b_\v $, and hence an extensive $q$, will be of main concern in this paper.

The dynamical mass centre of the minute yet extensive charge will be at position 
 $\Rb_q(X_q,Y_q,Z_q)$ at time $t$, assuming  along the $Z$ direction in an time interval under consdition.  $\Rb_q $ is displaced  from the equilibrium position, $\Rb_i$, 
by  ${\pmb \uscr}_q = \Rb_q-\Rb_i =Z_q -Z_i$.xxx   
The extensive distibution of the charge 
may be generally described by  a (normalised) probability density $\rho_q(z,t)=|\psi_q(z,t)|^2$. It will have a flow rate $j_q=\vel_q \rho_q$  at the velocity $\vel_q$,  along the $z$ direction here.   
 $\psi_q$ is a complex function because $|\psi_q(z,t)|^2$ will  be associated with the total energy  (\ref{eq-engq1}) below that is conserved in a conservative force  field
(see further e.g.  \cite{jxzjqmopt}). 
  $j_q$ may be alternatively described by a diffusion current   $j_q =-D_q[\psi_q^* \nabla \psi_q-(\nabla \psi_q^*) \psi_q ]$, where $D_q$
is an imaginary  diffusion constant (\ref{App-Diff}). The constant $D_q$, whence  diffusivity, is in inverse proportionality to the resistivity of the (vacuum) medium that will be identified to be measured by a dynamical mass $\Mcal_q$ later, 
 whence the usual relation   $D_q=\frac{i \hbar }{2 \Mcal_q}$. 
We will be mainly interested in the formation of stable particles, or particle states, as the proton, electrons are.
 This will only be ensured if  $\rho_q$ fulfils the continuity equation, 
$$\displaylines{
\refstepcounter{equation} \label{eq-eqsmotb}
\hfill  \pd_t \rho_q + \nabla (\rho_q \vel_q)=0 \quad {\rm or}
\ \pd_t \rho_q - D_q   [\psi_q^* \nabla^2 \psi_q-(\nabla^2 \psi_q^*)\psi_q]=0. 
 \hfill  (\ref{eq-eqsmotb})
}$$

The extesnive oscillatory charge $q$ constrained by (\ref{eq-eqsmotb})  
 will be found to move as a rigid object, and thus  
may be represented as a point particle located at its mass centre, of the coordinate $\Rb_q$ or $\uscr_q$ earlier. For this effective point particle, Newton's laws of motion are valid.   
Firstly, the spontaneiously moving charge will be  subject to  a spontaneous inertial force $\Fb_{ine} (\equiv \Fb_{app.0}) $ associated with $d^2_t\ubscrq (\equiv \frac{d^2\ubscrq}{dt^2})$.
This  force is  given according to  Newton's law of inertia  as 
$ \Fb_{ine} =\Mcal_q d^2_t\ubscrq$,  
where $\Mcal_q $ is a proportionality constant, or it  is the "(dynamical) inertial mass" of $q$. 

 In the vacuuonic vacuum, the motion of the charge $q$ will be resisted. 
This is as a consequence that the vacuuons in the vicinity of $q$ become polarised by $q$   and builds with $q$ an interaction potential,  $\Vvq (\ubscr_q) =\Vvqo+\sum_n  \frac{1}{n!}\nabla^n \Vvq(\ubscr_q) \ubscrqsq^n$, where $\Vvqo=\Vvq(0)$ is a constant.
$\Vvq$ is the superimposed result of the electrostatic interactions $V_{\v_j q}$ of $q$ with all of individual polarised vacuuons $j$  up to an intermediate range about $q$,
 $\Vvq (\ubscr_q) =\sum_j V_{\v_j q}$ (see further Sec. \ref{Sec-Vnps} and \ref{app-Vqv1}). 
The corresponding restoring force  is $\Fb_{res}= -\nabla V_{\v q}(\ubscr_q) -(-\nabla V_{\v q}(0)) $. 
There will be  a finite time interval during which the charge displacement $\uscr_q (< \frac{b_\v}{2})$  is about the fixed site $\Rb_i$ and along the $Z$ direction, hence $\uscr_q(=z)=Z_q-Z_i $.  We will  consider the dynamics   in  this time interval below.
$\uscr_q (< \frac{b_\v}{2})$ must  be relatively small, judging on the basis that  its radiation field obeys the linear Maxwell's equations and thus has an wave amplitude  that is relatively small. 
It thus suffices to retain the leading terms in $\Vvq (\ubscr_q)  $  to 
only. 
 The vacuum is isotropic,  
so  odd terms in  $\Vvq (\ubscr_q) $ must furthermore vanish.  The $\Vvq (\ubscr_q)  $ and  $\Fb_{res}$  are thus given as
$$\displaylines{
\refstepcounter{equation} \label{eq-Vvq}
\hfill 
V_{\v q}(\uscr_q)
=V_{\v q0} + \frac{1}{2}\beta_q \uscr_q^2, \quad \beta_q = \nabla^2 V_{\v q}; \quad 
 F_{res}=-\beta_q \uscr_q. 
\hfill (\ref{eq-Vvq}) 
}$$

Under the condition  (\ref{eq-eqsmotb}) and the action of the  forces above, and generally also in the presence of an external (total) force $\Fb_{ext}$, the equation of motion of the charge from time $t= 0$  is given according to Newton's second law  as   $\Fb_{ine}-(\Fb_{res} + \Fb_{ext})=0$,  or, 
$$\displaylines{
\refstepcounter{equation} \label{eq-eqsmota}
\hfill  d^2_t \uscr_q+ \w^2  \uscr_q -\Fb_{ext}/\Mcal_q=0,  
\quad 
\w=\left(\frac{\beta_q}{ \Mcal_q}\right)^{1/2}.
\hfill (\ref{eq-eqsmota})
}$$
\noindent

In an ordinary  environment there always present certain random radiation fields, 
which can statistically act (a)  a torque $\Fb_{ran} \times {\bf d}$ on the oscillating-charge dipole, and (b) a linear force $\Fb_{ran}$ on the charge's mass centre. Due to (a) and if no other external field present, $\uscr_q$ will  alter in orientation at every brief yet finite time interval and will explore all orientations over long time. Due to (b), the charge may be promoted to hop  over an energy barrier $\D_{\v i}$  to a neighbouring site, randomly in any  possible directions. An applied unidirectional force ($F_{u}$) acting on the oscillatory charge as a whole,  assuming here  $F_{u}=F_{u0}$ as  component of the initial  total force $\Fb_{appl.0}$ earlier and $F_{u}=F_{u0}$ is in the $X$ direction, will ordinate $q$ to hop from site to site in the $X$ direction. The motion has  a mean velocity given by $\vel=\frac{1}{N}\sum^N \frac{ X_{i+1}-X_{i}}{ \delta  t_i}$, $\delta  t_i$ being the dwelling time at site $i$. 
Owing to a Doppler effect associated with his source-charge motion, 
 $\w=\g \W$ is augmented  by a factor $\g$ from $\W$, and thus  $\beta_q= \g^2 \beta_q^0$, $\W$ and $\beta_q^0$ being the values  as measured when the charge oscillator is at rest ($\vel=0$) in the $X$ direction. 
$ \g= 1/\sqrt{1-\vel^2/c^2}$ as directly given by electromagnetic solutions 
\cite{Unif1},\cite{Unif1Schr}.
The $\Fb_{ran}$ above,  the $\Fb_{app}$ earlier, and  the $\Fb_{rad}$ of  \ref{App-tphi} later are all contributions to $\Fb_{ext}$.

We below consider first the  charge motion about the  fixed site $\Rb_i$ under no external force, i.e.  $F_{ext}=0$. Equation (\ref{eq-eqsmota}), to consider first, has  a general complex solution 
$$\displaylines{
\refstepcounter{equation} \label{eq-u}
\hfill
\uscr_q^c (t)=\Acal_q e^{-i (\w t+\a_o)}; 
\quad 
 \uscr_q(t)= {\rm Re}[\uscr_q^c(t)]=\Acal_q \cos(\w t+\a_o), \hfill (\ref{eq-u})}$$
where $\Acal_q$ is the amplitude and $\a_o$ an initial phase;  $\Acal_q=\Acal_q^0/\sqrt{\g}$, denotes the Lorentz    contracted quantity of the uncontracted value $\Acal_{q}^0$; and similarly $\uscr_q(=\uscr_q^0/\sqrt{\g})$, $\uscr_q^c(=\uscr_q^{c0}/\sqrt{\g})$.    
That is,   in the absence of applied force the minute charge as a whole executes a harmonic motion of displacement $ \uscr_q$, of a $\g$-augmented characteristic (or natural) angular frequency $\W$, $\w=\g\W$,  in the quadratic vacuum potential well $V_{\v q }$.
The corresponding kinetic  and elastic potential energies at any  time $t$  are thus  $\eng_{qk}(t)= \frac{1}{2}\Mcal_q \dot{\uscr}_q^2$ and  $\V_q(t)=V_{\v q }(\uscr_q(t))-V_{\v q 0}=\frac{1}{2}  \beta_q \uscr_q^2(t)$. The total mechanical energy, or Hamiltonian, is  
$$\displaylines{
\refstepcounter{equation} \label{eq-engq1}
\hfill \eng_q(t)(=\eng_{q.in})= \eng_{qk}(t)+ \V_{q}(t)
=\eng_q |e^{-i (\w t+\a_o)}|^2=\eng_q,  \quad 
\eng_q= \mbox{$\frac{1}{2}$} \Mcal_q \W^2 \Acal_{q}^2, 
\hfill (\ref{eq-engq1})
}$$
where $|e^{-i (\w t+\a_o)}|^2=\cos^2(\w t+\a_o)+\sin^2(\w t+\a_o)=1$. 
$\eng_q(t)$ given in (\ref{eq-engq1}) is constant in time and thus defines a stationary state of the harmonic charge oscillator.

The constraining equation (\ref{eq-eqsmotb}) decomposes into two conjugate second order differential equations for $\psi_q$ and $\psi^*_q$, that  are mathematically equivalent  to the Schr\"odinger equations for a harmonic  oscillator. 
The solution for  $\psi_q$ (and similarly $\psi^*$) follows therefore to be the standard hermit polynomial (e.g. \cite{Merzbacher}). And the solution for total energy  consists of quantised levels,   
$$ \displaylines{\refstepcounter{equation} \label{eq-eqn}
\hfill
 \eng_{qn} = n \hbar \w, \quad n=1,2, \ldots \hfill (\ref{eq-eqn})}$$
A solution  $\eng_{q0}=\frac{1}{2}\hbar \w$ is also mathematically permitted 
but has been discarded in (\ref{eq-eqn}), because it is judged as unphysical based on  a comparison with the empirical Plank energy equation for radiation.
(\ref{eq-eqn}) is a prediction of the Planck energy equation for the electromagnetic radiation  associated with the $\eng_q$ here, and hence the mass $m$ of the resulting IED particle to be specified below.
The total energy $ \eng_{qn} $ quantisation given  in (\ref{eq-eqn}) is the result of confinement of the minute extensive charge in the vacuum potential well at the scale $b_\v \sim 10^{-18}$ m.  
The thermal motion of the IED particle is on the other hand executed across a distance $A$ which contains (tremendously) many vacuuon spacings, $A>> b_\v$. 
Thermal energy quantisation  will be in question only when the IED particle is confined at a scale $A$, and this will not considered in this paper.

The  charge in oscillatory motion normally will  generate  electromagnetic waves, gradually and thus continuously, for the electromagnetic waves are propagated at the finite speed of light $c$ and are distributed in space.
The oscillating charge and its radiation field together make up our {\it  IED particle.} 
If the charge is restricted to emit radiation only and (re)absorb none, after a time $t_\vphi$ its entire $\eng_q$ will thus have been converted to the total energy $\eng''$, $=\eng_q$,
of the total electromagnetic wave.
In an open vacuum the electromagntic waves generated by the point source chareg here are  propgated in radial direction. With respect to their energies and linear momenta, the fields may be equivalently represented as two  Doppler-effected  effective plane waves $E^\dagsup,E^\ddagsup$  travelling oppositely at the velocities $+c,-c$ in the  $+X$  and $-X$ directions along a linear  vacuum path of a cross sectional area $s_0=\frac{8\pi (b_\v/2)^2}{3}$, of a  radiation electric field $E(X,t)=E^\dagger+ E^\ddagger=E_0 \cos [\w(X/c-t)]$, $E_0=qA_q\w^2/4\pi\ev_0 (b_\v/2)c^2$.  
                    %
$\eng''$ is given according to  electromagnetic theory  as $\eng''=\sqrt{\eng{''}^\dagger \eng{''}^{\ddagger}})=L_\vphi s \ev_0 \g |E|{}^2$, where $L_\vphi$ is the  geometric mean of the total lengths of the two wave trains.
If attributing the wave oscillations as the internal motions of the wave trains, 
the total wave motion thus reduces  to the rectilinear motion, at the speed of light $c$, of a total wave train as a rigid object, of a finite inertial mass $m''$ (which reflects the resistivity  against the motion of the wave train in the bulk vacuum continuum), and  linear momentum $p''=\eng''/c= \hbar k$, with $k=\w/c$. 
              %
The same $\eng''$ is thus now given \cite{Unif1,Unif1Schr,Unif1mass} as   the kinetic energy, $\eng_k= \frac{1}{2}m''c^2$, of the wave train plus an elastic potential energy equal to $\eng_k$, whence   
$\eng''= 2\times \eng_k''=m''c^2$;  $m''$ is thus also the relativistic mass 
 of the IED particle (see e.g. \cite{Unif1Schr}).

The $\eng''$ above and the Newtonian result  $\eng_q$ of (\ref{eq-engq1}), both being equal to $\eng_{qn}$, follow therefore to  be quantised each, in the inevitable way as  
$$\displaylines{
\refstepcounter{equation} \label{eq-engnb1}
\hfill
\eng'' \rightarrow \eng_n= L_\vphi s\ev_0 \g E_{n}^2= m_n c^2,  \quad 
{\rm with }  \ E''\rightarrow E_n=\sqrt{n}E_1, 
 \quad m''\rightarrow m_n=nm, \hfill  (\ref{eq-engnb1})
 \cr\refstepcounter{equation} \label{eq-engnb2}
\hfill
\eng_q \rightarrow \eng_{qn}^{newt}
= \frac{1}{2}  \beta_q\Acal_{qn}^2, \quad 
{\rm with }  \  \Acal_{q} \rightarrow \Acal_{qn}=\sqrt{n}\Acal_{q1},
\hfill (\ref{eq-engnb2})
 }$$
where  $\beta_q =\Mcal_q \w^2$ is as given by  (\ref{eq-eqsmota}b).
From the equalities $\eng_n=\eng_{qn}^{newt}=\eng_{qn}$,  
we obtain a few relevant relations for later use 
$$\displaylines{
\refstepcounter{equation}  \label{eq-engn2}
\hfill 
\w= \frac{mc^2}{\hbar}, \quad   
\Acal_{q1}= \left(\frac{2mc^2}{\beta_q}\right)^{1/2},
 \quad {\rm or}  \ 
  \beta_q
=\frac{2mc^2}{\Acal_{q1}^2},   
\quad  \Mcal_q =\frac{2mc^2}{\Acal_{q1}^2\w^2},
\hfill (\ref{eq-engn2})
}$$ 
where $m\equiv m_1=\g M$ is the relativistic mass and  $M=\lim_{\vel^2/c^2 \rightarrow 0} m$ the rest mass of the IED particle here formed of the charge $q$ alone (in the extreme case of no radiation) at the energy level  $n=1$ of excited state.

Any massive materials in ordinary conditions in the surrounding  will serve as non "absorbing" reflection walls to the wave of the  energy quanta $n\times \hbar \w$ which can only be "absorbed" through a pair annihilation  with its anti-particle, which is  rare so as to be deemed not to occur in a normal environment. 
So from the nearest such  walls the waves will be  reflected back to the charge, be  re-absorbed by it  and then re-emitted,  continuously and repeatedly.  The total energy $\eng_{tn} $ of the IED particle will in general be at any time  carried a fraction $a_1$ by the charge and a fraction $a_2$ by the radiation wave, with $0 \le a_1,a_2 \le 1$, $a_1+a_2=1 $.  So $\eng_{tn}=a_1\eng_{qn} +a_2 \eng_{n}$, which is similarly quantised. Accordingly,  the actual potential energy of the charge is at any time a fraction $a_1 $ of the total $V_{\v q n}$,  $a_1V_{\v q n}$;  and this, as one will readily obtain by combining with the solution of \ref{App-tphi}\cite{Unif1sho}, is as a result that  $\Acal_{q1}$  is scaled by $\sqrt{a_1}$ to  $\sqrt{a_1}\Acal_{q1}$. 
For the charge dynamics in the vacuum potential field as the major concern below,  unless specified otherwise we shall for simplicity return to  the extreme situation of $a_1=1$, $a_2=0$.

With $\Acal_q \rightarrow \Acal_{qn}$ of (\ref{eq-engnb2}b), we have $\uscr_q\equiv z\rightarrow \uscr_{qn}
\equiv z_n
=\sqrt{n}z$, $z=\Acal_{q1} \mbox{Re}[\theta_q]$; or   
$\zeta_n=\frac{z_n}{b_\v}=\sqrt{n} \zeta$,  $\zeta=\frac{z}{b_\v}$.  Placing  this $\uscr_{qn}$   in  (\ref{eq-Vvq}a), we obtain the quantised  vacuum potential energy of charge $q$
$$\displaylines{
\refstepcounter{equation} \label{eq-Vn}
\hfill V_{\v q n} (z)=V_{\v q 0}+ \frac{1}{2} \beta_{q} n z^2 =V_{\v q 0}+ \frac{1}{2}  \beta_{q}' n \zeta^2 \quad 
 \qquad (|z|< \frac{b_\v}{2})
\hfill (\ref{eq-Vn})
}$$
where $\beta_q'= b_\v^2\beta_q$. The $\beta_q$ and $V_{\v q0}$ are uniquely fixed for a fixed  $q$ value and the universal dielectric vacuum, if we disregard possible effects on  the local instantaneous vacuuon configurations from the variant sizes and frequencies 
of the charge. $V_{\v q n} $ thus is a
uniquely defined function for a specified  $q$ value. The two only isolatable, smallest or  unit charges $+e$ and $-e$ present in nature  therefore are associated with two unique vacuum potential energy functions. 
Considering that  the resistivity, $\propto \Mcal_q$, against a specified charge in a uniquely specified vacuum potential field also is uniquely fixed, then  (for $\vel=0$)  $\w|_{\vel=0}=\W=(\beta_q^0/\Mcal_q)^{1/2}$ given in (\ref{eq-eqsmota})  is a characteristic  quantity of the specified charge and the universal vacuum, i.e. $\W$ represents the natural angular frequency of the given charge and vacuum system.

If the charge oscillator is not restricted from radiation and is at present time in an equilibrated state of (re)emission and (re)absorption of radiation,  the total $\eng_q$ is thus distributed over a (large) $N_0$ number of radiation cycles. 
The (average) Hamiltonian of the charge oscillation of one cycle is thus $\frac{\eng_{qn}}{N_0}= \frac{1}{2}\beta_q |u_{qn}^c|{}^2$, where 
$$\displaylines{
\refstepcounter{equation} \label{eq-Aq}
\hfill 
 u_{qn}= N_0 A_{qn} e^{-i\w t}, \quad A_{qn}=\Acal_{qn}/\sqrt{N_0} 
\hfill (\ref{eq-Aq})
}$$
are the corresponding complex  oscillation displacement and amplitude. At any time $t$ the  Hamiltonian $\frac{\eng_{qn}}{N_0}$  is conveyed by the charge,   
and that  of the remaining $(N_0-1)$ cycles is conveyed by the radiation field.  
The dynamical variables $A_{qn}$, $V_{\v qn\tau}$,  etc. of  each cycle are quantised, inevitably against a fractional Planck constant $h$  (in a similar situation as  in \ref{App-tphi} and \cite{Unif1sho}); it is  the total  $\eng_{qn}$ only that is quantised 
against the whole $h$,  as given by (\ref{eq-eqn}). The notion involved here is  consistent with the physical origin of the Planck constant as recognised in \cite{Unif1Plnk}.

\section{Parameterisation
}
\label{sec-pot-para}
At the present we  lack  adequate input data,  the  vacuuon polarisability especially (see \ref{app-Vqv1}), for an  {\it ab initio} evaluation of  $V_{\v q}$. Instead, we shall below determine the parameters $\beta_q,V_{\v q 0}$ of 
the $V_{\v q}$ function  for charges  $+e,-e$ based on  experimental properties for particles, 
 the two most common particles  proton ($p$) and  electron ($e$) mainly. The parameterised potential energy functions will in the end be characteristically justified by comparison   with the direct electromagnetic interaction functions for an external $q$ and  individual  vacuuon given in  \ref{app-Vqv1} based on an 
arbitrary  value of polarisability.

 Observationally (e.g. \cite{ParticleDataGroup2010}),  the $p,e$ are   (I) of sharply-defined constant rest masses $M_p,M_e$, (II) of the smallest masses 
 (i.e. the $M_p,M_e$) among the particles which contain each one-unit  $+e$ or $-e$ and also possess  the properties   (III)-(IV) below,   (III) stable (i.e. of infinite lifetimes),   and  (IV)  free in the vacuum. That the $p$ and $e$ are free, point (IV), are said in the sense that they  are available for building the materials in our physical  world with no need of  extra energy for extraction. 
 The oscillating charges $+e$ and $-e$ composing the corresponding two IED particles are  therefore required to be (i)   stationary, i.e. being at one of the energy levels   $n=1,2, \ldots$ following (\ref{eq-eqn}),  (ii) factually at level $n=1$ in accordance to property (II),  (iii) of infinite lifetimes, and (iv) free in the vacuum.  (iii) is to be justified and (iv) to be furnished by positioning of the level $n=1$ in the vacuum potential well below.

From  (\ref{eq-time}),   \ref{App-tphi}, it follows that the time required for the (quasi) harmonic charge  oscillator $q$ at the initial state $n=1$ to have emitted its entire one energy quantum $\eng_{q1}-\eng_{q0}$, and transformed to the final state $n=0$, is 
$$\displaylines{
\refstepcounter{equation} \label{eq-time1}
\hfill t_{\vphi 1.0}=\infty.
\hfill (\ref{eq-time1})
}$$
By the usual  quantum mechanical principle, a transfer of only a fraction of a quantum $\hbar \w$ from one charged particle  $\a(q)$ to another charged particle $\a'(q')$ in their quantum states is  forbidden. Therefore a transition of  charge $q$ from level $n=1$ to 0 within a finite time is improbable. This verifies the (iii) above. This restriction however does not apply if $q$ is in an asymmetric potential field,  e.g. a field produced by the charge ($q'$) of an antiparticle at a very close distance.

We define "vacuum level", $V_{\v q\v}=V_{\v qn'}(z_\v)$, as the level  at which the pair of  vaculeons constituting a vacuuon are no longer attracted with one another, thus being (effectively) at an infinite separation; $n'$ is a specific value of $n$. At this level, an external charge $q$, oscillating at amplitude $z_{\v}$ about its equilibrium position $z=0$, thus just begins to be no longer attracted to the surrounding  vacuuons\footnote{
A  "negative" $V_{\v q 0}$ strictly  applies to $+q$ and has for $-q$ a relative meaning only due to the $V_{\v q}$ asymmetry over $+q$ and $-q$;  see further  \ref{app-Vqv1}.1.}, 
and it is subject to instantaneous collisions with the vacuuons only. Therefore, a charge $q$ at the vacuum level is free in the sense of (IV).  
The charges $+e,-e$ of the $p,e$ of the feature (iv) therefore lie at the vacuum level; and they are in turn in their $n=1$ stationary states as stated by  (ii)-(iii). That is,  for the charges of $p$ and $e$, the $n=1$ levels coincide with  the vacuum level $V_{\v q1}=V_{\v  q\v}=0$, and  $z_\v=\Acal_{q1}$, whence the property (iv) is furnished.  

\begin{figure}[tbh]
\vspace{-0.7cm}
\begin{center}\includegraphics[width=0.95\textwidth]{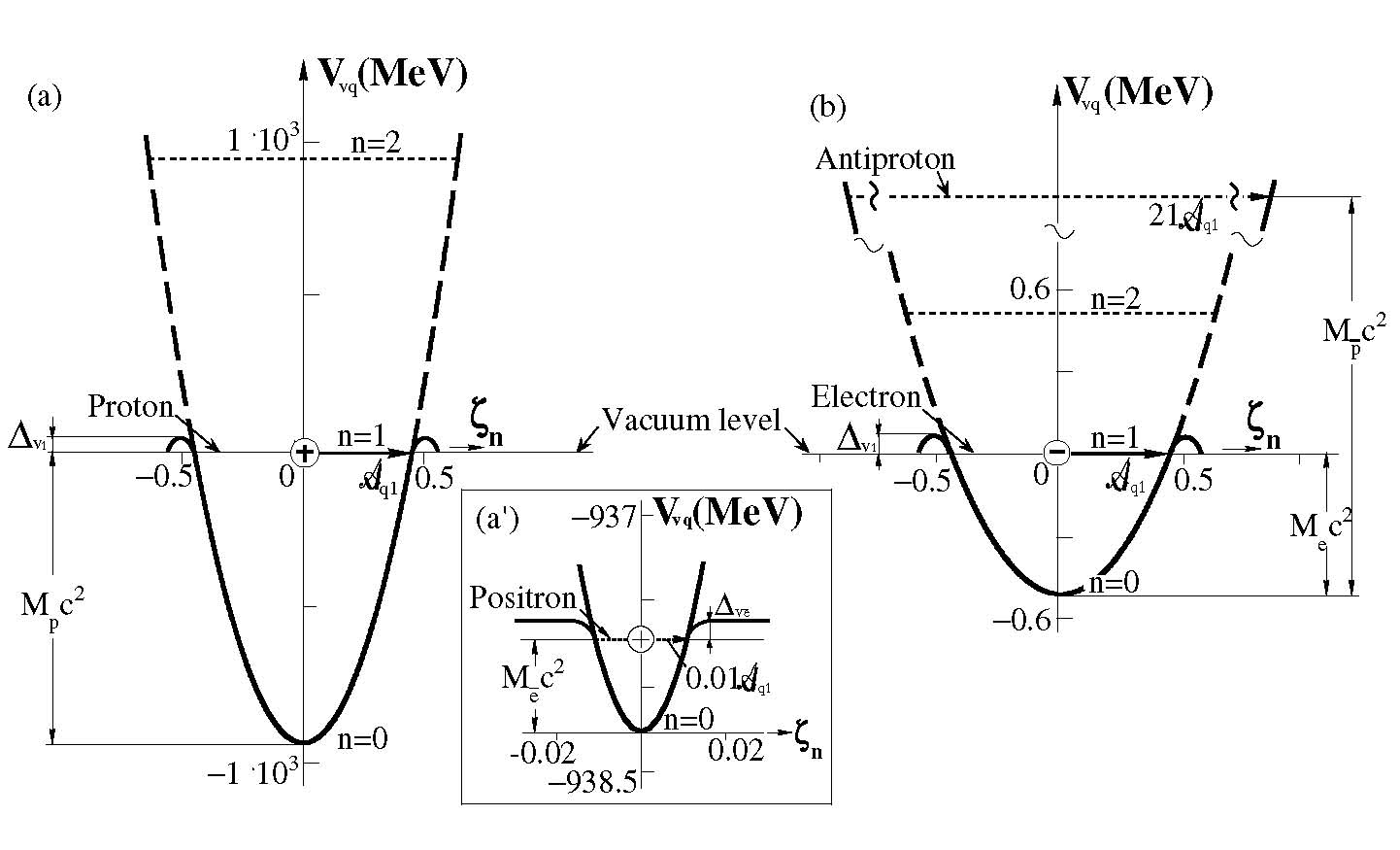}\end{center}  
             \vspace{-0.7cm}\caption{
Vacuum potential energy functions $V_{\v q} (\zeta_n)$ versus the centre-of-mass coordinate of a minute extensive charge $q$, $\zeta_n$, given by (\ref{eq-Vvqnpe}) for (a) $q=+e$, and (b)  $q=-e$. Used for the plot: $f_1 =0.9$.
}\label{fig-Vv2z-z.eps}
\vspace{-0.cm}                 
  \end{figure}

By the solution (\ref{eq-eqn}), in zero external field the harmonic state of the charge at level $n=1$ can only be promoted to  higher levels by a discrete amount at a time, i.e. $n\hbar \w$, $n=2, 3, \dots$. Or, it will not be altered at all.  The charge is however not restricted from being continuously promoted to higher  energies  if acted on by an unidirectional force $F_{u}$ and, assuming a sufficiently large $F_{u}$, may be  driven momentarily  to the mid point $z_{1/2}=(Z_{i+1}-Z_i)/2$ between site  $i$ and adjacent site $ i+1$ along a diffusion path. At $z_{1/2}$, it experiences  shortest  distances to the neighbouring vacuuons, and therefore  a maximum potential energy $V_{\v q 1 m}' =V_{\v q1}'(z_{1/2}) $. The potential energy difference $\D_{\v_1}(z)=V_{\v q 1}'(z) -V_{\v  q1} $ defines  an energy barrier which the charge $q$ must overcome to  hop to an adjacent site, see Figure \ref{fig-Vv2z-z.eps}. Its height $\D_{\v_1} (z_{1/2})=V_{\v q 1 m}' - V_{\v  q1}  $ and  width, $  \delta_1$, are both  dependent on the instantaneous interaction of $q$, while at $z_{1/2}$, with the vacuuons whose configuration fluctuates due to the influence of the random environmental fields and the instantaneous motion of the charge $q$. Their determination is beyond the scope of  this paper. 

In conformity with the observational vacuum, the vacuuons  are  densely packed in a disordered fashion and are, by virtue of their internal structure (see Sec. \ref{Sec-Vnps}),   in zero external field  electrically neutral and non interacting; they form  a perfect liquid\cite{Unif1}. 
The $V_{\v q n}$ is intermediate ranged (see \ref{app-Vqv1});  the vacuuons  thus to good approximation present  to $q$ with an average structure.
Then, assuming also disorder effect will be additionally included where in question (such as diffusion path),  the vacuuons may be in the simplest illustration represented as arranged on a simple cubic lattice of  spacing $b_\v$.   So   $z_{1/2}=\frac{b_\v}{2}$. The  oscillation amplitude of the charge $q$ at $n=1$ level thus  is (for $N_0=1$), 
$$\displaylines{
\refstepcounter{equation} \label{eq-Aqc1}
\hfill
\Acal_{q1}=\mbox{$\frac{1}{2}$}(b_\v-\delta_1).
\hfill (\ref{eq-Aqc1})
}$$
If $\eng_q$ is distributed over  $N_0$ radiation cycles, using   (\ref{eq-Aq}) in (\ref{eq-Aqc1}) gives  $A_q =\mbox{$\frac{1}{2N_0}$}(b_\v-\delta_1)$.
 With the  $ \Acal_{q1}$ above, and the experimental rest masses of $p$ and $e$, 
$M_p (= 938.27$ MeV) and  $M_e ( = 0.511$ MeV) for $M$ in  (\ref{eq-engn2}c),
 we obtain for the charges $+e$ and $-e$:
$$\displaylines{
\refstepcounter{equation} \label{eq-betaqpe}
                  \hfill
\beta_q=\frac{2M c^2}{(f_1b_\v/2)^2}, 
\quad  f_1=\frac{\Acal_{q1}}{b_\v/2}=1-\frac{ \delta_1}{b_\v} 
\qquad (q, M=+e, M_p; -e, M_e)
\hfill (\ref{eq-betaqpe})
}$$
Values of $\beta_q$ evaluated based on  (\ref{eq-betaqpe}) for the charges of the IED  proton $p$, electron $e$, antiproton $ \pbar$, and positron $\ebar$, together with  other  parameters involved in this section, $\W, \Acal_{q1} (\Acal_{q \a}), \Mcal_q, t_{\vphi 1.0}$, are tabulated in Table \ref{tab1}.
\begin{table}[h]         
\vspace{-0.5cm}
\begin{center} 
\caption{ } 
\label{tab1} 
\begin{tabular}[h]{
|p{0.4cm}  |p{1.2cm}                  |p{1.7cm}          |p{1.3cm}  |p{1.6cm}
|p{1.5cm} 
  |p{2.1cm} 
|p{1.39cm}| }
\hline
$q^{(a)}$         & \quad IED \quad Particle &  $ \hspace{0.4cm}  M^{(a)} \hspace{0.6cm} $   \qquad \hspace{-0.65cm} (MeV)  & $\quad \W^{(b)} $  ($10^{20}$r/s)  &     \hspace{0.cm}   $ \hspace{-0.1cm} 
\Acal_{q1}f_1^{-1}{}^{(c)}$ ($10^{-18}$ m)     
& 
\hspace{-0.1cm}
$\beta_q f_1^2{}^{(d)}$  ($10^{23}$N/m)  
& $\quad \Mcal_q f_1^2{}^{(e)}$ \qquad  \hspace{-0.3cm}   (kg)    \hspace{0.4cm}      
&Lifetime$^{(f)}$ \qquad\hspace{-0.4cm}(s)
\\
\hline
$+e$       &  \hspace{0.4cm} $p$                   & $(M_p) \ 938.27$         &$\hspace{0.1cm} 14280$ 
                                                                                                      &        $\hspace{0.2cm}0.5$
                                                                                         &  \hspace{0.4cm}$12020$
&                                                                 $5.896 \times 10^{-22}$       
                                                                                                                                &\quad $\infty$
 \\
$-e$       &  \hspace{0.4cm} $\bar{p} \ $               & $ (M_{\bar{p}}) \ 938.27$         & $\hspace{0.1cm}14280$ 
                                                                                                 & $\hspace{0.2cm}21.42$
&\hspace{0.4cm}$6.549$       
 & $3.218\times 10^{-25}$               
& \hspace{0.2cm}(short)
\\ 
\hline
$-e$     &   \hspace{0.4cm} $e$                  &     $  (M_{ e}) \ 0.511$           &
 $\hspace{0.1cm} 7.778 $ 
&        $\hspace{0.2cm}0.5 $
&\hspace{0.4cm}$6.549$ 
 &       $1.083\times 10^{-18}$         
& \quad$\infty$ 
\\
$+e$     &  \hspace{0.4cm}  $\bar{e}$                  &     $(M_{\bar{e}}) \ 0.511$           &
$\hspace{0.1cm} 7.778$          
               & $\hspace{0.2cm}0.0116$
&\hspace{0.4cm}$12020$        
         &       $1.989\times 10^{-15}$         
& $\quad(\infty)$ \\
\hline
\end{tabular}
\end{center}
\vspace{-0.1cm}
{\baselineskip 0.4cm
\footnotesize{(a) Experimental masses of $p,\pbar,e,\ebar$ \cite{ParticleDataGroup2010}.
(b) After  (\ref{eq-engn2}a).
(c) Given by (\ref{eq-Aqc1}) for $p,e$ and   (\ref{eq-Acal2}), (\ref{eq-Acal2b}) for $\pbar, \ebar$.
(d) After (\ref{eq-betaqpe}a). (e) After  (\ref{eq-engn2}d).          
(f) From (\ref{eq-time1}) for $p,e$ and the discussions before (\ref{eq-Acal2}) and after (\ref{eq-Acal2b}) 
for $\pbar$,$\ebar$.} 
} 
\end{table}

Taking (i)-(iv) together, the creation of particle $p$ or $e$ 
corresponds to the excitation  from energy level $n=0$ (the ground state) to $n=1$ (the first excited  state)  of the charge $q$ $=+e$ or $-e$ in its potential  well $V_{\v \pe n}$ or $V_{\v \m e n}$,  upon a minimum external energy supply  $\eng_{exc.m}(=\hbar \w_\g)=\eng_{q1} = M_pc^2$ or $M_ec^2$.  $\eng_{exc.m} $  is  thus used for overcoming the potential energy difference $V_{\v q 0}-V_{\v q 1} =V_{\v q 0} -0$, whence  
$$\displaylines{
\refstepcounter{equation} \label{eq-Vvqo}
\hfill 
V_{\v q 0} =-\eng_{exc.m}=-Mc^2  \qquad (q, M=+e, M_p; -e, M_e)
\hfill (\ref{eq-Vvqo})
}$$  
And the energy $\eng_{q1}$ gained by  the charge $+e$ or $-e$ 
corresponds to  the total energy associated with the rest mass $M_p$ or $M_e$ of the resulting IED particle $p$ or $e$.  

Placing (\ref{eq-betaqpe}a),(\ref{eq-Vvqo}) in (\ref{eq-Vn}), 
we obtain the parameterised quantised vacuum potential energy functions of the charges $+e$ and $-e$ respectively versus $ \zeta_n (=\sqrt{n}\zeta)$ across the interstice about $\zeta_n=0$,
$$\displaylines{
\refstepcounter{equation} \label{eq-Vvqnpe}
\hfill
V_{\v q n}(\zeta_n)=Mc^2 (-1 + \frac{4}{f_1^2} \zeta_n^2) \qquad (q, M=+e, M_p; -e, M_e;
\  |\zeta| < \mbox{$\frac{1}{2}$}).
\hfill (\ref{eq-Vvqnpe})
}$$
The function $V_{\v q n}$  is completely specified by (\ref{eq-Vvqnpe})
except  that $f_1$ depends on $\delta_1$ through (\ref{eq-betaqpe}b) and is yet to be determined. $f_1$  affects the steepness of the $V_{\v q n}$ well only; the energy levels of stable particle species (or particle states) therefore are completely specified by (\ref{eq-Vvqnpe}). 

Equations (\ref{eq-Vvqnpe}), see also the graphical plots in Figure \ref{fig-Vv2z-z.eps}a,b, show a strong asymmetry of $V_{\v q n}(\zeta_n)$ with respect to an external charge  $+e$ and $-e$:  $V_{\v\pe n}(\zeta_n)$ has a strongly negative depth $-M_pc^2=-938.27$ MeV (Figure \ref{fig-Vv2z-z.eps}a), and $V_{\v\m e n}$ has a  shallow "negative"  depth $-M_ec^2=-0.511$ MeV (Figure \ref{fig-Vv2z-z.eps}b). This very asymmetry will be directly demonstrated  in \ref{app-Vqv1}.1 through a formal evaluation of the electromagnetic interaction for  an external $q$ charge and vacuuon: an external positive charge $+q$  will be strongly attracted by the vacuum, while a negative charge $-q$ be strongly repelled  therein.  And this is a direct consequence  of the asymmetric structure of the vacuuon, of which   $-e$ envelops $+e$ (see Sec.\ref{Sec-Vnps}), combined with a "strong force" effect which onsets  at short interaction distances compared to the extension of the vacuuon. 

As already entered as an input for the $V_{\v q n}$ parameterisation, the proton lies at the first excited stationary state, i.e. energy level $n=1$, of charge $+e$  in the $V_{\v \pe n}$ well, and the electron at level $n=1$ of  $-e$ in the  $V_{\v \m e n}$ well, shown by the solid horizontal lines in Figure \ref{fig-Vv2z-z.eps}a and b. The very large mass ratio of $p$ over $e$, $M_p/M_e \approx 1836$, in retrospect, is a direct reflection of the asymmetry of the two vacuum potentials.

Based on  the solutions  (\ref{eq-eqn}),  there is no stationary state below the level $n=1$ for either charge. However, in a pair production out of a vacuuon (Sec. \ref{Sec-Vnps}) in the vacuum,  both its bound vaculeon charges $+e$ and $-e$ (assuming having been firstly disintegrated and now serving as two un-bound external charges) are by a resonance condition  (see the end of Sec. \ref{Sec-Vnps})  simultaneously excited with equal energies, provided a total energy $2\times \eng_{exc}=2 \times  \hbar \w_{\g} $ is externally supplied. If $\eng_{exc}$  is such that $-e$ is excited to its $n=1$ level in the  $V_{\v \m e n}$ well (Figure \ref{fig-Vv2z-z.eps}b), whence $\w_{\g}=M_ec^2/\hbar $  and  the creation of a stable electron $e$,  then  $+e$ is excited by the same quantum $\hbar \w_{\g}$  in the $V_{\v \pe n} $ well  (Figure \ref{fig-Vv2z-z.eps}a$'$), whence the creation of a positron $\ebar$. The $+e$ of $\ebar$ is at the level $V_{\v \pe \ebar} =V_{\v \pe n}(\Acal_{\pe \ebar})$ (dotted horizontal line in Figure \ref{fig-Vv2z-z.eps}a$'$) and has  an oscillation amplitude $\Acal_{\pe \ebar}$. This  $\ebar$ state is far below the $n=1$ (proton) level  in the $V_{\v\pe n}$ well, and is not a stationary state. But it would be virtually stable if, as is highly probable, the $e$  simultaneously created has  moved away and also no other electron presents nearby for annihilation. This $\ebar$ will be "hidden" in the vacuum and not "free" in the sense said in  (IV) earlier. 

On the other hand,  this excited  $+e$ of $\ebar$  is  free to travel from site to site  at its own constant potential energy level $V_{\v \pe \ebar}$, provided it has a sufficient kinetic energy to "hop" over a barrier  (cf Figure  \ref{fig-Vv2z-z.eps}a$'$), $\D_{\v_\ebar}=V_{\v \pe \ebar}' (z) -V_{\v \pe \ebar} (\Acal_{q \ebar})$,  crossing each two sites.  
The $\Acal_{\pe\ebar}$ of
$+e$ may be evaluated based on  the energy equation for $\ebar$, $\frac{1}{2}\beta_{\pe} \Acal_{\pe\ebar}^2= M_ec^2$, given by using $\beta_q =\beta_{\pe}$ for $+e$ in (\ref{eq-betaqpe}) but with the $\eng_{exc}$ equal to that of its opposite charge at level  $n=1$ (i.e. $M_ec^2$)
for $\eng_{q1}$, to be
$$\displaylines{
\refstepcounter{equation} \label{eq-Acal2}
\hfill 
 \Acal_{\pe\ebar}
 = \sqrt{2M_e c^2 / \beta_{\pe} } 
=\sqrt{(M_e/M_p)} \ \Acal_{\pe 1} = 0.0116 \Acal_{\pe 1}, 
\hfill (\ref{eq-Acal2})
}$$
which is exceedingly small.
The width of the barrier $\D_{\v_\ebar} $,   $\delta_\ebar= b_\v - 2 \times 0.01\Acal_{q 1} \sim b_\v $ (assuming $\delta_1 <<\frac{b_\v}{2}$), is thus  wide.
So after excited to above  the barrier $\D_{\v_\ebar}$, the charge will be translating across the large  distance $\delta_\ebar \sim b_\v $ before entering next 
$V_{\v \pe n}$  well. From  the experimental decay processes of the subatomic particles (e.g. \cite{ParticleDataGroup2010}), we observe  that, if disregarding the mediators $W^{\pm}$,   $\ebar $ is in fact the only non-composite particle formed of $+e$ which is below the $n=1$ level in the  $V_{\v \pe n}$ well. All of the other mass-deficit subatomic particles like  $\pi^+, K^+, \rho$, manifestly having one-unit charges $+e$'s,  are apparently composite particles built ultimately of a lepton $\mu$ and its neutrino, 
with $\mu$ being built of  charge $-e$ in the $V_{\v \m e n}$ well. 

If on the other hand $\eng_{exc}$ is such that $+e$ is excited to the $n=1$ level in the $V_{\v \pe n}$ well (Figure \ref{fig-Vv2z-z.eps}a), whence $\w_{\g}=M_pc^2/\hbar $ and 
the creation of a stable proton $p$, then similarly  by a  resonance condition  $-e$ is simultaneously excited by the same energy in the $V_{\v \m e n}$ well (Figure \ref{fig-Vv2z-z.eps}b), whence the creation of an antiproton $\pbar$. The charge $-e$ of $\pbar $ is at the potential energy level   $V_{\v \m e \pbar }=V_{\v \m e n }(\Acal_{\m\e \pbar}) $ (dotted horizontal line in Figure \ref{fig-Vv2z-z.eps}b) and has an oscillation amplitude
 $\Acal_{\m\e \pbar}$. Similarly from $\frac{1}{2}\beta_{\m e} \Acal_{\m e\pbar}^2= M_p c^2$ given by using  $\beta_q =\beta_{\m e}$ in  (\ref{eq-engnb2}) and $\eng_{exc}=M_pc^2$,  we formally  obtain 
$$\displaylines{
\refstepcounter{equation} \label{eq-Acal2b}
\hfill 
\quad  \Acal_{\m e\pbar} 
= \sqrt{       2M_p c^2/\beta_{\m e}  }
=\sqrt{        (M_p/M_e) } \ \Acal_{\m e 1}=21.42  \Acal_{\m e 1},  
\hfill (\ref{eq-Acal2b})
}$$
which is many times  larger than $\Acal_{\m e 1}$ of the $-e$ of an electron, as is an inevitable result for $V_{\v \m e \pbar } >>$ $V_{\v \m e  1}$. 

Since however the vacuum potential has a mean translation periodicity $b_\v$ along any diffusion path and thus is only quadratically well defined up to the vacuum level 
plus a $\D_{\v_1}$ about  $z=\frac{b_\v}{2}$, the charge $-e$ of $\pbar$ of the exceedingly large $\Acal_{\m e \pbar}$  factually traverses many  potential wells in each quart of its oscillation period. This motion is  no longer properly harmonic; and higher stationary levels than 1, i.e. $n=2,3, \dots $, become unphysical  except during charge--vacuuon  head-on collisions. The charge $-e$ of $\pbar$ accordingly will be so energetic as to translate swiftly across many sites in short time, meeting and scattering with other particles and losing its energy easily, until settling  down at the next and actually the only lower stationary level in the $V_{\v \m e n}$ well, which is the $n=1$ or electron state. That is, the resulting antiproton  is short-lived and  briefly will descend into a stable electron. This could explain the prominent "missing" of the antiprotons $\pbar$'s if all the protons present in nature indeed are produced in $p$---$\pbar$ pair productions.

The above scheme can similarly account for the short lifetimes of the other observational  heavier-mass, non-composite  subatomic particles made of one-unit charges, actually the  leptons $\mu, \tau$ only which are built of one-unit $-e$ in the $V_{\v \m e n}$ well,  if  disregarding the mediators, similarly based on the experimental decay processes of subatomic particles. All the other heavier-mass baryons such as  $\W^\m$, $\Lam$'s, $\Sigma$'s and mesons such as $\pi^\m, D^{\pm }$, etc.   having either one-unit   $-e$ or  (as earlier remarked)$+e$, are apparently composite particles ultimately built of $\mu$'s and their neutrinos. 

\section{Vacuuonic potentials. Pair productions}
\label{Sec-Vnps}
\begin{figure}[p]
\vspace{-0.5cm}
          \begin{flushleft}
\includegraphics[width=0.69\textwidth]{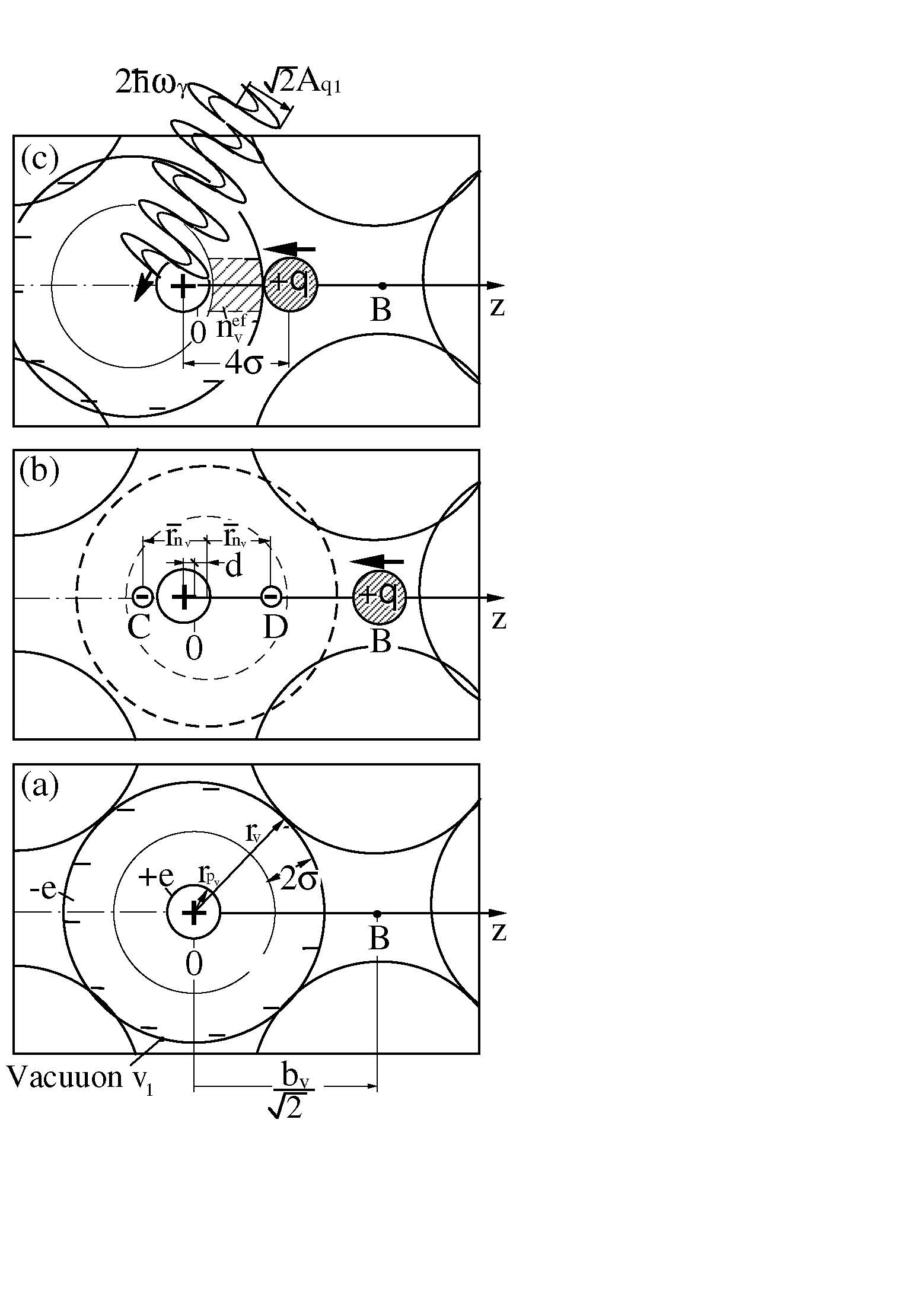}
\end{flushleft}  
\refstepcounter{figure}\label{fig2-Vpn_diagm.eps}
\vspace{-0.8cm}
\begin{flushright}
\vspace{-13.4cm}
\includegraphics[width=0.58\textwidth]{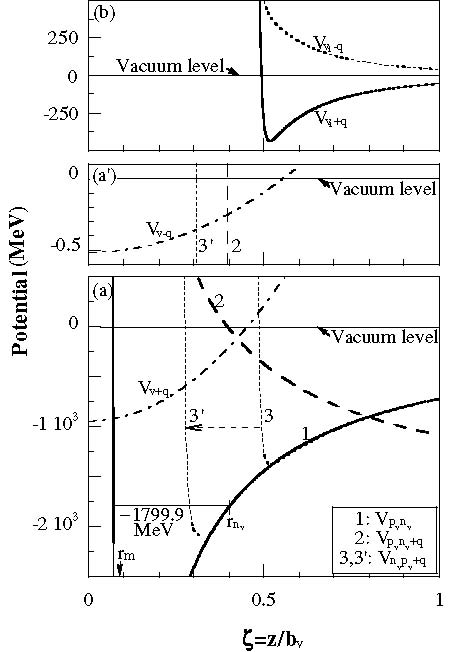} 
\end{flushright}
\begin{flushright}
\refstepcounter{figure} \label{fig2-Vpn.eps}\label{fig3-Vpn.eps}

\begin{minipage}{14.cm} \baselineskip 0.4 cm
{\footnotesize {\bf Figure \ref{fig2-Vpn_diagm.eps}} (left graphs). Vacuuons $\vrm_i$, with $\vrm_1$ at $z=0$,  arranged on  a simple cubic lattice  (a) in zero external field, and (b)--(c) in the field of an external charge $+q$ in the interstice $B$;   $+q$ is  moving toward $\vrm_1$ at a finite velocity.  In (c), $+q$  has collided with $\nv$ and in turn knocked $\nv$ into colliding with $\pv$ to their closed approaches each; at the same time, a $2\g$ wave of energy $2\hbar \w_\g (\ge 2 Mc^2)$ is incident on to $\vrm_1$.   }
\end{minipage}

\vspace{0.2cm}

\begin{minipage}{14.cm}  \baselineskip 0.4 cm {\footnotesize
{\bf Figure \ref{fig2-Vpn.eps}} (right graphs). (a) 
Solid curve: $\pv$---$\nv$ interaction potential energy $V_{\pv\nv}(\zeta)$ given by  (\ref{eqVpn}) for vacuuon $\vrm_1$ in zero external field (as in Figure \ref{fig2-Vpn_diagm.eps}a), $\zeta=r/b_\v$.  Dashed curve 2 and dotted curve 3 ($\zeta>r_\v$):  potential energies of $\pv$ and $\nv$ of  $\vrm_1$, $V_{\pv\nv\pq}(\zeta)$ and  $V_{\nv\pv\pq}(\zeta)$  given by (\ref{eq-Vpvnvq1})--(\ref{eq-Vpvnvq2}) in the field of external charge $+q$ as in Figure  \ref{fig2-Vpn_diagm.eps}b. The rapid rising part of curve 2 and curve 3$'$: the two potential energy functions $V_{\pv\nv\pq}(\zeta)$ and  $V_{\nv\pv\pq}(\zeta)$ when $+q$, $\nv$ and $\pv$ are as positioned in   Figure  \ref{fig2-Vpn_diagm.eps}c. Corresponding curves $2$, $3'$ for $-q$ are  shown in (a$'$).  Short-dot-dashed curves: the function $V_{\v \pe n}$ in (a) and $V_{\v \m e n}$ in (a$'$) given by (\ref{eq-Vvqo}). Used for the plots: $\sig=0.1b_\v$ (thus $u_1 =14400$ MeV), $\Nsm=12$, $d=0.01b_\v$. At $r_m=(g\Nsm)^{\frac{1}{\N-1}} \sig= 0.859 \sig$, $\frac{\pd V_{\pv\nv}}{\pd r}=0$ and $V_{\pv\nv m}(r_m) =- \frac{u_1}{2} (g\Nsm)^{-\frac{1}{\N-1}} [\frac{\Nsm-1}{\Nsm}] =-0.534 u_1.$ (b) Interaction potential energy function $V_{\v_1\pq} $ (solid curve), given by  (\ref{eq-Vpm-qpvnv}b), between vacuuon $\vrm_1$ and external charge $+q $ of position $\zeta$ as in  Figure \ref{fig2-Vpn_diagm.eps}b; and $V_{\v_1\m q}$ (dashed curve) between  $\vrm_1$  and $-q$. Values used for the plots are as in (a). 
}
\vspace{0.5cm}
\end{minipage} 
\vspace{-1.7cm}
\end{flushright}
                   \end{figure}

 A vacuuon (e.g. $\vrm_1$ in Figure  \ref{fig2-Vpn_diagm.eps}a) by construction\cite{Unif1,Unif1Vac} consists of a positive charge $+e$ seated on a minute sphere of radius $r_\pv$ at the centre, and a negative charge $-e$ on a concentric  spherical shell of thickness $2r_o$ and radius $r_\nv$ about $\pv$, termed as a p-vaculeon ($p_\v$)    and n-vaculeon ($n_\v$).  The  $\pv,\nv$ have  spins  $\frac{\hbar}{2}$ each; in their bound state in a vacuuon  their spin magnetic moments are  oriented in opposite directions in each others' magnetic fields. The vacuuon structure, as a building entity of the substantial vacuum, is constructed based on overall experimental indications, most directly the pair production and annihilation experiments in particular\cite{Unif1,Unif1Vac,jxzjied}; see further the discussion after (\ref{eq-pair1}) later.

The $r_\pv, r_\nv$ represent  the most probable radii of the practically  extensive $\pv, \nv$ (similarly as the single charge $q$ in Sec. \ref{Sec-VacPot}) at the scale $b_\v$,  and  $ r_o$ is said in a similar sense.  We presently lack experimental information  either on their direct  values or for their theoretical evaluation; although definitely they must be (much) smaller than  $\frac{b_\v}{2}$. For the illustration below we shall take  the $r_\pv,r_o$ values by their average, $\frac{1}{2}(r_\pv+r_o)=\sig$. And we set the vacuuon radius $r_\v$, $=r_\nv+r_o$, as the contact radius of the vacuuons on a simple cubic lattice (Sec. \ref{sec-pot-para}), so $r_\v=\frac{b_\v}{2}$, see Figure \ref{fig2-Vpn_diagm.eps}a. 
 The focus of our discussion  below will be  to demonstrate the characteristics of the interactions rather than to  perform an accurate  numerical calculation. 

In zero external  field, the two vaculeon charges  $+e$ and $-e$ of a vacuuon, say the $\vrm_1$ at $z=0$ in Figure  \ref{fig2-Vpn_diagm.eps}a, interact each other  by a Coulomb attraction,  $\V_{\pv\nv}^{coul}=- \frac{e^2}{4\pi\ev_0 r}= u_0\frac{r_\v}{r}=u_1 \frac{\sig}{r}$, and a short range repulsion, $\V_{\pv\nv}^{rep}=g u_1(\frac{\sig}{r})^\N$, 
where  $\uv_o=  \frac{e^2}{4\pi\ev_0 (b_\v/2)}=2879.9$  MeV for $r_\v =\frac{b_\v}{2}= 0.5 \times 10^{-18}$  m,  $\uv_1=\uv_o (r_\v/\sigma)$, and $g=\frac{\pi \sig^2}{4 \pi r_\nv^2} =\frac{ \sig^2}{4  r_\nv^2} $ is the fraction of charge of the segment, of size $\pi \sig^2$ on the extensive $\nv$ shell of an area $4 \pi r_\nv^2$, which makes direct contact with $\pv$. The $\Nsm, \sig$ values are to be determined.  The total $\pv$, $\nv$  interaction potential energy per vaculeon is  thus 
$$
\displaylines{\refstepcounter{equation}\label{eqVpn}
\hfill
V_{\pv \nv} (r)= \frac{1}{2}(\V_{\pv\nv}^{rep} (r) +\V_{\pv\nv}^{coul} (r)) =
\frac{u_1}{2} \left[
g 
\left(\frac{\sig}{r}\right)^{\N} - \frac{\sig}{r}\right]
         \hfill (\ref{eqVpn})
}$$
   See also  the graphical plot  of $V_{\pv\nv} (r) $ in Figure \ref{fig3-Vpn.eps}a   (solid curve 1),  where $\Nsm=12$ (Lennard-Jones' value) and  $\sig=0.1b_\v$ are used for the illustration.

At $r>>r_\pv$, $\nv$ is acted on by $\pv$ by (mainly) an attractive force $F_{\pv\nv}=-\frac{\pd \V_{\pv\nv}}{\pd r}$ $\approx -\frac{\pd \V_{\pv\nv}^{coul}}{\pd r} $, where  $\V_{\pv\nv} =2V_{\pv\nv} $. 
This, in the  zero mass representation, is counterbalanced by a magnetic force  $F_m$ on the spinning  $n$-vaculeon  charge on the spherical envelope 
 in the magnetic field produced by spinning motion of $p$-vaculeon  charge (Appendix A of \cite{Unif1}),
$F_m, = -F_{p_\vsub n_\vsub}$. The equality defines the equilibrium radius  $r_\nv(=\frac{b_\v}{2}-\sig)$, at which $\V_{\pv\nv}(r_\nv)\dot{=}-u_1(\sig/r_\nv)
=-3599.9 $ MeV, or $V_{\pv\nv}(r_\nv)=-1799.9$ MeV;
accordingly, $\nv$ has a spin kinetic energy $\Eng_{\nv k}
=-\frac{1}{2}\V_{\pv\nv} $ and Hamiltonian $\Eng_{\nv t}= \V_{\pv\nv}+ \Eng_{\nv k}=\frac{1}{2} \V_{\pv\nv}$. This $\V_{\pv\nv}$, of a GeV scale, is far too deep for the vaculeon pair to be disintegrated to the vacuum level,  by merely a  supply of  an excitation energy $2\eng_{exc.m}$ given by  (\ref{eq-Vvqo}), or 
$$\displaylines{
\refstepcounter{equation} \label{eq-exceng}
\hfill 2\times \eng_{exc} =2\times \hbar \w_{\g}  
\ge  2\times Mc^2
 =\frac{1}{2} \beta_q (\sqrt{2} \Acal_{q1})^2 \qquad (M=M_p,M_e; \ q=+e,-e)
\hfill (\ref{eq-exceng})
}$$
which are $ 2\times 938.27 $, $2\times $ 0.511  MeV  for the $p$---$\pbar $, $e$---$\ebar$ pair productions.  This $2\eng_{exc}$ is merely enough to impart masses to a pair of  dissociated vaculeon charges.

Inevitably, before the condition (\ref{eq-exceng}) becomes legible, an additional energy, as enormous as $2 \times (V_{\v q0}-V_{\pv\nv}(r_\nv)) \sim 3600$ MeV for $e$---$\ebar$ production or $1720$ MeV for $p$---$\pbar$ production, needs  firstly be  supplied so as  to  disassociate the pair of bound vaculeons of the $\vrm_1$ here to at or above the ground state of the charge, $V_{\v q 0}$. Such an enormous energy may be practically supplied if the two  vaculeons  are simultaneously approached by a charged particle  (e.g. a nucleon) $q$ at very short distance and thereby repelled to above $V_{\v q 0}$; 
an external $q$ thus needs be in the proximity (like the $+q$  in the interstice $B$ in Figure  \ref{fig2-Vpn_diagm.eps}b) and moving at an adequate  speed toward $\vrm_1$. A possible such process is  illustrated in Figure \ref{fig2-Vpn_diagm.eps}c. The corresponding  potential energies of $\pv$,$\nv$ in the presence of $+ q$, of coordinate $z$, and similarly of $-q$, $V_{\pv\nv\pm q} (z)$,$ V_{\nv\pv\pm q} (z)$ as functions of the position $z$ of $+q$ or $-q$ are given by (\ref{eq-Vpvnvq1})--(\ref{eq-Vpvnvq2}), \ref{app-Vqv1}.2. 
As the graphical plots, the dashed curves 2 and dotted curves 3 to $3'$  in Figure  \ref{fig2-Vpn.eps} a,a$'$ directly show, the two potential energy functions  rise each rapidly  to  above  $V_{\v q 0}$ at the closest approach between $+q,\nv$ and $\pv$ (Figure \ref{fig2-Vpn_diagm.eps}c), i.e. at $z\sim 3 \sig$.   The vaculeons $\pv$ and $\nv$ are now effectively no longer bound each other, being as if  separated infinitely apart.

If these, as soon as after their dissociation, are impinged by a $\g$ wave (see Figure \ref{fig2-Vpn_diagm.eps}c) of an energy $2\eng_{exc}=2\hbar \w_\g$ fulfilling (\ref{eq-exceng}), e.g. $\w_\g =m_p c^2/\hbar$,  then upon absorption of $2\eng_{exc}$ by  a "resonance condition"  (see below) the vaculeon charges $+e,-e$ will have been each endowed with an oscillation energy  $\hbar \w_p=m_pc^2$.  $+e$ is now promoted 
to the energy  level $n=1$ in  the $V_{\v \pe n}$ well at one site (short-dot-dashed curve in   Figure  \ref{fig2-Vpn.eps}a);  and $-e$ to the level of $\pbar$ in the $V_{\v \m e n}$ well, by a probable tendency, in another site located in the  opposite direction to the displacement of $+e$,   since the charges $+e,-e$ producing  (or absorbing) the same radiation $\Eb$ field  have opposite oscillation displacements. And similarly for $\w_\g =m_e c^2/\hbar$, with the charge $-e$ promoted to level $n=1$  in the $V_{\v \m e n}$ well (short-dot-dashed curve in   Figure  \ref{fig2-Vpn.eps}a$'$), and $+e$ to the level of $\ebar$ in the $V_{\v \pe n}$ well. These are the $p$---$\pbar$ and $e$---$\ebar$ pair productions of the reaction equations 
$$\displaylines{
\refstepcounter{equation} \label{eq-pair1}
\hfill 
2 \g \rightarrow p+\pbar, 
\quad 
 \quad 2 \g  \rightarrow e+\ebar. 
 \hfill (\ref{eq-pair1})
}$$ 
 The pair of particles produced will be at rest if  $\W=Mc^2/\hbar$ or will have a residual velocity $\vel=c \sqrt{1-1/\g^2}$ if $\w=\g\W >\W$, i.e. $\g >1$.

The reaction equations (\ref{eq-pair1}), together with the preceding energy criterion (\ref{eq-exceng}) and  the requirement for the presence of a nucleus (or nuclei) in a pair production, are in complete agreement with experiment.   Entirely as an experimental reaction equation, (\ref{eq-pair1}) are expressed such that they each inform explicitly all of "observables" before and after a pair production. In particular, (\ref{eq-pair1})  inform that both charges ($+e,-e$) and spins  ($\frac{\hbar}{2},\frac{\hbar}{2}$)  are present on their right-hand sides, but not the left-hand sides. And the external energy supply $2\eng_{exc} =2 Mc^2$  is only to ascribe  dynamical masses to the pair of vaculeon charges $+e,-e$ (which have zero rest masses), or equivalently, (dynamical) rest masses to the resulting IED particles in each reaction process, $e, \ebar$ or $p,\pbar$.  So the {\it charges} $+e,-e$ which carry a potential energy $V_{\pv\nv}(r_\nv)$ at the particles' production as given by (\ref{eqVpn}), and their {\it spins} $\frac{\hbar}{2}$'s which carry a kinetic energy, must {\it exist} in the  vacuum, whence the {\it vaculeons} composing  a vacuuon, so as to satisfy the requirement of energy conservation. Similar discussion was  made in terms of  the pair annihilation in   \cite{Unif1,Unif1Vac,jxzjied}.

Supplemental remarks regarding the pair production: (i) {\it  The resonance condition}. In mechanical terms, as follows from Sec. \ref{Sec-VacPot}, 
 the dielectric vacuum is  induced with an elasticity in the presence of an external charge ($q$) nearby. And the electromagnetic ($\g$) wave,  of a wavelength $\lam_\g\sim 1.3 \times 10^{-15} $ or $\sim 2.4 \times 10^{-12}$ m, is an elastic wave propagated in the vacuum by means of the elastic deformations of the vacuum, or in other terms,  of the oscillations of coupled oscillators each composed of (tremendously) many vacuuons  (of  size $\sim 10^{-18}$ m each). So relative to the extensive $\g$ wave, the pair of  vaculeons  $\pv,\nv$ in a  vacuuon (the $\vrm_1$ above)  are just a minute point on a large oscillator. They will respond to the $\g$ wave as one point, practically the only point in the large oscillator being in  the (internal) mode of resonance absorption to  the quanta $2 \hbar \w_r$ of the  $\g$ wave, assuming no other bound vaculeons in the large oscillator are dissociated to above level $V_{\v q 0}$.

(ii) The incident  $\g$  wave of  energy $2\eng_{exc}$  is an extensive electromagnetic  wave train (as schematically shown  in Figure \ref{fig2-Vpn_diagm.eps}c) of  length $L_\vphi$ and effective amplitude $A_{q2}=\frac{\sqrt{2}}{\sqrt{t_{\vphi 2.0} /(1/\w)       }} \Acal_{q1}$\cite{Unif1sho}. Accordingly, the "absorption of $2\eng_{exc}$" is a gradual, continuous process spanning  a total duration $t_{\vphi 1.0}$, in  which the wave train front runs at  the velocity of light $c$ on to the two vaculeon charges $+e$ and $-e$ of $\vrm_1$, and be thereby  absorbed  by them (by a certain fraction) continuously.
 Two new waves of the same $\w$, and of amplitude $A_{q1}$ each, are  subsequently continuously re-emitted by the two charges, and then, together with the transmitted fraction, re-absorbed after reflecting back from surrounding walls.

(iii) At the end of one $t_{\vphi 1.0}$, two  full wave trains (i.e. for the fraction $a_1+a_2=1$) maintain  the same $L_\vphi$, and same total $2\eng=2m c^2$, and $2p=2\eng/c=2mc$ (i.e. the linear momentum, which is conserved in this sense)  as the incident one. These two wave trains  have now become the respective (internal) components of the (IED) particle and antiparticle just produced.

\ack 
The author's research is privately funded by emeritus scientist P-I Johansson who has also given  continued moral support for the author's researches. 
A focused elaboration on the solution for the  vacuum potential in terms of the IED particle model presented  in this paper was motivated by one of a wide scope of  all essential questions put forward to the author at a seminar discussion during the author's  visit to Professor I Lindgren at his Atomic Physics Group, Gothenburg Univ., Feb, 2011. 
An introduction prior to the visit is indebted to Professor B Johansson (Uppsala Univ.)  The author expresses also thanks to a community of national and international distinguished  physicists for giving moral support for the author' recent-year research, and to the organising chairman Professor C Burdik, chairman Professor H-D Doebner,  the organisers and committee of the 7th Int Conf Quantum Theory and Symmetries (QTS 7) for facilitating the opportunity of communicating this research at the QTS 7, Prague, Aug., 2011.

\begin{appendix}

\section[\qquad \qquad \ \ Electromagnetic interaction]{Electromagnetic interaction 
}
\label{app-Vqv1}

\subsection*{A.1 Interactions at larger distances up to a closest approach }

As shown in Figure \ref{fig2-Vpn_diagm.eps}b (Sec. \ref{Sec-Vnps}), 
the vacuuon  $\vrm_1$ at $z=0$ is polarised by  the external charge $+q$  in the interstice $B$, moving from initial position say  $z=\frac{b_\v}{\sqrt{2}}$ toward $\vrm_1$ at a finite speed. We shall express the $\vrm_1 $---$+q$ and $\vrm_1$---$-q$  interaction potentials in electromagnetic terms below,  and shall do so  by situating ourselves in the frame where the mass centre of $\vrm_1$ is not moved during the  interaction. (This frame approximately corresponds to the frame fixed to the vacuum if $\vrm_1$ and its surrounding vacuuons can not move freely due to attachment to a fixed matrix of charged particles, but their configuration may be locally deformed under the dynamical impact of $q$.) In this frame, the  $\pv$ and $\nv$ vaculeons of the polarised vacuuon $\vrm_1$ are displaced  from the fixed position $z=0$ to $-\frac{d}{2}$   and $+\frac{d}{2}$. Since $r_\nv>>\sig$, we shall regard the $\pv$ and $q$  as point like and the $\nv$- spherical shell extensive in respect to their short range interactions. 

$+q$ interacts with a charge element $dq_1$ on the extensive spherical shell of $\nv$ by a Coulomb attraction  $d F = \frac{d q_1, q}{4\pi \ev_o \ell }$.   Integration over the entire shell gives the  total attraction of $+q$ and $\nv$  as \cite{Unif1}   $F\dot{=}- \frac{u_1\sig}{r_\nv^2}(\frac{r_\nv}{z })^{n+1} $, with $n=15.7$, which is strongly short ranged (whence a "strong force").  Accordingly $V^{coul}_{\nv \pq}= -\frac{1}{2}\int^\infty_z F d z \dot{=} - \frac{u_1 \sig}{2r_\nv n} (\frac{r_\nv}{z})^n $.
Because of the simple symmetry of the $\nv$-shell  with respect to $+q$, for a better physical transparency we below  express this interaction alternatively by representing $\nv$ effectively as two one-half charges $\frac{e}{2},\frac{e}{2}$ projected on the $z$ axis at $-\rbar_\nv, +\rbar_\nv$, with $\rbar_\nv=r_\v/2$\cite{Unif1}, as  
$$ \displaylines{
\refstepcounter{equation} \label{eq-Vpm-qpvnv1}
 \hfill
V_{\nv\pq}^{coul}=-\frac{u_1}{4}\left[\frac{\sig}{z-(\rbar_\nv+\frac{d}{2})} 
+\frac{\sig}{z+(\rbar_\nv-\frac{d}{2})} \right]
= -   \frac{u_1\sig\eta}{2(z-\frac{d}{2})}, \quad   
\eta=\frac{1}{1-(\rbar_\nv/(z- \frac{d}{2}))^2 }.  
\hfill
\cr
\hfill (\ref{eq-Vpm-qpvnv1})
}$$
In addition,  $+q$ interacts with $\nv$ similarly through $\nv$ of a fractional charge $gq$ as in (\ref{eqVpn})  by a short range repulsion,  $V_{\nv\pq}^{rep}= \frac{u_1}{2}g\left(\frac{\sig}{z-(r_\nv + \frac{d}{2})}\right)^\N$. And, with  the $\nv$-shell in between,   $+q$ interacts with  $\pv$ by a Coulomb repulsion only, 
$V_{\pv\pq}^{coul}=\frac{u_1}{2} \frac{\sig}{z+\frac{d}{2}}$ for $z\ge 4 \sig$. Adding the terms above, we obtain the interaction potential energy of $+ q$  with vacuuon $\vrm_1$, and similarly  of $-q$ with $\vrm_1$ after corresponding sign changes,  as 
$$ \displaylines{
\refstepcounter{equation} \label{eq-Vpm-qpvnv}
\hfill
 V_{\v_1\pm q} =
V_{\nv\pm q}^{rep} + 
V_{\nv\pm q}^{coul}+
V_{\pv\pm q}^{coul} 
= \frac{u_1}{2}\left[
g\left(\frac{\sig}{z-(r_\nv \pm \frac{d}{2})}\right)^\N
\mp \frac{\sig \eta}{(z\mp \frac{d}{2}) }
\pm \frac{\sig}{(z\pm \frac{d}{2})}
\right],
\hfill
\cr
\hfill V_{\v_1\pm q }\dot{=}  \frac{u_1}{2}\left[
g\left(\frac{\sig}{z-(r_\nv \pm \frac{d}{2})}\right)^\N
\mp \frac{\sig(\eta-1)}{z} 
- \frac{\sig d(\eta+1)}{2z^2}
\right], 
\quad   \eta=\frac{1}{1-(\rbar_\nv/(z\mp\frac{d}{2}))^2 },
\hfill
\cr
\hfill (\ref{eq-Vpm-qpvnv})  
}$$
 where (\ref{eq-Vpm-qpvnv}b) is given after expanding  the second and third terms of (\ref{eq-Vpm-qpvnv}a) in  $\frac{d}{2z}$ and retaining the respective  two first leading terms. The last term in (\ref{eq-Vpm-qpvnv}b), $ - \frac{u_1\sig d(\eta+1)}{4z^2}$, represents the  interaction energy of the $\nv$ vaculeon  dipole moment,  $\pb_\nv= e d \hat{z} $, with the static Coulomb field of charge $q$,  $\Eb_q=\frac{u_1\sig (\eta+1)}{2ez^2}\hat{z}$, and this may be directly obtained as  $V_{dip\pm q}= \frac{1}{2}\pb_\nv\cdot \Eb_q = -\frac{ u_1\sig d (\eta+1)}{4z^2} $. Since for small $d$ there is always $\eta >$( or $>>)1 $, at $z-\frac{3}{2} >\rbar_\nv$, $V_{dip\pm q} $ is thus an attraction for either $+q$ or $-q$. The second  term in $V_{\v_1\pm q }$ of (\ref{eq-Vpm-qpvnv}b), $\pm \frac{\sig}{(z\pm \frac{d}{2})}= V_{\nv\pm q }^{coul} $  is a main attraction term between $\vrm_1$ and $+q$, and is a repulsion between  $\vrm_1$ and $-q$. The sum of the interactions of $q$ with all surrounding vacuuons up to an intermediate range,  $\sum_i V_{ \v_i q} $, gives the $V_{\v q}$ of Sec. \ref{Sec-VacPot}.

As the graphical plots in Figure \ref{fig3-Vpn.eps} b directly show, from larger $z$ down to a closest approach at $z=r_\v$, the potential  $V_{\v_1\pq}$ (solid curve) for the positive charge $+q$  is strongly negative, while $V_{\v_1\m q}$ (dotted curve) for $-q$ is positive for a wide range of $d$ value ($d=0.01 b_\v$ for the plots).  

\subsection*{A.2 Dynamical interactions after $q,\mbox{$\nv$}$ closest approach}
\label{App-reppair}
At about $z=r_\v$, $+q$ and the segment $\nvef$ of the $\nv$-shell (cf Figure \ref{fig2-Vpn_diagm.eps}c)  are at closest approach.  And the $+q$---$\nv$ interaction potential, $V_{\nv +q}=V_{\nv +q}^{rep} + V_{\nv +q}^{coul}$ as given by the sum of first two terms in  (\ref{eq-Vpm-qpvnv}a), shown  by the dotted curve $3$  in Figure \ref{fig3-Vpn.eps}a,   
rises rapidly.

From $z=r_\v$ downward,   $+q$  continues to move toward $\pv$, now  together with  $\nv$  while impressing on the segment $\nvef $ (which has the coordinate  $z'=z-2\sig$) of $\nv$ a constant  repulsion  $V_{\nv\q}^{rep}(z-z')=g(\frac{\sig}{2\sig})^\N=g(\frac{1}{2})^\N$ (with the steep sector of the dotted curve 3 sweeping across the region,  ending at curve $3'$). In addition, $+q$ interacts with $n_\v$   by  a Coulomb potential $V_{\nv\pq}^{coul}(z)|_{z=r_\nv}=
- \frac{u_1\sig\eta}{2(r_\nv-d/2)}$ as given after (\ref{eq-Vpm-qpvnv1});   
and with $\pv$ by the $V^{coul}_{\pv\pq}=\frac{u_1\sig}{2(z+d/2) }$ as before.
 $\pv$ interacts with $\nv$, as a very crude approximation here,  by the constant Coulomb  potential $V_{\pv\nv}^{coul}(r_\nv)=-\frac{u_1}{2}\frac{\sig}{r_\nv}=-1799.9$ MeV, and with the segment $\nvef$ of $\nv$  by a short range repulsion  $V_{\pv\nv}^{rep}(z')=\frac{u_1}{2}g(\frac{\sig}{z-2\sig})^\N $.  
Adding the respective terms above, the total potentials of $\pv$ and $\nv$ as  functions of the  coordinate $z$ of $+q$ are  
$$
\displaylines{\refstepcounter{equation}\label{eq-Vpvnvq1}
\hfill V_{\pv\nv\pq}(z)= V_{\pv\nv}^{rep}(z') + V_{\pv\nv}^{coul}(r_\nv)  + V_{\pv\pq}^{coul}(z)
                            = \frac{u_1}{2} \left[g\left(\frac{\sig}{z-2\sig 
                                +\frac{d}{2}}\right)^\N 
                        - \frac{\sig}{r_\nv}
                          +\frac{\sig}{(z +\frac{d}{2})}\right], \hfill
\cr
\hfill (\ref{eq-Vpvnvq1})
\cr\refstepcounter{equation}\label{eq-Vpvnvq2}
\hfill V_{\nv\pv\pq}(z)
=V_{\nv\pv}^{rep}(z')+V_{\nv\pq}^{rep}(z-z') + V_{\pv\nv}^{coul} (r_\nv) 
     + V_{\nv\pq}^{coul} (z-z') \hfill 
\cr
\hfill\qquad \quad \ = \frac{u_1}{2} \left[g\left(\frac{\sig}{z-2\sig +\frac{d}{2}}\right)^\N  
+\left(\frac{1}{2}\right)^\N 
- \frac{\sig}{r_\nv} 
-\frac{\sig \eta}{z-\frac{d}{2}}
\right].
 \hfill (\ref{eq-Vpvnvq2})
}$$
These are plotted by the dashed  curve 2 and dotted curves 3--3$'$ in Figure \ref{fig3-Vpn.eps}a. When $+q$ is at $z=z'+2\sig=4\sig -\frac{d}{2}$,  $\nvef$ is at $z'=z-2\sig=2\sig-\frac{d}{2}$ and touches $\pv$, producing on $\pv$ a strong short range repulsion $V_{\pv\nv}^{rep}(z'=2\sig-\frac{d}{2})=\frac{u_1}{2}g(\frac{\sig}{z-2\sig +d/2})^\N$.

\section[\qquad \qquad \ \  Complex diffusion current]{Complex diffusion current} \label{App-Diff}

Let $\rho_\Asub(z,t)$ be the density of a real fluid in flow motion at velocity $\vel$ in $z$ direction with a flow rate $j_\Asub=\rho_\Asub \vel$. $j_\Asub$ may be  alternatively  written as a  diffusion current  $j_\Asub=-D_\Asub \nabla \rho_\Asub$ (Fick's first law), 
where $D_\Asub$ is a real diffusion constant; and $j_\Asub$ is positive    in the direction in which the density gradient decreases.  Let $\rho_\Asub$ be written as 
$\rho_\Asub= \f \f' $ where  $\f',  \f$ are two arbitrary differentiable real functions of $z,t$. Then  
$$\displaylines{
\refstepcounter{equation} \label{eq-app-j}
\hfill
j_\Asub= -D_\Asub [\f'  \nabla \f+(\nabla \f') \f].
\hfill (\ref{eq-app-j})
}$$ 

If now it is  a "complex" fluid of density $\rho_q=\psi_q^*\psi_q$, where $\psi_q(z,t)=e^{i\w t} \phi_q(z)$ and $\psi_q^*$ are  the complex functions as in Sec. \ref{Sec-VacPot}, and we want to write down a positive diffusion current  $j_q$ associated with $\rho_q$ on an equal footing with  (\ref{eq-app-j}),  certain transformations must be involved as we proceed as follows. Firstly, since $z(t)=\vel_q t$, $\vel_q$ being the flow velocity in $z$ direction,  thus $e^{i\w t}= e^{i\w z/\vel_q}$; accordingly  $\psi_q(z, t(z)) \rightarrow \psi_q(z)$, $\psi^* \rightarrow \psi^*_q(z)$, and $\rho_q \rightarrow \rho_q(z)$; i.e., $z$ is now an explicit independent variable of $\rho_q$ similarly as of $\rho_\Asub$ in  (\ref{eq-app-j}). We further define (for reason to become evident in the end) an imaginary diffusion constant, $D_q=i|D_q|$. We can now make three immediate substitutions of the corresponding variables of $\rho_q$
in (\ref{eq-app-j}), in such a way that each term is  ensured real and having a correct  sign so as to finally achieve a $j_q$  in accordance with the definition of (\ref{eq-app-j}):
$$\displaylines{
\refstepcounter{equation} \label{eDAq}
\hfill 
D_\Asub \rightarrow |D_q|= -i D_q, 
\quad \f' \rightarrow \psi^*,
\quad \f \rightarrow \psi. 
\hfill (\ref{eDAq})
}$$ 
The derivatives of $\psi_q^*$  and $\psi_q$ 
will however introduce an imaginary index $i$ and sign into the coefficients, as $\frac{1}{\psi^*} \nabla \psi^* =-ik$ and   $ \frac{1}{\psi} \nabla \psi =ik$. 
To obtain a "positive and real"  value for the term containing $\nabla \psi$  ($\psi$ represents a flow in positive direction) 
in the negative gradient of $\rho_q$, $-\nabla \rho_q$, 
and accordingly a "negative and real value" for  the term containing $\nabla \psi^*$, 
we rotate the two functions in the complex plane by angles 
 $-\frac{\pi}{2}$ and $+\frac{\pi}{2}$, thus 
$$\displaylines{
\refstepcounter{equation} \label{eq-difs}
\hfill
\nabla \f' \rightarrow  -i \nabla \psi^*; \quad  
\nabla \f \rightarrow  +i \nabla \psi, \quad 
 -\nabla \rho_\Asub  \rightarrow  - \nabla \rho_q =-[ \psi^* i\nabla \psi+ (-i\nabla \psi^*)\psi ]
\hfill (\ref{eq-difs})
}$$  
With  (\ref{eDAq}),(\ref{eq-difs})  in (\ref{eq-app-j}), we obtain 
$$\displaylines{
\refstepcounter{equation} \label{eq-app-j2}
\hfill
j_q (=-|D_q| |\nabla \rho_q|)= -(-i D_q) [\psi^*  (i\nabla \psi)+(-i\nabla \psi^*) \psi ]
=-D_q [\psi^* \nabla \psi-\psi (\nabla \psi^*)]. 
 \hfill (\ref{eq-app-j2})
}$$
\\
\footnotesize{Errata: In the first edition (arxiv:1111.3123v1) of this paper, the "positive real" value of $-\nabla \rho_q $ was ensured for the first of two  differential terms arranged in arbitrary order of sequence, rather than correctly for the term containing $\nabla \psi$.}

\section[\qquad \qquad \ \  Transition time]{Transition time
}\label{App-tphi}

Suppose that (i) the $F_{ext}$ in (\ref{eq-eqsmota}), Sec. \ref{Sec-VacPot}, is not zero but is equal to a radiation damping force, $F_{ext}=F_{rad}=- \w_r \Mcal_q d \uscr_q/dt$, where $\w_r$ is a radiation damping factor, (ii) $(\w_r/\w)^2<<1$, so the 
equations of motion and the solutions of Sec. \ref{Sec-VacPot}  continue to hold over a  finite time interval in which  damping of amplitude is negligible, whence 
a quasi stationary radiation, and (iii) we restrict  as before (Sec. \ref{Sec-VacPot}) 
 to the excitations which create matter particles only. Then, the energy solution for (\ref{eq-eqsmota}) combined with (\ref{eq-eqsmotb}) of the now quasi-harmonically oscillating charge is at any time $t_s$ given as,  dropping a 
term  $\frac{1}{2}\hbar \w$ similarly as in (\ref{eq-eqn}),
$$\displaylines{\refstepcounter{equation} \label{eq-engp-app}
\hfill
\eng'_{qn}(t_s)=e^{-\w_r t_s} \eng_{qn}, \quad 
\eng_{qn}=n\hbar \w, \quad n=1,2, \ldots
\hfill (\ref{eq-engp-app})
}$$

If at initial time $t_s=0$ the charge is  at level $n$ and just begins to emit radiation,  and after a time $t_s=t_{\vphi n.n-1}$ it has emitted one entire energy quantum $\D \eng_{qn.n-1}=n \hbar \w-(n-1)\hbar =\hbar \w $, whence transforming  to level $n-1$,  the energy reduction given after (\ref{eq-engp-app}) is  
$$\displaylines{
\refstepcounter{equation} \label{eq-Dengp-app}
\hfill
 \D \eng'_q (t_{\vphi n.n-1})= n\hbar \w(1 - e^{-\w_r t_{\vphi n.n-1}}). 
\hfill (\ref{eq-Dengp-app})
}$$ 
But $ \D \eng'_q (t_{\vphi n.n-1})=\D \eng_{qn.n-1} $;  or,  $n\hbar \w (1 - e^{-\w_r t_{\vphi n}})=\hbar \w$. This gives 
 $$\displaylines{
\refstepcounter{equation} \label{eq-time}
 \hfill t_{\vphi n.n-1} =-\frac{1}{\w_r}\ln \frac{n}{n-1 }. 
\hfill (\ref{eq-time})
}$$

\end{appendix}


\def\citePerkins1982Griffithsetal{{1}}
\def\citeWeinberg1967etal{{2}}
\def\rr{R}
\def\rrb{{\bf{\rr}}}
\def\rbav{\bar{\rrb}}
\def\rrbav{\bar{\rrb}}
\def\Fbav{\bar{\Fb}}

\def\rp{r}
\def\rpb{\mathbf{\rp}}
\def\rbp{\rpb}

\def\ka{\kappa}
\def\xib{\pmb{\xi}}
\def\Vel{u}
\def\Velb{{\mathbf{\Vel}} }
\def\rsub{{{_o}}}
\def\vtheta{\vartheta}
\def\pbf{{\bf{p}}}
\def\neu{{\mbox{\tiny{Neu}} }}
\def\orb{{\mbox{\tiny{orb}} }}

\def\life{{\mbox{\tiny{life}} }}
\def\max{{\mbox{\tiny{max}} }}
\def\min{{\mbox{\tiny{min}} }}

\def\GWS{{\mbox{\tiny{GWS}} }}
\def\Xsub{{\mbox{\tiny{X}} }}
\def\Isub{{\mbox{\tiny{I}} }}
\def\Rsub{{\mbox{\tiny{R}} }}
\def\Lsub{{\mbox{\tiny{L}} }}

\def\Rsub{{\mbox{\tiny{R}} }}
\def\Lsub{{\mbox{\tiny{L}} }}
\def\HRsub{{\mbox{\tiny{H}} }}

\def\osup{{\mbox{\tiny{$0$}} }}
\def\inTo{{\mbox{\tiny{in$T_o$}} }}
\def\Lamsub{{\mbox{\tiny{Larm}} }}

\def\exto{{o}}
\def\ex{{eq}}
\def\eq{{eq}}
\def\exti{{1}}

\def\sigb{{\pmb{\sigma}}}
\def\scat{{\rm{scat}}}
\def\tot{{     \mbox{\tiny{tot}} }}
\def\ef{{\rm{ef}}}
\def\rar{\rightarrow}
\def\Lrw{\Longrightarrow}
\def\nubar{{\bar{\nu}}}
\def\nubarem{{\bar{\nu}_{e}}}
\def\nuep{\nu_{e^+}}
\def\orb{{orb}}
\def\Mcm{M}
\def\mcm{M}
\def\velcm{U}

\def\gbar{{\overline{\g}}}
\def\gpbar{ {\overline{\g'}} }
\def\gbarb{ { \overline{\overline{\g}} } }

\def\Vol{{V\hspace{-0.3cm}^{_{\mbox{-}}} \hspace{0.13cm}}}

\def\l{l}
\def\mn{{m}}
\def\lsub{{     \mbox{\tiny{$l$}} }}
\def\isub{{     \mbox{\tiny{$1$}} }}

\def\sb{{\bf{s}}}

\def\velsp{\vel_{s_p}}
\def\velse{\vel_{s_e}}

\def\Db{{\bf{D}}}
\def\velop{\velavp}
\def\veloe{\velave}

\def\velsp{\vel^s_{p}}
\def\velse{\vel^s_{e}}
\def\velavp{\bar{\vel}^s_{p}}
\def\velsavp{\bar{\vel}^s_{p}}
        \def\velspav{\bar{\vel}^s_{p}}  
        \def\velave{\bar{\vel}^s_{e}}
\def\velsave{\bar{\vel}^s_{e}}
         \def\velseav{\bar{\vel}^s_{e}}

\def\velbavp{\bar{\velb}^s_{p}}
\def\velbave{\bar{\velb}^s_{e}}

\def\velbaveb{\bar{\velb}^s_{e}}

\def\velav{\bar{\vel}}
           \def\velavb{\bar{\bf{\vel}}}
\def\velbav{\bar{\pmb{\vel}}}

\def\aavp{\bar{a}_{p}}
\def\apav{\bar{a}_{p}}
\def\aeav{\bar{a}_{e}}
\def\aave{\bar{a}_{e}}
\def\aav{\bar{\hspace{0.1cm}\hspace{-0.1cm}a }}
\def\aavb{\bar{{\bf{a}}}}

\def\aop{\bar{a}_{p}}
\def\aoe{\bar{a}_{e}}
\def\amp{a_{m_p}}
\def\ame{a_{m_e}}
\def\aqp{a_{p}}
\def\aqe{a_{e}}
\def\ws{\w^s}
\def\wsp{\w^s_{p}}
\def\wse{\w^s_{e}}
\def\wsa{\w^s_\a}

\def\musa{\mu^s_\a}

\def\velsa{\vel^s_\a}

\def\Ls{L^s}
\def\Lsp{L^s_{p}}
\def\Lse{L^s_{e}}
\def\mus{\mu^s}
\def\musp{\mu^s_{p}}
\def\muse{\mu^s_{e}}
\def\Lorb{L}
\def\worb{\w}
\def\muorb{\mu}

\def\Mu{\mathscr{M}}
\def\gb{{\bf{g}}}
\def\zuni{\hat{z}}
\def\Zuni{\hat{Z}}

\def\runi{\hat{r}}
\def\thuni{\hat{\phi}}
\def\Runi{\hat{R}}
\def\Thuni{\hat{{\mit\Phi}}}

\def\Bbext{\Bb_{ext}}
\def\Fbext{\Fb_{ext}}
\def\mrm{{\mathbin{\mu\mkern-5.6mu\mbox{-}}}}

\def\rvec{\vec{r}}
\def\Rvec{\vec{R}}

\def\suf{{}_{}}
\def\labsup{{     \mbox{\tiny{IL}} }}
\def\lab{\labsup}
\def\Lab{{\mbox{\tiny{$L$}}}}
\def\Labi{     {\mbox{\tiny{       $L_1$      }}} }

\def\arssup{\star}
\def\arsup{\star}
\def\ars{\arsup}



\def\FbIL{{\bf{F}}}
\def\BbIL{{\bf{B}}}
\def\BfbIL{{\bf{B}}}
\def\MIL{M}
\def\RIL{{\sf{R}}}
\def\RbIL{{\bf{R}}}

\def\Fbicm{{\pmb{\sf{F}}}}
\def\Bbicm{{\pmb{\sf{B}}}}
\def\Bfbicm{{\pmb{\sf{B}}}}
\def\Ebicm{{\pmb{\sf{E}}}}
\def\Lbicm{{\pmb{\sf{L}}}}
\def\Sbicm{{\pmb{\sf{S}}}}
\def\Lbicm{{\pmb{\sf{L}}}}
\def\Vicm{{\sf{V}}}
\def\Hicm{{\sf{H}}}
\def\Eicm{{\sf{E}}}
\def\Ticm{{\sf{T}}}
\def\Ficm{{\sf{F}}}
\def\Licm{{\sf{L}}}

\def\Wb{{\bf{W}}}

\def\Vast{V}                                                                                                                                            \def\Hast{H}                                                                                                                                            \def\East{E}                                                                                                                                            \def\Tast{T}                                                                                                                                            \def\Fast{F}                                                                                                                                            \def\Last{L}                                                                                                                                            \def\Sast{S}                                                                                                                                            \def\Bast{B}                                                                                                                                            \def\Mast{M}                                                                                                                                            \def\Jast{J}                                                                                                                                            \def\Uast{U}                                                                                                                                            \def\Gast{G}                                                                                                                                            \def\Gam{{\mit{\Gamma}}}

\def\Fbast{{\mathbf{\Fast}}}
\def\Bbast{{\mathbf{\Bast}}}
\def\Bfbast{{\mathbf{\Bast}}}
\def\Ebast{{\mathbf{\East}}}
\def\Efbast{{\mathbf{\East}}}
\def\Lbast{{\mathbf{\Last}}}

\def\Sbast{{\mathbf{\Sast}}}
\def\Jbast{{\mathbf{\Jast}}}


\def\Vdag{{V^\dag}}                                                                                                                                            

\def\Hdag{{H^\dag}}                                                                                                                                            \def\Edag{{E^\dag}}                                                                                                                                            \def\Tdag{{T^\dag}}                                                                                                                                                        \def\Mdag{{\sf{M}}}                                                                                                                                            \def\Fdag{{F^\dag}}                                                                                                                                            \def\Ldag{{L^\dag}}                                                                                                                                            \def\Sdag{{S^\dag}}                                                                                                                                            \def\Bdag{{B^\dag}}                                                                                                                                            \def\Mdag{{M^\dag}}                                                                                                                                            \def\Jdag{{J^\dag}}                                                                                                                                            \def\Udag{{U^\dag}}                                                                                                                                            

\def\Fbdag{ {{\mathbf F}^\dag} }
\def\Bbdag{{{\mathbf B}^\dag} }
\def\Bfbdag{{{\mathbf B}^\dag} }
\def\Ebdag{{{\mathbf E}^\dag} }
\def\Efbdag{{{\mathbf E}^\dag} }
\def\Lbdag{{{\mathbf L}^\dag} }
\def\Sbdag{{{\mathbf S}^\dag} }
\def\Lbdag{{{\mathbf L}^\dag} }
\def\Sbdag{{{\mathbf S}^\dag} }
\def\Jbdag{{{\mathbf J}^\dag} }

\def\astsup{{}}
\def\ast{\astsup}

\def\astn{{}}  

\def\pp{{}}

\def\ab{{\bf{a}}}
\def\abf{{\pmb{\mbox{\it{a}}}}}
\def\acb{{\bf{\mbox{\sf{a}}}}}



\def\mr{{\mathscr{M}}}
\def\mrrr{{\mathscr{M}^*}}
\def\mrr{{\mathscr{M}^*}}
          \def\gar{\g^*}
\def\garr{\g^*{}}
\def\velrr{\vel^\ddagger{}}
        \def\velr{\vel^\ddagger{}}
     \def\velbrr{\velb^\ddagger{}}

     \def\trr{t^*{}}
\def\tr{{t^*}}
     \def\rbrr{{\bf{r}^*}}
\def\rbr{{\bf{r}^*}}

\def\rvecrr{{\vec{r}^*}}
 \def\rvecr{{\vec{r}^*}}

\def\mr{{\mathscr{M}}}
\def\mrdag{{\mathscr{M}^*\dagger}}
\def\mrlab{{\mathscr{M}}}
\def\gadag{\g^\dagger}
\def\gdag{\g^\dagger}
\def\veldag{\vel^\dagger{}}
       \def\velbdag{\velb^\dagger{}}
\def\velrr{\vel^\ddagger{}}
     \def\velbrr{\velb^\ddagger{}}

\def\rbdag{{\bf{r}^\dagger}}

\def\tdag{t^\dagger{}}
\def\rbdag{{\bf{r}^\dagger}}

\def\rlab{r}
\def\glab{\g}
\def\galab{\g}
\def\velblab{\velb}
\def\vellab{\vel}
\def\mrlab{{\mathscr{M}}}
\def\rblab{{\bf{r}}}
\def\rveclab{{\vec{r}}}
\def\rveclab{{\vec{r}}}

\def\rb{{\bf r}}
\def\tlab{t}
\def\garcmf{\g_{\lab:\cmsub}}

\def\aba{{\bar{a}}}
\def\rbar{{\bar{r}}}
\def\m{{\mbox{-}}}

\def\ubscr{\pmb{\mathscr{U}}}
        \def\ubscrq{{\pmb{\mathscr{U}}\hspace{-0.14cm}}_q}
         \def\ubscrqsq{{\pmb{\mathscr{U}}\hspace{-0.14cm}}_q{\hspace{-0.05cm}}}

\def\N{\mathfrak{N}}
\def\Vvqo{V_{\v q0}}
\def\Vvq{V_{\v q}}
\def\mub{\pmb{\mu}}
\def\taub{\pmb{\tau}}
\def\thetab{\pmb{\theta}}
\def\phib{\pmb{\phi}}
\def\Phimb{\pmb{\Phim}}

\def\wb{\pmb{\w}}
\def\mb{\mathbf{m}}

\def\lep{l}
\def\Rbb{\mathbb{R}}
\def\Kbb{\mathbb{P}}
\def\R{r_{max}}
\def\Pscr{\mathscr{P}}
\def\Hscr{\mathscr{H}}
\def\Vscr{\mathscr{V}}
\def\Tscr{\mathscr{T}}

\def\Ds{\mathscr{D}}

\def\Xim{{\mit{\Xi}}}

\def\rw{\rightarrow}
\def\jm{{j\mu}}
\def\kp{{j'}}
\def\muk{{\n'}}
\def\Nssk{\Nss_1}
\def\p{{\mbox{\scriptsize{$+$}} \hspace{-0.03cm}}}
\def\pe{\p e}
\def\pq{\p q}
\def\H{{a_{\Sigsub} \hspace{-0.05cm}}}
\def\Hn{a_{n\Sigsub}}
\def\Ia{\mathcal{A}}
\def\hpbar{\abar}
\def\hpbars{\hpbar}
\def\hp{a}
               \def\abar{{a\hspace{-0.18cm}\mbox{{\small $^{_{_{-}}}$}}\hspace{-0.07cm}}}
                \def\abars{{\hspace{-0.03cm}a\hspace{-0.1cm}\mbox{\small{-}}\hspace{0.0cm}}}
\def\abar{{a\hspace{-0.18cm}\mbox{{\small $^{_{_{-}}}$}}\hspace{-0.07cm}}}
\def\Qcal{\mathbin{{Q}\mkern-8.5mu^{_{\mbox{\small{$\dash$}}}}\hspace{-0.04cm} }}

\def\Sa{{\mathfrak{S}}}
\def\nfrak{{\mathfrak{N}}}

\def\h{h}
\def\taubar{\mathbin{{\tau}\mkern-10.3mu_{^{{}^{{\mbox{\tiny{$-$}}}}\hspace{-0.10cm}}} }}
\def\ho{\eta_0}
\def\hjn{\hp_{\jn}}
\def\Wst{{\mathbin{\Omega\mkern-4.1mu^{_{\mbox{\footnotesize{-}}}}}\hspace{-0.04cm}}}
\def\Wsts{{\mathbin{\Omega\mkern-4.2mu^{_{\mbox{\scriptsize{-}}}}}\hspace{-0.06cm}}}
\def\Wstsup{{\mathbin{\Omega\mkern-5.mu^{_{\mbox{\scriptsize{-}}}}}}}
\def\Wstt{{\mathbin{\Omega\mkern-3.5 mu^{_{\mbox{\scriptsize{-}}}}}}}  


\def\Nstat{{\mathcal{N}}}

\def\Nsts{{\Wsts}}
\def\Nst{{\Wst}}
\def\Nss{{\Wst}}
\def\Nstsup{{\Wstsup}}

\def\nsig{{\Sigsub\n}}
\def\nPi{{\n_{^{_\Pi}}}}

\def\pjn{p_{j\n}}
\def\jn{{jn}}
\def\xjn{{j\n}}
\def\Pcaln{\Pcal_{n}}
\def\Pcalbarn{ \overline{\Pcal_\n}}

\def\Pcalens{\Pcal_{\ens}}
\def\Pens{\Pcal_{\ens}}
\def\Pensm{\Pcal_{\ens,max}}
\def\Nstgm{\Nst_{i.g.m}}
\def\Nstam{\Nst_{i.a.m}}
\def\quadd{\ \ }
\def\la{\langle}
\def\ra{\rangle}
\def\mrsub{{\mbox{{\scriptsize{$\mr$}}}}}
\def\Msub{{        {_{\mbox{{\tiny{M}}}}} }}

\def\cmsub{{{_{\mbox{{\tiny{CM}}}}}}}

\def\cmsub{   {\mbox{{\tiny{cm}}}} }

\def\icmsub{{\mbox{{\tiny{ICM}}}}}

\def\ncmsub{{\mbox{{\tiny{NCM}}}}}
\def\Neusub{{\mbox{{\tiny{Neut}}}}}

\def\pmsub{{\mbox{{\tiny{$\pm$}}}}}
\def\Hsub{{\mbox{{\tiny{H}}}}}

\def\Fsub{{\mbox{{\tiny{$F$}}}}}
\def\Larmsub{{\mbox{{\tiny{$Larm$}}}}}

\def\Lsub{{\mbox{{\tiny{$L$}}}}}
\def\Sigsub{{\mbox{{\tiny{$\Sigma$}}}}}
\def\Wsub{{\mbox{{\tiny{$W^\pmsub$}}}}}
\def\Tsub{{\mbox{{\tiny{$T$}}}}}
\def\wsub{{\mbox{\tiny{w}}}}

\def\wsubi{{\mbox{\tiny{$W_1$}}}}
\def\wpmsub{{\mbox{\tiny{$W^\pmsub$}}}}
\def\wsubpm{{\mbox{\tiny{$W^\pmsub$}}}}
\def\wpmsub{{\mbox{\tiny{$W^\pmsub$}}}}
\def\wssub{\wsub}

\def\wsubo{{\mbox{\tiny{$0$}}}}
\def\wsubthi{{\mbox{\tiny{$\theta_1$}}}}

\def\Nsub{{\mbox{{\tiny{$N$}}}}}
\def\rep{{rep}}
\def\Sig{\Sigma}
\def\bi{b^{i}}
\def\i{i}
\def\n{\nu}
\def\uscr{\mathscr{U}}
\def\uscrdotbar{\bar{\dot{\mathscr{U}}}}
\def\vac{{\rm{vac}}}
\def\Vcal{{\mathscr{V}}} 
 \def\V{V} 
\def\Bsub{{\mbox{\tiny{{\rm B}}}}}
\def\Nsub{{{\mbox{\tiny${N}$}}}}
\def\Hsub{{{\mbox{\tiny${H}$}}}}

\def\Hbar{\bar{H}}
\def\pbar{\bar{p}}

\def\exc{{\rm exc}}
\def\ext{{{\rm ext}}}
\def\mini{0}
\def\Pcal{{\mathcal{P}}}
\def\bav{{\bar{b}}}
\def\v{{\rm v}}
\def\vrm{\vel_{t}{}}
\def\vit{\vrm}
\def\vrmb{{\bf{v}}}

\def\Hbar{\bar{H}}
\def\D{\Delta}
\def\bcal{b}
\def\bbar{\mathbin{{b}\mkern-9.5mu^{{\mbox{\tiny{$-$}}}}\hspace{-0.00cm} }}
\def\nstat{\nu}
\def\nst{\nu}
\def\engbar{\bar{\eng}}
\def\engobar{\bar{\eng}_0}
\def\psias{\psi}
\def\Phimas{\Phim}
\def\fas{f}
\def\rbb{\as}

\def\La{L}
\def\Ja{J}
\def\as{p}
\def\ioii{{\mbox{\normalsize${\frac{1}{2}}$}}}
\def\Rb{{\bf R}}

\def\xb{{\bf x}}
\def\Xb{{\bf X}}

\def\ub{{\bf{u}}}
\def\hatu{\hat{u}_q}
\def\Nsub{{{\mbox{\tiny${N}$}}}}
\def\Pisub{{{\mbox{\tiny${\mit{\Pi}}$}}}}

\def\q{\bar{q}}
\def\xdot{\dot{x}}
\def\exc{{\rm ex}}
\def\ens{{ens}}

\def\Gcal{\Gast}


\def\Lcal{\Last}
\def\Cbar{\hspace{0.02cm}C\hspace{-0.4cm}^{{  \atop  -}\hspace{0.08cm}}{}   }

\def\Lscr{\Lcal\hspace{-0.3cm}^{{  \atop  -} }{}   }
\def\Lbscr{\mathbf{\Lcal}\hspace{-0.32cm}^{{  \atop  -} }{} }
\def\Lscrb{\mathbf{\Lcal}\hspace{-0.31cm}^{{  \atop  -} }{} }

\def\Lbcal{\mathbf{\Lcal}}
\def\Lcalb{\Lbcal}

\def\Scal{\Sast}
\def\Jcal{\Jast}
\def\Sbcal{\mathbf{\Scal}}
\def\Jbcal{\mathbf{\Jcal}}
\def\Hcal{\Hast}

\def\Kcal{{\mathcal{K}}}
\def\Xcal{{\mathcal{X}}}
\def\Tcal{\Tast}

\def\Ocal{\mathcal{O}}
\def\Rcal{\mathcal{R}}
\def\Ycal{\mathcal{Y}}

\def\Wvel{\Omegavel}
\def\Ncal{{\mathcal{N}}}
\def\Omegavel{\mathbin{{\mit\Omega}\mkern-13.mu^{_{\mbox{$-$}}}\hspace{-0.08cm}{}_d }}

\def\omegavel{{\w\mbox{\hspace{-0.38cm} \vspace{0.15cm}$-$\hspace{-0.02cm}}}}
\def\wvel{\omegavel_d}

\def\Ucal{\bar{\eng}_{0}}
\def\Omegavel{\mathbin{{\mit\Omega}\mkern-13.mu^{_{\mbox{$-$}}}\hspace{-0.08cm}{}_d }}
\def\Wvel{\Omegavel}

\def\q{\mathbin{q\mkern-11mu-}}
\def\PE{\mbox{\tiny{{\rm P.E.}}}}
\def\ME{\mbox{\tiny{{\rm M.E.}}}}
\def\QM{\mbox{\tiny{{\rm QM}}}}
\def\Psub{\mbox{\tiny{{\rm P}}}}
\def\Bsub{{\mbox{\tiny{{\rm B}}}}}
\def\TP{{\mbox{\tiny{{\rm T}}}}}

\def\SM{{\mbox{\tiny{{\rm SM}}}}}
\def\MT{{\mbox{\tiny{{\rm MT}}}}}

\def\ev{\epsilon}

\def\Ci{1}
\def\betamt{{\bf{b}}}
\def\kb{{\bar{k}}}
\def\kbf{{\bf{k}}}
\def\Kb{{\bf{K}}}
\def\cb{{\bf{c}}}

\def\pb{{\bar{p}}}
\def\pbf{{\bf{p}}}
\def\Pbf{{\bf{P}}}
\def\Mbf{{\bf{M}}}
\def\pbf{{\bf{p}}}
\def\Acal{\mathscr{A}}

\def\Bcal{\mathcal{B}}
\def\Bbcal{\pmb{\mathcal{B}}}

\def\Ecal{{\mathcal{E}}}
\def\Ebcal{\pmb{{\mathcal{E}}}}

\def\Fcal{{\mathcal{F}}}
\def\Fbcal{\pmb{{\mathcal{F}}}}

\def\Ccal{{\cal{C}}}
\def\Vp{V}
\def\Ccal{{\cal{C}}}
\def\p{{{}_{+\hspace{-0.1cm}}}}

\def\psipi{\psi_{\p}(1)}
\def\psipii{\psi_{\p}(2)}
\def\psimi{\psi_{\m}(1)}
\def\psimii{\psi_{\m}(2)}

\def\ai{\alpha(1)}
\def\aii{\alpha^{'}(2)}
\def\bi{\beta^{'}(1)}
\def\bii{\beta(2)}

\def\fa{f_r}
\def\fb{f_\ell}

\def\Ca{C_a}
\def\Cb{C_b}
\def\fbf{{\bf{f}}}
\def\Ocal{{\cal{O}}}
\def\psib{{\pmb{\psi}}}
\def\alphab{{\pmb{\alpha}}}
\def\sigmab{{\pmb{\sigma}}}
\def\sig{\sigma}
\def\Eb{{\bf E}}
\def\Bb{{\bf B}}
\def\ke{\kappa}
\def\nabb{{\pmb{\nabla}}}
\def\nablab{{\pmb{\nabla}}}
\def\vir{{\rm vir}}

\def\psitot{\psi}
\def\jb{{\bf{j}}}
\def\vel{upsilon}

\def\vels{{\hspace{0.1cm}\breve{\hspace{-0.1cm}\vel}}}
\def\velsb{{\breve{\velb}}}
\def\vb{{\bf{v}}}
\def\Imtr{I}
\def\citeUnif{4?}
\def\App{}
\def\Qcal{{\mathcal{Q}}}
\def\Cross{Q}

\def\vphilim{f}
\def\ft{{\mathcal{B}}}
\def\vphibar{\mathbin{\varphi\mkern-12.5mu-}}
\def\vphi{\varphi}
\def\med{{\med}}
\def\Mcal{{\mathfrak{M}}}
\def\Sb{{\bf{S}}}
         \def\xia{{\mathcal{A}}}
\def\tha{\theta}

\def\nb{\bf{n}}
\def\zb{{\bf{z}}^0}

\def\ph{{ph}}
\def\phiv{\varphi}
\def\Lb{{\bf{L}}}
\def\velsub{_{\vel}}
\def\Jb{{\bf{J}}}
\def\Pb{{\bf{P}}}
\def\Mb{{\bf{M}}}
\def\Zo{{Z^0}}
\def\nablab{{\pmb{\nabla}}}
\def\velb{{\pmb{\vel}}}
\def\velbecmast{\velb_e^\astn}

\def\Db{{\bf{D}}}

\def\Ab{{\bf{A}}_a}
\def\Abb{{\bf{A}}}

\def\vel{\upsilon}
\def\Thm{\vartheta}
\def\Thetam{{\mit{\Theta}}}
\def\Thetamb{ \pmb{{\mit{\Theta}}}}

\def\lb{{\bf l}}
\def\ldb{{\pmb{\ld}}}
\def\ld{\ell}
\def\ellb{{\pmb{\ell}}}
\def\vb{\velb_{t}{}}

\def\Rb{{\bf R}}
\def\pd{\partial}
\def\vphi{\varphi}

\def\psitot{\varphi}
\def\psiR{\widetilde{\psi}}
\def\psiL{\widetilde{\psi}^{{\rm vir}}}
\def\Phim{{\mit{\Phi}}}
\def\PhimR{\widetilde{ {\mit \Phi}}}
\def\PsimR{\widetilde{ {\mit \Psi}}}
\def\PsimL{{\widetilde{ {\mit \Psi}}}^{{\rm vir}}}
\def\a{\alpha}
\def\uav{\bar{u}}
\def\D{\Delta}
\def\th{\theta}
\def\r{{\mbox{\tiny${R}$}}}
\def\re{{\mbox{\tiny${R}$}}}
\def\Fmed{F_{{\rm a.med}}}
\def\med{{\rm med}}
\def\Lw{L_{\varphi}}
\def\Fb{{\bf{F}}}

\def\Efb{{\bf{E}}}
\def\Bfb{{\bf{B}}}
\def\Bf{B}
\def\Ac{ \varphi}
\def\Xsub{{\mbox{\tiny${X}$}}}
\def\Ysub{{\mbox{\tiny${Y}$}}}
\def\Zsub{{\mbox{\tiny${Z}$}}}
\def\MTsub{{}}

\def\Ksub{{\mbox{\tiny${K}$}}}
\def\W{{\mit \Omega}}
\def\Wd{\W_d{}}
\def\Nu{{\cal V}}
\def\Nud{\Nu_d{}}
\def\Eng{{\cal E}}
\def\eng{{\varepsilon}}
\def\vep{\varepsilon}
\def\Kmscr{{\mathscr{K}}}
           \def\engk{\Kcal}
\def\Acuni{\Ac_{{\Ksub}^\dagsup}^{\dagsup}}
\def\unduni{\Ac_{{\Ksub}^\dagger}^{\dagsup}}
\def\Acauni{\Ac_{{\Ksub}^\ddagsup}^{\ddagsup}}
\def\Acunim{{\Ac_{{\Ksub}^\dagsup}^{\dagsup *}}}
\def\undunim{{\Ac_{{\Ksub}^\dagsup}^{\dagsup}}^*}
\def\Acaunim{{\Ac_{{\Ksub}^\ddagsup}^{\ddagsup *}}}
\def\pd{\partial}
\def\Ad{ {\mit \psi}}
\def\psim{ {\mit \psi}}
\def\Kd{K_d{}}
\def\Lam{{\mit \Lambda}}
\def\lam{\lambda}

\def\dagsup{\m}
\def\ddagsup{{\mbox{\tiny{+}}}}

\def\dagsupa{{\mbox{\tiny{$\dagger$}}}}

\def\psimKdK{\psim_{\Ksub,\Kdsub}}
\def\wk{1}

\def\w{\omega{}}
\def\wrm{{\rm{w}}}
\def\wit{\w_{t}{}}
\def\witb{{\pmb{\it{w}}}}

\def\wdlow{\omega_d }
\def\g{\gamma{}} 
\def\Phimc{{\mathcal C}}
\def\Psim{{\mit \Psi}}
\def\arm{{\rm a}}
\def\brm{{\rm b}}
\def\crm{{\rm c}}
\def\drm{{\rm d}}
\def\erm{{\rm e}}
\def\frm{{\rm f}}
\def\grm{{\rm g}}
\def\hrm{{\rm h}}
\def\lf{\left}
\def\rt{\right}
\def\Kdsub{{\mbox{\tiny${K_d}$}}}
\def\psimkd{\psim_{\kdsub}}
\def\psimKd{\psim_{\Kdsub}}
\def\hquad{ \ \ } 
\def\Taum{{\mit \Gamma}}

\def\mrm{{\mathbin{\mu\mkern-5.6mu\mbox{-}}}}



\eject
\noindent
{\large{Part B} (Published in  {\it J. Phys.: Conf. Ser.} {\bf 670} 012056, 2016.)}
\vspace{-1.cm}
\vspace{1.cm}

\title[Microscopic Theory of the Neutron]{A Microscopic Theory of the Neutron}
\author{J.X. Zheng-Johansson}
\address{
Institute of Fundamental Physics Research} 

\setcounter{equation}{0}
\setcounter{section}{0}
\setcounter{figure}{0}



\def\citePerkins1982Griffithsetal{{1}}
\def\citeWeinberg1967etal{{2}}

\def\point{{\rm{point}}}
\def\j{j}
\def\jsub{{\mbox{\tiny{$j$}}}}

\def\hfpp{{^1\hspace{-0.06cm}\mbox{\tiny{/}}\hspace{-0.05cm}{}_2}}
\def\hfp{{\mbox{\tiny{1/2}}}} 

\def\hfp{\hf}
\def\hf{{\mbox{\tiny{$\frac{1}{2}$}}}}
\def\hfb{{}^{{\mbox{\tiny{$\frac{1}{2}$}}}}}
\def\J{J}
\def\Jscr{J}

\def\Jtot{\mbox{$\mathscr{J}$}}

\def\Jb{{\mathbf{J}}}
\def\Jscrb{{\mathbf{J}}}
\def\Jbscr{{\mathbf{J}}}

\def\Jorb{{\mathcal{J}}}
\def\Jtr{{\mathcal{J}}}
\def\Jtrb{{\pmb{\mathcal{J}}}}

\def\rr{R}
\def\rrb{{\bf{\rr}}}
\def\rbav{\bar{\rrb}}
\def\rrbav{\bar{\rrb}}
\def\Fbav{\bar{\Fb}}

\def\rp{r}
\def\rpb{\mathbf{\rp}}
\def\rbp{\rpb}

\def\ka{\kappa}
\def\xib{\pmb{\xi}}
\def\Vel{u}
\def\Velb{{\mathbf{\Vel}} }
\def\rsub{{{_o}}}
\def\vtheta{\vartheta}
\def\pbf{{\bf{p}}}
\def\neu{{\mbox{\tiny{Neu}}}}
\def\orb{{\mbox{\tiny{orb}} }}

\def\life{{\mbox{\tiny{life}} }}
\def\max{{\mbox{\tiny{max}} }}
\def\min{{\mbox{\tiny{min}} }}

\def\GWS{{\mbox{\tiny{GWS}} }}
\def\Xsub{{\mbox{\tiny{X}} }}
\def\Isub{{\mbox{\tiny{I}} }}

\def\Rsub{{\mbox{\tiny{$R$}} }}
\def\Lsub{{\mbox{\tiny{$L$}} }}
\def\HRsub{{\mbox{\tiny{H}} }}

\def\osup{{\mbox{\tiny{$0$}} }}
\def\inTo{{\mbox{\tiny{in$T_o$}} }}
\def\Lamsub{{\mbox{\tiny{Larm}} }}

\def\exto{{o}}
\def\ex{{eq}}
\def\eq{{eq}}
\def\exti{{1}}

\def\sigb{{\pmb{\sigma}}}
\def\scat{{\rm{scat}}}
\def\tot{{     \mbox{\tiny{tot}} }}
\def\ef{{\rm{ef}}}
\def\rar{\rightarrow}
\def\Lrw{\Longrightarrow}
\def\nubar{{\bar{\nu}}}
\def\nubarem{{\bar{\nu}_{e}}}
\def\nuep{\nu_{e^+}}
\def\orb{{orb}}
\def\Mcm{M}
\def\mcm{M}
\def\velcm{U}

\def\gbar{{\overline{\g}}}
\def\gpbar{ {\overline{\g'}} }
\def\gbarb{ { \overline{\overline{\g}} } }

\def\Vol{{V\hspace{-0.3cm}^{_{\mbox{-}}} \hspace{0.13cm}}}

\def\l{l}
\def\mn{{m}}
\def\lsub{{     \mbox{\tiny{$l$}} }}
\def\isub{{     \mbox{\tiny{$1$}} }}

\def\sb{{\bf{s}}}

\def\velsp{\vel_{s_p}}
\def\velse{\vel_{s_e}}

\def\Db{{\bf{D}}}
\def\velop{\velavp}
\def\veloe{\velave}

\def\velsp{\vel^s_{p}}
\def\velse{\vel^s_{e}}
\def\velavp{\bar{\vel}^s_{p}}
\def\velsavp{\bar{\vel}^s_{p}}
        \def\velspav{\bar{\vel}^s_{p}}  
        \def\velave{\bar{\vel}^s_{e}}
\def\velsave{\bar{\vel}^s_{e}}
         \def\velseav{\bar{\vel}^s_{e}}

\def\velbavp{\bar{\velb}^s_{p}}
\def\velbave{\bar{\velb}^s_{e}}

\def\velbaveb{\bar{\velb}^s_{e}}

\def\velav{\bar{\vel}}
           \def\velavb{\bar{\bf{\vel}}}
\def\velbav{\bar{\pmb{\vel}}}

\def\aavp{\bar{a}_{p}}
\def\apav{\bar{a}_{p}}
\def\aeav{\bar{a}_{e}}
\def\aave{\bar{a}_{e}}
\def\aav{\bar{\hspace{0.1cm}\hspace{-0.1cm}a }}
\def\aavb{\bar{{\bf{a}}}}

\def\aop{\bar{a}_{p}}
\def\aoe{\bar{a}_{e}}
\def\amp{a_{m_p}}
\def\ame{a_{m_e}}
\def\aqp{a_{p}}
\def\aqe{a_{e}}
\def\ws{\w^s}
\def\wsp{\w^s_{p}}
\def\wse{\w^s_{e}}
\def\wsa{\w^s_\a}

\def\musa{\mu^s_\a}

\def\velsa{\vel^s_\a}

\def\Ls{L^s}
\def\Lsp{L^s_{p}}
\def\Lse{L^s_{e}}
\def\mus{\mu^s}
\def\musp{\mu^s_{p}}
\def\muse{\mu^s_{e}}
\def\Lorb{L}
\def\worb{\w}
\def\muorb{\mu}

\def\Mu{\mathscr{M}}
\def\gb{{\bf{g}}}
\def\zuni{\hat{z}}
\def\Zuni{\hat{Z}}

\def\runi{\hat{r}}
\def\thuni{\hat{\phi}}
\def\Runi{\hat{R}}
\def\Thuni{\hat{{\mit\Phi}}}

\def\Bbext{\Bb_{ext}}
\def\Fbext{\Fb_{ext}}
\def\mrm{{\mathbin{\mu\mkern-5.6mu\mbox{-}}}}

\def\rvec{\vec{r}}
\def\Rvec{\vec{R}}

\def\suf{{}_{}}

\def\Lab{{\mbox{\tiny{lab}}}}

\def\labsup{\Lab}
\def\lab{\labsup}
\def\Labi{     {\mbox{\tiny{       $L_1$      }}} }

\def\arssup{\star}
\def\arsup{\star}
\def\ars{\arsup}



\def\FbIL{{\bf{F}}}
\def\BbIL{{\bf{B}}}
\def\BfbIL{{\bf{B}}}
\def\MIL{M}
\def\RIL{{\sf{R}}}
\def\RbIL{{\bf{R}}}

\def\Fbicm{{\pmb{\sf{F}}}}
\def\Bbicm{{\pmb{\sf{B}}}}
\def\Bfbicm{{\pmb{\sf{B}}}}
\def\Ebicm{{\pmb{\sf{E}}}}
\def\Lbicm{{\pmb{\sf{L}}}}
\def\Sbicm{{\pmb{\sf{S}}}}
\def\Lbicm{{\pmb{\sf{L}}}}
\def\Vicm{{\sf{V}}}
\def\Hicm{{\sf{H}}}
\def\Eicm{{\sf{E}}}
\def\Ticm{{\sf{T}}}
\def\Ficm{{\sf{F}}}
\def\Licm{{\sf{L}}}

\def\Wb{{\bf{W}}}

\def\Vast{V}                                                                                                                                            \def\Hast{H}                                                                                                                                            \def\East{E}                                                                                                                                            \def\Tast{T}                                                                                                                                            \def\Fast{F}                                                                                                                                                       
\def\Sast{S}                                                                                                                                            \def\Bast{B}                                                                                                                                            \def\Mast{M}                                                                                                                                            \def\Jast{J}                                                                                                                                            \def\Uast{U}                                                                                                                                            \def\Gast{G}                                                                                                                                                         
\def\Gam{K}

\def\Fbast{{\mathbf{\Fast}}}
\def\Bbast{{\mathbf{\Bast}}}
\def\Bfbast{{\mathbf{\Bast}}}
\def\Ebast{{\mathbf{\East}}}
\def\Efbast{{\mathbf{\East}}}
\def\Lbast{{\mathbf{\Last}}}

\def\Sbast{{\mathbf{\Sast}}}
\def\Jbast{{\mathbf{\Jast}}}


\def\Vdag{{V^\dag}}                                                                                                                                            

\def\Hdag{{H^\dag}}                                                                                                                                            \def\Edag{{E^\dag}}                                                                                                                                            \def\Tdag{{T^\dag}}                                                                                                                                                        \def\Mdag{{\sf{M}}}                                                                                                                                            \def\Fdag{{F^\dag}}                                                                                                                                            \def\Ldag{{L^\dag}}                                                                                                                                            \def\Sdag{{S^\dag}}                                                                                                                                            \def\Bdag{{B^\dag}}                                                                                                                                            \def\Mdag{{M^\dag}}                                                                                                                                            \def\Jdag{{J^\dag}}                                                                                                                                            \def\Udag{{U^\dag}}                                                                                                                                            

\def\Fbdag{ {{\mathbf F}^\dag} }
\def\Bbdag{{{\mathbf B}^\dag} }
\def\Bfbdag{{{\mathbf B}^\dag} }
\def\Ebdag{{{\mathbf E}^\dag} }
\def\Efbdag{{{\mathbf E}^\dag} }
\def\Lbdag{{{\mathbf L}^\dag} }
\def\Sbdag{{{\mathbf S}^\dag} }
\def\Lbdag{{{\mathbf L}^\dag} }
\def\Sbdag{{{\mathbf S}^\dag} }
\def\Jbdag{{{\mathbf J}^\dag} }

\def\astsup{{}}
\def\ast{\astsup}

\def\astn{{}}  

\def\pp{{}}

\def\ab{{\bf{a}}}
\def\abf{{\pmb{\mbox{\it{a}}}}}
\def\acb{{\bf{\mbox{\sf{a}}}}}



\def\mrdag{{\mathscr{M}^*\dagger}}
\def\mrlab{{\mathscr{M}}}
\def\mrrr{{\mathscr{M}^*}}
\def\mrr{{\mathscr{M}^*}}
 \def\mr{{\mbox{$\mathscr{M}$}}}

          \def\gar{\g^*}
\def\garr{\g^*{}}
\def\velrr{\vel^\ddagger{}}
        \def\velr{\vel^\ddagger{}}
     \def\velbrr{\velb^\ddagger{}}

     \def\trr{t^*{}}
\def\tr{{t^*}}
     \def\rbrr{{\bf{r}^*}}
\def\rbr{{\bf{r}^*}}

\def\rvecrr{{\vec{r}^*}}
 \def\rvecr{{\vec{r}^*}}

\def\gadag{\g^\dagger}
\def\gdag{\g^\dagger}
\def\veldag{\vel^\dagger{}}
       \def\velbdag{\velb^\dagger{}}
\def\velrr{\vel^\ddagger{}}
     \def\velbrr{\velb^\ddagger{}}

\def\rbdag{{\bf{r}^\dagger}}

\def\tdag{t^\dagger{}}
\def\rbdag{{\bf{r}^\dagger}}

\def\rlab{r}
\def\glab{\g}
\def\galab{\g}
\def\velblab{\velb}
\def\vellab{\vel}
\def\mrlab{{\mathscr{M}}}
\def\rblab{{\bf{r}}}
\def\rveclab{{\vec{r}}}
\def\rveclab{{\vec{r}}}

\def\rb{{\bf r}}
\def\tlab{t}
\def\garcmf{\g_{\lab:\cmsub}}

\def\aba{{\bar{a}}}
\def\rbar{{\bar{r}}}
\def\m{{\mbox{-}}}

\def\ubscr{\pmb{\mathscr{U}}}
        \def\ubscrq{{\pmb{\mathscr{U}}\hspace{-0.14cm}}_q}
         \def\ubscrqsq{{\pmb{\mathscr{U}}\hspace{-0.14cm}}_q{\hspace{-0.05cm}}}

\def\N{\mathfrak{N}}
\def\Vvqo{V_{\v q0}}
\def\Vvq{V_{\v q}}
\def\mub{\pmb{\mu}}
\def\taub{\pmb{\tau}}
\def\thetab{\pmb{\theta}}
\def\phib{\pmb{\phi}}
\def\Phimb{\pmb{\Phim}}

\def\wb{\pmb{\w}}
\def\mb{\mathbf{m}}

\def\lep{l}
\def\Rbb{\mathbb{R}}
\def\Kbb{\mathbb{P}}
\def\R{r_{max}}
\def\Pscr{\mathscr{P}}
\def\Hscr{\mathscr{H}}
\def\Vscr{\mathscr{V}}
\def\Tscr{\mathscr{T}}

\def\Ds{\mathscr{D}}

\def\Xim{{\mit{\Xi}}}

\def\rw{\rightarrow}
\def\jm{{j\mu}}
\def\kp{{j'}}
\def\muk{{\n'}}
\def\Nssk{\Nss_1}
\def\p{{\mbox{\scriptsize{$+$}} \hspace{-0.03cm}}}
\def\pe{\p e}
\def\pq{\p q}
\def\H{{a_{\Sigsub} \hspace{-0.05cm}}}
\def\Hn{a_{n\Sigsub}}
\def\Ia{\mathcal{A}}
\def\hpbar{\abar}
\def\hpbars{\hpbar}
\def\hp{a}
               \def\abar{{a\hspace{-0.18cm}\mbox{{\small $^{_{_{-}}}$}}\hspace{-0.07cm}}}
                \def\abars{{\hspace{-0.03cm}a\hspace{-0.1cm}\mbox{\small{-}}\hspace{0.0cm}}}
\def\abar{{a\hspace{-0.18cm}\mbox{{\small $^{_{_{-}}}$}}\hspace{-0.07cm}}}
\def\Qcal{\mathbin{{Q}\mkern-8.5mu^{_{\mbox{\small{$\dash$}}}}\hspace{-0.04cm} }}

\def\Sa{{\mathfrak{S}}}
\def\nfrak{{\mathfrak{N}}}

\def\h{h}
\def\taubar{\mathbin{{\tau}\mkern-10.3mu_{^{{}^{{\mbox{\tiny{$-$}}}}\hspace{-0.10cm}}} }}
\def\ho{\eta_0}
\def\hjn{\hp_{\jn}}
\def\Wst{{\mathbin{\Omega\mkern-4.1mu^{_{\mbox{\footnotesize{-}}}}}\hspace{-0.04cm}}}
\def\Wsts{{\mathbin{\Omega\mkern-4.2mu^{_{\mbox{\scriptsize{-}}}}}\hspace{-0.06cm}}}
\def\Wstsup{{\mathbin{\Omega\mkern-5.mu^{_{\mbox{\scriptsize{-}}}}}}}
\def\Wstt{{\mathbin{\Omega\mkern-3.5 mu^{_{\mbox{\scriptsize{-}}}}}}}  


\def\Nstat{{\mathcal{N}}}

\def\Nsts{{\Wsts}}
\def\Nst{{\Wst}}
\def\Nss{{\Wst}}
\def\Nstsup{{\Wstsup}}

\def\nsig{{\Sigsub\n}}
\def\nPi{{\n_{^{_\Pi}}}}

\def\pjn{p_{j\n}}
\def\jn{{jn}}
\def\xjn{{j\n}}
\def\Pcaln{\Pcal_{n}}
\def\Pcalbarn{ \overline{\Pcal_\n}}

\def\Pcalens{\Pcal_{\ens}}
\def\Pens{\Pcal_{\ens}}
\def\Pensm{\Pcal_{\ens,max}}
\def\Nstgm{\Nst_{i.g.m}}
\def\Nstam{\Nst_{i.a.m}}
\def\quadd{\ \ }
\def\la{\langle}
\def\ra{\rangle}
\def\mrsub{{\mbox{{\scriptsize{$\mr$}}}}}
\def\Msub{{        {_{\mbox{{\tiny{M}}}}} }}

\def\cmsub{{\mbox{{\tiny{M}}}}}
\def\Msub{{\mbox{{\tiny{M}}}}}
\def\cmsublab{{\mbox{{\tiny{cm}}}}}

\def\icmsub{{\mbox{{\tiny{ICM}}}}}

\def\ncmsub{{\mbox{{\tiny{NCM}}}}}
\def\Neusub{{\mbox{{\tiny{Neut}}}}}

\def\pmsub{{\mbox{{\tiny{$\pm$}}}}}
\def\Hsub{{\mbox{{\tiny{H}}}}}

\def\Fsub{{\mbox{{\tiny{$F$}}}}}
\def\Larmsub{{\mbox{{\tiny{$Larm$}}}}}

\def\Sigsub{{\mbox{{\tiny{$\Sigma$}}}}}
\def\Wsub{{\mbox{{\tiny{$W^\pmsub$}}}}}

\def\Tsub{{\mbox{{\tiny{$TP$}}}}}

\def\wsub{{\mbox{\tiny{w}}}}

\def\wsubi{{\mbox{\tiny{$W_1$}}}}
\def\wpmsub{{\mbox{\tiny{$W^\pmsub$}}}}
\def\wsubpm{{\mbox{\tiny{$W^\pmsub$}}}}
\def\wpmsub{{\mbox{\tiny{$W^\pmsub$}}}}
\def\wssub{\wsub}

\def\wsubo{{\mbox{\tiny{$0$}}}}
\def\wsubthi{{\mbox{\tiny{$\theta_1$}}}}

\def\Nsub{{\mbox{{\tiny{$N$}}}}}
\def\rep{{rep}}
\def\Sig{\Sigma}
\def\bi{b^{i}}
\def\i{i}
\def\n{\nu}
\def\uscr{\mathscr{U}}
\def\uscrdotbar{\bar{\dot{\mathscr{U}}}}
\def\vac{{\rm{vac}}}
\def\Vcal{{\mathscr{V}}} 
 \def\V{V} 
\def\Bsub{{\mbox{\tiny{{\rm B}}}}}
\def\Nsub{{{\mbox{\tiny${N}$}}}}
\def\Hsub{{{\mbox{\tiny${H}$}}}}

\def\Hbar{\bar{H}}
\def\pbar{\bar{p}}

\def\exc{{\rm exc}}
\def\ext{{{\rm ext}}}
\def\mini{0}
\def\Pcal{{\mathcal{P}}}
\def\bav{{\bar{b}}}
\def\v{{\rm v}}
\def\vrm{\vel_{t}{}}
\def\vit{\vrm}
\def\vrmb{{\bf{v}}}

\def\Hbar{\bar{H}}
\def\D{\Delta}
\def\bcal{b}
\def\bbar{\mathbin{{b}\mkern-9.5mu^{{\mbox{\tiny{$-$}}}}\hspace{-0.00cm} }}
\def\nstat{\nu}
\def\nst{\nu}
\def\engbar{\bar{\eng}}
\def\engobar{\bar{\eng}_0}
\def\psias{\psi}
\def\Phimas{\Phim}
\def\fas{f}
\def\rbb{\as}

\def\La{L}
\def\Ja{J}
\def\as{p}
\def\ioii{{\mbox{\normalsize${\frac{1}{2}}$}}}
\def\Rb{{\bf{R}}}

\def\xb{{\bf x}}
\def\yb{{\bf y}}
\def\zb{{\bf z}}
\def\Xb{{\bf X}}

\def\ub{{\bf{u}}}
\def\hatu{\hat{u}_q}
\def\Nsub{{{\mbox{\tiny${N}$}}}}
\def\Pisub{{{\mbox{\tiny${\mit{\Pi}}$}}}}

\def\q{\bar{q}}
\def\xdot{\dot{x}}
\def\exc{{\rm ex}}
\def\ens{{ens}}

\def\Gcal{\Gast}



\def\Ltr{\mathscr{L}}
\def\Ltrb{\pmb{\Ltr}}
\def\Htr{\mathscr{H}}

\def\Ltr{\mathcal{L}}
\def\Ltrb{\pmb{\Ltr}}
\def\Htr{\mathcal{H}}
\def\Ttr{\mathcal{T}}
\def\Vtr{\mathcal{V}}

\def\gstru{\hspace{0.02cm}\g\hspace{-0.155cm}_{^-}}
\def\gbar{\hspace{0.02cm}\g\hspace{-0.155cm}_{^-}}

\def\gd{\underline{\g}}
\def\gu{\g}
\def\uu{\mathcal{U}}
\def\uub{\pmb{\mathcal{U}}}

\def\Cbar{\hspace{0.02cm}C\hspace{-0.365cm}^{{  \atop  -}\hspace{0.08cm}}{}   }

\def\Last{L}                                                                                                                                            
\def\Lcal{\Last}


\def\Lbcal{\mathbf{\Lcal}}
\def\Lcalb{\Lbcal}
\def\Lscr{\Lcal\hspace{-0.3cm}^{{  \atop  -} }{}   }
\def\Lbscr{\mathbf{\Lcal}\hspace{-0.32cm}^{{  \atop  -} }{} }
\def\Lscrb{\mathbf{\Lcal}\hspace{-0.31cm}^{{  \atop  -} }{} }

\def\Lscr{L}
\def\Lbscr{\mathbf{L}}
\def\Lscrb{\mathbf{L}}

\def\Scal{\Sast}
\def\Jcal{\Jast}
\def\Sbcal{\mathbf{\Scal}}
\def\Jbcal{\mathbf{\Jcal}}
\def\Hcal{\Hast}

\def\Kcal{{\mathcal{K}}}
\def\Xcal{{\mathcal{X}}}
\def\Tcal{\Tast}

\def\Ocal{\mathcal{O}}
\def\Rcal{\mathcal{R}}
\def\Ycal{\mathcal{Y}}

\def\Wvel{\Omegavel}
\def\Ncal{{\mathcal{N}}}
\def\Omegavel{\mathbin{{\mit\Omega}\mkern-13.mu^{_{\mbox{$-$}}}\hspace{-0.08cm}{}_d }}

\def\omegavel{{\w\mbox{\hspace{-0.38cm} \vspace{0.15cm}$-$\hspace{-0.02cm}}}}
\def\wvel{\omegavel_d}

\def\Ucal{\bar{\eng}_{0}}
\def\Omegavel{\mathbin{{\mit\Omega}\mkern-13.mu^{_{\mbox{$-$}}}\hspace{-0.08cm}{}_d }}
\def\Wvel{\Omegavel}

\def\q{\mathbin{q\mkern-11mu-}}
\def\PE{\mbox{\tiny{{\rm P.E.}}}}
\def\ME{\mbox{\tiny{{\rm M.E.}}}}
\def\QM{\mbox{\tiny{{\rm QM}}}}
\def\Psub{\mbox{\tiny{{\rm P}}}}
\def\Bsub{{\mbox{\tiny{{\rm B}}}}}
\def\TP{{\mbox{\tiny{{\rm T}}}}}

\def\SM{{\mbox{\tiny{{\rm SM}}}}}
\def\MT{{\mbox{\tiny{{\rm MT}}}}}

\def\ev{\epsilon}

\def\Ci{1}
\def\betamt{{\bf{b}}}
\def\kb{{\bar{k}}}
\def\kbf{{\bf{k}}}
\def\Kb{{\bf{K}}}
\def\cb{{\bf{c}}}

\def\pb{{\bar{p}}}
\def\pbf{{\bf{p}}}
\def\Pbf{{\bf{P}}}
\def\Mbf{{\bf{M}}}
\def\pbf{{\bf{p}}}
\def\Acal{\mathscr{A}}

\def\Bcal{\mathcal{B}}
\def\Bbcal{\pmb{\mathcal{B}}}

\def\Ecal{{\mathcal{E}}}
\def\Ebcal{\pmb{{\mathcal{E}}}}

\def\Fcal{{\mathcal{F}}}
\def\Fbcal{\pmb{{\mathcal{F}}}}

\def\Ccal{{\cal{C}}}
\def\Vp{V}
\def\Ccal{{\cal{C}}}
\def\p{{{}_{+\hspace{-0.1cm}}}}

\def\psipi{\psi_{\p}(1)}
\def\psipii{\psi_{\p}(2)}
\def\psimi{\psi_{\m}(1)}
\def\psimii{\psi_{\m}(2)}

\def\ai{\alpha(1)}
\def\aii{\alpha^{'}(2)}
\def\bi{\beta^{'}(1)}
\def\bii{\beta(2)}

\def\fa{f_r}
\def\fb{f_\ell}

\def\Ca{C_a}
\def\Cb{C_b}
\def\fbf{{\bf{f}}}
\def\Ocal{{\cal{O}}}
\def\psib{{\pmb{\psi}}}
\def\alphab{{\pmb{\alpha}}}
\def\sigmab{{\pmb{\sigma}}}
\def\sig{\sigma}
\def\Eb{{\bf E}}
\def\Bb{{\bf B}}
\def\ke{\kappa}
\def\nabb{{\pmb{\nabla}}}
\def\nablab{{\pmb{\nabla}}}
\def\vir{{\rm vir}}

\def\psitot{\psi}
\def\jb{{\bf{j}}}
\def\vel{\upsilon}

\def\vels{{\hspace{0.1cm}\breve{\hspace{-0.1cm}\vel}}}
\def\velsb{{\breve{\velb}}}
\def\vb{{\bf{v}}}
\def\Imtr{I}
\def\citeUnif{4?}
\def\App{}
\def\Qcal{{\mathcal{Q}}}
\def\Cross{Q}

\def\vphilim{f}
\def\ft{{\mathcal{B}}}
\def\vphibar{\mathbin{\varphi\mkern-12.5mu-}}
\def\vphi{\varphi}
\def\med{{\med}}
\def\Mcal{{\mathfrak{M}}}
\def\Sb{{\bf{S}}}
         \def\xia{{\mathcal{A}}}
\def\tha{\theta}

\def\nb{\bf{n}}

\def\ph{{ph}}
\def\phiv{\varphi}
\def\Lb{{\bf{L}}}
\def\velsub{_{\vel}}
\def\Jb{{\bf{J}}}
\def\Pb{{\bf{P}}}
\def\Mb{{\bf{M}}}
\def\Zo{{Z^0}}
\def\nablab{{\pmb{\nabla}}}
\def\velb{{\pmb{\vel}}}
\def\velbecmast{\velb_e^\astn}

\def\Db{{\bf{D}}}

\def\Ab{{\bf{A}}_a}
\def\Abb{{\bf{A}}}

\def\Thm{\vartheta}
\def\Thetam{{\mit{\Theta}}}
\def\Thetamb{ \pmb{{\mit{\Theta}}}}

\def\lb{{\bf l}}
\def\ldb{{\pmb{\ld}}}
\def\ld{\ell}
\def\ellb{{\pmb{\ell}}}
\def\vb{\velb_{t}{}}
\def\pd{\partial}
\def\vphi{\varphi}

\def\psitot{\varphi}
\def\psiR{\widetilde{\psi}}
\def\psiL{\widetilde{\psi}^{{\rm vir}}}
\def\Phim{{\mit{\Phi}}}
\def\PhimR{\widetilde{ {\mit \Phi}}}
\def\PsimR{\widetilde{ {\mit \Psi}}}
\def\PsimL{{\widetilde{ {\mit \Psi}}}^{{\rm vir}}}
\def\a{\alpha}
\def\uav{\bar{u}}
\def\D{\Delta}
\def\th{\theta}
\def\r{{\mbox{\tiny${R}$}}}
\def\re{{\mbox{\tiny${R}$}}}
\def\Fmed{F_{{\rm a.med}}}
\def\med{{\rm med}}
\def\Lw{L_{\varphi}}
\def\Fb{{\bf{F}}}

\def\Efb{{\bf{E}}}
\def\Bfb{{\bf{B}}}
\def\Bf{B}
\def\Ac{ \varphi}
\def\Xsub{{\mbox{\tiny${X}$}}}
\def\Ysub{{\mbox{\tiny${Y}$}}}
\def\Zsub{{\mbox{\tiny${Z}$}}}
\def\MTsub{{}}

\def\Ksub{{\mbox{\tiny${K}$}}}
\def\W{{\mit \Omega}}
\def\Wd{\W_d{}}
\def\Nu{{\cal V}}
\def\Nud{\Nu_d{}}
\def\Eng{{\cal E}}
\def\eng{{\varepsilon}}
\def\vep{\varepsilon}
\def\Kmscr{{\mathscr{K}}}
           \def\engk{\Kcal}
\def\Acuni{\Ac_{{\Ksub}^\dagsup}^{\dagsup}}
\def\unduni{\Ac_{{\Ksub}^\dagger}^{\dagsup}}
\def\Acauni{\Ac_{{\Ksub}^\ddagsup}^{\ddagsup}}
\def\Acunim{{\Ac_{{\Ksub}^\dagsup}^{\dagsup *}}}
\def\undunim{{\Ac_{{\Ksub}^\dagsup}^{\dagsup}}^*}
\def\Acaunim{{\Ac_{{\Ksub}^\ddagsup}^{\ddagsup *}}}
\def\pd{\partial}
\def\Ad{ {\mit \psi}}
\def\psim{ {\mit \psi}}
\def\Kd{K_d{}}
\def\Lam{{\mit \Lambda}}
\def\lam{\lambda}

\def\dagsup{\m}
\def\ddagsup{{\mbox{\tiny{+}}}}

\def\dagsupa{{\mbox{\tiny{$\dagger$}}}}

\def\psimKdK{\psim_{\Ksub,\Kdsub}}
\def\wk{1}

\def\w{\omega{}}
\def\wrm{{\rm{w}}}
\def\wit{\w_{t}{}}
\def\witb{{\pmb{\it{w}}}}

\def\wdlow{\omega_d }
\def\g{\gamma{}} 
\def\Phimc{{\mathcal C}}
\def\Psim{{\mit \Psi}}
\def\arm{{\rm a}}
\def\brm{{\rm b}}
\def\crm{{\rm c}}
\def\drm{{\rm d}}
\def\erm{{\rm e}}
\def\frm{{\rm f}}
\def\grm{{\rm g}}
\def\hrm{{\rm h}}
\def\lf{\left}
\def\rt{\right}
\def\Kdsub{{\mbox{\tiny${K_d}$}}}
\def\psimkd{\psim_{\kdsub}}
\def\psimKd{\psim_{\Kdsub}}
\def\hquad{ \ \ } 
\def\Taum{{\mit \Gamma}}

\def\mrm{{\mathbin{\mu\mkern-5.6mu\mbox{-}}}}

\begin{abstract}
\def\mrm{{\mathbin{\mu\mkern-5.6mu\mbox{-}}}}
\def\wsub{{\mbox{\tiny{$W$}}}}


A microscopic theory of the neutron, which consists in a neutron model constructed using key relevant experimental observations as input information and the first principles solutions for the basic properties of the model neutron,  is proposed within a framework consistent with the Standard Model. The neutron is composed of an electron $e$ and a proton $p$ that are separated at a distance $r_1$ of the order $10^{-18} $ m, and are in relative orbital angular motion and Thomas precession highly relativistically, with their reduced mass moving along a quantised  circular orbit $l=1, j=\frac{1}{2}$ of  radius vector $\rb_{1 \hf}=r_1 \hat{r}_{1 \hf}$ about their mass centre.
The associated rotational energy flux
has a spin 
$\frac{1}{2}$ and 
resembles a confined antineutrino. The particles $e,p$ are attracted with one another predominantly by a central magnetic force produced as result of the particles' relative precessional-orbital and intrinsic angular motions. The interaction force (resembling the weak force), potential (resembling the Higgs' field), and a corresponding excitation Hamiltonian ($\Hast_I$), among others, are derived based directly on first principles laws of electromagnetism, quantum mechanics and relativistic mechanics within a unified framework. In particular, the equation for $\frac{4}{3}\pi r_1^3 \Hast_I $, which is directly comparable with the Fermi constant  $G_\Fsub$, is predicted as $G_\Fsub=\frac{4}{3}\pi r_1^3 \Hast_I =A_o C_{0\hf} /\g_e^\astn \g_p^\astn$, where $A_o=e^2 \hbar^2/12\pi\ev_0 m_e^\osup m_p^\osup c^2$, $m_e^\osup, m_p^\osup$ are the $e,p$  rest masses, $C_{0\hf}$  is a geo-magnetic factor, and  $\g_e^\astn, \g_p^\astn$ are the Lorentz factors. Quantitative solution for a stationary meta-stable neutron is found to exist at the extremal point $r_{1m}=2.537 \times 10^{-18} $ m, at which the $G_\Fsub$ is a minimum (whence the neutron lifetime is a maximum) and is equal to the experimental value. Solutions for the magnetic moment, effective spin ($\frac{1}{2}$), fine structure constant, and intermediate vector boson masses of the neutron are also given in this paper.

\end{abstract}

\setcounter{section}{0}
\section{Introduction}\label{Sec-Intro}

The neutron is a  building particle of matter, as the proton and electron are. The neutron distinguishes yet from the proton and electron prominently in its undergoing weak decay with a notable non-conservative parity.  In inverse proportion to the weak interaction strength 
represented by the Fermi constant $G_\Fsub$, the lifetime of a free neutron is of a finite 12 minutes only. The Fermi constant $G_\Fsub$ combined with  the Heisenberg relation indicates moreover a weak interaction distance of an order $10^{-18} m$. Weak decay is  
a common property of all of the other several hundred  elementary matter particles observed in the laboratory except for  the proton and electron, by virtue of which process these particles are unstable, short lived. The basic properties of the weak processes, foremost the neutron $\beta$ decay, have been experimentally  studied extensively over the past eight decades or so, and summarised under the Standard Model for elementary particles \cite{citePerkins1982Griffithsetal}. Theoretically, the weak decay of neutron and other particles has been accounted  satisfactorily for, most notably  in quantitative prediction of the branching ratio, by the unified renormalisable theories of weak interaction. The Glashow-Weinberg-Salam (GWS) electroweak theory [\citeWeinberg1967etal{}a-c] based on group $SU(2) \times U(1)$ is one of these. This theory in particular also predicts the charged and neutral intermediate  vector bosons $W^\m$,$W^+$ and  $Z^\osup$  which were confirmed by the experiments at CERN;  its renormalisability was proven by  t'Hooft in 1971 [\citeWeinberg1967etal{}d].
      %
All of the current field theories of the neutron are essentially focused with the neutron $\beta$ decay, and are rested on the original hypothesis of Fermi[\citeWeinberg1967etal{}e].  
Namely that, in a neutron ($n$) $\beta$ decay reaction  $n \rightarrow p + e + \nubar_e$,            
the matter particles proton $p$ and electron $e$,   
and the antineutrino $ \bar{\nu}_e$ do not exist until the neutron $n$ decays. And upon neutron decay, these particles are envisaged as simply emitted by the neutron (as a point entity) in an analogous way to an accelerated point charge emitting electromagnetic radiation. The current theory of the neutron remains as a phenomenological one. There remain certain outstanding questions yet to be resolved.  In particular, the nature and origin of the weak interaction force are not yet well understood, an equation of the weak force accordingly is yet to be derived, and the Fermi constant $G_\Fsub$ is yet to be derived based on the interaction force. At a similar significant level, the nature and the origins of the (anti)neutrino, the intermediate vector bosons, the Weinberg weak mixing angle, and the Higgs mass are not yet fully well understood. One common feature suggestive of the nature of the weak phenomena however is readily recognisable directly from observations, namely that the weak phenomena  present  with (precede) only the electrons and protons emitted from the baryon ($n$, $\Lam$, etc) and meson ($\pi$, $K$, etc.) disintegrations or conversely (succeed) ones upon the productions of the $n$, $\Lam$,  $\pi$, $K$, etc., but not with the same electrons and protons in free-particle or  bound atomic processes. Weak phenomenon has thus to do with the internal structure of the weak emitting particles.
For a more comprehensive understanding of the nature of the weak phenomena, a microscopic theory would be indispensable. The purpose of this paper is to develop a microscopic theory of the neutron, serving as a prototype of the weak interaction 
(meta-)stabilised systems, 
 based firstly on a realistic real-space model construction of the neutron,  
such that the fundamental weak force and the variety of weak-interaction related properties and phenomena can be predicted  based on first principles solutions within a unified  framework of electromagnetism, quantum mechanics, and relativistic mechanics.

\begin{figure}[h]     
\begin{center}
\includegraphics[width=0.95 \textwidth]{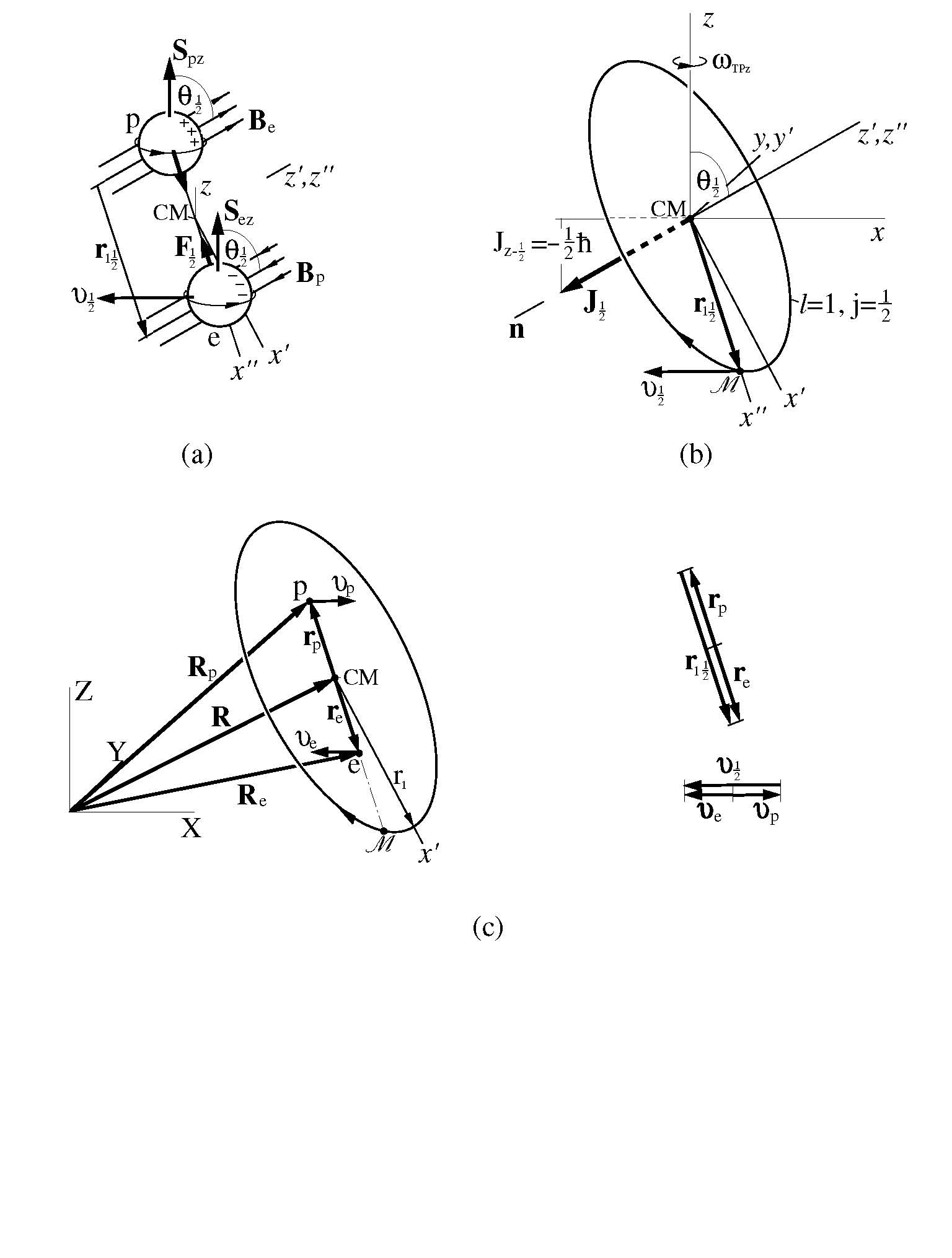}
\end{center}  
             \vspace{-0.49cm}
\vspace{-4.5 cm}
\caption{
Schematic of the model neutron composed of an electron $e$ and a proton $p$. 
(a) The $e,p$ are separated by a distance $\rb_{1 \hfp}= r_1 \hat{r}_{1 \hf}$
 and are in relative angular motion and a Thomas precession at  velocity $\velb_\hfp$ under a magnetic interaction force $\Fb_{\hfp}$ in the  magnetic fields $\Bb_p,\Bb_e$  of $p,e$ at $e,p$; their spins $\Sb_{ez},\Sb_{pz}$ (in units $\hbar$) are aligned parallel, in the $+z$ direction for the $m_j=-j$ magnetic state shown, and antiparallel to $\Jscr_{z\m\hf}$ of graph (b).
 (b) The reduced mass $\mr$ of $e,p$ moves at velocity $\velb_\hfp$ about the CM along a $l=1,j=\frac{1}{2}$ circular orbit of radius vector $\rb_{1 \hfp}$ and normal $\nb$ at angle $\pi-\theta_\hfp$ to the $z$ axis; it has a $z$-component angular momentum $\Jscr_{z\m\hf}$. 
(c) Left: The $e,p$ are located at positions $\rb_e,\rb_p$, moving at velocities $\velb_e,\velb_p$,  relative to the CM in the CM frame (coordinates $x,y,z$ in graph b), and at $\Rb_e,\Rb_p$ in the lab frame (coordinates $X,Y,Z$).
Right: vector relations between $\rb_e,\rb_p$ and $\rb_{1 \hf}$, and 
 $\velb_e,\velb_p$ and $\velb_\hf$.  The drawings are made for $m_e \simeq m_p$.
} \label{fig-neutron.eps} \label{fig-neutron1.eps}
  \vspace{-.1cm}
  \end{figure}
Using several key relevant experimental facts, in particular the neutron beta decay reaction equation $n \rightarrow p + e + \nubar_e$,  the neutron spin ($\frac{1}{2}$),  the order of magnitude of the Fermi constant $G_\Fsub$ and the so implied 
weak interaction distance $\sim 1 \cdot 10^{-18} $ m 
 as direct input information, we propose at the outset of the theory development a real-space ($e,p$-) neutron model as follows: The {\it neutron} is composed of an electron $e$ and a proton $p$ separated at a distance $r (=r_1)$ of an order $ 10^{-18}$ m; see Fig \ref{fig-neutron.eps}a. The $e,p$ are in relative orbital angular motion and in addition a Thomas precession at a velocity approaching the velocity of light $c$, under a central force of an electromagnetic origin. The central force is in effect predominantly an attractive magnetic force produced  by the magnetic fields ($ \Bb_p,\Bb_e$) of $p,e$ at $e,p$ as result of their intrinsic spin and relative motions. The $z$-components ($\Scal_{ez}$, $\Scal_{pz}$) of the $e,p$ spin angular momenta are aligned parallel to each other and antiparallel to that of their relative motion ($\Jscr_{z\m\hf}$, Figs \ref{fig-neutron1.eps}b), so that the magnetic interaction force is maximally attractive. The $e,p$ relative motion is in such a way that  their reduced mass ($\mr$) moves at a velocity ($\velb_\hfp $) accordingly approaching  $c$ along a (quantised $l=1,j=\frac{1}{2}$) circular orbit of radius $r(=r_1)$ about their (the $e,p$) common centre of mass (CM), with a normal  ($\nb$) at a precession-modified quantised angle ($\pi-\theta_\hfp $ for $j$ spin down state) to the $z$ axis; see Fig \ref{fig-neutron.eps}b. The relative precessional--orbital angular momentum projected in $z$ direction ($\Jscr_{z\m\hf}$) will show to be a negative half-integer quantum $\Jscr_{z\m\hf}=-\frac{1}{2}\hbar$. The corresponding neutral rotational energy flux, or vortex, along the $l=1, j=\frac{1}{2}$ circular orbit, of accordingly a $z$-component angular momentum $\Jscr_{z\m\hf}$, resembles a "confined antineutrino" ($\nubar_e$).

It is commented that, the proposed $e,p$-neutron model suggests also a scheme for the strong force similarly on a unified basis with electromagnetism: A proton $p$ would  be attracted with a neutron  $n(e,p)$ (mainly) through an electrostatic attraction with the electron $e$ of the neutron at short range; in the same order of the short-range electrostatic interaction, two protons will repel, but never attract with one another. Such characteristics are in accordance with the observational fact that no nucleus exists which is made of more than one protons and  protons only without neutrons. The author's more recent research (internal work) has further shown that a microscopic representation of the muon and the "muon-emitting" composite elementary particles may be achieved within a consistent scheme with the neutron model. The system of the so-represented elementary particles furthermore is in conformity with the quark model, in a manner  that the (internal) spin states of the (composite) elementary particles are in one-to-one correspondences with the configurations of the (observationally-never-isolatable) quarks. The internal spin states of the model neutron under reversed signs (which represents  the neutron effective spin, see Sec \ref{Sec-mag-mmt}), $-\frac{1}{2} ,-\frac{1}{2}, \frac{1}{2}$ (i.e.  "down, down, up"), for example, directly correspond with the $ddu$ quarks.   The (free) proton, as another example, is a non-composite particle with only one spin state, assigned as $+\frac{1}{2}$ (spin up) by convention. But  this may be translated into a  systematic  three-spin states representation as $\frac{1}{2},\frac{1}{2},-\frac{1}{2}$ (i.e. "up, up, down"), by adding  two dummy spins $\frac{1}{2},-\frac{1}{2}$ without changing the original spin $\frac{1}{2}$; the three spin states correspond directly to the $uud$ quarks.

The remainder of this paper gives a first-principles mathematical representation of the  model neutron, mainly in respect to the internal relativistic kinematics, dynamics  (Secs \ref{Sec2.1.Eq-Mot}), magnetic structure (Sec \ref{Sec-mag-mmt}), and a derivation of the internal $e,p$ interaction force (Sec \ref{Sec-force}) of the neutron in stationary state, the dynamics upon the neutron $\beta$ decay (Secs \ref{Sec-excit}), and a quantitative evaluation of the dynamical variables (Sec \ref{Sec-num}). The (quantitative) predictions of the basic  properties of the neutron are subsequently subjected to comparisons with, or constraints by, the available experimental data where in question, so that critical checks and controls of the viability of the neutron model are made as far as possible. Other basic aspects, including the parity associated with the $\beta$ decay, a direct derivation of the intermediate vector boson masses and Weinberg mixing angle of the neutron, and a corresponding dynamic scheme for the other (composite) elementary particles participating weak interaction, will be elucidated in separate papers.

\section{Equations of motion. Coordinate transformations. Solutions }\label{Sec-eom} \label{Sec2.1.Eq-Mot}

{\it \quad \ref{Sec2.1.Eq-Mot}.1. Transformed Newtonian equations of motion of the mean and instantaneous positions of $e,p$. Representation in $\rb,\Rb$ coordinates }
Consider that an electron $e$ and a proton $p$ comprising a neutron are at time $t^\astn$ located with the probability densities $|\psi_\a(\rrb_\a^{\astn},t^\astn)|^2$  ($\a=e,p$) at positions $\rrb_e^{\astn \pp}, \rrb_p^{\astn \pp}$ 
relative to coordinates $X,Y,Z$ fixed in the laboratory (lab) frame; see  Fig \ref{fig-neutron.eps}c. (The usual statistical point-particle picture suffices and is referred to here.)
 The $e,p$ are in relative motion at a velocity to prove high compared to $c$  (Sec \ref{Sec-num}) under a mutual interaction force $\Fb^{\astn}{}^\pp$ and gravity $\gb^\astn{}^\pp$; no applied  force presents. Their mean positions, $\rbav_\a^\astn{}^\pp  =\int \rrb_\a^\astn{}^\pp  | \psi_\a |^2 d^3 \rr_\a^{\astn }$, evolve according to the transformed Newtonian  equations of motion, $ \frac{ d(m_\a^\astn{}^\pp  (d \rrbav_\a^\astn{}^\pp /dt^\astn) ) }{dt^\astn} = \int (   m_\a^\astn{}^\pp  \gb \pm \Fb^\astn{}^\pp )| \psi_\a |^2 d^3 \rr_\a^{\astn }$ (the correspondence principle), where $m_e,m_p$ are the $e,p$ masses. 
The $e,p$ are assumed to form a bound stationary system until Sec \ref{Sec-excit} and hence feasibly move circularly at constant (tangential) velocities ($\ub_\a^\astn{}^\pp  =d \rrb_\a^\astn{}^\pp /dt^\astn$). The equations of motion thus reduce to 
$$\displaylines{
\refstepcounter{equation} \label{eq-EqMt} \label{eq-EqMt2}   
\hfill  
m_e^\astn{}^\pp  \frac{d^2 \rrb_e^\astn{}^\pp }{d t^{\astn 2}}
=  m_e^\astn{}^\pp  \gb^\astn{}^\pp   +\FbIL^\astn{}^\pp, 
\quad 
m_p^\astn{}^\pp   \frac{d^2 \rrb_p^\astn{}^\pp }{d t^{\astn 2}}
=  m_p^\astn{}^\pp  \gb^\astn{}^\pp   -\FbIL^\astn{}^\pp.
\quad
{\rm Or} \ \
\mcm^\astn{}^\pp \frac{d^2\Rb^\astn{}^\pp}{d t^{\astn 2}}=\mcm^\astn{}^\pp \gb^\astn{}^\pp, \quad
\mr^\ast  \frac{d^2 \rb^\ast}{dt^{\astn 2}}= \FbIL^\astn{}^\pp,
\hfill
(\ref{eq-EqMt})
\cr
{\rm where} 
\hfill
\cr
\hfill \Rb^\astn{}^\pp =\frac{m_e^\astn{}^\pp  \rrb_e^\astn{}^\pp + m_p^\astn{}^\pp  \rrb_p^\astn{}^\pp }{ \mcm^\astn{}^\pp }, 
\ \ M^\astn{}^\pp=m_e^\astn{}^\pp +m_p^\astn{}^\pp, 
\ \ 
\rb^\ast{} =\rrb_e^\astn{}^\pp  -\rrb_p^\astn{}^\pp=\rpb_e^{\astn}{}^\pp  -\rpb_p^{\astn}{}^\pp, \ \
\mr^\ast=\lf(\frac{1}{m_e}+\frac{1}{m_p}\rt)^{-1}
=\frac{m_e^\astn{}^\pp  m_p^\astn{}^\pp }{M^\astn{}^\pp};
\hfill
\cr
\refstepcounter{equation} \label{eq-EqMt3}\label{eq-EqMt3-b}
\hfill
\rrb_e^\astn{}^\pp=\Rb^\astn{}^\pp+\frac{m_p^\astn{}^\pp  }{\mcm^\astn{}^\pp  } \rb^\ast,  
\quad
\rrb_p^\astn{}^\pp  =\Rb^\astn{}^\pp-\frac{m_e^\astn{}^\pp }{\mcm^\astn{}^\pp} \rb^\ast;
\quad
\rpb_e^{\astn }{}^\pp =\rrb_e^\astn{}^\pp - \Rb^\astn{}^\pp= \frac{m_p^\astn{}^\pp }{\mcm^\astn{}^\pp} \rb^\ast,  
\quad
\rpb_p^{\astn }{}^\pp =\rrb_p^\astn{}^\pp  -\Rb^\astn{}^\pp
=-\frac{m_e^\astn {}^\pp  }{\mcm^\astn{}^\pp} \rb^\ast.
\hfill (\ref{eq-EqMt3})  
}$$
$\Rb^\astn{}^\pp$ is the position of the centre of mass, CM;  $\mcm$ is the total mass located at $\Rb$; $\rb$ is the relative position; $\mr$ is the reduced mass (of a fictitious particle) located at $\rb$; and $\rpb_e^{\astn }{}^\pp,\rpb_p^{\astn }{}^\pp$ are the $e,p$ positions relative to $\rrb^\astn$. Eqs (\ref{eq-EqMt}c,d) are given for the masses $\mcm^\astn{}^\pp$ and $\mr$ travelling accordingly circularly at constant velocities 
($\ub_\cmsublab = d \Rb /dt$ relative to the lab frame and $\velb$ about the CM). 
A common time $t$ measured by a clock fixed at the CM has been used in order to facilitate the direct  transformation of Eqs (\ref{eq-EqMt}a,b) to (\ref{eq-EqMt}c,d). 
Corresponding directly with the dynamic effect of $d^2\Rb_\a/dt^2$ on the left of Eqs (\ref{eq-EqMt}a,b), this  $t$ enters as an independent variable of $\gb,\Fb$: $\gb=\gb(t)$, $\Fb=\Fb(t)$; 
 the relativistic masses $m_e,m_p$ may remain as (implicit) functions of the local times ($t_e,t_p$) at $\rb_e,\rb_p$ (Sec  \ref{Sec2.1.Eq-Mot}.2) in so far as the same masses are used through the equations. For the $e,p$ relative motions internal of a neutron are of major concern in this paper, unless specified otherwise 
we shall work in the CM frame, i.e. immediately in terms of the relative positions $\rb_e,\rb_p, \rb$ measured with respect to a set of relative coordinate axes  $x,y,z$ parallel with the $X,Y,Z$ axes, and with an origin fixed at the CM (cf Fig \ref{fig-neutron1.eps}b). This in more general terms means that  we shall work with the unsuperscripted variables, including $\Rb, M, \Rb_e, m_e, t$ etc,
which we hereafter reserve to explicitly refer to ones measured in the CM frame. We shall refer to their counterparts for example measured in the lab frame  by $\Rb^\Lab, M^\Lab, \Rb_e^\Lab, m_e^\Lab, t^\Lab$, etc. where in question.

The partial--relative and relative velocities of the $e,p$, and the corresponding           
 rotational angular momenta  in the CM frame, in terms of the time $t^\astn$,  follow as  
$$\displaylines{ 
 \refstepcounter{equation} \label{eq-vs}
\hfill
\velb_e^\astn
=\frac{d\rbp_e^\astn}{dt^\astn}
=  \frac{m_p^\astn}{\mcm^\astn} \velb,
\quad 
\velb_p^\astn
=\frac{d \rbp_p^\astn }{d t^\astn}
= -\frac{m_e^\astn}{\mcm^\astn} \velb,
\quad 
\velb=\frac{d \rb}{dt}=\velb_e^\astn-\velb^\astn_p;
\hfill (\ref{eq-vs})
\cr
 \refstepcounter{equation} \label{eq-Lddg}
\hfill
\Jb_{e}
= \rbp_e^\astn   \times (m_e^\astn  \velb_{e}^\astn ) 
= \frac{m_p^\astn }{\mcm^\astn}\Jb^\astn,  
\quad
\Jb_{p}
=\rbp_p^\astn   \times (m_p^\astn   \velb_{p}^\astn  ) 
=\frac{ m_e^\astn  }{\mcm^\astn } \Jb^\astn,  
\quad
\Jb^\astn
=\Jb_{e}+\Jb_{p}
=\rb\times (\mr \velb) 
\hfill (\ref{eq-Lddg})
}$$
From the relations (\ref{eq-EqMt3-b}g,h)  between the distances $\rb_e,\rb_p$ of $e,p$  to the CM, and $\rb$ of $e$ to $p$ it follows that, by virtue how time  in essence is defined, the local times $t_e^\astn,t_p^\astn$ and the time $t$ for light to traverse the distances $\rb_e, \rb_p, \rb$ at a constant velocity $c$ are related as  $t_e^\astn = (m_p^\astn /\mcm^\astn) t^\astn$, $t_p^\astn = (m_e^\astn/\mcm^\astn) t^\astn$.
The partial-relative velocities in terms of $t_e^\astn,t_p^\astn$ are
$$\displaylines{
\refstepcounter{equation} \label{eq-velsp}
\hfill
\velb_e^{\astn \prime}=d \rbp_e^\astn /d t_e^\astn  =\velb, \quad 
\velb_p^{\astn \prime}=d \rbp_p^\astn /d t_p^\astn  =-\velb. \quad
\hfill(\ref{eq-velsp})
}$$
Denote $f_t(e)=\frac{t}{t_e}= \frac{|d \rb_e /d t_e| }{ |d \rb_e/d t| }=\frac{\vel_e'}{\vel_e } $. 
Substituting $ t= f_t(e) t_e $ in (\ref{eq-EqMt}a), setting $\Rb=0$, we have $m_e \frac{d^2 \rb_e}{d (f_t^2(e) t^2_e) } =m_e \gb(t) + \Fb(t)$, or $m_e \frac{d^2 \rb_e}{d t^2_e } =m_e f_t^2(e)\gb(t) + f_t^2(e) \Fb(t) = m_e \gb (t_e)+ \Fb(t_e)$, recovering the original form of (\ref{eq-EqMt}a)  expressed by its local time $t_e$ provided $ \Fb(t_e)=f_t^2(e) \Fb(t)$, $\gb (t_e)=f_t^2(e)\gb(t)$. Similarly  a factor $f_t(p)= \frac{t}{t_p}=\frac{\vel_p'}{\vel_p}$ will project (\ref{eq-EqMt}b)  to its original form expressed in $t_p$. The same projection factors, in the form of geometric mean $f_t^2= (f_t^2(e)  f_t^2(p))^{1/2}$, will be obtained  through direct derivation of the magnetic force  in Sec \ref{Sec-force}.

{\it \ref{Sec2.1.Eq-Mot}.2. Lorentz-Einstein transformations }
The instantaneous rest frame fixed to each rotating particle, $e$, $p$, $\mr$ or $ \mcm$, may be regarded as an inertial frame for each differential rotation which is essentially linear. (For a complete macroscopic rotation, non-inertial frame effects present and will be included separately, see Eqs (\ref{eq-Tinv}) vs (\ref{eq-Tinvp}) below and in turn Sec \ref{Sec2.1.Eq-Mot}.4). Subsequently, the differentials of the space and time coordinates $\rb_e$, $t_e$; $\rb_p$, $t_p$; $\rb$, $t$; $\Rb$, $\bar{t}_{ep}$ in the CM frame, and their counterparts $\rb_e^\osup$, $t_e^\osup$; $\rb_p^\osup$, $t_p^\osup$; $\rb^\osup$, $t^\osup$; $\Rb^\osup$, $\bar{t}_{ep}^\osup$ in the respective (instantaneous) rest frames are related by the Lorentz-Einstein transformations, 
$$\displaylines{
\refstepcounter{equation} \label{eq-Eq-transf-Ax1.1} \label{eq-Eq-transf-Ax1} 
\label{eq-Eq-transf-Ax2} 
\hfill
\g_e (d\rb_e -\velb_e'  dt_e)= d\rb_e^\osup,
\ \ 
\g_e(dt_e - \frac{ \velb_e' \cdot d\rb_e}{ {c'}^2 }) =dt_e^\osup;
\ \
\g_p (d\rb_p -\velb_p'  dt_p)= d\rb_p^\osup,
\ \
\g_p(dt_p -  \frac{\velb_p' \cdot d\rb_p}{ {c'}^2 }) =dt_p^\osup;
\hfill
\cr
\hfill
\g(d\rb-\velb dt^\astn)= d\rb^\osup,
\ \
\g(dt^\astn -  \frac{\velb\cdot d\rb}{ c^{_\astn 2}} ) =dt^\osup;
\ \ 
\g_\Msub (d \Rb- \velb_\Msub d \bar{t}_{ep})=d \Rb^\osup,  
\ \
\g_\Msub(d \bar{t}_{ep} -  \frac{\velb_\Msub \cdot d\Rb }{ {c'}^2 }) =d \bar{t}_{ep}^\osup
\hfill (\ref{eq-Eq-transf-Ax2})
}$$
where $\g_\a=(1- {\vel_\a'}^2/c'{}^{2})^{-1/2}$ ($\a=e,p$), $\g=(1- \vel^2/c^{ 2})^{-1/2}$, $\g_\Msub= (1- \vel_{\Msub}^{2}/c^{ 2})^{-1/2}$; 
$c' = d \rb_{\a_{pht}} / d t_\a =c=d \rb_{pht} / d t^\astn$ is the light speed measured in the CM frame; $\g_\Msub, \g$ are the (effective) Lorentz factors of the fictitious particles of masses $\mcm, \mr$ moving effectively at the  velocities  $\velb_\Msub, \velb$, 
such that their dynamical consequence is the same as  that due to the motions of $m_e,m_p$ relative to the CM. In particular, $\velb_\Msub$ needs be  thought of as the speed of the CM relative to the $e,p$, i.e. 
$\velb_\Msub=\frac{d \Rb}{d \bar{t}_{ep} }$ given in terms of  a mean local time $ \bar{t}_{ep}$ of $e,p$; the CM is not moving relative to itself. 

Transformations from the scalar distances $\rp_e^\astn, \rp_p^\astn, R^\astn,r$ to  $\rp_e^\osup, \rp_p^\osup, R^\osup,r^\osup$ at fixed $t^\astn$ (hence $t_e,t_p,\bar{t}_{ep}$), from the time $t^\astn$ to $ t^\osup$ at fixed $\rb$, and from the CM-frame masses $m_e^\astn, m_p^\astn, \mcm^\astn, \mr^\astn$ to their respective rest-frame counterparts $m_e^\osup, m_p^\osup, \mcm^\osup (=m_e^\osup+m_p^\osup), \mr^\osup(=m_e^\osup m_p^\osup/\mcm^\osup)$ 
follow as 
$$\displaylines{ 
\refstepcounter{equation} \label{eq-Eqtrans-xB1x} \label{eq-Sca-trans}
\hfill
\g_e^\astn{}^\pp \rp_e^\astn{}^\pp=\rp_e^\osup, 
\quad  \g_p^\astn{}^\pp \rp_p^\astn{}^\pp=\rp_p^\osup, 
\quad \g_\Msub^\astn  R^\astn= R^\osup, 
\quad \g r=r^\osup; 
\quad \g t^\astn =t^\osup; 
\hfill 
(\ref{eq-Sca-trans}.1)
\cr
\hfill
 m_e^\astn{}^\pp=\g_e^\astn{}^\pp m_e^\osup, 
\quad  m_p^\astn{}^\pp=\g_p^\astn{}^\pp m_p^\osup, 
\quad
 \mcm^{\astn \pp} =m_e+m_p= \g_\Msub^\astn \mcm^\osup, 
\quad   \mr=\g \mr^\osup.
\quad
\hfill 
(\ref{eq-Sca-trans}.2)
}$$
Using Eqs  (\ref{eq-Sca-trans}.2) for $m_e^\astn, m_p^\astn, \mcm^\astn, \mr$ in (\ref{eq-EqMt3}b),(d) gives (\ref{eq-EqMt3-c-Y1}), and solving gives (\ref{eq-EqMt3-c-Y}) below: 
$$\displaylines{
\refstepcounter{equation} \label{eq-EqMt3-c-Y1} 
\hfill
\g_\Msub^{\astn \pp}M^\osup
=\g_e^\astn{}^\pp m_e^\osup+\g_p^\astn{}^\pp m_p^\osup,
\quad
\g_\Msub^{\astn \pp} \g
= \g_e^\astn{}^\pp \g_p^\astn{}^\pp; 
\quad {\rm or}
\quad
\mcm^\osup=m_e^\dagsupa+ m_p^\dagsupa,
\quad {\rm where} 
\hfill(\ref{eq-EqMt3-c-Y1}.1)
\cr
\hfill
m_e^\dagsupa=\frac{m_e^\astn }{ \g_\Msub^\astn}=\g_e^\dagsupa m_e^\osup,
\ \
m_p^\dagsupa
=\frac{m_p^\astn}{\g_\Msub^\astn}
=\g_p^\dagsupa m_p^\osup,
\quad
\g_e^\dagsupa =\frac{\g_e^\astn{}^\pp}{\g_\Msub^\astn},
\ \ 
\g_p^\dagsupa =\frac{\g_p^\astn{}^\pp}{\g_\Msub^\astn}; 
\quad
\g_e^\dagsupa \g_p^\dagsupa
=\frac{\g_e^\astn \g_p^\astn }{\g_\Msub^\astn{}^2}
=\frac{\g }{\g_\Msub^\astn}
= \g^\dagsupa 
\hfill (\ref{eq-EqMt3-c-Y1}.2)
\cr
\refstepcounter{equation} \label{eq-EqMt3-c-Y} 
\cr
\hfill
\g_e^\astn{}^\pp=  \frac{\g_\Msub^\astn (\mcm^\osup \pm \Gam)
}{2m_e^\osup}, 
\quad
\g_p^\astn{}^\pp= \frac{ \g_\Msub^\astn{}^\pp (\mcm^\osup \mp \Gam)
}{2m_p^\osup}, 
\quad
\Gam =\sqrt{ (\mcm^\osup)^2 -4 m_e^\osup m_p^\osup  \g^\dagsupa  }.
\hfill (\ref{eq-EqMt3-c-Y})
}$$
For (\ref{eq-EqMt3-c-Y}) to have real solutions requires $(\mcm^\osup)^2 -4 m_e^\osup m_p^\osup  \g^\dagsupa \ge 0$, or $\g^\dagsupa \le (\g^\dagsupa)_{\max}=(\mcm^\osup)^2 / 4 m_e^\osup m_p^\osup  =459.556,
$ where $\g^\dagsupa=(\g^\dagsupa)_{\max} $ if $\Gam=0$, in which case $\g_e^\astn= \g_\Msub^\astn \mcm^\osup/2m_e^\osup$, $\g_p^\astn= \g_\Msub^\astn \mcm^\osup/2 m_p^\osup \simeq \frac{1 }{2}\g _\Msub^\astn$, $m_e=m_p$. In general  $m_e$ and $m_p$ may  not be equal. Let $m_e=k m_p $; this combined with  (\ref{eq-EqMt3-c-Y}a) gives $(m_e=)k m_p= \g_e m_e^\osup=\frac{1}{2} \g_\Msub(\mcm^\osup+\Gam) $. Dividing it by  (\ref{eq-EqMt3-c-Y}b) times $m_p^\osup$, i.e. $(m_p=)\g_p m_p^\osup= \frac{1}{2} \g_\Msub(\mcm^\osup-\Gam) $ gives  (\ref{eq-gamdag}a,b), and re-arranging (\ref{eq-EqMt3-c-Y}c) gives (\ref{eq-gamdag}c) below, 
$$\displaylines{
\refstepcounter{equation} \label{eq-gamdag}
 \hfill
k=\frac{\mcm^\osup+\Gam}{ \mcm^\osup-\Gam},
 \ \ {\rm or} \ \ 
\Gam=\frac{ (k-1)\mcm^\osup}{k+1}; 
\quad
  \g^\dagsupa = \frac{ (\mcm^\osup)^2-\Gam^2}{4 m_e^\osup m_p^\osup} 
=\frac{k (\mcm^\osup)^2}{(k+1)^2 m_e^\osup m_p^\osup}
\hfill (\ref{eq-gamdag})
}$$ 
Substituting  in these  $k=m_e/m_p=1.3165$ from the solution for neutron magnetic moment (Sec \ref{Sec-mag-mmt}) gives  $\Gam=\frac{(1.3165-1)}{1.3165+1}  938.78(3)= 128.26(5)$ GeV, and $\g^\dagsupa = 450.96(0)$.
Eqs (\ref{eq-EqMt3-b}g),(h) and  (\ref{eq-vs}a),(b) for this case become $\rb_e^\astn=\frac{ \rb}{k+1} =0.43 \rb $, $\rb_p=-\frac{k\rb}{k+1} =-0.57 \rb$; $\velb_e =0.43 \velb$, $\velb_p=-0.57 \velb$  (cf Fig. \ref{fig-neutron1.eps}c, right graph). $k {> \atop \sim } 1$ implies $\g_e,\g_p >>1$.

Multiplying $\frac{\g_e m_e}{\g_e+1}$ to the quadratics of Eq (\ref{eq-vs}a), and $\frac{\g_p m_p}{\g_p+1}$ to that of  (\ref{eq-vs}b), adding, we obtain on the left side the total kinetic energy $T_e+T_p$ of $e$,$p$ measured in the CM frame and in time $t$,
$$\displaylines{ 
 \refstepcounter{equation} \label{eq-Tinv}
\hfill
(T_e+T_p\equiv )
\frac{\g_e m_e \vel_e^2}{\g_e+1} +\frac{\g_p m_p \vel_p^2}{\g_p+1}
= \lf[ \frac{\g_e m_p}{(\g_e+1) \mcm} + \frac{\g_p m_e}{(\g_p+1) \mcm} \rt] \mr \vel^2 
(\equiv T)
\hfill (\ref{eq-Tinv}.1)
\cr
\hfill 
\mbox{for} \ \g_e,\g_p >>1:\quad 
m_e \vel_e^2 + m_p \vel_p^2  =  \mr \vel^2
\hfill (\ref{eq-Tinv}.2)
}$$
The right side of (\ref{eq-Tinv}.1) or (\ref{eq-Tinv}.2) expresses the kinetic energy $T$ of the reduced mass relative to the CM. Eq (\ref{eq-Tinv}.1) or  (\ref{eq-Tinv}.2) expresses invariance of kinetic energy under the  $\rb_e,\rb_p$ to $\rb,\Rb$ coordinate transformation as described in the CM frame and in time $t$. Performing similar operations to Eqs (\ref{eq-velsp}a,b) instead we obtain on the left side the total  kinetic energy $T_e'+T_p'$ of $e$,$p$ measured in the CM frame but in their local times $t_e,t_p$,
$$\displaylines{ 
 \refstepcounter{equation} \label{eq-Tinvp}
\hfill
(T_e'+T_p'\equiv )
\frac{\g_e m_e }{\g_e+1} \vel_e'{}^2 +\frac{\g_p m_p}{\g_p+1} \vel_p'{}^2
= \lf[ \frac{\g_e m_e}{(\g_e+1) \mcm } + \frac{\g_p m_p}{(\g_p+1) \mcm} \rt] \mcm \vel^2 
(\equiv T')
\hfill (\ref{eq-Tinvp}.1)
\cr
\hfill 
\mbox{for} \ \g_e,\g_p >>1:\quad 
m_e \vel_e'{}^2 + m_p \vel_p'{}^2  = \mcm \vel^2
\hfill (\ref{eq-Tinvp}.2)
}$$
The right side of (\ref{eq-Tinvp}.1) or (\ref{eq-Tinvp}.2) represents in effect the kinetic energy $T'$ of the total mass at the CM relative to the $e,p$ local space and time coordinates. Since $ \mcm>\mr$, so $T_e'+T_p' >T_e+T_p$.  The difference $(T_e'+T_p' )-(T_e+T_p)$ apparently represents a kinetic energy contribution from the non-inertial frame motion at $\rb_e,\rb_p$ relative to the CM.

Unless specified otherwise we shall hereafter suppose for simplicity the $e,p$ system as a whole, i.e. its CM, to be at rest in the lab frame. Provided further setting the coordinate origins of the CM and lab frames the same, hence $\Rb=0$, the relativistic effects in the two frames are the same.

{\it \ref{Sec2.1.Eq-Mot}.3. Total mass as measured in the lab frame. Neutron mass }
For the centre of mass CM of the $e,p$ system assumed at rest in the lab frame,  naturally an observer in the lab frame will measure a rest total mass  $\mcm^\Lab =\mcm^\osup =m_e^\osup+m_p^\osup $ of the model neutron. In more elaborate terms, a measurement of the neutron mass in the laboratory is typically made in a specified say $X$ direction over a macroscopic time interval $\D t $ which is $>>2\pi r/\vel$, the rotation period of $\mr$ (Secs \ref{Sec2.1.Eq-Mot}.4, \ref{Sec-num}). During $\D t$, 
$\velb_e,\vel_p$ explore
all directions each with a  zero  average projection in the $X$ direction. Hence the relativistic augments in the masses of $e,p$ as measured along the instantaneous directions $\velb_e,\vel_p$ in the CM frame do not enter the mass $\mcm^\Lab (=m_e^\Lab+ m_p^\Lab)$ measured in the lab frame. (This mass augment however evidently enters the interaction force or potential of Sec \ref{Sec-force}, which has a constant magnitude
 so as to manifestly effectuate a bound $e,p$ in stationary state irrespective of the direction of the $e,p$ separation.) 

A dually relevant example here is electron scattering by a target neutron. In respect to the internal dynamics of a target neutron, an incident electron $e$ travelling in a fixed direction is as a (moving) observer in the lab frame. The incident $e$  thus will see the rest (as contrasted to relativistic) masses of the $e,p$ of the neutron. Moreover, the $e,p$ of the neutron are fast rotating along circles of similar radii about their CM and thus about equally exposed to the incident $e$. So in terms of exposure frequency, the $e,p$ would equally probably scatter with the incident $e$,  through electro and magnetic potentials and naturally at their contracted radii $a_e,a_p$. The scattering potential from the proton $p$ of the neutron, on the other hand, would dominate  because of its much heavier rest mass, which for the electrostatic part at least is attractive. Incidentally, the experimentally measured electron-neutron  scattering length is negative and suggests an attractive scattering potential.

The (very large) $e,p$ interaction potential fields within the neutron, on the other hand, are liable to (considerably) modify the vacuum potential surrounding the $e,p$ charges; the effect would be particularly large given the $e,p$ separation  distance  $10^{-18}$ m (Sec \ref{Sec-num}) here is comparable with the inter-vacuuon distance based on the  "vacuuonic vacuum structure"$^a$.  This would consequently further modify the $e,p$ particles' rest masses,  in terms of the IED model$^b$, produced as their generating charges move through this modified vacuum. ($a,b$: see the author's earlier published work). The above gives a qualitative account for the (order of MeV) larger  neutron rest mass over the sum of the $e,p$ rest masses; this  difference is relatively small and is ignored where in question throughout this paper.

{\it \ref{Sec2.1.Eq-Mot}.4. Eigenvalue equations. Orbital and precessional angular momenta. Antineutrino }
In the absence of applied force and omitting the very weak gravity,  $\mcm^\astn$ is free and hence not directly subject to quantisation condition. We thus need only to establish the relativistic Schr\"odinger or Klein-Gordon equation (KGE) for the reduced mass $\mr$, in terms of the spherical polar coordinates $r,\vartheta, \phi$ transformed directly from $x, y, z$.
The KGE has the usual form $[ ( (E_{tot})_{op}  -V)^2- {\mr^\osup}^2c^4 - (p^2)_{op} c^2]\psi_{tot}=0$, where $(E_{tot})_{op}  -V= \mr c^2$;
 the associated non-inertial frame effect is not contained in it and will be included separately.
 Since the mass $\mr$ under consideration is moving at velocity exceedingly close to $c$ such that its rest-mass energy is negligibly small compared to its kinetic energy (Sec \ref{Sec-num}), more relevant here is the square-root (SQR) form of the KGE: $\Htr_{op} \psi =  \Htr \psi $, where $\Htr_{op} = \lf( (E_{tot})_{op} -V \rt) -\mr^\osup c^2+V = \Ttr_{op}+V$, and $\Ttr_{op}=\mr c^2-\mr^\osup c^2 =(\g-1) \mr^0 c^2
= \frac{(\g-1) \g (p^2)_{op}}{ \g^2 (\vel/c)^2  \mr  } 
=\frac{\g (p^2)_{op}}{(\g+1)\mr}$ (with $\g^2 (\frac{\vel}{c})^2=\g^2-1$) are the 
  Hamiltonian and kinetic energy operators associated with the kinetic motion of $\mr$; 
$(p^2)_{op} = (p_r^2)_{op} + \frac{(\Jtr^2)_{op}}{  r^2} $; $(p_r^2)_{op}$ and $(\Jtr^2)_{op} $  are the squared radial and orbital angular momentum operators.  For the $e$,$p$ interaction potential $\Vast $ being  central (Sec \ref{Sec-force}), hence $\Vast (\rb)=\Vast(r)$, the wave function of $\mr$, $\psi(r^\ast, \vartheta^\ast,\phi^\ast)$, may be written as $\psi^\ast = \Rcal^\ast (r^\ast) \Ycal_{}^\ast(\vartheta^\ast,\phi^\ast)$. And  the SQR-KGE separates into two eigenvalue equations,
$$\displaylines{
\refstepcounter{equation} \label{eq-Sch1-R}
\hfill
\lf[-\frac{\g \hbar^2}{(\g+1)\mr r^2}\frac{\pd }{\pd r}(r^2 \frac{\pd}{\pd r})
+ \frac{\g l(l+1) \hbar^2}{(\g+1) \mr r^2} +V(r)
\rt]\Rcal(r) =\Htr \Rcal(r) \hfill (\ref{eq-Sch1-R})
\cr
\refstepcounter{equation} \label{eq-Sch1-b}  \label{eq-eignev-eqL2}
\hfill
(\Jtr^\ast{}^2)_{op} \Ycal^\ast{}(\vartheta^\ast{},\phi^\ast{})
= \Jtr^\ast{}^2  \Ycal^\ast{}(\vartheta^\ast{},\phi^\ast{}),
\quad
(\Jtr^\ast{}^2)_{op} =-\hbar^2 \lf(\frac{\pd^2}{\pd \vartheta^2}+\cot \vartheta \frac{\pd }{\pd \vartheta}+\frac{1}{\sin^2\vartheta }\frac{\pd^2}{ \pd \phi^2}\rt). 
\hfill (\ref{eq-eignev-eqL2})
}$$
(\ref{eq-eignev-eqL2}) may be solved without $\Vast^\ast(r)$ being explicitly known. The eigen functions  are the spherical harmonics, $\Ycal_\l^{m_\lsub} =C_{\lsub}^{m_\lsub} P_\l^{m_\lsub}(\cos \vartheta^\ast) e^{i m_\lsub \phi^\ast} $. The square-root eigen values and their $z$ components are
 $$\displaylines{\refstepcounter{equation} \label{eq-eigenLcal} 
\hfill
\Jtr^{\ast}_\lsub =|\rb_\lsub^\ast \times (\mr^\ast  \vb_\lsub^\ast)| =\sqrt{l(l+1)} \ \hbar, 
\ \ \Jtr_{z\mn_\lsub}= \Jtr_\lsub \cos \vartheta_{m_\lsub} = \mn_\lsub \hbar,
\ \  l=0,\,1, \,\ldots; 
 \mn_\lsub=0, \,
\ldots, \,\mp l. \ 
\hfill (\ref{eq-eigenLcal})
}$$
 Based on the semiclassical expression $\rb_\lsub^\ast \times (\mr^\ast  \vb_\lsub^\ast)$, the particle of mass $\mr$ in $l$th state executes an orbital angular motion along a circular orbit $l$ of radius vector $\rb_\lsub $ at a tangential velocity $\velb_{t\lsub} =d \rb_\lsub/dt =\wb_o \times \rb_\lsub$, $\wb_o= \rb_\lsub \times \velb_{t \lsub}/r_\lsub^2 $; $\rb_\lsub \equiv \rb_{n}$ for all $l (=0,1, \ldots, n-1)$ values of the same  principal quantum number $n$.
The normal of the orbital plane or the axis of rotation $\nb_{o}$ passes through the CM and is at a quantised angle $ \vartheta_{m_\lsub}$  to the $z$ axis.

Owing to their having a finite acceleration $\acb^\astn_\lsub=-|d^2 \rb_\lsub /d t^\astn{}^2| (\rb_\lsub/r_\lsub) $ in radial direction here, as a well-known non-inertial frame effect the $e,p$ in addition execute a Thomas precession (TP), with an instantaneous angular velocity  denoted by $\wb_\Tsub$ and thus angular momentum  $\Jb_{\Tsub}= r_\lsub^2 \mr \wb_\Tsub$. 
 $\wb_\Tsub^\astn = \frac{\g^2}{(\g+1) }  \frac{ \acb^\astn_\lsub \times \velb_{ t \lsub}^\astn }{ c^\astn{}^2}$ 
according to LH Thomas (1927),  
as may be alternatively derived directly based on (transformed) infinitesimal Newtonian inertial-frame and hence linear motion combined with 
acceleration in infinitesimal time 
(internal work). $\wb_\Tsub^\astn $ is in the instantaneous direction $\acb^\astn_\lsub \times \velb_{ t \lsub}^\astn \propto -\rb_\lsub \times \velb_{t \lsub} =-\Jtrb_{\lsub}/ \mr $, i.e. opposite to $\Jtrb_{\lsub}$, describing an instantaneous rotation in opposite sense to the orbital angular motion underlining $\Jtrb_{\lsub}$. For a quantum system as the bound $e,p$ here, the $z$ component of $ \Jb_\Tsub$, $ J_{\Tsub z}$, will be necessarily constrained such that both the space quantisation conditions  (\ref{eq-eigenLcal}) above and (\ref{eq-Lzb}) below are met.

The total, precessional-orbital angular momentum $J_\jsub 
 =|\Jtrb^\ast_\lsub -\Jb^\astn_{\Tsub }|
$ and its $z$ component $\Jscr_{z m_j}=  \Jtr^\ast_{z m_\lsub }-\J^\astn_{\Tsub z }
= \Jscr_\jsub \cos \theta_{m_\j} $ 
are given according to the quantum vector addition model as 
$$\displaylines{
\refstepcounter{equation} \label{eq-Lzb}                      \label{eq-velnta}
\hfill
 \Jscr_\jsub  
=|\rb^\ast_{\lsub \jsub}  \times (\mr^\ast \velb^\ast_\jsub)|
=r_\lsub \mr \vel_j
= \sqrt{j(j+1)}\ \hbar, 
\quad
j = l-l_\Tsub=0,\, \frac{1}{2},\, 1, \,\frac{3}{2}, \,\ldots
 \hfill (\ref{eq-Lzb}.1)
\cr
\hfill
\Jscr_{z\mn_j}^\ast = \Jscr_\jsub \cos \theta_{\mn_j} 
= \mp \Jscr_{\jsub  }^\ast \cos \theta_{\jsub}
= \mn_j \hbar, \quad \mn_j=  \mp \j,  \hfill (\ref{eq-Lzb}.2)
}$$
where the permitted $\j$ values are results of the general solutions of the quantum commutation relation for the angular momentum $\Jscrb$  here, $\Jscrb \times \Jscrb =i\hbar \Jscrb$. $\rb_{\lsub \jsub}=r_\lsub \hat{r}_{\lsub \jsub} $ is the quantised radius vector of the instantaneous circular orbit $l$ of a normal or axis of rotation  $\nb$;  
$\nb$ is at a fixed angle $\theta_{m_\jsub} =\arccos {(J_{z m_\jsub}/J_j)}$ to the $z$ axis and rotates about the $z$ axis at the angular velocity $\wb_{\Tsub z}$ in opposite sense to 
that of the $\w_o$-orbital angular motion projected in the $xy$ plane, 
whence the Thomas precession; see Fig \ref{fig-neutron1.eps}b. The magnitude of $\rb_{\lsub \jsub} $, $|\rb_{\lsub \jsub}|= r_\lsub
$ is unchanged subjected to the radial eigenvalue equation (\ref{eq-Sch1-R}) but immediately to the quantum equation (\ref{eq-eigenLcal}a) here. $\velb_\j =d \rb_{\lsub \jsub}/dt= \wb \times \rb_{\lsub \jsub}$; $\wb= \rb_{\lsub \jsub} \times \velb_\j /r_\lsub^2=|\wb_o -\wb_{\Tsub }| \nb     $ is the precessional-orbital angular velocity. For facilitating later discussion we attach as in Fig \ref{fig-neutron1.eps}b the axes $x',y',z'$ to the instantaneous rest frame  of the precessional orbit $l$, with their origin coinciding with that of $x,y,z$, i.e. 
fixed at the CM. So the $x',y',z'$ axes precess about the $z$ axis at the angular velocity $\w_{\Tsub z}$, in counterclockwise sense for the $m_j=-j$ state in the figure, 
in such a way that the $z'$ axis maintains at fixed angle $\theta_\j$ to the $z$ axis, the $x'$ axis at fixed angle $\theta_\j $ to (its projection $x'_{xy}$ in) the $xy$ plane and along the $\rb_{\lsub \jsub}$ direction, and the $y'$ axis  in the $xy$ plane.
And we attach the $x'',y'',z''$ axes to the instantaneous rest frame of the precessing-orbiting particles $e,p$ with an origin fixed at the CM too, the $z''$ axis coinciding with $z'$, and the $x''$ axis lying along the line joining the $e,p$; see Fig \ref{fig-neutron1.eps}a,b. So the $x'',y''$ axes rotate in the $x'y'$ plane and about the $z'$ ($z''$) axis, in clockwise sense for the $m_j=-j$  state, at the angular velocity $\w_o+\w_\Tsub$  relative to the $x',y'$ axes.

From Eq  (\ref{eq-Lzb}.1b), it follows that the permitted $l_\Tsub$ values are uniquely specified once $l, \j$ are specified according to (\ref{eq-eigenLcal}),(\ref{eq-Lzb}):  For $l=0$, only  $l_\Tsub=0$ is permitted;  and   $\j =l=0$. This gives  an $e,p$  system not magnetically bound at a separation of the order $10^{-18}$ m (see further Sec \ref{Sec-force});  a bound system would in principle be obtainable at much larger separation as a hydrogen  only.
For any non-zero integer $l$ values,
 $l_\Tsub=0$ is  permitted formally by  (\ref{eq-Lzb}.1) but is  however unphysical because of the so implied absence of Thomas precession. $l_\Tsub=\frac{1}{2}$ is therefore the smallest finite and hence physical value which is also permitted  based on the quantum solutions for $l,\j$. Moreover,  $l_\Tsub=\frac{1}{2}$ is itself a solution for $\Jb_{\Tsub z}$ to separately satisfy the quantum commutation relation $\Jb_{\Tsub z} \times \Jb_{\Tsub z}=i\hbar \Jb_{\Tsub z}$; this establishes a condition for the carrier of  $ \Jb_{\Tsub z}$, the neutral rotational energy flux (to be identified as the antineutrino) to be created or emitted as a quantum particle.
Higher  half-integer or integer $l_\Tsub$ values satisfying (\ref{eq-Lzb}.1) and (\ref{eq-eigenLcal}) are permitted in theory but 
are not liable for a neutron candidate  
and will not be considered.  
 For the permitted $l$ and  $l_\Tsub=\frac{1}{2}$, Eqs (\ref{eq-Lzb}) are written as
$$\displaylines{
\hfill\Jscr_\jsub  
=r_\lsub \mr \vel_j
=\sqrt{j(j+1)}\, \hbar
= \mbox{$  \frac{\sqrt{(4\l^2-1) }  \ \hbar     }{2}$}, 
\quad
\mbox{ $\j =l-\frac{1}{2}=\frac{1}{2}, \frac{3}{2}, \ldots$};  \hfill (\ref{eq-Lzb}.1)'
\cr
\hfill \Jscr_{z\mn_j}^\ast =\Jscr_\jsub \cos \theta_{m_j} 
= \mn_j \hbar=( \mp |m_\lsub| \pm \mbox{$\frac{1}{2}$})\hbar, 
\quad
\mn_\jsub=\mp \j=\mp|m_\lsub| \pm \mbox{$\frac{1}{2}$} =\mp 
\mbox{$\frac{1}{2}$}, \ldots,  \mp \j, \hfill (\ref{eq-Lzb}.2)'
}$$
where $\cos \theta_{m_\jsub} = \frac{\Jscr_{z \mn_j}}{\Jscr_\jsub} = \frac{\mp 2 l \pm1}{\sqrt{4 l^2-1}}$. For $\j=\frac{1}{2}$, $\mn_j= \mp \frac{1}{2}$:
$$\displaylines{
\refstepcounter{equation} \label{eq-Lzb-p} 
\hfill
\Jscr_{\hfp}^\ast=|\rb_{1\hf} \times (\mr \velb_\hfp^\ast) |
= r_1 \mr \vel_\hfp
= \mbox{$\frac{\sqrt{3} \ \hbar }{2}$},
\quad
\Jscr_{z\mp\hf}^\ast= r_1 \mr \vel_\hfp^\ast \cos \theta_{\mp \hf}
 = \mp \mbox{$\frac{\hbar}{2}$};  \hfill
 (\ref{eq-Lzb-p})
}$$
$\cos \theta_\hfp= \J_{z\hf}/\J_\hfp= 1/  \sqrt{3}$; $\mbox{$ \theta_\hfp^\ast =\arccos (1/\sqrt{3}) = 54.7 ^o$}$. The $\j=\frac{1}{2}$ ($l=1$) states describe a ground-state neutron  (Secs \ref{Sec-mag-mmt}, \ref{Sec-force}). Eq (\ref{eq-Lzb-p}a) thus gives the $e,p$ relative precessional--orbital angular momentum  internal of the neutron, and (\ref{eq-Lzb-p}b)  the two possible $z$ components associated with a  minimum-energy ($ \mn_j=-\frac{1}{2}$) and excitation ($\mn_j= \frac{1}{2}$) state in an applied magnetic field in the $+z$ direction (see Sec \ref{Sec-mag-mmt}). The precessing circular orbit of the mass $\mr$ has the quantised  radius vector  $\rb_{1\hf}$  about the CM in the $x'y'$ plane. For the $\mn_j=-\frac{1}{2}$ state,  the normal $\nb$ of the rotation plane, and hence $\Jbscr_{ \hfp}^\ast $,  is at angle $\theta_{\m \hfp}=\pi-\theta_{\hfp}$ to the $z$ axis, the rotation being in clockwise sense, as shown in Fig \ref{fig-neutron1.eps}b.
 And conversely for $\mn_j=\frac{1}{2}$. For the next  orbital,  $\j=\frac{3}{2}$ ($\l=2$), $\Jscr_{z\frac{3}{2}}/\Jscr_{\frac{3}{2}}=3/\sqrt{15}
$, $\theta_{\frac{3}{2}} =39.2^o$.


For the neutron existing (in zero applied field) only in a single non-degenerate  state $\j = \l-\frac{1}{2}=\frac{1}{2}$ and presuming that, in terms of the SQR-KGE here, energies of different $l$ and same $n$ are degenerate, then  $N$ (the radial degree of freedom)$=0$ and $n=N+l+1 =0+l+1 =2$. So $\Ttr^\astn_{r \lsub} |_{\lsub=n-1=1}=0$; and the total kinetic energy of $\mr$ is, with $ \Jscr_\hfp^2 (=(\Jtrb_1-\Jb_{\Tsub })^2)$ for $ \Jtr_1^2$ and $T$ in place of $\Ttr$, 
$T_\hfp = T_{t\hf}=\frac{\g }{(\g+1)} \mr \vel_{\hf}^2
=\frac{    \g  }{ (\g+1)    } (\Jscr_{\hf}^{2} /  \mr r_1^2)
=\frac{3 \g \hbar^2 \mcm }{ 4 (\g+1) m_e m_p  r_{1}^{2} }
$. Accordingly, the eigen energy (Hamiltonian) $H_j=T_j+V_j$, in place of $\Htr_{\lsub =n-1} = \Ttr_{\lsub =n-1} + V_{\lsub =n-1} $ (where $\Htr_{\lsub =n-1} \equiv \Htr_n$, etc.), may be directly computed from the sum of the $ T_\jsub$ given above 
and the $ V_{\jsub} $  to be derived in Sec \ref{Sec-force}, without formally solving the radial differential equation  (\ref{eq-Sch1-R}).


The precessional--orbital motion (propagation) of the $e,p$ particle waves $\psi_e,\psi_p$ relative to one another, or alternatively of the matter wave $\psi$ of $\mr$ relative to the CM, 
is associated with a neutral precessing-orbiting --- simply rotational kinetic energy flux, or vortex, on disregarding their charges and also their rest-mass energies. 
This vortex entity carries a spin angular momentum with a $z$ component equal to one unit of half-integer quantum $\Jscr_{z \mp \hf}=\mp \frac{1}{2} \hbar$,  has apparently a positive helicity, and therefore resembles directly an antineutrino ($\nubar_e$) confined within the neutron; see further Sec \ref{Sec-excit}. Accordingly the $z$-component spin angular momenta of $\nubar_e$ are 
$$\displaylines{\refstepcounter{equation} \label{eq-Snubar}
\hfill
\Scal_{\nubar_e z}^\ast
=\Jscr_{z\mp\hf}^\ast 
=\mp \mbox{$\frac{1}{2}$} \hbar 
=\mp s_{\nubar_e} \hbar ,
\quad  
s_{\nubar_e}
=\mbox{$\frac{1}{2}$}. 
\hfill (\ref{eq-Snubar})
}$$


 \section{$e,p$ system magnetic structure. Neutron magnetic moment, (effective) spin}\label{Sec-mag-mmt}

 \setcounter{equation}{18}

{\it \qquad \ref{Sec-mag-mmt}.1 $e,p$ spins and magnetic moments. Neutron internal spin configurations. Total spin }  Either particle $\a=e$ or $p$ has an intrinsic spin $s_\a=\frac{1}{2}$, spin angular momentum $S_\a=\sqrt{s_\a (s_\a+1)} \ \hbar =\frac{\sqrt{3} }{2} \hbar $ about a spin axis $\nb_\a$ passing through $\rb_\a$, and $z$ component  $S_{\a z} =S_\a \cos \theta^s_\a =\pm s_\a  \hbar$. For the spin up state as in Fig \ref{fig-neutron1.eps}a, $ \cos \theta^s_\a = S_{\a z} /S_{\a }  = 1/\sqrt{3}$; the spin axis $\nb_\a$ is at fixed angle $ \theta^s_\a= 54.73^o$ 
to a $z_\a^s$ axis parallel with $z$ and passing through $\rb_\a$. 
Certain external, random environmental in the case of zero applied, magnetic field would always present and thus define the (instantaneous) $z$ direction here. For computing the magnetic field produced by  the spin motion of one particle $\a$ at the other particle ($\a'$), $\Bb_\a^s$ (Eq \ref{eq-Bp1b}b for $\a=p$,  Sec \ref{Sec-force}) at a separation $r_\lsub$ comparable to the sizes of their charges  (Secs. \ref{Sec-force}, \ref{Sec-num}), it is appropriate to treat the  $e,p$ charges as extended objects, firstly simply as spheres of radii $a_{e}, a_{p}$ with specific mass and charge distributions.
 We assume that the mass $m_\a$ of either particle $\a$ is distributed predominantly within its charge (and thus by a negligible amount in its wave field), with a volume density $\rho_{\a}(\xib_{\a})$ at a distance $\xib_\a$ from $\rb_\a$;  $d m_\a= m_\a \rho_{\a}(\xib_{\a }) d^3 {\xib_{\a}}$ is a mass element at $\xib_\a$. So $S_\a$ formally is given rise to by the angular motion of the sphere about $\nb_\a$ at the angular velocity $\w^s_\a=d \phi_\a^s /d t_\a $,  tangential velocity $\vel_\a^{s} =a_\a \w^s_\a$ as, 
$$\displaylines{
\refstepcounter{equation} \label{eq-spinLs3}
\hfill
\mbox{
$\Sast_\a
=m_\a \int |\xib_{\a} \times \velb^{s }_\a|  
\rho_{\a}(\xib_{\a})
d^3{\xib_{\a }}
=\frac{1 }{g_\a} a_\a^2 \w^s_\a m_\a 
=\frac{\sqrt{3} }{2}\ \hbar$}; 
\quad
\Sast_{\a z}
=\mbox{ $ \frac{1}{g_\a} $} a_\a^2  m_\a \w^s_\a \cos \theta^s_\a 
= \mbox{ $\frac{1}{2} \hbar$}
\hfill
(\ref{eq-spinLs3})
}$$
where $g_\a$ is the Lande $g$ factor of particle $\a$. And the charge $q_\a$ of either particle $\a$ is distributed along a circular loop of radius $a_{\a}$ with a normal parallel with $\nb_\a$. The spin (dipole) magnetic moment $\mu_\a^{s} $ of particle $\a$ is accordingly produced by the current loop of charge $q_\a$, area $\pi a_\a^2$, and angular velocity $\w^s_\a$ in the $\mp S_\a$ direction for $q_\a =\mp e$.  The $z$ components, written for $e,p$ explicitly, are
$$\displaylines{\refstepcounter{equation} \label{eq-muez}
\hfill
\mu_{ez}^{s \ast }
=e\frac{\w_e^{s\astn }       }{2\pi} \pi  a_e^2 \cos (\pi-\theta_e^{s\astn})
=- \frac{g_e e  \Scal_{ez} }{2 m_e^\astn}
=-\frac{g_e e \hbar }{4 m_e^\astn}; \quad 
\mu_{pz}^{s \ast }= \frac{g_p e  \Scal_{pz} }{2 m_p^\astn}
=\frac{g_p e \hbar }{4 m_p^\astn}
\hfill  (\ref{eq-muez})
}$$

For either particle $\a$, besides the $z_\a^s$ above we further specify the local axes $x_\a^s,y_\a^s$ to be parallel with the $x,y$ but with an origin located at $\rb_\a$. 
Its spin axis $\nb_\a$ as projected in this $x_\a^s y_\a^s$ plane is  unspecified in orientation according to the uncertainty principle, and  in the external magnetic field in $z$ direction would rotate about the $z_\a^s$ axis. Accordingly the direction of $\velb_\a^s{}$ at any fixed point on the current loop  varies over time as the current loop precesses. Since the $z$ component  $S_{\a z}$ of $S_\a$  is a constant (Eq \ref{eq-spinLs3}b), the projection of $\velb_\a^s$ in the $x_\a^s y_\a^s$ plane is a constant: $\vel^s_{\a_{xy}} = (a_\a \cos \theta^s_\a) \w^s_\a= \vel_\a^s \cos \theta^s_\a$; this gives also  the time average of $\velb_\a^s$, for the projection of $\velb_\a^s$ in $z$ direction averages  to  zero over time. For deriving an effective algebraic equation for the magnetic field produced by the spin current loop (Sec \ref{Sec-force}), the distinct symmetry property of the system will be utilised to further reduce the system to a two point half-charge representation (\ref{App-red-geom}).

{\it \ref{Sec-mag-mmt}.2 $e,p$ system spin configuration. Total spin angular momentum }  For the $e,p$ to be in a bound, minimum (internal magnetic) energy state (Sec \ref{Sec-force}), apart from the specific $\j =\frac{1}{2}$ and $m_j=\mp \frac{1}{2}$ values (Sec \ref{Sec2.1.Eq-Mot}.4),  $\Scal_{ez},\Scal_{pz}$ need be parallel mutually  and each antiparallel to $\Jscr_{z\mn_j}$ (Figs \ref{fig-neutron.eps}a,b;  \ref{fig-Sn.eps}a,b). $ \Scal_{ez}, \Scal_{pz}, \Jscr_{z\mn_j}$ may therefore assume two possible configurations
$$\displaylines{\refstepcounter{equation} \label{eq-anglmmtx1} 
\hfill
\mbox{$
(i) 
\ \   \Scal_{ez} =\frac{1}{2} \hbar, 
\ \  \Scal_{pz}=\frac{1}{2} \hbar,
\ \  \Jscr_{z \m \hf} =-\frac{1}{2} \hbar; 
$}
\quad
\mbox{$
(ii) 
\ \  \Scal_{ez}=-\frac{1}{2} \hbar, 
\ \  \Scal_{pz}=-\frac{1}{2} \hbar,
\ \  \Jscr_{z \hf}=\frac{1}{2} \hbar 
$}
\hfill
(\ref{eq-anglmmtx1})
}$$
as shown in Figs \ref{fig-Sn.eps}a,b. 
\begin{figure}[h] 
             \vspace{-0.1cm}

\begin{center}
\setcounter{figure}{1}
\includegraphics[width=0.96 \textwidth]{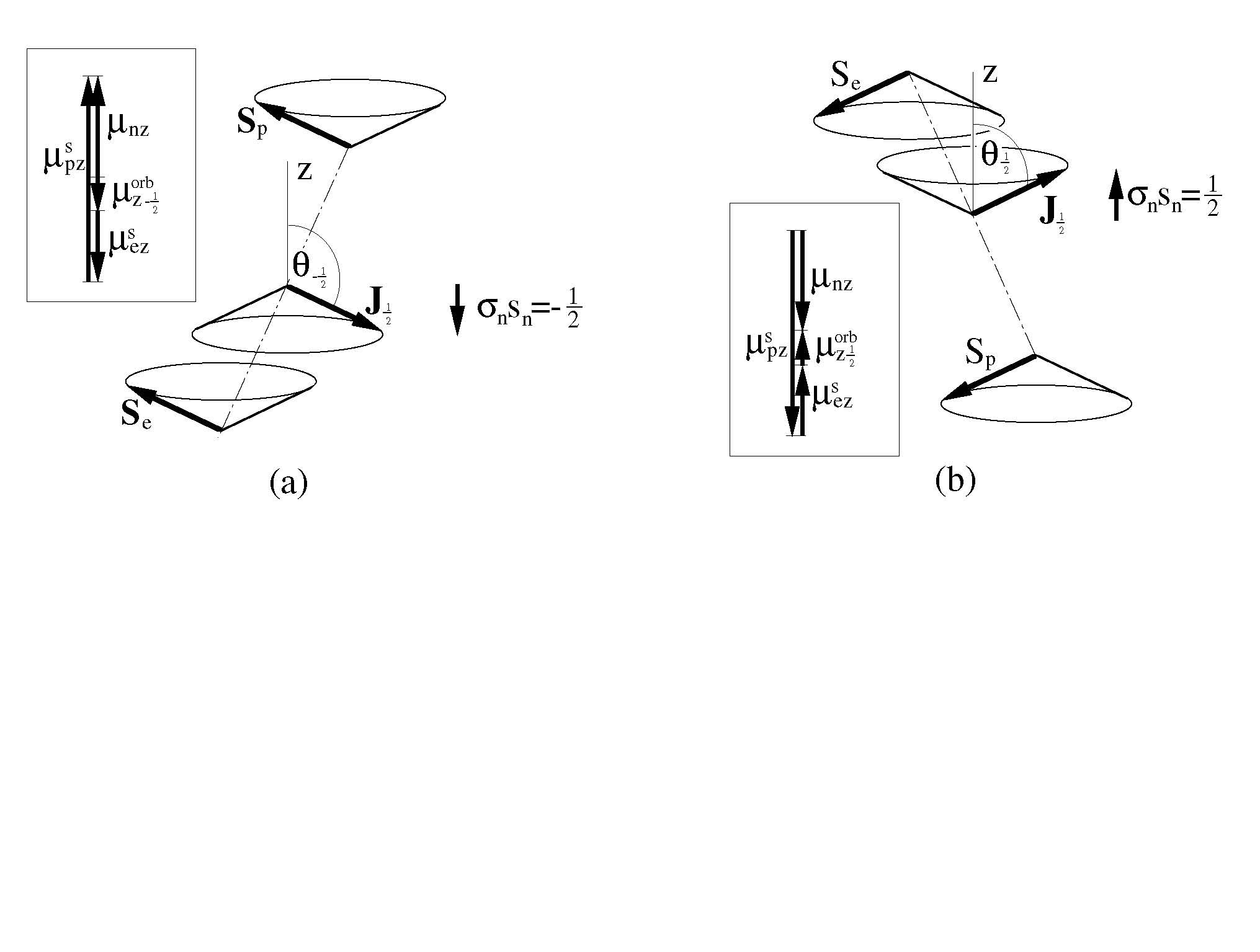}
\end{center}  
\vspace{-5.5 cm}

\caption{
Internal angular momenta $S_e,S_p,J_{m_j}$  of the model neutron 
with the internal spin configurations (a) $s_e,s_p,m_j = \frac{1}{2},\frac{1}{2},-\frac{1}{2} $ in effective spin down state $s_n \sig_n =m_j=-\frac{1}{2}$,  and  (b) $s_e,s_p,m_j= -\frac{1}{2},-\frac{1}{2},\frac{1}{2} $ in effective 
up state $\sig_ns_n =m_j=\frac{1}{2}$. Insets in (a),(b): 
schematics of the corresponding $z$-component neutron magnetic moments 
$\mub_{nz} $ as the respective vector sums of the internal component moments along $z$ direction.
}\label{fig-Sn.eps}
\vspace{0.1cm}
\end{figure}

Based on the usual vector addition model, the total angular momentum of the $e,p$ system, denoted by $\Jtot_{\jsub_n}$, in the precessional-orbital state $j=\frac{1}{2}$ are
$$\displaylines{
\refstepcounter{equation} \label{eq-totangm-z}\label{eq-totangm}
\hfill 
\Jtot^\ast_{j_n}
= \sqrt{j_n(j_n+1)} \hbar     
= \mbox{$\frac{\sqrt{3}}{2}  \hbar $},    \quad j_n=(s_e+s_p) - \j =(\frac{1}{2}+\frac{1}{2})- \frac{1}{2} =\frac{1}{2} 
\hfill (\ref{eq-totangm-z}) 
}
$$
Its $z$ components $\Jtot_{z \mn_{j_n}}$ for the $m_{j_n}= \pm j_n$
states are 
$$\displaylines{
\refstepcounter{equation} \label{eq-totangm-zz}
\hfill \Jtot_{z  \pm \hf }    
=(\Scal_{p z}^\ast + \Scal_{ez}^\ast)_{{up \atop down}}+ \Jscr_{ z \mp \hfp}^\ast  
\mbox{$
=[\pm (\frac{1}{2} + \frac{1}{2}) \mp \frac{1}{2}  ] \hbar
=\pm \frac{1}{2} \hbar.
$}
\hfill (\ref{eq-totangm-zz}) 
}$$

{\it \ref{Sec-mag-mmt}.3  Total magnetic moment }  The differing $g$ factors and potentially also asymmetric relative positions of $e,p$ suggest an asymmetric internal magnetic structure of the $e,p$ system. One thus expects a total magnetic moment  which is not simply related to  $\Jtot_{z m_{j_n}}$ of  Eqs (\ref{eq-totangm-zz}) as for a simple particle; nor would it be zero as implied by its zero net charge as both given by the direct sum of  the $e,p$ charges here and from experimental observation. We shall find the total magnetic moment  based directly on vector addition of the  individual magnetic moments along the $z$ direction below. For the spin configuration (i) of Eqs (\ref{eq-anglmmtx1}), the $z$-component total magnetic moment of the $e,p$ spins is the vector sum 
$$\displaylines{
\refstepcounter{equation} \label{eq-mun-1a}
\hfill
\mu^s_{z}= \mu_{pz}^s+ \mu_{ez}^s 
= \frac{g_pe S_{pz}}{2 m_p} - \frac{g_ee S_{ez}}{2 m_e}
= (g_p -\frac{g_e}{k})\frac{e S_{pz}}{2 m_p},
\hfill (\ref{eq-mun-1a}) 
}$$ 
where the last of Eqs  (\ref{eq-mun-1a}) is given after substituting $m_e=k m_p$ (as in Sec \ref{Sec2.1.Eq-Mot}.2) and $S_{ez}=S_{pz}$. $k$ may in general differ from 1. So the relative precessional-orbital motion may contribute a finite moment given by the vector sum, for the case (i) or $m_j= -\frac{1}{2}$,
$$\displaylines{
\refstepcounter{equation} \label{eq-mun-2}
\hfill
\mu_{z\m\hf}^{orb}
= \frac{e  \J_{pz\m \hf}}{ 2 m_p }-\frac{e \J_{ez\m \hf}}{ 2 m_e } 
= \frac{e  (\frac{m_e}{\mcm}) \J_{z\m \hf}}{ 2 m_p }-\frac{e  (\frac{m_p}{\mcm})\J_{z\m \hf}}{ 2 m_e } 
= \lf(\frac{k-1}{k}\rt) \frac{e \J_{z\m \hf}}{ 2 m_p }
\hfill (\ref{eq-mun-2})
}$$
where $J_{\a z m_j}$ are the $z$-projections of $J_{\a j}$ given after  Eqs (\ref{eq-Lddg}a,b); 
a $g$ factor equal to $1$ is assumed. 
The total $z$-component magnetic moment of the $e,p$ system is the vector sum
$$\displaylines{
\refstepcounter{equation} \label{eq-mun-3}
\hfill
\mu_{z \m\hf \phi}
= \mu_{z}^s +  \mu_{z\m\hf}^{orb}
= (g_p -\frac{g_e}{k})\frac{ e (-   \J_{z\m \hf})}{2 m_p} 
+\lf(\frac{k-1}{k}\rt) \frac{e \J_{z\m \hf}}{ 2 m_p }
=-  \frac{g_n e \J_{z\m \hf} }{ 2 m_p }, \hfill (\ref{eq-mun-3}a)
\cr
\hfill
g_n= g_p -\frac{g_e}{k}-\frac{k-1}{k} 
\hfill (\ref{eq-mun-3}b)
}$$
where in the expression of $\mu_{z}^s $ we substituted $S_{pz} =- \J_{z\m\hf}$; $\J_{z m_j}$, instead of $\Jtot_{z m_{j_n}} $, is used for reason to be explained below. The subscript $\phi$ indicates that the total $\mu_{z\m\hf\phi}$, as the $ \mu_{pz}^s$, $ \mu_{ez}^s$ and $\mu_{z\m\hf}^{orb}$, is as directly probed by a detector placed at the CM relative to which the $e,p$ are moving in the   $\velb_e, \velb_p$ or $\phi$ directions, in which case the relativistic masses $m_e,m_p $ in Eq  (\ref{eq-mun-3}a) remain physical.

An experimenter (as an external observer) in the laboratory on the other hand  commonly probes the neutron magnetic moment by applying a magnetic field to  turn the moment 
 along the $\theta $ direction, typically at a speed $u_\theta<<c$. This is to turn the $e,p$ system as a whole here,  or manifestly the $e,p$ precessional-orbital plane about the $y'$ axis along the $\theta $ direction in the $x'z'$ plane, i.e. in a direction perpendicular to $\velb_e, \velb_p$.  So immediately for the proton $\g_{p \theta} =(1- u_{p\theta}^2/c^2   )^{-1/2}\simeq 1$, $m_p (u_{p\theta}) = \g_{p \theta} m_p^\osup \simeq m_p^\osup$; the proton dominates the turning process because of its much larger rest-mass moment of inertia. 
The electron can only be turned in the same rigid precessional-orbital plane as the proton in a bound relativistic dynamic process, its mass (hence moment of inertia)  must therefore manifestly be weighed by the same factor $k$ as in this relativistic  process, i.e. as  $m_e=k m_p$, not as $m_e^\osup$. The only means of correctly carrying the factor $k$ through to the experimenter's result (Eq \ref{eq-mu-4} below) is to convert $m_e $ to $k m_p$ before transforming to Eq  (\ref{eq-mu-4}), as has been done in (\ref{eq-mun-3}a). Substituting  $m_p^\osup$ in place of $m_p$ in the last of Eqs  (\ref{eq-mun-3}a), and accordingly $\mu_{z\m\hf }$ 
in place of $\mu_{z\m\hf \phi} $,  gives therefore the $z$-component magnetic moment of the model neutron, $\mu_{nz}$, as probed by the experimenter, for the $m_j=-\frac{1}{2}$ state
$$\displaylines{
\refstepcounter{equation} \label{eq-mu-4}
\hfill
\mu_{nz}(\mbox{$m_j=-\frac{1}{2}$})=\mu_{z\m\hf }  =  -\frac{ g_n e \J_{z\m \hf} }{2 m_p (u_{p\theta}) }        
=  -\frac{ g_n e \J_{z\m\hf}  }{2 m_p^\osup }     
=  \frac{ g_n e \hbar }{4 m_p^\osup } =  \frac{ 1 }{2  } g_n \mu_\Nsub, 
\hfill (\ref{eq-mu-4})
}$$
where $ \mu_\Nsub= \frac{e\hbar }{2 m_p^\osup}$ (the nuclear magneton). Similarly for the $m_j=\frac{1}{2}$ state we obtain $\mu_{nz}(m_j=\frac{1}{2})=\mu_{z\hf } 
= - g_n e \J_{z\hf}  /2 m_p^\osup 
=-  \frac{ 1 }{2  } g_n \mu_\Nsub$. $g_n$  represents  the $g$ factor of the model neutron. Equating $g_n$ of (\ref{eq-mun-3}b) with the experimental value   $g_n^{exp}= 3.8261$, and using the experimental $g$ values of $e,p$, $g_p=5.5857$, $g_e=2$, numerical solution for $k$ is obtained as $k=1.3165$. According to (\ref{eq-mu-4}) or (\ref{eq-mun-3}a), $\mu_{z \m \hf} >0$, i.e. $\mu_{z \m \hf} $ points in the $+z$ direction, and is in the opposite direction to $ \J_{z\m \hf}  $ (Fig \ref{fig-Sn.eps}a).  And $\mu_{z  \hf} <0$ (Fig \ref{fig-Sn.eps}b). Clearly, the magnetic moment of the model neutron is dominated by the proton spin magnetic moment because of the asymmetrically much larger  $ g_p$ over $g_e$. From  $k>1$ and hence $m_e>m_p$, $ r_p>r_e$ (by a small amount each), it follows  that $\mu_{z \m \hf}^{orb}$ points in the $-z$ direction, adding a negative but small  term to $\mu_{z \m \hf} $.

{\it \ref{Sec-mag-mmt}.4  Effective spin of the neutron }  In an applied magnetic field say $\Bb_0$ in $+z$ direction, the magnetic-interaction energy of the $e,p$ system with the field is $U_{j} =- \mub_j \cdot \Bb_0 = -\mu_{z m_j} B_0$. $U_j<0$ for the spin configuration (i) of Eqs  (\ref{eq-anglmmtx1}) or the $m_j=-\frac{1}{2}$ state, and $U_j>0$ for (ii) or the  $m_j=\frac{1}{2}$ state. That is, (i) or $m_j=-\frac{1}{2}$ corresponds to the minimum-energy state and (ii) or $m_j=\frac{1}{2}$ the excited state in the applied field. A transition from the minimum-energy to excited state corresponds to a flip of the spin-configuration (i) to (ii) of  the bound $e,p$ system as a whole, which is dictated, and thus manifested by the flip of the processional-orbital plane from  the $m_j=-\frac{1}{2}$  state to $m_j=+\frac{1}{2}$. In other terms, $\mu_{z_{\m\hfp}}$ is as if produced by a negative charged current loop in spin up state, or alternatively by a positively charged  current loop in spin down state. In so far as the total magnetic moment of the $e,p$ system as a whole, whence the model neutron, is probed, it is therefore physical to assign to it an  effective spin $s_n$ and spin vector $\sig_n$, corresponding directly to the $  j$ and $m_j =\mp j$ values  (rather than the $j_n$ and $m_{j_n}$). So the effective neutron spin angular momentum $S_n  $,  its $z$ components $S_{nz}$, and accordingly $\mu_{z m_j} $ in relation with $S_{nz}$, given by substituting $S_{nz}$ for $J_{z m_j}$ in (\ref{eq-mu-4}), are
$$\displaylines{
\refstepcounter{equation}  \label{eq-sn}
\hfill
S_n  = \sqrt{s_n (s_n+1) } \hbar \equiv \J_j =\frac{\sqrt{3}}{2}\hbar, 
\quad
s_n=j =\frac{1}{2}; 
\quad
S_{nz}= \sig_n s_n \hbar \equiv J_{z_{m_j}}=m_j \hbar = \mp \frac{1}{2} \hbar, 
\hfill
\cr
\hfill 
\sig_n=\mp 1  \mbox{ (for $m_j=\mp \frac{1}{2}$}); \quad 
\mu_{nz} 
= -\frac{g_ne S_{nz} }{2 m_p^\osup} 
= -\frac{g_n \sig_n s_n e \hbar }{2 m_p^\osup} 
\equiv \mu_{z_{m_j}} 
=\pm \frac{g_ne \hbar }{4 m_p^\osup} 
\hfill (\ref{eq-sn})
}$$
Notice that for either spin configuration the $S_n$, $S_{nz}$ are equal to  $\Jtot_{j_n}$, $\Jtot_{z m_{j_n}}$ in magnitudes but opposite in directions. 
The  magnitude of  $S_n$ or  $\Jtot_{z m_{j_n}}$, being a one-half integer quantum, 
 has the absolute significance that it identifies the neutron as a fermion in accordance with observation.
The assignment of the effective spin parameters $s_n, \sig_n$ above is in direct conformity with the role of the Standard Model neutron spin with respect to the experimental determination of 
neutron magnetic monument based on magnetic resonance method [\citePerkins1982Griffithsetal{}f].


\section{$e,p$ electromagnetic interaction. Weak interaction force} \label{Sec-force}
\setcounter{equation}{28}

We shall below derive for the electron $e$ and proton $p$ comprising the model neutron in stationary state their interaction force $\Fbast  $, the corresponding potential  $V^\astn $ and Hamiltonian $H$ based on first principles laws of electromagnetism and (the solutions of Sec \ref{Sec-eom} of) relativistic kinematics and quantum-mechanics. We shall continue to  work in the CM frame, i.e. in terms of the unsuperscripted mass and space-time variables which will directly enter the electromagnetic interactions below, and for simplicity the time $t^\astn$ instead of $t_e^\astn, t_p^\astn $;  the local time $t_e^\astn, t_p^\astn $ effect  will be included in the end by a projecting factor ($f_t^2$). In this section, $\rb$ or $\rb_{\lsub \jsub} $ refers to the $e,p$ separation distance starting at $\rb_p$, ending at $\rb_e$, as in Figs \ref{fig-neutron1.eps}a, c; its magnitude is equal to that of $\rb$ or $\rb_{\lsub \jsub}$ of  Sec \ref{Sec2.1.Eq-Mot}.4, Fig \ref{fig-neutron1.eps}b.

Consider the $e,p$ system in a precessional-orbital state $j=l-l_\Tsub$, 
$m_j=- j$, with the $e,p$ spins  $S_{pz}$, $S_{ez}$ in the $+z$ direction, i.e. antiparallel with $\Jscr_{z\m j}$ (as in Figs \ref{fig-neutron.eps}a,b or Fig \ref{fig-Sn.eps}a for $j=1-\frac{1}{2} =\frac{1}{2}$, $m_j=- \frac{1}{2}$). Firstly, the proton of a charge $+e$  produces at  the electron at $\rb_{\lsub \jsub}$ apart a (transformed) Coulomb field $\Ebast_{p} (r_\lsub) = (e /4\pi\ev_0 r^2_\lsub) \ \hat{\rb}_{\lsub \jsub}$ (in SI units here and below); $\hat{r}_{\lsub \jsub}=\rb_{\lsub \jsub}/r_\lsub$ is a unit vector pointing from $p$ to $e$. 
$|\Ebast_{p }^\ast| $ is amplified  from its rest-frame value $E^\osup_p$ by a factor $\propto  (1/r^2    )    /        (           1/ (r^\osup)^2           )= \g^2=1/f_c$ and hence has a narrowed profile at a point $\rb$ perpendicular to its motion $\phi$ direction by an inverse  factor,   $f_c$. And  so are the magnetic fields below. Furthermore, the proton is in relative precessional--orbital motion in clockwise sense at the tangential velocity $\velb_p^\astn $ about the CM in the $x'y'$ plane, and in spin motion at the tangential velocity $\velb_{p_{xy}}^s$ in the $x_p^sy_p^s$ plane (Sec \ref{Sec-mag-mmt}.1). The latter, 
after a projection on to the $x'y'$ and hence $x''y''$ plane, may be effectively represented 
(Eq \ref{eq-app-SpzRp}, \ref{App-red-geom}) as two point half-charges located at $-\aav, \aav$ from $\rb_p$ 
on the $x''$ axis and moving oppositely at  velocities $- \velavb_p^s{}'',+\velavb_p^s{}''  $ in $-y'',+y''$ directions. So $p$ produces at $e$ magnetic fields $\Bbast_{p}^{\orb \ast} (=-\velb_p  \times \Ebast_p/c^2)$ and $\Bbast_{p}^{s \ast} (r_\lsub \pm \aav)  (= \pm |\velavb_p^s{}'' \times \Ebast_p (r_\lsub \pm \aav)|/c^2 )$ along the $z'$ direction given as (the transformed Biot-Savart law),
$$\displaylines{\refstepcounter{equation}  \label{eq-Bp1orb}
\hfill
\Bb^{orb \ast }_{p } (r_\lsub,\theta_j)
=\frac{e \velb_p^{\astn }  \times \rb_{\lsub \jsub} }{ 4\pi \ev_0 c^2 r_\lsub^3 }
=-\frac{e \rb_{\lsub \jsub} \times (-\frac{m_e^\astn  m_p^\astn  }{\mcm^\astn}) \velb_\jsub
}{4\pi \ev_0 m_p^\astn c^2 r_\lsub^3}
=-\frac{\sqrt{4 l^2-1 } \ e\hbar  \ \hat{z}' }{8\pi \ev_0 m_p^\astn  c^2 r_\lsub^3};
\hfill (\ref{eq-Bp1orb})
\cr
\refstepcounter{equation}  \label{eq-Bp1a} \label{eq-Bp1b}
\hfill
\Bbast_{p}^{s \ast }(r_\lsub \pm \aav, \theta_j) 
= \frac{ \mp \frac{1}{2}e   \velbav_p^s{}'' \times (\rb_{\lsub \jsub}/r_\lsub) }{4\pi \ev_0 c^2  (r_\lsub \pm \aav )^2},
\quad
\Bbast_{p}^{s } (r_\lsub,\theta_j)
=\Bbast_{p}^{s \ast}(r_\lsub-\aav, \theta_j)+\Bbast_{p}^{s \ast}(r_\lsub+\aav,\theta_j) =
\hfill
\cr
\hfill
=\frac{e \aav  \velbav_p^s{}'' \times  (\rb_{\lsub \jsub}/r_\lsub)     }{2\pi \ev_0 c^2       r^{ 3}_\lsub (1-\frac{\aav ^{2}  }{ r^{2}_\lsub})^2 } 
= \frac{  -  \eta^2 g_p  e \hbar \cos \theta_\jsub \hat{z}' }{ 4\pi \ev_0   m_p^\astn c^2  r^\ast{}^3_\lsub  C_{1\lsub} },
\quad
C_{1\lsub}\mbox{$=\lf(1-\frac{ \aav^{2}}{r^{2}_\lsub}\rt)^2$}  
\hfill
 (\ref{eq-Bp1b})
}$$
The last of Eqs (\ref{eq-Bp1orb})
 is given after substituting (\ref{eq-vs}b) for $\velb_p$ and  (\ref{eq-Lzb}.1)$'$ for $\rb_{\lsub \jsub} \times (\frac{m_e^\astn  m_p^\astn}{\mcm^\astn}) \velb_\jsub$, given for $l_\Tsub=\frac{1}{2}$. The last of 
Eqs (\ref{eq-Bp1b}b)
 is given after substituting (\ref{eq-app-aav}a), (\ref{eq-app-SpzRp}b) for $\aav$, $\velavb_p^s{}''$, $a$ for $a_p$, and (\ref{eq-spinLs3}) for $a \vel_p^{s} m_p^\astn  \cos \theta_p^s/g_p(= \frac{1}{2}\hbar)$. The negative signs in the two final results indicate $\Bb_p^\orb $, $\Bb_p^s$ to be in the $-z'$ direction each.

In the $\Efbast_{p }^\ast$ and $\Bfbast_p^{orb \ast} + \Bfbast_{p}^{s \ast} =\Bfbast_p$ fields of the proton (cf Fig \ref{fig-neutron1.eps}a), the electron at $\rb_{\lsub \jsub}$ apart, with an effective charge $q_e=-f_c e$,  and in precessional--orbital and spin motions at the tangential velocities $\velb_e$ about the CM and $ \velavb_e^s{}''$ about $\rb_e$  in clockwise and counter-clockwise senses in the $x'y'$ plane, is acted by an electromagnetic force along the $\rb_{\lsub \jsub}$ direction according to the Lorentz force law,
$$\displaylines{
\refstepcounter{equation} \label{eq-Fpe1}
\hfill 
\Fbast^\ast_{pe} (r^\ast_\lsub, \theta_j)
=-f_c e \Ebast_{p}^\ast (r^\ast_\lsub)
+ f_t^2 [\Fbast_{pe,m}^{orb-orb \ast}{} (r^\ast_\lsub, \theta_j, t^\astn)
+ \Fbast_{pe,m }^{s-orb \ast}{}(r^\ast_\lsub, \theta_j, t^\astn)
+\Fbast_{pe,m}^{s-s \ast}{} (r^\ast_\lsub, \theta_j,t^\astn)], 
\hfill  (\ref{eq-Fpe1})
\cr
\refstepcounter{equation} \label{eq-Fmo2} \label{eq-Forb-orb}
\refstepcounter{equation} \label{eq-Fmo1} 
\refstepcounter{equation} \label{eq-Fms1}
 {\rm where} \hfill
\cr
\hfill
\Fbast^{\orb-\orb }_{pe,m}
=-e \velb_e^\astn \times \Bfbast_{p}^{\orb\ast } 
=-\frac{ e |\rb_{\lsub \jsub}\times (\frac{m_e^\astn  m_p^\astn }{\mcm^\astn}  \velb_\jsub) |  |\Bb_{p}^{orb}| \ \hat{r}_{\lsub \jsub}
}{m_e^\astn r_\lsub  } 
= -  \frac{  (4l^2-1)   e^2 \hbar^2 \ \hat{r}_{\lsub \jsub} 
}{ 16\pi \ev_0 m_e^\astn  m_p^\astn c^2 r_\lsub^4 }, 
\hfill (\ref{eq-Fmo2})
\cr
\hfill
\Fbast_{pe,m}^{s-\orb}{} 
=-e \velb_e^\astn \times \Bbast_{p}^s 
=-\frac{e |\rb_{\lsub \jsub} \times (\frac{m_e^\astn m_p^\astn }{\mcm^\astn}\velb_\jsub) | |\Bbast_{p}^s| \ \hat{r}_{\lsub \jsub} }{m_e^\astn r_\lsub}  
=-\frac{(2l-1) \eta^2 g_p e^2 \hbar^2  \ \hat{r}_{\lsub \jsub} }{16 \pi \ev_0 m_e^\astn m_p^\astn c^2 r_\lsub^4 C_{1\lsub}}, 
\hfill (\ref{eq-Fmo1})
\cr
\hfill
\Fbast_{pe,m}^{s-s \ast}
= - \frac{\pd \Vast^{s-s}_{pe,m}  }{ \pd r_\lsub}   \ \hat{r^\ast}_{\lsub \jsub}  
=  |\mu_{ez}^{s}  |  \frac{\pd |\Bbast_{p }^s  |  }{\pd r_\lsub } \cos \theta_{\jsub} \ \hat{r^\ast}_{\lsub \jsub} 
=-\frac{  3 \eta^2 g_e g_p e^2  \hbar^2                 \cos^2 \theta_{\jsub}   \ \hat{r}_{\lsub \jsub}  }{
16 \pi \ev_0 m_e^\astn m_p^\astn c^2 r^\ast{}^4_\lsub C_{1\lsub} }. 
\hfill (\ref{eq-Fms1})
}$$
$f_c$ is the fraction of the $e$-charge sphere  momentarily facing the narrowed $\Ebast_p$ profile at $\rb_{\lsub \jsub}$. Eq (\ref{eq-vs}a) for $\velb_e$ and again (\ref{eq-Lzb}.1)$'$  for $r_\lsub (\frac{m_e^\astn  m_p^\astn }{\mcm^\astn})  \vel_\jsub $   are used in (\ref{eq-Fmo2}),(\ref{eq-Fmo1}). In (\ref{eq-Fms1}), $\Vast^{s-s\ast}_{pe,m} 
=-\mub_e^s \cdot \Bb_p^s 
= - | \mu_{ez}^{s \astn }|   | \Bbast_{p}^{s \ast}| \cos \theta_{\jsub}$ is the magnetic potential of the spin-spin interaction; $\mu_{ez}^{s \astn } $ is an intensive quantity at $\rb_e$, hence not affected by the $\Bast_p$ profile narrowing, and is given by Eq (\ref{eq-muez}a). $\Fbast_{m 0}^{s-s \ast} =-\int^{2\pi}_0 e \velb_e^{s\astn} \times \Bbast_{pz}^{s}(r_\lsub) d \phi_e^s =0$; $\frac{\pd |\Bbast_{p}^{s}| }{\pd r_\lsub }=- \frac{3 \Bast_{p}^{s} }{r_\lsub}$. $f_t^2$  projects the  product term $\vel_e^\astn \vel_p^\astn $ contained in each component magnetic force  to  $\vel_e^{\astn \prime} \vel_p^{\astn  \prime}$ which actually enters the  interaction;  $\Fbast^\ast_{pe} (r^\ast_\lsub, \theta_\j) \equiv
\Fbast^\ast_{pe} (r^\ast_\lsub, \theta_\j, t^\astn_e,t_p^\astn)$ is implicitly meant. $\vel_e^{\astn \prime} \vel_p^{\astn  \prime} 
= (\vel_e^\astn \mcm^\astn /m_p^\astn)   (\vel_p^\astn \mcm^\astn/m_e^\astn) =(\mcm/\mr)\vel_e\vel_p =f_t^2 \vel_e^\astn\vel_p^\astn $ (Sec \ref{Sec2.1.Eq-Mot}.1), so $f_t^2 = \mcm/\mr$. A repulsion $\Fbast^{rep \ast}_{pe} =A_{rep} \hat{r}_{\lsub \jsub} /r^{\Nsub+1}_\lsub$ at short range, relative to the 
magnetic interaction strength at the distance $r \sim 10^{-18}$ m, 
 may generally also present but is omitted for the intermediate range of interest here. Given the presumed $\Sast_{ez}, \Sast_{pz}, \Jscr_{z\m j}=\frac{1}{2}, \frac{1}{2}, -j$  
configuration in units $\hbar $ here,  all the three component magnetic forces (for $l > 0$ for $F^{orb-orb}_{pe,m}, F^{s-orb}_{pe,m}$) acted by $p$ on $e$ above are optimally in the $-\rb_{\lsub \jsub}$ direction and hence attractive. $ \Fbast_{pe}$ is therefore in the  $-\rb_{\lsub \jsub}$ direction and maximally attractive. Any alteration of the relative orientations between   $\Sast_{ez}, \Sast_{pz}, \Jscr_{z\m j}$ will render some or all of the component magnetic forces repulsive. An alteration of $\Sast_{ez}, \Sast_{pz}, \Jscr_{z\m j}$ as a whole, i.e. from the configuration (i) to (ii) of Eqs (\ref{eq-anglmmtx1}) or from Fig \ref{fig-Sn.eps}a to \ref{fig-Sn.eps}b for $j=\frac{1}{2}$, retains all component magnetic forces attractive, and hence a total force $ \Fbast_{pe}$ the same as given in (\ref{eq-Fmo2}), or $F_j$ same as given in (\ref{eq-Fpe2}) below.

Similarly, $e$ produces at $p$ at  $-\rb^\ast_{\lsub \jsub}$ apart the electromagnetic fields $\Efbast^\ast_e$ and $\Bfbast_e^\ast$, and  forces given as $f_c e \Ebast_e$,  $f_t^2 \Fbast_{ep,m}^{orb-orb }=- f_t^2\Fbast_{pe,m}^{orb-orb }$, $f_t^2\Fbast_{ep,m}^{s-orb }=-f_t^2 \Fbast_{pe,m}^{s-orb } \frac{g_e }{g_p }$, $f_t^2\Fbast_{ep,m}^{s-s }=- f_t^2\Fbast_{pe,m}^{s-s } $. The action and reaction forces for the $e,p$ in equilibrium must be equal in magnitude and opposite in direction (Newton's third law); the magnitude may be here represented by the geometric mean as 
$F =\sqrt{|\Fb_{pe}||\Fb_{ep}|} 
=\sum_{\lam, \lam'} \sqrt{ |\Fb_{pe}^\lam ||\Fb_{ep}^{\lam'}| } \, \delta_{\lam \lam'} 
$, where $\lam,\lam'$ indicate the different component forces; $\delta_{\lam \lam'} $ is the Kronecker delta. The last equation needs to hold for the action and reaction to maintain detailed balance upon any small variation of the independent variables such as $\rb_{\lsub \jsub}$. The final total (attractive) force of $p$ on $e$ in equilibrium in the $j=l-l_{\Tsub}, m_j=-j$ state is therefore, suffixing $j$ after $\Fbast$ explicitly, 
$\Fbast_\j (r_\lsub,\theta_\j)
=- [f_c e \sqrt{|\Ebast_p|| \Ebast_e|} 
+f_t^2 \sum_\lam \sqrt{ |\Fbast_{pe,m}^{\lam}| |\Fbast_{ep,m}^{\lam}| }] \hat{r}       
= -f_c e \Ebast_p
+ f_t^2 [ \Fbast_{pe,m}^{orb-orb}
+ \Fbast_{pe,m}^{s-orb}  \frac{ \sqrt{g_e g_e}  }{g_p }
+ \Fbast_{pe,m}^{s-s}]  
$. 
Substituting Eqs (\ref{eq-Fmo2})--(\ref{eq-Fms1}) into the foregoing we obtain this force in explicit and scalar form (and explicitly for $l_\Tsub=\frac{1}{2}$), 
$$\displaylines{
\refstepcounter{equation} \label{eq-Fpe2} 
\hfill 
\Fast^{\ast }_\jsub(r_{\lsub}^\ast, \theta_j) 
=-\frac{ e^2}{           4\pi \ev_0 r^2_\lsub}(f_c+f_{m})
\simeq -\frac{ e^2 f_m }{ 4\pi \ev_0 r^2_\lsub}
=-\frac{   f_t^2 e^2  \hbar^2 \Cbar_{0\j}
 }{16 \pi \ev_0  m_e^\astn m_p^\astn  c^2  r^\ast{}^4_\lsub},
\hfill (\ref{eq-Fpe2})
\cr
\refstepcounter{equation} \label{eq-fwsub}\label{eq-fwsub-C0}
\hfill
 f_m= \frac{  f_t^2  \hbar^2 \Cbar_{0j }}{4 m_e^\astn m_p^\astn  c^2 r_\lsub^2},
\quad
\Cbar_{0j} =
                (4l^2-1)   
              +\frac{  (2l-1) \eta^2 \sqrt{g_eg_p } }{ C_{1\lsub} } 
              +\frac{3 g_eg_p\eta^2 \cos^2\theta_{\jsub}
}{ C_{1\lsub}}.
\hfill (\ref{eq-fwsub-C0})
}$$
The negative sign in Eqs (\ref{eq-Fpe2}) 
indicates that $F_\jsub$ is attractive. The approximation  is given for $f_{m} >> f_c=1/\g^2 $. For $j=\frac{1}{2}$ ($\l=1$), using the solution values from Sec \ref{Sec-num} gives $f_{m}= \hbar^2 c^2 \Cbar_{0 \hf} /m_e^\astn m_p^\astn c^4 r_{1m}^2 \simeq 6.9 $ which indeed is $ >>$ $f_c=1/\g^2 \simeq 5.7 \times 10^{-11}$.

$\j=0$ yields $\Jast_0=0$, $\Bast^{orb}_p =0$, and hence zero orbit-orbit and orbit-spin  interactions.  The resultant system, even if possible to also form a bound state by the finite spin-spin interaction only,  is not a viable candidate of the neutron, at least because it does not contain a confined antineutrino. For $j \ge \frac{1}{2}$, the three component magnetic forces are each finite and attractive. $\j= \frac{1}{2}$ therefore is the lowest possible (eigen) state of the $e,p$ bound by an attractive magnetic force at the separation scale $\sim 10^{-18}$ m (Sec \ref{Sec-num}), has a confined antineutrino,  and has the correct spin  $\frac{1}{2}$ (Sec \ref{Sec-mag-mmt}). 
                %
The $\j=\frac{1}{2} $ state  is therefore a liable candidate for (the ground state of) the neutron. For $\j=\frac{1}{2} $ ($l=1$), hence $\cos \theta_\hfp=1/\sqrt{3}$, and $m_e = km_p=1.3165 m_p$  (Sec \ref{Sec-mag-mmt}), hence $f_t^2=\frac{(m_e+m_p)^2}{m_e m_p}=\frac{(k+1)^2}{k}\dot{=}4$, Eq (\ref{eq-Fpe2}),  the corresponding interaction potential $\Vast_\hfp$ and Hamiltonian $\Hast_\hfp$ are written as, with Eqs (\ref{eq-Sca-trans}.2a,b) for $m_e^\astn,m_p^\astn$, and $T_\hfp$ given in Sec \ref{Sec2.1.Eq-Mot}.4,
$$\displaylines{
\refstepcounter{equation} \label{eq-Fdag}\label{eq-FLab} \label{eq-Al}
\hfill
\Fast_\hfp(r_1,\theta_{\hfp})  
=-\frac{3A_o \Cbar_{0\hf}}{ \g_e^\astn\g_p^\astn r^4_1},
\quad
A_o=\frac{  \    e^2 \hbar^2 }{ 
          12\pi\ev_0   m_e^\osup m_p^\osup c^2},
\quad  
\Cbar_{0\hf}=
                3
               +\frac{\eta^2 \sqrt{g_eg_p} }{ C_{11}}
              +\frac{\eta^2 g_eg_p}{C_{11} };
\hfill (\ref{eq-Fdag})
\cr
\refstepcounter{equation} \label{eq-Vast} \label{eq-Fpe2-bVa} 
\hfill
\Vast^\ast_\hfp (r^\ast_1,\theta_\hfp)
=-\int_\infty^{r_1} \Fast_\hf (r^\ast,\theta_\hf)d r^\ast
=\frac{r^\ast_{1}  \Fast_\hf (r^\ast_{1},\theta_\hf)}{ 3}
=-\frac{A_o\Cbar_{0\hf}}{\g_e^\astn\g_p^\astn r_1^3}
=-\frac{e^2 \hbar^2 \Cbar_{0\hf} }{
12 \pi\ev_0  m_e^\astn m_p^\astn  c^2 r^\ast{}^3_{1}  };
\hfill (\ref{eq-Fpe2-bVa}) 
\cr
\refstepcounter{equation} \label{eq-H} \label{eq-HNCM}
\hfill
T_{\hfp} =-C_{k\hfp}V_\hfp, 
\quad 
C_{k\hf} 
= \frac{  \g 9 \pi \ev_0 \mcm c^2 r_{1}^\ast }{(\g+1) e^2\Cbar_{0\hf}}; 
\quad
\Hcal_\hfp(r_1^\ast, \theta_\hfp)
=T_\hfp +\Vast_\hfp 
= \Vast_\hfp( 1 - C_{k\hfp}).
\hfill 
(\ref{eq-HNCM})
}$$
In terms of the $e,p$-neutron model, $F_\hfp$ represents the weak interaction force, $V_\hfp$ the corresponding interaction potential, and $H_\hfp$ Hamiltonian of  neutron.

\section{$e,p$ disintegration. Neutron $\beta$ decay } 
\label{Sec-excit}
\setcounter{equation}{39}
Suppose that an afore-described (free) neutron, being initially in stationary state of the Hamiltonian $\Hast_\hfp$ at a time earlier, is now perturbed by an excitation or external-interaction Hamiltonian $\Hast_{I} =\Hast_I^\osup+\Hast_I^1  =\Hast_I^1$ given in the CM frame; evidently $\Hast_I^\osup=0$. So the bound $e,p$  are in the final ($f$) state disintegrated into free $e,p$ separated at an effective infinite distance 
$r_\infty$ such that $ \Vast_{\hfp f} (r_\infty) =0$. The removal of the central force, say acted by $p$ on $e$ in the $-\rb_{1 \hf}$ direction, 
subjects $e$ to a deceleration along that direction and subsequently deceleration or Bremsstrahlung radiation. Provided no exertion of external torque on the neutron, the angular momentum must  be conserved before (being a quantum $S_{\nubar_e}=-\frac{1}{2}\hbar$ in $-z$ direction) and after the deceleration radiation. The electromagnetic radiation emitted is therefore necessarily in the form of  a precessing-orbiting or simply rotational energy flux so as to  convey the same angular momentum quantum $-\frac{1}{2}\hbar$ in the $z$ direction, and the same rotational kinetic energy $T_{\nubar_e}=T_{\hf } $ provided also no exchange of the kinetic energy with the surrounding. The rotational radiation energy flux emitted resembles directly an antineutrino, $\nubar_e$, which is now free. The equation of the foregoing (disintegration) reaction straightforwardly is 
$$n\rightarrow p +e+\nubar_e$$ 
 i.e. the $e,p$ disintegration resembles a neutron $\beta$ decay. The final-state total Hamiltonian has the general form $\Hast_{\hf f} = \Vast_{\hf f} (r_\infty) + \Tcal_{\hf f}^\astn =0+\Tcal_{\hf f}^\astn$. The emitted particles would convey a certain translational kinetic energy  $T_{tr} $ as converted from the total mass difference before (assuming the $n$ being at rest) and after the neutron decay.
 $T_{tr} $ is of a MeV scale (a scale as is also known from $\beta$ decay experiment) which is $<<  T_\hf$ of GeV scale. Omitting this  $T_{tr} $, in the case of $T_{\nubar_e}=T_{\hfp } $, we have  $\Tcal_{\hfp f}^\astn = \Tcal_{\hfp } ^\astn+ T_{tr} \simeq  \Tcal_{\hfp}^\astn$, and $\Hast_{\hfp f} = 0 + \Tcal_{\hfp f}^\astn \simeq  \Tcal_\hfp^\astn$.

The energy condition for the neutron $\beta$ decay to occur is $\Hast_{I} =\Hast_{\hfp f}-\Hast_\hfp$. Substituting in it  the equation for $\Hast_{\hfp f}$ above and (\ref{eq-HNCM}c) for $\Hast_\hfp$ gives
$$\displaylines{
\refstepcounter{equation} \label{eq-HI} \label{eq-GFtrans-a}
\refstepcounter{equation} \label{eq-HI} \label{eq-GFtrans}
\hfill  
\Hast_I 
= \Tcal_\hfp^\astn - (\Vast_\hfp+ \Tast_\hfp^\astn)
=-\Vast_\hfp
=\frac{A_o \Cbar_{0\hf}}{ \g_e^\astn \g_p^\astn r_1^3} \quad {\rm or} \hfill (\ref{eq-GFtrans-a})
\cr
\hfill  
\Gast_\Fsub
=\Hast_I \lf(\frac{4 }{3}\pi r_1^3\rt)  
=\frac{A_o C_{0\hf} }{\g_e^\astn \g_p^\astn}
=\frac{e^2 \hbar^2 C_{0\hf} }{
12 \pi\ev_0  m_e^\astn m_p^\astn  c^2  }
= \frac{e^2\hbar^2 C_{0\hf}}{48 \pi \ev_0 \mr^2 c^2}, 
\quad C_{0\hf}=\frac{4 \pi \Cbar_{0\hf}}{3},
\hfill
(\ref{eq-HI})
}$$ 
where $\lf(\frac{4 }{3} \pi r_1^3\rt)$ is the volume in which the electron is confined about the proton; the last of Eqs (\ref{eq-HI}a) is given after substituting the relation $m_em_p=\mr \mcm =\mr (\frac{(k+1)^2}{k}\mr) \dot{=} 4 \mr^2$ given for  $m_e=km_p=1.3165 m_p$.
By virtue of its physical significance, the product term $\Gast_\Fsub =\Hast_{I} (\frac{4}{3}\pi r_{1}^3)$ in (\ref{eq-HI}b) is directly identifiable with  the CM-frame counterpart of the Fermi constant  $G^{\Lab}_\Fsub$.


$G^{\Lab}_\Fsub$ is experimentally determined (as $G^{exp}_\Fsub$)  from the neutron lifetime, denoted by $\tau^\Lab$ here as is usually measured in the lab frame, on the basis of the quantum theoretical relation $G^{\Lab}_\Fsub \propto 1/\sqrt{\tau^\Lab}$.  The neutron under consideration may be generally in motion,  say at a velocity $u_\cmsublab^\lab $ in a fixed $X$  direction. The (model) neutron mass  in this direction is  (cf Sec \ref{Sec2.1.Eq-Mot}.3) $m_n^\lab\dot{=} \mcm^\lab= \g_\cmsublab^\lab \la \mcm \ra= \g_\cmsublab^\lab \mcm^\osup$, where $ \g_\cmsublab^\lab =(1- (u_\cmsublab^\lab)^2/c^2)^{-1/2}$. Its component masses are formally $m_e^\lab =   (\g_\cmsublab^\lab{})^\ka m_e$, $m_p^\lab = (\g_\cmsublab^\lab)^\ka m_p$, i.e. each in effect augmented by a factor $(\g_\cmsublab^\lab)^\ka $, where  $\ka$ is a certain (positive) exponent  resulting from the mapping of $u_\cmsublab^\lab$ onto the instantaneous interaction direction of $e,p$. So Eqs (\ref{eq-HI}a), here re-written from its original form as 
$G_\Fsub 
=\frac{e^2 \hbar^2 C_{0\hf} }{12 \pi\ev_0  m_e^\astn m_p^\astn  c^2  }
=\frac{A_1}{ m_em_p}$, $A_1= \frac{e^2 \hbar^2 C_{0\hf} }{12 \pi\ev_0    c^2  }$, transformed to the lab frame is formally
$$\displaylines{
\refstepcounter{equation} \label{eq-GFlab}
\hfill
G_\Fsub^\lab 
= \frac{A_1}{m_e^\lab m_p^\lab  } = \frac{A_1}{ (\g_\cmsublab^\lab)^{2\ka} m_e m_p}  
=\frac{G_\Fsub}{ (\g_\cmsublab^\lab)^{2 \ka} }
\propto \frac{1}{ (\g_\cmsublab^\lab)^{2 \ka}  \sqrt{\tau^\osup }}
=\frac{1}{  \sqrt{   \tau^\lab    }  } 
\hfill (\ref{eq-GFlab})
}$$
where $\tau^\osup$ denotes the lifetime of a neutron at rest. (\ref{eq-GFlab}) suggests that for a fast moving neutron such that $u_\cmsublab^\lab{}^2/c^2 >0$ and  $\g_\cmsublab^\lab>1 $ appreciably each, the neutron life time $ \tau^\lab =(\g_\cmsublab^\lab)^{2 \ka} \tau^\osup $ will appear appreciably "dilated", as the result of a reduced internal (magnetic) interaction strength, or reduced Fermi constant. For a neutron at rest of major concern in this paper, $G_\Fsub $ identifies with $G_\Fsub^\lab$ as measured for a rest or slow-moving neutron. We shall continue to speak of $G_\Fsub$.

Multiplying $\frac{ 137 \times 12 \mr^2 c^2}{\hbar^3 c C_{0 \hf}}$ on its both sides, rearranging, the last of Eqs (\ref{eq-HI}a) is written as
$$\displaylines{
\refstepcounter{equation} \label{eq-alp-1}
\refstepcounter{equation} \label{eq-alp}
\hfill
\frac{G_\Fsub (137 \times 12   \mr^2  /C_{0\hf}) c^2 }{  \hbar^3 c }
 =  \frac{137  e^2 }{4\pi \ev_0 \hbar c}; \quad  {\rm or}  
\hfill (\ref{eq-alp-1})
\cr
\hfill
\frac{G_\Fsub M_{\ef}^2 c^2 }{\hbar^3 c } =\frac{g_{neu}^2 }{ \hbar c},
\quad
M_{\ef} = \lf(\frac{137 \times 12   \mr^2 }{C_{0 \hf}}\rt)^{1/2}
= \frac{40.546 \mr}{\sqrt{C_{0\hf}}} \dot{=}\frac{23.043 m_p}{\sqrt{C_{0\hf}} },  
 \quad 
g_{neu}^2=   \frac{137  e^2 }{4\pi \ev_0 }; 
\hfill (\ref{eq-alp})
}$$ 
or $G_\Fsub=\frac{g_{neu}^2 (\hbar c)^2     }{M_\ef^2 c^4}$. $M_\ef$ is an effective mass;  for the last of Eqs (\ref{eq-alp}b) 
$m_e=1.3165 m_p$ is used.
$\frac{e^2 }{4\pi \ev_0 \hbar c}=\frac{g_\Hsub^2}{\hbar c} =\frac{1}{137} =\a_\Hsub(=\frac{\vel_{1\Hsub} }{c}) $ corresponds to the fine structure constant of the hydrogen atom (and $\vel_{1\Hsub}$  the orbiting velocity of electron relative to proton thereof). 
So the right side of (\ref{eq-alp-1}) is unity, $ \frac{137 \times e^2 }{4\pi \ev_0 \hbar c} =1$. 
The fine structure constant for the model neutron is accordingly defined by  $\a_{neu}=\vel_\hfp/c$; using $\vel_\hfp\dot{=} c$ from Sec \ref{Sec-num} in it 
gives $\a_{neu} \dot{=}1 $. Based on its unity value,  and on the physical significance of the dynamical variables, say $\frac{g_{neu}^2 }{ \hbar c}$ on its right side compared to $\frac{g_\Hsub^2}{\hbar c} $, Eq (\ref{eq-alp}a) is immediately identifiable as the equation for $\a_{neu}$,
$$\displaylines{
\refstepcounter{equation} \label{eq-alp-3}
\hfill
\a_{neu}=\frac{\vel_\hf}{c}\equiv 137 \a_\Hsub(\dot{=}1)= \frac{g_{neu}^2 }{ \hbar c} =\frac{G_\Fsub M_{\ef}^2 c^2 }{\hbar^3 c } 
\hfill (\ref{eq-alp-3})
}$$ 
In the GWS theory, $G_\Fsub$ is given  the  formula $G_\Fsub^{\GWS} =   \frac{ g_\wssub^2\sqrt{2} (\hbar c)^2 }{M_{\wsub}^2 c^4}$, where $g_\wssub^2 =\frac{e^2}{8  \ev_0 \sin^2\theta_\wsub }$.
           %
         %
         %
Equating $G_\Fsub^{\GWS}$ with $G_\Fsub$ of (\ref{eq-GFtrans}a) gives a first-principles microscopic expression for $M_\wsub$, accordingly $M_{z}$,
$$\displaylines{\refstepcounter{equation} 
\label{eq-Mw-p}
\hfill  
M_\wsub
=\lf(  \frac{ 3\sqrt{2} \  \pi m_e^\astn  m_p^\astn}{
                2  C_{0\hf} \sin^2 \theta_\wsub} \rt)^{1/2}
=\lf(     \frac{ 3\sqrt{2} \  \pi k  }{2C_{0\hf} \sin^2 \theta_\wsub       }   \rt)^{1/2}  m_p^\astn; 
\quad
\quad M_{z}= M_\wsub/\cos \theta_\wsub
\hfill
(\ref{eq-Mw-p})
}$$
It is well appreciated in the literature that, whilst the $G_\Fsub$  value is absolutely determined by the  lifetime of the neutron in question, the $M_{\wsub}$ value (or $M_\ef$ in Eq \ref{eq-alp-3})  is dependent on the definition or choice of the coupling constant $g_{\wsub}^2$ (or $g^2_{neu}$ in Eq \ref{eq-alp-3}); $g_{neu}^2$ in  (\ref{eq-alp-3}) is uniquely specified for $\vel_\hfp$ is separately known. In terms of the $e,p$-neutron model, $M_\wsub$, or $M_{\ef}$, represents essentially the (reduced) mass of the $e,p$ particles in the binding  and hence manifestly resistive potential field $V_\hfp$.  $V_\hfp$ resembles the Higgs field. The $M_\wsub$, or $M_{\ef}$, of a neutron is highly relativistically augmented (Sec \ref{Sec-num}) over that of a hydrogen, primarily as the result of the relative velocity of $e,p$ within a neutron being so high as to approach $c$. Moreover, the $e,p$ interaction in a neutron is predominately magnetic, and in a hydrogen electrostatic.

\section{Numerical evaluation}\label{Sec-num}
\setcounter{equation}{46}
Equations (\ref{eq-Fdag})--(\ref{eq-GFtrans}) are specified effectively by four independent variables $\aav$,  $r_{1}$, $\vel_\hfp$, and $\g(\vel_\hfp, c^\astn)(=\g_e^\astn\g_p^\astn/ \g_\cmsub^\astn )$ ($\g_\cmsub$ is given if $\g$ is given), to be determined each. We need four independent constraints for quantitatively determining  these and subsequently the remaining dynamical variables. Equation (\ref{eq-EqMt2}d) would ordinarily serve as one basic constraint: it describes a stable state condition under which the central force $\Fb$ on mass $\mr$ counterbalances with the inertial (or here centrifugal) force $\mr \frac{d^2 \rb}{dt^2}$. It may be checked (App \ref{App-Stable-state}) that at a $r_1$ value satisfying Eq (\ref{eq-EqMt2}d), the lifetime of the $e,p$ system however is not an optimum.
This suggests that the neutron candidate $e,p$ system, if opted for a maximum lifetime, is not in stable state. We shall choose the maximum lifetime condition here on the basis that 
a real free neutron indeed  is  "meta" stable only, with a relatively short lifetime 12 m.

In overall view of the basic solutions  from preceding sections, the discussion just made above, and the available key experimental data such as to realistically identify the neutron, we employ 
(i) the quantisation condition (\ref{eq-Lzb-p}a)  for $\Jscr_\hfp$\footnote{ 
The eigen value solution (\ref{eq-Lzb-p}a) represents directly a Heisenberg relation for $\Jscr_\hfp$ and the angular interval $2\pi$, or alternatively in theory the Maupertuis-Jacobi's action integral $2\Tast_\hfp=C_{k\hf}H_\hfp/(C_{k\hf}-1)$ and $\D t_\hfp$; $\D t_\hfp=2\pi r_1/\vel_\hf$. The excitation Hamiltonian $H_I$ is not necessarily conjugated with the $\D t_\hfp$, but generally with some other time interval subjecting to a Heisenberg relation depending on the excitation scheme.
},
(ii) a maximum neutron lifetime,  hence a minimum $G_\Fsub$, 
and 
(ii) the experimental value of the Fermi constant, $G_\Fsub^{exp}$, 
as three basic constraints. 
These are (re-) written as, on dividing (\ref{eq-Lzb-p}a) by $ r_1 \mr^\osup \vel_\hfp$ for (i), and denoting by $r_{1m}$ the extremal value of $r_1$ at which $G_\Fsub$ is a minimum,
$$\displaylines{
\refstepcounter{equation} \label{eq-veldagx}
(i): \hfill  \g =\g_\cmsub^\astn \g^\dagsupa =\frac{\sqrt{3} (\hbar c) c }{ 2 \mr^\osup c^2 r_1 \vel_\hfp}= \frac{D_o c }{r_1  \vel_\hfp },
\quad D_o=  \frac{\sqrt{3}\hbar c }{2 \mr^\osup c^2} 
\hfill (\ref{eq-veldagx})
\cr
             \refstepcounter{equation} \label{eq-r1HR} \label{eq-r1ex2} 
(ii): \hfill  G_\Fsub (r_{1m})= G_{\Fsub. min} 
\hfill  (\ref{eq-r1HR})  
\cr
\refstepcounter{equation} \label{eq-GFsubx1}
(iii):\hfill
G_\Fsub (r_{1m})=G_\Fsub^{exp} 
=1.43585(37) \times 10^{-62} \mbox{ Jm}^3\quad  \mbox{(data from [\citePerkins1982Griffithsetal{}e]})
\hfill (\ref{eq-GFsubx1})
}$$ 
Since (\ref{eq-veldagx}a) suggests that $\g>>1$ for any $\vel_\hfp$ value not too 
far below $c$, and  $c^\astn=c^\Lab$  by the standard assumption, so $\vel_\hfp=c \sqrt{\g^2-1}/\g \simeq c=c^\Lab$, which serves as the fourth constraint here. With this $\vel_\hfp$ value in (\ref{eq-veldagx}a), we obtain (\ref{eq-go-GF}a,b)  below; further with (\ref{eq-EqMt3-c-Y1}.1b) for  $\g_e^\astn \g_p^\astn (=\g_\cmsub^\astn \g=\g_\cmsub^{\astn 2}  \g^\dagsupa)$ and the resultant $\g_\Msub$ from (\ref{eq-go-GF}b) in (\ref{eq-GFtrans}a), with $\g^\dagsupa = 450.96$ given in Sec  \ref{Sec2.1.Eq-Mot}.2 (for $m_e=1.3165 m_p$), we obtain (\ref{eq-go-GFb}) below,
$$\displaylines{
\refstepcounter{equation} \label{eq-g-cmsub-p}\label{eq-go-GF}
\hfill
\g= \frac{D_o}{r_{1}},
\quad
\g_\Msub^\astn= \frac{\g}{ \g^\dagsupa}= \frac{D_o}{ \g^\dagsupa r_1}, 
\hfill (\ref{eq-go-GF})
\cr
\refstepcounter{equation} \label{eq-g-cmsub-p}\label{eq-go-GFb}
\hfill
G_\Fsub
=\frac{A_oC_{0\hf}}{ \g_\Msub^{\astn ^2} \g^\dagsupa }
=\frac{\g^\dagsupa A_o C_{0\hf}r_1^2  }{D_o^2}
=\frac{450.96 A_o C_{0 \hf }r_1^2 }{D_o^2 }.
\hfill (\ref{eq-go-GFb})
}$$
$D_o (=3.3462\times 10^{-13}$ m) and $A_o (=6.2455 \times 10^{-57}$ Jm$^3$) are constants. 
              For evaluating $C_{0 \hf}$ (Eqs \ref{eq-HI}b, \ref{eq-Fdag}c), we shall use the experimental $g$ values of $e,p$, $g_e=2$, $g_p=5.5857$, and $\eta=1/\sqrt{2}$ (\ref{App-red-geom}).
 $G_\Fsub$ of (\ref{eq-go-GFb}) is then solely dependent on $r_1$,$\aav$.
  \begin{figure}[tbhpbhtbh]
\setcounter{figure}{2}
  \vspace{0.  cm}
\begin{center}
\includegraphics[width=0.99\textwidth]{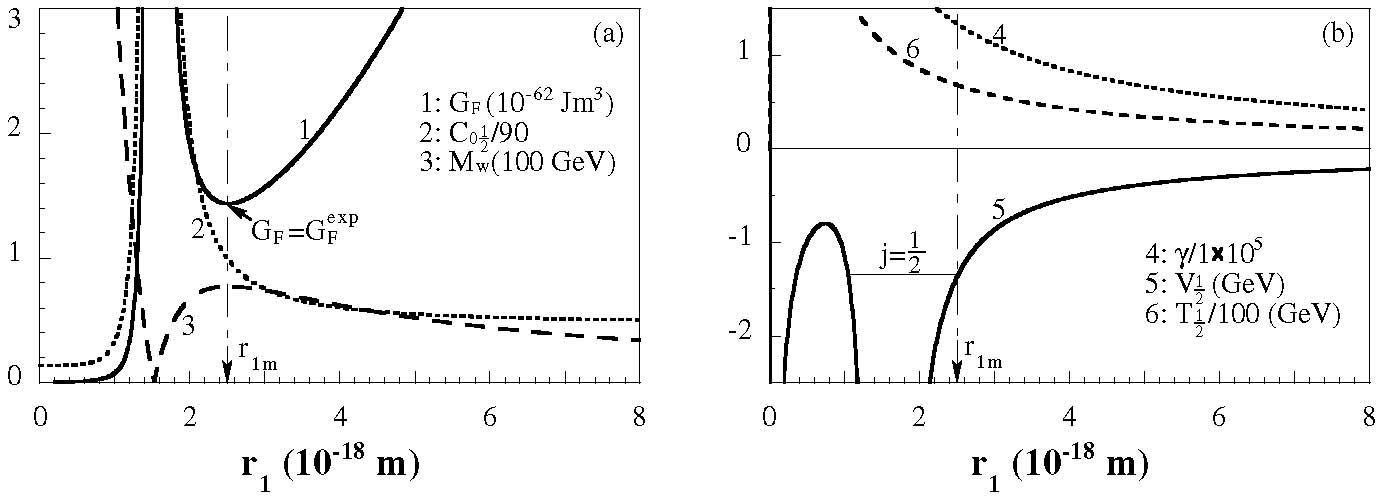}
     \vspace{-.6  cm}
\caption{(a) $G_\Fsub=H_I  r_1^3$, $C_{0\hf}$,   $M_\wsub $ (curves 1,2,3), and (b) $\g$, $V_\hfp= - H_I$, $T_\hfp$ (curves 4,5,6) as functions of $r_1$ computed from Eqs  (\ref{eq-go-GFb}), (\ref{eq-Fdag}c), (\ref{eq-Mw-p}), (\ref{eq-go-GF}a), (\ref{eq-Fpe2-bVa}), (\ref{eq-HNCM}a) for $\aav =1.5391(8) \times 10^{-18}$ m. At $r_1=r_{1m}=2.5369(5) \times 10^{-18}$ m, $G_\Fsub =G_\Fsub^{exp}=1.43585(37) \times 10^{-62}$ Jm$^3$.
}                   
\label{fig2a-Vr3-r1.eps}
\vspace{-.3  cm}
\end{center}
\end{figure}
Characteristically, for a specified $\aav$ value, the $G_{\Fsub }(r_1)$ vs $r_1$  function presents an extremal point at a (uniquely specified) $r_1$, $r_{1m} $, at which $G_{\Fsub }(r_{1m})$ is a minimum satisfying Eq (\ref{eq-r1HR}), as in Fig \ref{fig2a-Vr3-r1.eps}a, although this is not generally equal to $G_\Fsub^{exp}$. $G_{\Fsub }(r_{1m})$ increases monotonically with $\aav$. Computing $G_{\Fsub }(r_{1m})$ as a function of $\aav$ over a range of $\aav$ values, a unique  $\aav$ is found at  $\aav = 1.5391(8) \times 10^{-18}$ m 
 at which $G_\Fsub (r_{1m})=G_\Fsub^{exp} $ satisfying Eq (\ref{eq-GFsubx1}), 
$r_{1m} =2.5369(5)  \times 10^{-18}$ m, $\g= 1.3190 \times 10^5$ (Eq \ref{eq-go-GF}a), 
and $C_{0\hf}=\lf(\frac{4\pi}{3}\rt) \times 3 (1+1.3952+4.6633)=88.69$ (Eqs \ref{eq-HI}b,   \ref{eq-Fdag}c). Note that the $\aav$ value obtained is in accordance with the order of magnitude of the neutron charge radius, $\sim 1.4 \times 10^{-18}$ m, measured by electron-neutron scattering experiment (see also Sec \ref{Sec2.1.Eq-Mot}.3).

With the $\aav$, $r_{1m}$ (hence $C_{0 \hf}$), $ \vel_\hfp, \g $ values obtained, all the remaining dynamical variables and functions may be evaluated. For the fixed  $\aav = 1.5392  \times 10^{-18}$ m value, the $G_\Fsub$,  $C_{0\hf}$, $M_\wsub$ (using the average experimental value $\sin^2 \theta_\wsub =0.23$),
$\g$, $\Vast_\hfp (=-\Hast_\hfp)$, and $T_{\hfp}$ vs. $r_1$ functions, computed from Eqs (\ref{eq-go-GFb}), (\ref{eq-HI}b,\ref{eq-Fdag}c), (\ref{eq-Mw-p}a), (\ref{eq-veldagx}a), (\ref{eq-Fpe2-bVa}), (\ref{eq-HNCM}a) are as shown in Figs \ref{fig2a-Vr3-r1.eps}a,b (curves 1--6). $r_1=r_{1.\min}$ lies as expected in the region where $-\pd V_\hfp(r)/  \pd r=F_\hfp (r)<0$, and $V_\hfp(r)<0$. At $r_1=r_{1m}$, $\Vast_\hfp=-H_I= - 1.310$ GeV, $T_\hfp^\astn(\simeq \mr c^2) =67.36$ GeV,  $H_\hfp (\simeq E_{tot.\hfp} ) = 66.05$ GeV, $M_\wsub=77.23$ GeV. Furthermore specifically, with the $\g$ value in (\ref{eq-go-GF}b),(\ref{eq-EqMt3-c-Y}a),(b),  we obtain  $\g_\Msub^\astn=292.48(3)$, 
$\g_e^\astn= \g_\Msub^\astn \frac{(\mcm^\osup+\Gam)}{2 m_e^\osup}=3.0537(6) \times 10^5$, and $\g_p^\astn= \g_\cmsub^\astn \frac{(\mcm^\osup-\Gam)}{2 m_p^\osup} 
=126.33$, which are $>>1$ each. So the particles $e,p$ within the neutron are travelling at velocities $\vel_e', \vel_p' \simeq c$ measured in their local space and time coordinates $\rb_e,t_e, \rb_p,t_p$ (Eqs \ref{eq-velsp}) in the (non-inertial) CM frame; and so is the total mass $\mcm$ at the CM  relative to $e,p$. The total kinetic energy of $e,p$ in these absolute terms is given by substituting the $\vel_e',\vel_p', m_e(=km_p), m_p$ values in (\ref{eq-Tinvp}.2) as $T_e'+T_p' \dot{=} (k+1) m_p c^2 =2.3165 \g_p m_p^\osup c^2=2.3165  \times  118.53 = 2 \times 137.29$ GeV. Substituting $m_e=k m_p$ into Eqs (\ref{eq-vs}a,b) gives the $e,p$ velocities measured in time $t$, $\vel_e=\frac{m_p }{\mcm} \vel =\frac{c}{k+1}= 0.43 c $, $\vel_p=-\frac{m_e}{\mcm}  \vel =- \frac{kc}{k+1}= -0.57 c $; and in turn the above values into Eq (\ref{eq-Tinv}.2) gives the corresponding total kinetic energy $T_e^\astn+T_p^\astn = km_p ( \frac{c}{k+1})^2 + m_p (\frac{kc}{k+1})^2 = \frac{k}{k+1} m_p c^2=0.56831 \times 118.53 =67.36$ GeV, equal to the solution value for $T_{\hf}=\mr c^2 $ earlier.
 The non-inertial frame motion contributes an amount $(T_e'+T_p' )-(T_e+T_p) =(k+1) m_pc^2 -  \frac{k}{k+1} m_p c^2
=\frac{k^2+k+1}{k+1 } m_p c^2= 1.75 m_p c^2$. The exceedingly large kinetic energy apparently is mainly consumed to contract the size of the system.

The author expresses thanks to emeritus scientist P-I Johansson for his private financial support of the author's research, to Kissemiss Johansson for his joyful companion when the unification researches were carried out, and to Professor C Burdik for providing the opportunity of presenting this wok at the 23rd International  Conference on Integrable Systems and Quantum Symmetries (ISQS23), Tech Univ, Prague, June 2015, during which the author also very much enjoyed interesting discussions with a number of participants.

\begin{appendix}
\appendix

\section[\qquad \qquad \ \  Mapping of the spin current loop on to reduced geometries]{Mapping of the spin current loop on to reduced geometries}
\label{App-red-geom}

Consider here the circular spin current loop of proton 
as projected in the $x_p^s y_p^s$ plane, spinning at tangential velocity $\vel^s_{p_{xy}}$ 
about the  $z_p^s$ axis passing through $\rb_p$ in counterclockwise sense,  i.e. in spin up state as in Sec  \ref{Sec-mag-mmt}.1.  $d q_p= \rho_{p xy} a_p d \phi_p^s$ is a charge element  at $\xib_p(\phi^s_p)$ on it. The magnetic field produced by this spin current loop at a distance $\rb$ from $\rb_p$ has the general form (Biot-Savart law) $\Bb_p^s(\rb)=  \int d \Bb_{p  \theta}^s (\rb') = \int \frac{d q_p \velb^s_{p_{xy}} \times \rb'  }{4\pi \ev_0 c^2 r'{}^3}$,  where $\rb'= \rb + \xib_p(\phi_p^s)$. The integration in algebraic terms is a complex problem. We below reduce the current loop to simpler geometries to facilitate an effective algebraic expression for the field (Eq \ref{eq-Bp1b}, Sec \ref{Sec-force}).

Consider first this field produced at a point at distance $x_p^s=|\rb| $ from $\rb_p$
on the positive $x_p^s$ axis, $\Bb_p^s(x_p^s=|\rb|)$; 
 $x_p^s$  is on the right side to the $z_p^s$ axis as plotted for the parallel $x,y,z$ axes  of $x_p^s,y_p^s,z_p^s$ in Fig \ref{fig-neutron1.eps}b.
The problem has the obvious symmetry that the left-half current loop produces at $x_p^s$ a magnetic field $\Bb_{p \Lsub}^s (\propto \velb_{p xy}^s \times \xb_p^s) >0$ in $+z$ direction, and the right-half a field $\Bb_{p \Rsub}^s  <0$ in $-z$ direction. The total field $\Bb_{p \Rsub}^s - \Bb_{p \Lsub}^s = \Bb_{p }^s (x_p^s=|\rb|)$ is in $-z$ direction. Furthermore, on either half loop the differential $d \Bb_p^s$ fields produced by all charge elements as associated with the $y$-component velocities add up, and with the $x$-component velocities cancel out. So, in so far as the same $\Bb_{p \Lsub}^s, \Bb_{p \Rsub}^s$ are in effect produced,  the left- and right- half spin current loops may be further reduced to two point half-charges $+\frac{1}{2}e,+\frac{1}{2}e$ located at effective distances $-\aav, \aav$ from $\rb_p$ on the $x_p^s$ axis and moving oppositely at velocities $-\velavb_{p xy}^s{}, \velavb_{p xy}^s{}$ in the $-y_p^s,+y_p^s$ directions, where 
$$\displaylines{
\refstepcounter{equation} \label{eq-app-aav}
\hfill
\aav = \eta a, 
\quad 
\velav_{p xy}^s= \velav_p^s \cos \theta_p^s, \quad 
\velav_p^s= \aav \w_p^s =\eta  a   \w_p^s =\eta \vel_p^s,  \quad
\vel_p^s =a \w_p^s;
\hfill (\ref{eq-app-aav})
}$$
$\eta$ is a coefficient to be determined.
We have set $a= a_p=a_e=\frac{1}{2}(a_e+a_p)$ here; so $\aav$ is effective also in its being  scaled by $\eta$ from the average radius $a$ of the $e,p$ charges.
In analogy to $r= r^\osup/\g$, $a$ is contracted\footnote{Contractions in charge radius  and in the wavelength of matter wave  may be comprehended on a common physical ground as follows. Charges and matter waves are distributed energy entities in space each. Two charges or two matter waves by this nature will  inevitably repel with one another when attempting to occupy  same space. Higher velocities facilitate two charges or two matter waves to counterbalance such repulsion to a larger extent, manifesting consequently as the contraction in their dimensions.
}
from its rest value $a^\osup$ formally according to $a=a^\osup/\g_a$; $\g_a$ is a factor analogous
 to $\g$, being an increasing function with $(\vel_p/c)^2$.

By virtue of its physical significance, $\aav$ should be such that the moments of inertia of the right and left point half-charges about the $z_p^s$ axis, $I_{\Rsub}^{point} $, $I_{\Lsub}^{point} (= I_{\Rsub}^{point} )$, are equal to those of the half current loops about $z_p^s$, $I_{\Rsub}^{loop }$, $I_{ \Lsub}^{loop} (=I_{\Rsub}^{loop})$. Let  $m_p^*$ be the effective mass uniformly distributed along the full current and in turn on the full charge, and hence a mass $\frac{1}{2}m_p^*$ on each half loop and in turn at each half point charge. The moments of inertia of one point half charge and one half loop are known as, written for the right  ones,
$$\displaylines{
\refstepcounter{equation} \label{eq-app-Is}
\hfill
I_{\Rsub}^{point} = \frac{1}{2}m_p^*  \aav^2, 
\quad
I_{\Rsub}^{loop }= \int^\pi_0 x^2 d m_p^* 
= 2 \int^{\pi/2}_0  (a_p \cos \phi)^2 \frac{(m_p^*/2)}{\pi} d \phi
= \frac{m_p^* a^2}{4}
\hfill (\ref{eq-app-Is})
}$$ 
 The equality $I_{\Rsub}^{loop } =I_{\Rsub}^{\point }$ gives $\frac{1}{2} \aav^2 (=\frac{1}{2} (\eta a)^2 )= \frac{1}{4}a^2$, so $\eta= 1/\sqrt{2}$. 
Accordingly,  the right point half-charge is associated with a spin angular momentum  
$S_{p \Rsub}
=I_{\Rsub}^{\point } \w_p^s 
(=I_{\Rsub}^{loop } \w_p^s )
=  \frac{1}{2}m_p^*  \aav^2 \w_p^s
= \frac{1}{2}m_p^*  \aav \velav_p^s 
$, 
so that the $z$ component is $S_{p\Rsub z} =S_{p\Rsub}  \cos \theta_p^s = \frac{1}{2}m_p^*  \aav \velav_{p_{xy}}^s $.

Now with respect to a point at a distance $\xb_p^s{}'' $ from $\rb_p$ on the $x''$ axis (Sec \ref{Sec2.1.Eq-Mot}.4) lying in a plane whose normal is along the $z'$ or $z''$ direction at angle $\theta_j$ to the $z$ axis, with $|\xb_p^s{}''| =|\xb_p^s|=|\rb|$, the component spin angular momentum of the right point half-charge perpendicular to $x''$ is the  projection of $S_{p\Rsub z}$ onto the $z''$ or $z'$ direction,
$$\displaylines{\refstepcounter{equation} \label{eq-app-SpzRp}
\hfill
S_{p\Rsub z}'' 
= S_{p \Rsub z} \cos \theta_j
= \frac{1}{2}m_p^*  \aav \velav_{p_{xy}}^s  \cos \theta_j
= \frac{1}{2}m_p^*  \aav  \velav_p^s{}'', 
\quad  
\velav_p^s{}''=\velav_{p_{xy}}^s  \cos \theta_j = \velav_p^s \cos \theta_p^s  \cos \theta_j 
\hfill (\ref{eq-app-SpzRp})
}$$

\section[\qquad \qquad \ \  Stable-state solution for $r_\lsub$]{Stable-state solution for $r_\lsub$} \label{App-Stable-state}
Substituting $\frac{d^2\rb_{\lsub \jsub}}{d t^2}= -\frac{\vel_\jsub^2}{r_\lsub}\hat{r}_{\lsub \jsub}$, $\vel_j$ from (\ref{eq-Lzb}.1)$'$, and $\Fb_{\jsub} $ from (\ref{eq-Fpe2})  into (\ref{eq-EqMt}d) gives $-\mr \frac{(4l^2-1)\hbar^2 \,\hat{r} }{4 \mr^2 r_\lsub^3} = -\frac{f_t^2 e^2 \hbar^2 \Cbar_{0 j} \,\hat{r} }{16\pi \ev_0 m_e m_p c^2 r_\lsub^4}$. Cancelling common factors on both sides, sorting, with $f_t^2=4$ for $m_e=1.3165 m_p$ (Sec \ref{Sec-force}) and $\mcm=\g_{\Msub}^{ (j)} \mcm^\osup$ for $j$th state, we obtain 
$$\displaylines{
\refstepcounter{equation} \label{eq-r1stab} 
\hfill   
r_\lsub
=  \frac{ e^2 \Cbar_{0 j}}{(4l^2-1) \pi \ev_0 \g_\Msub^{ (j)} \mcm^\osup c^2}.
\quad {\rm For} \ j=\frac{1}{2},\ l=1: \ \ 
r_1=  \frac{e^2 \Cbar_{0 \hf}}{3\pi \ev_0 \g_\Msub^{ (\hf)}    \mcm^\osup  c^2};
\hfill 
(\ref{eq-r1stab})
}$$
for $j=\frac{3}{2}$, $l=2$, $r_2=  \frac{e^2 \Cbar_{0 \frac{3}{2} }}{15 \pi \ev_0 \g_\Msub^{(3/2)} {\mcm^\osup} c^2}$, 
$\Cbar_{0 \frac{3}{2}}= 15 + \frac{3 \sqrt{g_eg_p}}{2 C_{12}} +\frac{9 g_e g_p}{10 C_{12}}$
given after Eq (\ref{eq-fwsub-C0}b).
For the $j=\frac{1}{2}$ state, with (\ref{eq-Fdag}b) for $\Cbar_{0 \hf}$, 
(\ref{eq-Bp1b}c) 
for $C_{11}$, and  $\aav=1.53918 \times 10^{-18}$ m, the right side of (\ref{eq-r1stab}b) may be computed as a function of $r_1$. Two numerical solutions 
are found at $r_1= 1.440 \times 10^{-18}$, $ 1.661(7) \times 10^{-18}$ m, 
at which the two sides of (\ref{eq-r1stab}b) are equal. 

\end{appendix}

\vspace{3cm}


\noindent
{\large{Part C}  (Published in {\it  J. Phys.: Conf. Ser.} {\bf 1416} 012043, 2019)} 
\vspace{-1.cm}
\vspace{1.cm}

\title[Quantum Electromagnetic Theory of the Pions, Muons and Their Emitting Particles (I)]{A Quantum Electromagnetic Theory of the Pions, Muons and Their Emitting Particles (I) }

\setcounter{equation}{0}
\setcounter{section}{0}
\setcounter{figure}{0}



\author{J.X. Zheng-Johansson}
\address{Institute of Fundamental Physics Research
 (October, 2019)
} 

\def\Kh{{K_h}}
\def\rhohat{\hat{\rho}}
\def\vehat{\hat{\ve}}

\def\nc{n}
\def\lc{{\mbox{\small{$l$}}}}
\def\lcsub{{\mbox{\tiny{$l$}}}}
\def\lcfoot{{\mbox{\footnotesize{$l$}}}}

\def\nclc{{\nc\lcsub}}

\def\ka{\kappa}
\def\Htr{\mathscr{H}}
\def\Hcal{\mathcal{H}}
               \def\Hcal{\mathscr{E}}

\def\vebar{\bar{\ve}}

\def\jbar{\bar{j}}

\def\rbbar{\bar{\rb}}
\def\Mbar{\bar{M}}
\def\mrbar{\bar{\mr}}
\def\Rbbar{\bar{\Rb}}
\def\onebar{{\bar{1}}}
\def\tobar{{\bar{2}}}
\def\fbar{{\bar{4}}}
\def\tobarsub{{_{\bar{2}}}}
\def\onebarsub{{_{\bar{1}}}}
\def\osup{{\mbox{$\tiny{0}$}}}

\def\etaav{\bar{\eta}}
\def\etd{\tilde{e}}

\def\hm{h}
\def\tm{t}
\def\hy{t}

\def\rlar{\leftrightarrow}

\def\astt{{\mbox{{\tiny{$\wedge$}}}}}
\def\astt{{\mbox{{\tiny{$\bigtriangleup$}}}}}

\def\pn{p_n}
\def\pbn{\pb_n}
\def\En{E_n}
\def\Mn{\mathcal{M}}

\def\Ombar{{\Om\hspace{-0.3cm}^{_{\mbox{-}}} \hspace{0.13cm}}}

\def\pstru{\mathbin{p\mkern-9.2mu\mbox{\scriptsize{$-$}}}}
\def\Pstru{\mathbin{P\mkern-12.5mu^{_{\mbox{$-$}}}}}
\def\pbstru{\mathbin{\pb\mkern-9.2mu\mbox{\scriptsize{$-$}}}}
\def\pstrub{\mathbin{\pb\mkern-9.2mu\mbox{\scriptsize{$-$}}}}

\def\Pbstru{\mathbin{\Pb\mkern-12.5mu^{_{\mbox{$-$}}}}}
\def\Pstrub{\mathbin{\Pb\mkern-12.5mu^{_{\mbox{$-$}}}}}

\def\Pbar{{P\hspace{-0.2cm}^{{\mbox{-}}} \hspace{0.13cm}}}

\def\k{{\mbox{{\scriptsize{$\ve$}}}}}
\def\Tsub{{\mbox{{\tiny{$TP$}}}}}

\def\dagsup{\dagger}
\def\qu{c}
\def\M{M}

\def\excm{\hm^\m{}}

\def\eexcm{\hm^\m{}}

\def\eexc{{\hm}}

\def\ph{{p_{_h}}}
\def\pho{{ph}}

\def\eh{h}
\def\ehxy{{{h}_{xy}}}
\def\hxy{{{h}_{xy}}}
\def\ehxym{{h^\m_{xy}}}
\def\hxym{{h^\m_{xy}}}

\def\ehsub{{h}}

       \def\ehm{h^\m}

\def\ehmsub{{h^\m}}

\def\xysub{{{xy}}}
\def\xy{{{xy}}}
\def\em{{e^\m}}
\def\ep{{e^\p}}

\def\uq{u_q}
\def\uqb{{\bf{u}}_q}

\def\uqr{\xi_{q}}
\def\uqrb{{\pmb{\xi}}_q}
\def\uqz{\zeta_q}
\def\uqzb{{\pmb{\zeta}}_q}

\def\pim{{\pi^\m}}
\def\pip{{\pi^\p}}
\def\pio{{\pi^0}}
\def\ein{{$e^{\m}_{in}$ }}
\def\einsub{{\mbox{$e^{\m}_{(in)}$}}}

\def\str{\star}
\def\astr{{a^\star}}
\def\bstr{{b^\star}}

\def\mbstr{m_b^\str}
\def\mastr{m_a^\str}
\def\MAsub{M^\Asub}
\def\mrAsub{\mr^\Asub}
\def\MBsub{M^\Bsub}
\def\mrBsub{\mr^\Bsub}

\def\Cbar{\hspace{0.02cm}C\hspace{-0.365cm}^{{  \atop  -}\hspace{0.08cm}}{}   }
\def\Cbarp{\hspace{0.02cm}C\hspace{-0.34cm}^{{  \atop  -}\hspace{0.08cm}}{}   }
\def\Ups{\mit{\Upsilon}}
\def\ebar{\bar{e}}
\def\b{\beta}
\def\qt{{q_t}}
\def\mj{{m_j}}
\def\chat{\hat{c}}

\def\nhat{\hat{n}}
\def\uhat{\hat{u}}
\def\rhat{\hat{r}}
\def\xihat{\hat{\xi}}

\def\zetahat{\hat{\zeta}}

\def\ehat{\hat{e}}
\def\zhat{\hat{z}}
\def\yhat{\hat{y}}
\def\xhat{\hat{x}}
\def\phihat{\hat{\vphi}}
\def\vq{\v}
\def\vphihat{\hat{\vphi}}
\def\thetahat{\hat{\theta}}
\def\xphat{\hat{x'}}
\def\xpphat{\hat{x''}}

\def\Jcal{{\mathcal{J}}}

\def\Jcalb{{\pmb{\mathcal{J}}}}

\def\Jorb{{\mathcal{J}}}
\def\Jtr{{\mathcal{J}}}
\def\Jtrb{{\pmb{\mathcal{J}}}}
\def\Tsub{{\mbox{{\tiny{$TP$}}}}}

\def\Asub{{\mbox{{\tiny{$A$}}}}}
\def\Bsub{{\mbox{{\tiny{$B$}}}}}

\def\AA{\mbox{$A${\hspace{-0.25cm}}{}$^{^o}$\hspace{0.2cm}}}

\def\rhob{\pmb{\rho}}

\def\hfmbox{\mbox{$\frac{1}{2}$}}
\def\hfp{{\mbox{\tiny{$1/2$}}}}

\def\hf{{\mbox{\tiny{$\frac{1}{2}$}}}}

\def\hfpp{{^1\hspace{-0.06cm}\mbox{\tiny{/}}\hspace{-0.05cm}{}_2}}
\def\hfb{{}^{{\mbox{\tiny{$\frac{1}{2}$}}}}}

\def\pbar{\bar{p}}
\def\nbar{\bar{n}}

\def\tr{{\mbox{\tiny{$I$}}}}

\def\rarno{\mbox{$\rightarrow \hspace{-0.45cm}/$\hspace{0.45cm}}}

\def\rar{\rightarrow}
\def\lar{\leftarrow}
\def\sear{\searrow}
\def\hrar{\hookrightarrow}

\def\uarsup{{\mbox{\tiny{$\uparrow\hspace{-0.09cm}$}}}}
\def\darsup{{\mbox{\tiny{$\downarrow\hspace{-0.09cm}$}}}}

\def\uar{{\mbox{$\hspace{-0.05cm}\uparrow$}}}
\def\dar{{\mbox{$\hspace{-0.05cm}\downarrow$}}}

\def\point{{\rm{point}}}
\def\j{j}
\def\jsub{{\mbox{\tiny{$j$}}}}

\def\J{J}
\def\Jscr{J}

\def\Jtot{\mbox{$\mathscr{J}$}}

\def\Jb{{\mathbf{J}}}
\def\Jscrb{{\mathbf{J}}}
\def\Jbscr{{\mathbf{J}}}

\def\Jorb{{\mathcal{J}}}
\def\Jtr{{\mathcal{J}}}
\def\Jtrb{{\pmb{\mathcal{J}}}}

\def\Tsub{{\mbox{{\tiny{$TP$}}}}}

\def\db{\pmb{\delta}}

\def\dob{d}
\def\Cb{{\bf{C}}}

\def\d{\delta}
\def\GWS{{\mbox{\tiny{GWS}} }}
\def\Xsub{{\mbox{\tiny{X}} }}
\def\Isub{{\mbox{\scriptsize{I}} }}
\def\Rsub{{\mbox{\tiny{R}} }}
\def\Lsub{{\mbox{\tiny{L}} }}
\def\Ssub{{\mbox{\tiny{S}} }}

\def\rsub{{{_o}}}
\def\vtheta{\vartheta}
\def\pbf{{\bf{p}}}
\def\neu{{\mbox{\tiny{Neu}} }}
\def\rbav{\bar{\rb}}
\def\Fbav{\bar{\Fb}}
\def\fb{{\bf{f}}}

\def\Rsub{{\mbox{\tiny{R}} }}
\def\Lsub{{\mbox{\tiny{L}} }}
\def\HRsub{{\mbox{\tiny{H}} }}

\def\in{{in}}
\def\osup{{\mbox{\tiny{$0$}} }}
\def\inTo{{\mbox{\tiny{in$T_o$}} }}

\def\Ksub{{\mbox{\tiny{$K$}} }}
\def\Lamsub{{\mbox{\tiny{$\Lam$}} }}
\def\exto{{o}}
\def\ex{{eq}}
\def\eq{{eq}}
\def\exti{{1}}
\def\ext{{ext}}

\def\sigb{{\pmb{ \sigma }}}

\def\scat{{\rm{scat}}}
\def\tot{{     \mbox{\tiny{tot}} }}
\def\ef{{\rm{ef}}}
\def\Lrw{\Longrightarrow}
\def\nubar{{\bar{\nu}}}
\def\nubarmu{{\bar{\nu}_{\mu}}}

\def\numu{{\nu_{\mu}}}
\def\numup{{\nu_{\mu}'}}

\def\nubarem{{\bar{\nu}_{e}}}
\def\nuepr{{\nu_e'}}
\def\nuep{{\nu_{e^+}}}
\def\nue{{\nu_{e}}}
\def\nubare{{\nubar_e}}
\def\nubarep{{\nubar_e'}}

\def\orb{{orb}}
\def\Mcm{M}
\def\mcm{M}
\def\velcm{U}
\def\Lambar{\overline{\Lam}}

\def\gbar{\gamma}

\def\Vol{{V\hspace{-0.25cm}^{_{\mbox{-}}} \hspace{0.01cm}}_o}

\def\l{l}
\def\mn{{m_\lsub}}
\def\lsub{{     \mbox{\tiny{$l$}} }}
\def\isub{{     \mbox{\tiny{$1$}} }}

\def\Ocal{\mathcal{O}}
\def\Rcal{\mathcal{R}}
\def\Ycal{\mathcal{Y}}
\def\Tcal{\mathcal{T}}

\def\Rbcal{{\pmb{\mathcal{R}}}}

\def\sb{{\bf{s}}}

\def\velsp{\vel_{s_p}}
\def\velse{\vel_{s_e}}

\def\Db{{\bf{D}}}
\def\velop{\velavp}
\def\veloe{\velave}

\def\velsp{\vel^s_{p}}
\def\velse{\vel^s_{e}}
\def\velavp{\bar{\vel}^s_{p}}
\def\velsavp{\bar{\vel}^s_{p}}
        \def\velspav{\bar{\vel}^s_{p}}  
        \def\velave{\bar{\vel}^s_{e}}
\def\velsave{\bar{\vel}^s_{e}}
         \def\velseav{\bar{\vel}^s_{e}}

\def\velbavp{\bar{\velb}^s_{p}}
\def\velbave{\bar{\velb}^s_{e}}

\def\velbaveb{\bar{\velb}^s_{e}}

\def\velav{\bar{\vel}}
\def\velavb{\bar{\pmb{\vel}}}
\def\velbav{\bar{\pmb{\vel}}}

\def\aavp{\bar{a}_{p}}
\def\apav{\bar{a}_{p}}
\def\aeav{\bar{a}_{e}}
\def\aave{\bar{a}_{e}}
\def\aav{\bar{a}}
\def\aavb{\bar{{\bf{a}}}}

\def\aop{\bar{a}_{p}}
\def\aoe{\bar{a}_{e}}
\def\amp{a_{m_p}}
\def\ame{a_{m_e}}
\def\aqp{a_{p}}
\def\aqe{a_{e}}
\def\ws{\w^s}
\def\wsp{\w^s_{p}}
\def\wse{\w^s_{e}}
\def\wsa{\w^s_\a}

\def\mubar{\bar{\mu}}
\def\mumu{\mu\mu}
\def\mum{{\mu^\m}}
\def\mup{{\mu^\p}}

\def\musub{\mu}
\def\musa{\mu^s_\a}

\def\velsa{\vel^s_\a}

\def\Ls{L^s}
\def\Lsp{L^s_{p}}
\def\Lse{L^s_{e}}
\def\mus{\mu^s}
\def\musp{\mu^s_{p}}
\def\muse{\mu^s_{e}}
\def\Lorb{L}
\def\worb{\w}
\def\muorb{\mu}

\def\Mu{\mathscr{M}}
\def\gb{{\bf{g}}}
\def\zuni{\hat{z}}
\def\Zuni{\hat{Z}}

\def\runi{\hat{r}}
\def\thuni{\hat{\phi}}
\def\Runi{\hat{R}}
\def\Thuni{\hat{{\mit\Phi}}}

\def\Bbext{\Bb_{ext}}
\def\Fbext{\Fb_{ext}}
\def\mrm{{\mathbin{\mu\mkern-5.6mu\mbox{-}}}}

\def\rvec{\vec{r}}
\def\Rvec{\vec{R}}

\def\suf{{}_{}}
\def\labsup{{     \mbox{\tiny{IL}} }}
\def\lab{\labsup}
\def\Lab{     {\mbox{\tiny{L}}} }
\def\Labi{     {\mbox{\tiny{       $L_1$      }}} }

\def\arssup{\star}
\def\arsup{\star}
\def\ars{\arsup}



\def\FbIL{{\bf{F}}}
\def\BbIL{{\bf{B}}}
\def\BfbIL{{\bf{B}}}
\def\MIL{M}
\def\RIL{{\sf{R}}}
\def\RbIL{{\bf{R}}}

\def\Fbicm{{\pmb{\sf{F}}}}
\def\Bbicm{{\pmb{\sf{B}}}}
\def\Bfbicm{{\pmb{\sf{B}}}}
\def\Ebicm{{\pmb{\sf{E}}}}
\def\Lbicm{{\pmb{\sf{L}}}}
\def\Sbicm{{\pmb{\sf{S}}}}
\def\Lbicm{{\pmb{\sf{L}}}}
\def\Vicm{{\sf{V}}}
\def\Hicm{{\sf{H}}}
\def\Eicm{{\sf{E}}}
\def\Ticm{{\sf{T}}}
\def\Ficm{{\sf{F}}}
\def\Licm{{\sf{L}}}

\def\Fbast{{\pmb{\sf{F}}}}
\def\Bbast{{\pmb{\sf{B}}}}
\def\Bfbast{{\pmb{\sf{B}}}}
\def\Ebast{{\pmb{\sf{E}}}}
\def\Efbast{{\pmb{\sf{E}}}}
\def\Lbast{{\pmb{\sf{L}}}}
\def\Sbast{{\pmb{\sf{S}}}}
\def\Lbast{{\pmb{\sf{L}}}}
\def\Sbast{{\pmb{\sf{S}}}}
\def\Jbast{{\pmb{\sf{J}}}}

\def\Wb{{\bf{W}}}

\def\Vast{{\sf{V}}}
\def\Hast{{\sf{H}}}
\def\East{{\sf{E}}}
\def\Tast{{\sf{T}}}
\def\Fast{{\sf{F}}}
\def\Last{{\sf{L}}}
\def\Sast{{\sf{S}}}
\def\Bast{{\sf{B}}}
\def\Mast{{\sf{M}}}
\def\Jast{{\sf{J}}}
\def\Uast{{\sf{U}}}
\def\Gast{{\sf{G}}}

\def\Vdag{{V^\dag}}
\def\Hdag{{H^\dag}}
\def\Edag{{E^\dag}}
\def\Tdag{{T^\dag}}
\def\Fdag{{F^\dag}}
\def\Ldag{{L^\dag}}
\def\Sdag{{S^\dag}}
\def\Bdag{{B^\dag}}
\def\Mdag{{M^\dag}}
\def\Jdag{{J^\dag}}
\def\Udag{{U^\dag}}

\def\Fbdag{ {{\mathbf F}^\dag} }
\def\Bbdag{{{\mathbf B}^\dag} }
\def\Bfbdag{{{\mathbf B}^\dag} }
\def\Ebdag{{{\mathbf E}^\dag} }
\def\Efbdag{{{\mathbf E}^\dag} }
\def\Lbdag{{{\mathbf L}^\dag} }
\def\Sbdag{{{\mathbf S}^\dag} }
\def\Lbdag{{{\mathbf L}^\dag} }
\def\Sbdag{{{\mathbf S}^\dag} }
\def\Jbdag{{{\mathbf J}^\dag} }

\def\Ecal{{\mathcal{E}}}

\def\astsup{{}}
\def\ast{\astsup}

\def\astn{{*}}
\def\astp{{*}}

\def\ab{{\pmb{\mbox{$a$}}}}
\def\acb{{\bf{\mbox{\sf{a}}}}}
\def\ac{{\mbox{\sf{a}}}}

\def\dagsup{\dagger}


\def\mr{{\mathscr{M}}}
\def\mrb{{\pmb{\mathscr{M}}}}

\def\mrsub{{\mbox{{\tiny{$\mr$}}}}}

\def\mrrr{{\mathscr{M}^*}}
\def\mrr{{\mathscr{M}^*}}
          \def\gar{\g^*}
\def\garr{\g^*{}}
\def\velrr{\vel^\ddagger{}}
        \def\velr{\vel^\ddagger{}}
     \def\velbrr{\velb^\ddagger{}}

     \def\trr{t^*{}}
     \def\rbrr{{\bf{r}^*}}
\def\rbr{{\bf{r}^*}}

\def\rvecrr{{\vec{r}^*}}
 \def\rvecr{{\vec{r}^*}}

\def\mr{{\mathscr{M}}}
\def\mrdag{{\mathscr{M}^*\dagger}}
\def\mrlab{{\mathscr{M}}}
\def\gadag{\g^\dagger}
\def\gdag{\g^\dagger}
\def\veldag{\vel^\dagger{}}
       \def\velbdag{\velb^\dagger{}}
\def\velrr{\vel^\ddagger{}}
     \def\velbrr{\velb^\ddagger{}}

\def\rbdag{{\bf{r}^\dagger}}

\def\tdag{t^\dagger{}}
\def\rbdag{{\bf{r}^\dagger}}

\def\rlab{r}
\def\glab{\g}
\def\galab{\g}
\def\velblab{\velb}
\def\vellab{\vel}
\def\mrlab{{\mathscr{M}}}
\def\rblab{{\bf{r}}}
\def\rveclab{{\vec{r}}}
\def\rveclab{{\vec{r}}}

\def\rb{{\bf r}}
\def\tlab{t}
\def\garcmf{\g_{\lab:\cmsub}}

\def\aba{{\bar{a}}}
\def\rbar{{\bar{r}}}
\def\m{{\mbox{-}}}
\def\mpr{{\mbox{-}' }}

\def\ubscr{\pmb{\mathscr{U}}}
        \def\ubscrq{{\pmb{\mathscr{U}}\hspace{-0.14cm}}_q}
         \def\ubscrqsq{{\pmb{\mathscr{U}}\hspace{-0.14cm}}_q{\hspace{-0.05cm}}}

\def\N{{\mbox{{\small{$N$}}}}}

\def\Nsub{{\mbox{{\tiny{$N$}}}}}

\def\Vvqo{V_{\v q0}}
\def\Vvq{V_{\v q}}
\def\mub{\pmb{\mu}}
\def\taub{\pmb{\tau}}
\def\tbar{\bar{t}}
\def\thetab{\pmb{\theta}}
\def\phib{\pmb{\phi}}
\def\Phimb{\pmb{\Phim}}

\def\wb{\pmb{\w}}
\def\mb{{\mathbf{m}}}

\def\lep{l}
\def\Rbb{\mathbb{R}}
\def\Kbb{\mathbb{P}}
\def\R{r_{max}}
\def\Pscr{\mathscr{P}}
\def\Hscr{\mathscr{H}}
\def\Vscr{\mathscr{V}}
\def\Tscr{\mathscr{T}}
\def\Lscr{\mathscr{L}}

\def\Ds{\mathscr{D}}

\def\Xim{{\mit{\Xi}}}

\def\rw{\rightarrow}
\def\jm{{j\mu}}
\def\kp{{j'}}
\def\muk{{\n'}}
\def\Nssk{\Nss_1}
\def\p{{\mbox{\footnotesize{$+$}} \hspace{-0.03cm}}}
\def\psc{{\mbox{\scriptsize{$+$}} \hspace{-0.03cm}}}
\def\pe{\p e}
\def\pq{\p q}
\def\H{{a_{\Sigsub} \hspace{-0.05cm}}}
\def\Hn{a_{n\Sigsub}}
\def\Ia{\mathcal{A}}
\def\hpbar{\abar}
\def\hpbars{\hpbar}
\def\hp{a}
               \def\abar{{a\hspace{-0.18cm}\mbox{{\small $^{_{_{-}}}$}}\hspace{-0.07cm}}}
                \def\abars{{\hspace{-0.03cm}a\hspace{-0.1cm}\mbox{\small{-}}\hspace{0.0cm}}}
\def\abar{{a\hspace{-0.18cm}\mbox{{\small $^{_{_{-}}}$}}\hspace{-0.07cm}}}
\def\Qcal{\mathbin{{Q}\mkern-8.5mu^{_{\mbox{\small{$\dash$}}}}\hspace{-0.04cm} }}

\def\Sa{{\mathfrak{S}}}
\def\nfrak{{\mathfrak{N}}}

\def\h{h}
\def\taubar{\mathbin{{\tau}\mkern-10.3mu_{^{{}^{{\mbox{\tiny{$-$}}}}\hspace{-0.10cm}}} }}
\def\ho{\eta_0}
\def\hjn{\hp_{\jn}}
\def\Wst{{\mathbin{\Omega\mkern-4.1mu^{_{\mbox{\footnotesize{-}}}}}\hspace{-0.04cm}}}
\def\Wsts{{\mathbin{\Omega\mkern-4.2mu^{_{\mbox{\scriptsize{-}}}}}\hspace{-0.06cm}}}
\def\Wstsup{{\mathbin{\Omega\mkern-5.mu^{_{\mbox{\scriptsize{-}}}}}}}
\def\Wstt{{\mathbin{\Omega\mkern-3.5 mu^{_{\mbox{\scriptsize{-}}}}}}}  


\def\Nstat{{\mathcal{N}}}

\def\Nsts{{\Wsts}}
\def\Nst{{\Wst}}
\def\Nss{{\Wst}}
\def\Nstsup{{\Wstsup}}

\def\nsig{{\Sigsub\n}}
\def\nPi{{\n_{^{_\Pi}}}}

\def\pjn{p_{j\n}}
\def\jn{{jn}}
\def\xjn{{j\n}}
\def\Pcaln{\Pcal_{n}}
\def\Pcalbarn{ \overline{\Pcal_\n}}

\def\Pcalens{\Pcal_{\ens}}
\def\Pens{\Pcal_{\ens}}
\def\Pensm{\Pcal_{\ens,max}}
\def\Nstgm{\Nst_{i.g.m}}
\def\Nstam{\Nst_{i.a.m}}
\def\quadd{\ \ }
\def\qe{{q}}

\def\la{\langle}
\def\ra{\rangle}
\def\mrsub{{\mbox{{\scriptsize{$\mr$}}}}}
\def\Msub{{_M}}

\def\cmsub{{        {_{\mbox{{\tiny{CM}}}}} }}

\def\cmsub{   {\mbox{{\tiny{cm}}}} }

\def\icmsub{{\mbox{{\tiny{ICM}}}}}

\def\ncmsub{{\mbox{{\tiny{NCM}}}}}
\def\Neusub{{\mbox{{\tiny{Neut}}}}}

\def\pmsub{{\mbox{{\tiny{$\pm$}}}}}
\def\Hsub{{\mbox{{\tiny{H}}}}}

\def\TPsup{{\mbox{{\tiny{$TP$}}}}}
\def\TPsub{{\mbox{{\tiny{$TP$}}}}}

\def\Hsub{{\mbox{{\tiny{$H$}}}}}

\def\Fsub{{\mbox{{\tiny{$F$}}}}}
\def\Larmsub{{\mbox{{\tiny{$Larm$}}}}}

\def\Lsub{{\mbox{{\tiny{$L$}}}}}
\def\Sigsub{{\mbox{{\tiny{$\Sigma$}}}}}
\def\Wsub{{\mbox{{\tiny{$W^\pmsub$}}}}}
\def\wsub{{\mbox{\tiny{w}}}}

\def\wsubi{{\mbox{\tiny{$W_1$}}}}
\def\wpmsub{{\mbox{\tiny{$W^\pmsub$}}}}
\def\wsubpm{{\mbox{\tiny{$W^\pmsub$}}}}
\def\wpmsub{{\mbox{\tiny{$W^\pmsub$}}}}
\def\wssub{\wsub}

\def\wsubo{{\mbox{\tiny{$0$}}}}
\def\wsubthi{{\mbox{\tiny{$\theta_1$}}}}

\def\Nsub{{\mbox{{\tiny{$N$}}}}}
\def\rep{{rep}}
\def\Sig{{\mit\Sigma}}
\def\bi{b^{i}}
\def\i{i}
\def\n{\nu}
\def\uscr{\mathscr{U}}
\def\uscrdotbar{\bar{\dot{\mathscr{U}}}}
\def\vac{{\rm{vac}}}
\def\DV{V_{aq}{}}
\def\Vd{\widetilde{V}}
\def\DF{F_{aq}{}}
\def\DFb{\Fb_{aq}{}}

\def\Vcal{{\mathscr{V}}} 
 \def\V{V} 

\def\Bsub{{\mbox{\scriptsize{B}}}}

\def\Csub{{\mbox{\scriptsize{C}}}}
\def\Csubti{{\mbox{\tiny{C}}}}
\def\Dsub{{\mbox{\scriptsize{D}}}}

\def\Nsub{{{\mbox{\tiny${N}$}}}}
\def\Hsub{{{\mbox{\tiny${H}$}}}}

\def\Hbar{\bar{H}}
\def\pbar{\bar{p}}

\def\exc{{\rm exc}}
\def\ext{{{\rm ext}}}
\def\mini{0}

\def\Pcal{{\mathcal{P}}}
\def\Pcalb{{\pmb{{\mathcal{P}}}}}

\def\bav{{\bar{b}}}
\def\v{{\rm v}}
\def\vrm{\vel_{t}{}}
\def\vit{\vrm}
\def\vrmb{{\bf{v}}}

\def\Hbar{\bar{H}}
\def\D{\Delta}
\def\bcal{b}
\def\bbar{\mathbin{{b}\mkern-9.5mu^{{\mbox{\tiny{$-$}}}}\hspace{-0.00cm} }}
\def\nstat{\nu}
\def\nst{\nu}
\def\engbar{\bar{\eng}}
\def\engobar{\bar{\eng}_0}
\def\psias{\psi}
\def\Phimas{\Phim}
\def\fas{f}
\def\rbb{\as}

\def\La{L}
\def\Ja{J}
\def\as{p}
\def\ioii{{\mbox{\normalsize${\frac{1}{2}}$}}}
\def\Rb{{\bf R}}

\def\xb{{\bf x}}
\def\Xb{{\bf X}}

\def\ub{{\bf{u}}}
\def\hatu{\hat{u}_q}
\def\Nsub{{{\mbox{\tiny${N}$}}}}
\def\Pisub{{{\mbox{\tiny${\mit{\Pi}}$}}}}

\def\q{\bar{q}}
\def\xdot{\dot{x}}
\def\ens{{ens}}
\def\Lcal{\mathcal{L}}
\def\Lcalb{\pmb{\mathcal{L}}}
\def\Tcal{\mathcal{T}}
\def\Kcal{{\mathcal{K}}}
\def\Xcal{{\mathcal{X}}}

\def\Wvel{\Omegavel}
\def\Ncal{{\mathcal{N}}}
\def\Omegavel{\mathbin{{\mit\Omega}\mkern-13.mu^{_{\mbox{$-$}}}\hspace{-0.08cm}{}_d }}
\def\Om{{{\mit{\Omega}}}}
\def\omegavel{{\w\mbox{\hspace{-0.38cm} \vspace{0.15cm}$-$\hspace{-0.02cm}}}}
\def\wvel{\omegavel_d}

\def\Ucal{\mathcal{U}}

\def\Omegavel{\mathbin{{\mit\Omega}\mkern-13.mu^{_{\mbox{$-$}}}\hspace{-0.08cm}{}_d }}
\def\Wvel{\Omegavel}

\def\q{\mathbin{q\mkern-11mu-}}
\def\PE{\mbox{\tiny{{\rm P.E.}}}}
\def\ME{\mbox{\tiny{{\rm M.E.}}}}
\def\QM{\mbox{\tiny{{\rm QM}}}}
\def\Psub{\mbox{\tiny{{\rm P}}}}
\def\TP{{\mbox{\tiny{{\rm TP}}}}}

\def\SM{{\mbox{\tiny{{\rm SM}}}}}
\def\MT{{\mbox{\tiny{{\rm MT}}}}}

\def\ev{\epsilon}

\def\Ci{1}
\def\betamt{{\bf{b}}}
\def\kb{{\bf{k}}}
\def\qb{{\bf{q}}}
\def\kbf{{\bf{k}}}
\def\Kb{{\bf{K}}}
\def\cb{{\bf{c}}}

\def\orar{\overrightarrow}

\def\pbf{{\bf{p}}}
\def\pb{\pbf}

\def\Pbf{{\bf{P}}}
\def\Mbf{{\bf{M}}}
\def\pbf{{\bf{p}}}
     \def\Acal{{\cal{A}}}
\def\Bcal{{I_{{\rm{ex}}}}}
\def\Ccal{{\cal{C}}}
\def\Vp{V}
\def\Ccal{{\cal{C}}}

\def\psipi{\psi_{\p}(1)}
\def\psipii{\psi_{\p}(2)}
\def\psimi{\psi_{\m}(1)}
\def\psimii{\psi_{\m}(2)}

\def\ai{\alpha(1)}
\def\aii{\alpha^{'}(2)}
\def\bi{\beta^{'}(1)}
\def\bii{\beta(2)}

\def\fa{f_r}

\def\Ca{C_a}
\def\fbf{{\bf{f}}}
\def\fb{{\bf{f}}}

\def\Ocal{{\cal{O}}}
\def\psib{{\pmb{\psi}}}
\def\alphab{{\pmb{\alpha}}}
\def\sigmab{{\pmb{\sigma}}}
\def\sig{\sigma}
\def\Eb{{\bf E}}
\def\Bb{{\bf B}}
\def\ke{\kappa}
\def\nabb{{\pmb{\nabla}}}
\def\nablab{{\pmb{\nabla}}}
\def\vir{{\rm vir}}

\def\psitot{\psi}
\def\jb{{\bf{j}}}
\def\vel{upsilon}
\def\ve{\upsilon}

\def\vels{{\hspace{0.1cm}\breve{\hspace{-0.1cm}\vel}}}
\def\velsb{{\breve{\velb}}}
\def\vb{{\bf{v}}}
\def\Imtr{I}
\def\citeUnif{4?}
\def\App{}
\def\Qcal{{\mathcal{Q}}}
\def\Tcal{{\mathcal{T}}}
\def\Cross{Q}

\def\vphilim{f}
\def\ft{{\mathcal{B}}}
\def\vphibar{\mathbin{\varphi\mkern-12.5mu-}}
\def\vphi{\varphi}
\def\vphib{{\pmb{\varphi}}}

\def\med{{\med}}
\def\Mcal{{\mathfrak{M}}}
\def\Mfrak{{\mathfrak{M}}}
\def\Mca{{\mathcal{M}}}

\def\Sb{{\bf{S}}}
         \def\xia{{\mathcal{A}}}
\def\tha{\theta}

\def\xib{\pmb{\xi}}
\def\zetab{\pmb{\zeta}}
\def\nb{\bf{n}}
\def\zb{{\bf{z}}^0}

\def\phiv{\varphi}
\def\Lb{{\bf{L}}}
\def\velsub{_{\vel}}
\def\Jb{{\bf{J}}}
\def\Pb{{\bf{P}}}
\def\Mb{{\bf{M}}}
\def\Zo{{Z^0}}
\def\nablab{{\pmb{\nabla}}}
\def\velb{{\pmb{\vel}}}
\def\veb{{\pmb{\ve}}}

\def\Db{{\bf{D}}}

\def\Ab{{\bf{A}}_a}
\def\Abb{{\bf{A}}}

\def\vel{\upsilon}
\def\ve{\upsilon}
\def\Thm{\vartheta}
\def\Thetam{{\mit{\Theta}}}
\def\Thetamb{ \pmb{{\mit{\Theta}}}}

\def\lb{{\bf l}}
\def\ldb{{\pmb{\ld}}}
\def\ld{\ell}
\def\lp{{\ell_p}}
\def\Lcal{\mathcal{L}}
\def\ellb{{\pmb{\ell}}}
\def\vb{\velb_{t}{}}

\def\Rb{{\bf R}}
\def\pd{\partial}
\def\vphi{\varphi}

\def\psitot{\varphi}
\def\psiR{\widetilde{\psi}}
\def\psiL{\widetilde{\psi}^{{\rm vir}}}
\def\Phim{{\mit{\Phi}}}
\def\PhimR{\widetilde{ {\mit \Phi}}}
\def\PsimR{\widetilde{ {\mit \Psi}}}
\def\PsimL{{\widetilde{ {\mit \Psi}}}^{{\rm vir}}}
\def\a{\alpha}
\def\ap{{\kappa}}

\def\uav{\bar{u}}
\def\D{\Delta}
\def\th{\theta}
\def\r{{\mbox{\tiny${R}$}}}
\def\re{{\mbox{\tiny${R}$}}}
\def\Fmed{F_{{\rm a.med}}}
\def\med{{\rm med}}
\def\Lw{L_{\varphi}}
\def\Fb{{\bf{F}}}

\def\Ef{{\sf{E}}}
\def\Bf{{\sf{B}}}

\def\Efb{{ \pmb{ \Ef} }}
\def\Bfb{{ \pmb{ \Bf} }}
\def\Yb{{\bf{Y}}}

\def\Ac{ \varphi}
\def\Xsub{{\mbox{\tiny${X}$}}}
\def\Ysub{{\mbox{\tiny${Y}$}}}
\def\Zsub{{\mbox{\tiny${Z}$}}}
\def\MTsub{{}}

\def\Ksub{{\mbox{\tiny${K}$}}}
\def\W{{\mit \Omega}}
\def\Wd{\W_d{}}
\def\Nu{{\cal V}}
\def\Nuscr{{\mathscr{V}}}
\def\Nud{\Nu_d{}}
\def\Eng{{\cal E}}
\def\eng{{\varepsilon}}
\def\vep{\varepsilon}
\def\Kmscr{{\mathscr{K}}}
\def\Lscr{{\mathscr{L}}}
           \def\engk{\Kcal}
\def\Acuni{\Ac_{{\Ksub}^\dagsup}^{\dagsup}}
\def\unduni{\Ac_{{\Ksub}^\dagger}^{\dagsup}}
\def\Acauni{\Ac_{{\Ksub}^\ddagsup}^{\ddagsup}}
\def\Acunim{{\Ac_{{\Ksub}^\dagsup}^{\dagsup *}}}
\def\undunim{{\Ac_{{\Ksub}^\dagsup}^{\dagsup}}^*}
\def\Acaunim{{\Ac_{{\Ksub}^\ddagsup}^{\ddagsup *}}}
\def\pd{\partial}
\def\Ad{ {\mit \psi}}
\def\psim{ {\mit \psi}}
\def\Kd{K_d{}}
\def\Xim{{\mit \Xi}}
\def\Sigm{{\mit \Sigma}}

\def\Lam{{{\mit \Lambda}}}

\def\lam{\lambda}
\def\dagsup{{\mbox{\tiny${\dagger}$}}}
\def\ddagsup{{\mbox{\tiny${\ddagger}$}}}
\def\psimKdK{\psim_{\Ksub,\Kdsub}}
\def\wk{1}

\def\w{\omega{}}

\def\om{\omega{}}
\def\omb{\pmb{\omega{}}}
\def\Omb{\pmb{\Omega}}
\def\ombar{\omega{}\hspace{-0.4cm}-\hspace{-0.1cm}}

\def\wrm{{\rm{w}}}
\def\wit{\w_{t}{}}
\def\witb{{\pmb{\it{w}}}}

\def\wdlow{\omega_d }
\def\g{\gamma{}} 
\def\Phimc{{\mathcal C}}
\def\Psim{{\mit \Psi}}
\def\arm{{\rm a}}
\def\brm{{\rm b}}
\def\crm{{\rm c}}
\def\drm{{\rm d}}
\def\erm{{\rm e}}
\def\frm{{\rm f}}
\def\grm{{\rm g}}
\def\hrm{{\rm h}}
\def\lf{\left}
\def\rt{\right}
\def\Kdsub{{\mbox{\tiny${K_d}$}}}
\def\psimkd{\psim_{\kdsub}}
\def\psimKd{\psim_{\Kdsub}}
\def\hquad{ \ \ } 
\def\Taum{{\mit \Gamma}}
\def\Gam{{\mit \Gamma}}

\def\dagsup{{\mbox{\tiny${\dagger}$}}}
\def\ddagsup{{\mbox{\tiny${\ddagger}$}}}




\begin{abstract}

\def\mrm{{\mathbin{\mu\mkern-5.6mu\mbox{-}}}} 
In direct accordance to the overall relevant experimental demonstrations, we represent the charged pion $\pi^\m$ as a heavy electron $h^\m$ in precessional-orbital (P-O) motion at essentially the light speed $c$ about 
 $\nubar_e$-orbit
 of a normal at quantised angle $\pi-\theta_\hfp=-\arccos (1/\sqrt{3})$ to the $z$ axis.  $h^\m$ is the level $N=1$ oscillation of charge $-e$ and its electromagnetic radiation originally generated in the weak potential field of another particle.
 The P-O kinetic energy current and two additional opposite ones  created upon $\pi^\m$ decay represent confined  neutrinos $\nubar_e$, $\nubar_\mu$, $\nu_\mu$. The muon $\mu^\m$ is a $xy$-projected $h^\m$ in two superposing P-O motions along 
$\nubar_e$-, $\nu_\mu$-orbits 
of normals at angles $\pi-\theta_\hfp$, $\theta_\hfp$ to $z$. The $\mu^\m$ (rest) mass is 
a geometric projection of the reduced $\pi^\m$ mass, $M_\mu \dot{=} (M_\pi - M_\numu)\sqrt{\cos \theta_\hfp} = 105.86$ MeV.  The $\mu^\m$ mass is fundamentally predetermined by the mixed two states $m_\lcsub=- 1,+1$  of level $n=2$ above the vacuum in the CM frame of a double heavy positronium produced in a  relativistic $e^\m,e^\psc$ collision, and is {\it ab initio} predicted to be 
$M_\mum= (\frac{3}{4} \frac{ 2 }{ \a}+1)M_e=105.549$ MeV, where $\a=e^2/4\pi\ev_0 \hbar c$. The un-projected $\nc=2$ level gives the bound $\pi^\m$ mass $M_\pi +\Ocal_\eta=  (\frac{2}{\a} +1)M_e  =140.525$ MeV before subtracting a friction term $\Ocal_\eta$.
Their antiparticles $\pi^\psc, \mu^\psc$ and the tauons $\tau^\mp$ can be similarly represented. The remaining unstable elementary particles  can be constructed as composites of two or more 
single charged ones 
in certain spatial quantised P-O motions.
 \end{abstract}

\section{Introduction}

In the Standard Model (SM) description\cite{perkins-20,FrauenfelderHenley91}, elementary particles are divided into $2 \times 6$  leptons, baryons / mesons   made of $2 \times 6$ quarks,  and  five or so force mediators.
Regarding their weak decays, such as the neutron $\beta$ decay $n \rar e^\m p \nubar_e$, the SM quark model assumes that 
the decay product particles  $e^\m, p, \nubar_e$ do not exist until  $n$ decays, and that they are  instantaneously produced and emitted upon the $n$ decay. 
The SM has been successful in reproducing the charges, spins, C, P and T symmetries of the hundreds of observational elementary  particles, and predicting their decay 
rates -absolute or comparative. But the SM has unnatural aspects, concerning the quarks and weak decays in particular. The baryons and mesons are made of quarks rather than the particles they decay into,   which is an abrupt departure from the atomic and nuclear descriptions. Free quarks are never observed in experiment. The SM weak interaction Hamiltonian $H=G/(\frac{4\pi}{3} r^3)$ is a phenomenological construction.  The weak force is not predicted.  The weak, strong, gravitational and electromagnetic forces are not unified.  The basic questions common to the regular particles $e^\m,p$ (being the two only permanent particles - why?) remain outstanding, including the  origin of mass,  the nature of matter  waves, and the cause of gravity.

The Internally Electrodynamic (IED)  particle model 
(see a review and the original references given in \cite{jxzj-ied}) 
 is a complementary approach to  the SM and is beyond the SM. The IED model was developed using overall observational properties of particles as input information, and aimed from the beginning to achieve a unification of the 
         three basic mechanics 
and of the four fundamental forces. In terms of this model, the electron $e^\m$ and the proton $p$ are composed of oscillatory charges $-e$ and $+e$  and their electromagnetic  radiation (EMR) fields in a polarisable dielectric vacuum filled of vacuuons\cite{jxzj-dielec-vac,jxzj-unifbook}. Their distinct masses are  uniquely determined \cite{jxzj-vacpot}  by the quantum oscillation  levels of $-e,+e$  in their vacuum potentials that are asymmetric to $-e,+e$. 
The masses of their antiparticles $e^\m, \pbar$ are  determined because of pair productions.
The remaining single charged particles, including the six  manifestly free forms  $\pi^\mp$, $\mu^\mp$, $\tau^\mp$ and ones in (confined) higher excited  states 
 such as comprising $\rho^0(770)$, are the main subject of investigation of this paper.  We shall show  that,  in accordance to the overall experimental demonstrations, 
 these can be generally each described as a charge $- e$ or $ +e$ oscillation and the resulting EMR originally generated in the weak potential field of another particle, that are as a whole  in certain precessional orbital (P-O) motions. The remaining unstable elementary matter  particles such as $n$, $\Lam$, $\pi^0$, $\rho(770)$ are composites of two or more of  the single charged ones. A formal $e^\m,p$  model of the neutron $n$ and  first principles solutions for spin, magnetic -hence  weak- interaction force, coupling constant, and (unpublished) mixing angle have been achieved in  \cite{jxzj-neutron}. The $\Lam$-hyperon is the simplest baryon emitting a pion $\pim$, but  otherwise a direct analogue of $n$, and  will serve a prototype system for the model constructions in this paper.

The formal representation divides in three sections. In Sec \ref{Sec-pion}, using the  relevant observations as input information we propose the existence of mass states "heavy electrons (or positrons)", originally generated by charges $-e$'s (or $+e$'s) in the weak potential fields of other particles. In terms of this we propose  the structures of the  pions $\pi^\mp$ and  muon $\mu^\mp$. In Sec \ref{Sec-IED-gen}, we present a generalised  first principles description of the masses of the single charged particles such as $\pi^\mp, \mu^\mp$, $e^\m$, $p$, and the composite particles  such as $n,\Lam$. In Sec \ref{Sec-pi-mass-abinitio}, we predict  the existence of a double heavy positronium preceding the rudimentary productions of $\mu^\mp, \pi^\mp$, etc. and, based on its eigen level $n=2$  solutions, we {\it ab initio} predict the  masses of $\mu^\mp$, $\pi^\mp$. 
 Representation of the remaining  composite unstable elementary particles will be described in a separate paper.

\section{Structures of the charged pions and muons}\label{Sec-pion}\label{Sec-muon-model}
\begin{figure}[bptbh]
\refstepcounter{figure} 
\vspace{-1.3cm}
\vspace{0.3cm}
\begin{flushleft}
\includegraphics[width=0.77\textwidth]{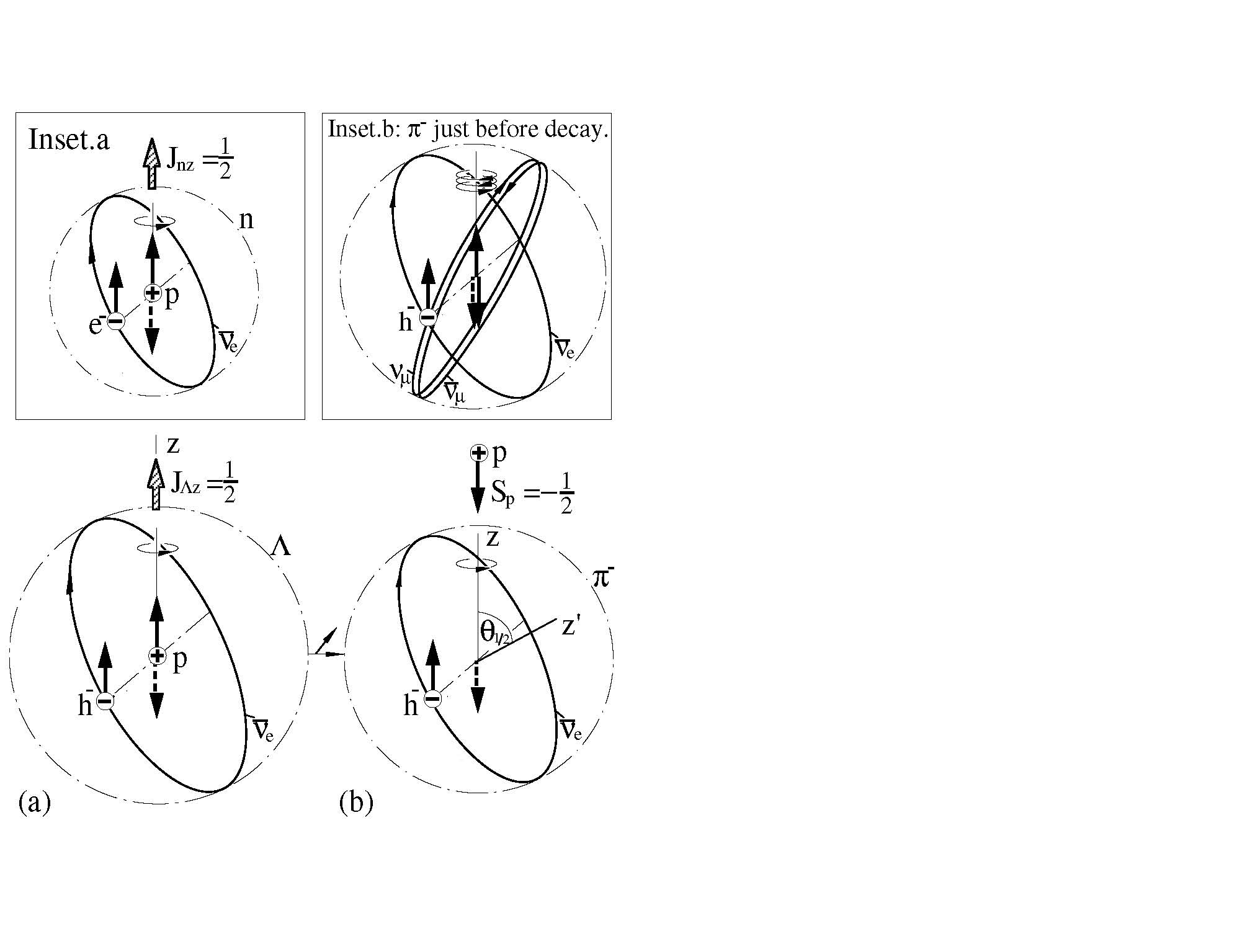}
\end{flushleft}  
\vspace{-2.2cm}
\vspace{0.6cm}
\begin{flushleft}\begin{minipage}[here]{5.7cm} 
{\footnotesize
{\bf Figure \ref{fig-pi-decay-to-mu.eps}.  }
Schematic structures of (a) $\Lam$ in analogy to $n$ (Inset.a), except in place of $e^\m$, in $\Lam$ is a heavy electron $h^\m$ in P-O motion  shown in the $p$ rest frame, and (b) stationary $\pi^\m$; (Inset.b) shows $\pim$ just before decay.
}
\label{fig-pi-decay-to-mu.eps}
\vspace{-0.5cm}
\end{minipage}
\end{flushleft}

\vspace{1.9cm}
\refstepcounter{figure} 
\vspace{-0.5cm}
\vspace{2.7cm}
\vspace{-3.1cm}
\vspace{-9.9cm}
\begin{flushright}
\includegraphics[width=0.60\textwidth]{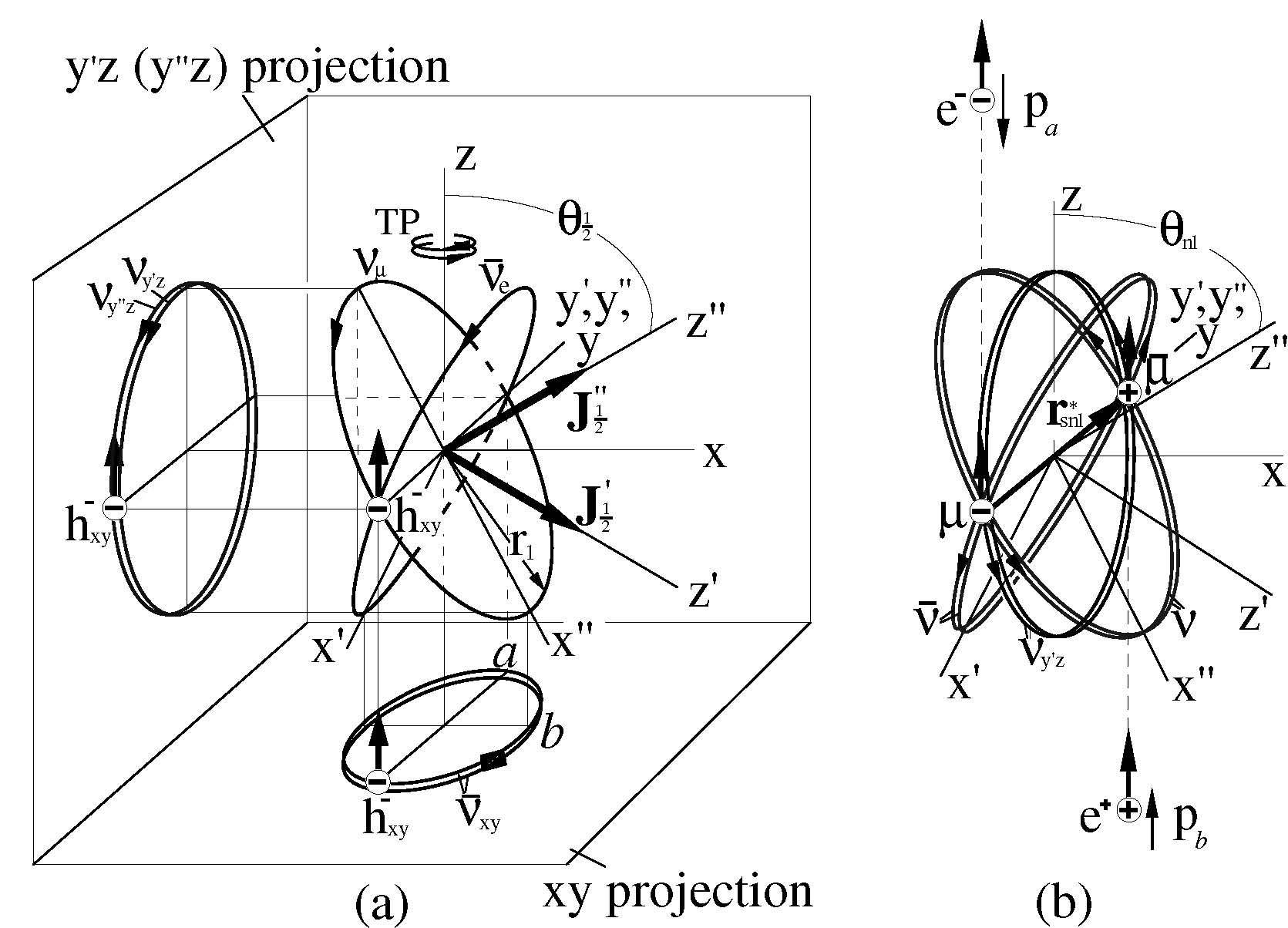}
\vspace{-0.15cm}
          \end{flushright} 
\vspace{-0.49cm}
\vspace{0.5cm}
\begin{flushright}
\hspace{-0.9cm}
\begin{minipage}[here]{8.3cm} 
{\footnotesize {\bf Figure \ref{fig2.mu_orb_mu-+pr.eps}. }
(a) Schematic structure of  $\mu^\m$, which is a $\ehxym$ of   charge $-e$, spin $\frac{1}{2}$, and in two superposing P-O motions along  orbits $\nu_\mu, \nubar_e $ of normals $z',z''$ at  angles $\pi-\theta_{1/2}$, $\theta_{1/2}$ to the $z$ axis. 
(b) $e^\m, e^\p$ near head-on collision at separation $r^\astp_{2,1}$ to form a double heavy positronium, preceding to $\mu^\m \mu^\p$ production. 
}
\label{fig2.mu_orb_mu-+pr.eps}
\label{fig-mu.eps} \label{fig-mu_orbits.eps}
\label{fig-mu_orbits-b.eps} \label{fig-mu-mu+fr_e-e+.eps}
\end{minipage}
\end{flushright}

\vspace{-0.1cm}
\vspace{-0.6cm}
\end{figure}

{\it \ref{Sec-muon-model}.1 The charged pions $\pi^\m, \pip$ \ }  
The simplest composite baryon emitting a charged pion $\pi^\m$, and secondarily a muon $\mu^\m$, is the $\Lam$ hyperon. Observationally, $\Lam$ has an identical charge $0$, spin $\frac{1}{2}$, and analogous decay reaction $\Lam \rar \pi^\m p \rar \mu^\m \nubar_\mu p $ to those of the neutron $n$  ($n\rar e^\m \nubar_e p$), but has a heavier mass 1115.6 MeV. 
These suggest an analogous structure of $\Lam$ 
to $n$ \cite{jxzj-neutron} (Fig \ref{fig-pi-decay-to-mu.eps}a, Inset), except  in place of  $\em$ of  $n$, $\Lam$ has   a "heavy electron" ($h^\m$) in P-O motion relative to $p$ (Fig \ref{fig-pi-decay-to-mu.eps}a).
In accordance to the implication of $\Lam$,  the observational $\pi^\m$ charge $-e$,  spin $s_\pi=0$,  mass $M_\pi=139.569$ MeV, decay reaction $\pi^\m \rar \mu^\m  \nubar_\mu \rar (e^\m \nubar_e \nu_\mu) \nubar_\mu  $   
and rudimentary direct production reaction $e^\m e^\p \rar \rho(770) \rar \pi^\m \pi^\p $, we propose a two-step description of the $\pi^\m$ structure:
(1) There presents a (confined) particle state called  heavy electron, $h^\m$, 
that has a charge $-e$,  spin $\frac{1}{2}$ as the electron $e^\m$, 
and yet is a heavier mass state, the charge $-e$ oscillation and its resultant EMR of a  mass $M_h$, (originally) generated in the weak potential field $V_\hf$ of another particle. 
(2) The pion $\pi^\m$ is a heavy electron $\ehm$, of  spin $S_{\eh z } =\frac{1}{2} \hbar $  in $z$ direction, in  P-O (precessional-orbital) motion at 
essentially the light  speed, $\ve_\hf \dot{=}c$,  along 
$\nubar_e$-orbit 
 of a normal ($-z'$) at quantised angle $\pi- \theta_\hf$ to the $z$ axis (Fig \ref{fig-pi-decay-to-mu.eps}b), in the absence of other particle(s). The P-O kinetic energy current along $\nubar_e$ resembles a confined antineutrino $\nubar_e$, of an apparent rest mass $M_\nue$. Two equal but opposite P-O momentum  currents along  $\nubar_\mu$ $\nu_\mu$ 
(hence confined neutrinos) are generated  momentarily upon  $\pi^\m$ decay in an explosive collision  (Fig \ref{fig-pi-decay-to-mu.eps}b, Inset), these do not contribute to the $\pim$  dynamical variables. By virtual of its anti-symmetric properties, $\pi^\p$ is a heavy positron $h^\p$ in P-O motion along orbit $\nu_e$. The rest mass of $\pi^\m$ (or $\pi^\p$) is  formally  $M_\pi = M_\eh +M_{\nubar_e} $. 

The $z$ component  and the total angular momenta of the P-O motion of $h^\m$ are given by the eigen solutions of bound state $n=2$, $\lcfoot=1$, $j=\lcfoot-\lcfoot^\TPsup=\frac{1}{2}$  in the original $V_\hf$ field (e.g. of $p$ as in $\Lam$) in the lab frame similarly as for $n$ (\cite{jxzj-neutron}),
$J_{\hf z } = J_{\hf}  \cos (\pi-\theta_\hf) =-j\hbar =-\frac{1}{2} \hbar$, 
$J_{\hf }=| \rb_{\hf \pi} \times (m_\pi \veb_{\hf } )| 
(= | \rb_{\hf \Lamsub} \times (\mr_\Lamsub \veb_{\hf \Lamsub} )|)= \sqrt{j(j+1)}|_{j=1/2} \hbar =\frac{\sqrt{3}}{2}\hbar $, 
where $ \cos (\pi-\theta_\hf) =- \frac{1}{\sqrt{3}}$, $m_\pi= \g_\pi  M_\pi$,  $\g_\pi=(1-(\ve_\hf /c)^2)^{-1/2} $ ($\mr_\Lamsub $, is the reduced mass and $r_{1 \Lamsub}$ the orbit radius  of $\pi^\m,p$ comprising $\Lam$).
Those of the total system  $\pi^\m (e_h^\m [\nubar_e])$ are
 $S_{\pi z}=J_{\hf z} +S_{\eh z }  = (-\frac{1}{2} +\frac{1}{2})\hbar  =s_{\pi}\hbar =0 $, $s_{\pi}=0$, and  $S_{\pi}=\sqrt{s_{\pi}(s_{\pi}+1)} \hbar =0$.
The kinetic energy is $T_{\pi}= \frac{\g}{\g+1}  m_\pi \ve_\hf^2 =\frac{\g }{\g+1} J_\hf c/r_1$.

{\it \ref{Sec-muon-model}.2 The muons $\mu^\m, \mup$ \ } In direct accordance to the  observational charge $-e$,  spin $s_\mu=\frac{1}{2}$, mass $M_\mu=105.658$ MeV and decay reaction $\mu^\m \rar e^\m \nubar_e \nu_\mu$ of $\mu^\m$, to its specific  production from $\pi^\m\rar \mu^\m \nubar_\mu$ and  the $\pi^\m$ structure of Sec \ref{Sec-pion}.1, we propose: By an apparent  loss of mass energy of its motion in the $y'z$ plane(s),  $\eh^\m$ transforms to a  $xy$-projected  mass state  $\ehxym$ of mass $M_{\ehxy}$, and of an unchanged charge $-e$ and spin 
$s_{h_\xy}=\frac{1}{2} = S_{h_\xy z}/\hbar$ 
in $+z$ direction here. The muon $\mu^\m$ is a $\ehxym$ in rotational motion along two coinciding  elliptics $\nu_{y'z}$, $\nu_{y''z}$ projected from the $\nubar_e$-,  $\nu_\mu$-orbits in a 
$y'z$ (or $y''z$) plane. $\nu_{y'z}$ is the result of supposition of  two P-O motions, at speed $\ve_\hf \dot{=}c$ each, along  $\nubar_e$-, $\nu_\mu$-orbits of equal radius $r_{1\mu}$, and of  normals $z',z''$ at quantised angles $\pi-\theta_\hf$ and  $\theta_\hf$ to the $z$ axis (Fig \ref{fig-mu_orbits.eps}a).

The projections of the $\nubar_e$-,$\nu_\mu$- orbits in  the $y'z $ plane, $\nu_{y'z}$,$\nu_{y''z}$, or in the $xy$ plane, $\nu_{xy}$'s, necessarily coincide. The component kinetic motions  of $-e$  along $\nu_{xy}$'s are thus equal and opposite and  cancel out, leaving $\hxym$ at rest on $\nu_{xy}$ at a random angle $\varphi$ to the $x$ axis,  and those along $\nu_{y'z}$,$\nu_{y''z}$ add up. Similarly as for $n$\cite{jxzj-neutron} and $\pi^\m$ (Sec \ref{Sec-pion}.1), the angular momenta of the P-O motions along $\nubar_e,\nu_\mu$ are each given by the eigen solutions in  the states  $n=2$, $\lcfoot=1$, $j= \lcfoot - \lcfoot^\TPsub= \frac{1}{2} $ in the lab frame: 
$J_{\hf} =r_{1\mu} \g_\mu M_\mu c = \sqrt{j(j+1)} \hbar =\frac{\sqrt{3}}{2} \hbar$.
The  projections along $z$  (with $m_j',m_j''=- \frac{1}{2}, \frac{1}{2}$)
 are $J_{\hf z}'  =-J_{\hf z}'' =  J_\hf \cos (\pi-\theta_\hf) =- \frac{1}{2} \hbar $, and in the $xy$ plane are  
$J_{\hf xy}'  =J_{\hf xy}'' = J_\hf  \sin \theta_\hf   
= \frac{\sqrt{2}}{2} \hbar $, where  $\cos \theta_\hf = 1/\sqrt{3}$, $\sin \theta_\hf =\sqrt{2/3}$. 
$J _{\hf xy,\mu } =J_{\hf xy}'+J_{\hf xy}''= \sqrt{2} \hbar$ at angle $\varphi$ to $x$ in the $xy$ plane,  of  a time average  $\la J _{\hf xy,\mu }  \ra =0$.
The spin (angular momentum) of 
the total system $\mu^\m(h^\m_\xy[\nubar_e,\nu_\mu])$ is 
$ S_{\mu z}
=  S_{\hxy,z} + J_{\hf z, \mu} =\frac{1}{2}\hbar$ in $+z$ direction. 

\section{A generalised IED model extending to weak potential field }
\label{Sec-IED-gen}

In the IED picture, the mass of a single charged permanent  particle $\a$  (e.g.  $\em$) is generated by the oscillation of its charge $q (=-e)$, of a displacement $u_q$ (of the CM of the minute but extensive $q$),  in the vacuum potential field $V_\v$  (mainly due to $q$ and vacuuon-dipoles' electrostatic interaction) \cite{jxzj-ied,jxzj-vacpot}. In absence of other particles, 
$V_{\v } = \Vd_\v +V_{\v  0} \dot{=} \frac{1}{2} \beta_{\v}  u_q^2 +V_{\v  0} $, 
where $\beta_{\v}= \Mcal_\v \Om_\v^2$ is the force constant,  $ \Mcal_\v$  is the (sub-vacuum) mass of the charge $q$, and  $\Om_\v$ is the angular frequency. Assume $q$ oscillates about a fixed position and hence  $\a$ is at rest. $ \Mcal_\v$  represents a friction against  the $q$ motion ordinarily in its own $V_{\v } $ field.   If another particle $\a' $ of charge $q'$ presents in the vicinity and exerting on the vacuum  a potential $\Vd_\v'$, then $\Mcal_\vq$ ($\propto$ friction), hence $\beta_\vq$, $\Vd_\v$ of $q$ of $\a$ would in general be modified, by a factor $\eta_\v$ to $\Mcal_\vq^\dagsup$,  $\beta_\vq^\dagsup$, and 
$$\displaylines{
\refstepcounter{equation} \label{eq-DV.Fp.b-p} 
 \label{eq-DV.Fp.c}  
\hfill
\mbox{$
\Vd_\v^\dagsup(u_q)
=\frac{1}{2} \beta_\vq^\dagsup u_q^2 
=\frac{1}{2} \beta_\vq u_q^2 + \frac{1}{2}\Ocal_\eta c^2,
\quad
\beta_\vq^\dagsup
= \beta_\vq (1+ \frac{\Ocal_\eta c^2}{ 2\la \Vd_{\vq} \ra}), \quad
\Ocal_\eta=  \frac{\eta_\v D \la \Vd_{\v'} \ra  }{D' c^2}, 
$}
\hfill (\ref{eq-DV.Fp.b-p})
}$$
where $D,D'$ are the dimensions of the $q,q'$ oscillations; $\la \ \ra$ indicates the time average.

Consider a particle $\a$ of charge $q$ is in P-O motion along orbit $\nubar$  (of a normal $z'$)
relative to  $ \a'$ in the  central magnetic or weak potential field of $\a'$
$V_\hf (r_1) = -G/(\frac{4}{3}\pi r_1^3)$, at an (equilibrium) separation $r_1 (\sim 10^{-18}$ m),  $G$ being the coupling constant. Attach the local co-ordinate axes $\xi,\zeta$ to the moving $\a$, with an origin fixed at $r_1$ and $\xi,\zeta$ parallel to  $r$, $z'$ (the radius and normal of orbit $\nubar$).
The oscillation of (the CM of) $q$, of displacement  $\ub_q =\rb-\rb_1=\xib_q+\zetab_q$,  in  $\Vd_\v$ is now in the (instantaneous)   $\xi,\zeta$ plane of dimension $D=2$, at a random angle $\vtheta$ to  $\zeta$, and 
 of a rest  amplitude $\Acal_{\v \vtheta}  = \Acal_{\v } |\sin \vtheta \xihat + \cos \vtheta  \zetahat |  =   |\Acal_{\v\xi}    \xihat + \Acal_{\v\zeta} \zetahat | $;  $\vtheta$ 
varies randomly in $2\pi $ over time. Consider further $q$ is at initial time  driven by an external force ($\Fb_{ext}$)  into displacement $\xi_q$ from $r_1$ along radial $r$- or $\xi$- direction. After a relaxation time $q$  maintains an oscillation of displacement  $\xi_q(t) =r-r_1 =\Acal_{h} \psi_q(t)$, 
expressed for  a free $\a$ at rest here, 
under the restoring force $\Fb_{\hm} =-\frac{\pd V_{\hm}}{\pd \xi_q} \xihat $   produced by one half of the difference weak potential, 
$$\displaylines{
\refstepcounter{equation} \label{eq-DV.Fp.b}
\hfill
 V_{\hm} (\xi_q)
=  \frac{1}{2} [V_{\hf } (r_1 + \uqr)-  V_{\hf } (r_1 ) ] 
\dot{=} \frac{9 G |\xi_q|}{ 8\pi r_1^4 } 
\approx \frac{9 G \xi_q^2}{ 8\pi r_1^4  \Acal_{\hm}} 
=  \frac{1}{2 } \beta_{\hm} \uqr^2, 
\hfill\cr\hfill
\beta_{\hm} = \frac{9G }{ 4 \pi r_1^4  \Acal_{\hm} }
=\frac{2 \Kh c^2}{\Acal_h}, 
\hfill
 (\ref{eq-DV.Fp.b})
}$$ 
where the second of Eqs (\ref{eq-DV.Fp.b}a) is given for  $\xi/r_1<<1$;
the third is a quadratic approximation but gives  the exact energy level at $\xi_q=\Acal_{\hm}$. The other half $V_{\h'}(\xi_{q'})=V_{h}(\xi_q)$ is shared by $\a'$.  
$\a'$  similarly presents to $V_h$ a friction term 
 $\Ocal_\eta =\frac{ \eta_\v D_h \la \Vd_{\v'}\ra }{ D' c^2 } $, $V_h^\dagsup= V_h+\frac{1}{2} \Ocal_\eta c^2$, $\beta_h^\dagsup=\beta_h(1+\frac{\Ocal_\eta c^2 }{ 2\la V_h\ra })$.


Setting the friction effect aside, the total  restoring forces  on $q$  in $\xi$, $\zeta$ directions are  thus $F_{}(\xi_q)= F_{\vq}(\xi_q) + F_{h} (\xi_q) = - (\beta_\vq + \beta_{h} )\xi_q $,   $F_{}(\zeta_q)=  F_{\vq} (\zeta_q)=-\beta_\vq \zeta_q$.  The Newtonian equations of motion (eom's) of the CM of $q$ along $\xi$, $\zeta$  are thus $- (\beta_\vq +\beta_{ \hm}) \xi_q =(\Mcal_\vq+  \Mcal_{ \hm} )  \frac{\pd^2 \xi_q }{\pd t ^2} $,  $- \beta_\vq \zeta_q =(\Mcal_\vq) \frac{\pd^2 \zeta_q }{\pd t ^2} $. 
The general solutions are $\xi _q^{(c)}=| \Acal_\vq e^{i \Om_\vq t} \ehat_0 +\Acal_{ \hm} e^{i \Om_{ \hm} t} \ehat_h|$, $\zeta_q^{(c)} = \Acal_\vq  e^{i \Om_\vq t}$, where $\Om_\vq^2= \beta_\vq/\Mcal_\vq$,  $\Om_{\hm} ^2 =\beta_{\hm}/\Mcal_{\hm}$;  $\ehat_0, \ehat_h$ are unit vectors for the regular and  excited $h$ mass states and  assumed orthogonal. $\Acal_\vq $ is the time average of $\Acal_{\v \vtheta} $, $\Acal_\vq =\frac{ \Acal_{\vq \vartheta}}{2\pi} =\frac{1}{2} \Acal_{\vq \xi} $. 
The total mechanical energies of the $q$ oscillations in  
$\Vd_{\vq}^\dagsup$, $V_{h}^\dagsup$ including  friction   are 
$\Eng_{\vq}^\dagsup =\Eng_\vq + \Ocal_\eta c^2$, 
$\Eng_\vq= \frac{1}{2}\Mcal_{\v } \dot{u}_q^2+ \Vd_{\vq} = \frac{1}{2}\beta_{\v } |u_q^{(c)}|^2 = \frac{1}{2}\beta_{\v } \Acal_{\v }^2$, 
$\Eng_{h}^\dagsup=  \Eng_{h } + \Ocal_\eta c^2$, 
$\Eng_{h } = \frac{1}{2}\Mcal_{h} \dot{\xi}_q^2+ V_{ h} = \frac{1}{2}\beta_{h} |\xi_q^{(c)}|^2  = \frac{1}{2} \beta_{h}\Acal_{h}^2$. 
For simplicity of discussion we consider the extreme case that no (internal) radiation is being generated by the charge oscillation;  $\Eng_\v^\dagsup$,$\Eng_{h }^\dagsup $  thus equal 
 the originally imparted  energies.

The charge $q$ is point like to its EMR  field and yet is extensive at the scale $10^{-18}$ m and hence in $V_\hf (r)$\cite{jxzj-vacpot} and $V_\v$\cite{jxzj-neutron}. Denote by $\rho_c=|\psi_c(\ub,t)|^2$ the  dimensionless charge density;   $\psi_c$ is a complex function, $\ub= \rb-\rb_1=\xib+ \zetab$ is the displacement  a volume element  on $q$ makes at time $t$. The $\rho_c$ current is thus
$j_{\qu }=  \rho_\qu \frac{d \ub}{d t} = -D_{q } [\psi_{\qu }^*(\nabla \psi_{\qu}) -  (\nabla \psi_{\qu}^*)\psi_{\qu}  ]$;
$D_{ } = \frac{i \hbar }{2 \Mcal (\ub_q)}$  is an imaginary diffusion constant. 
 The  stationary state is described by the continuity equation $\pd_t \rho_{\qu }  + \nabla j_{\qu} + \Ocal \rho_{\qu } =0$, where $\Ocal = \frac{ V_{ }(\ub)-V_{ }(\ub)  }{i\hbar } \equiv 0$. This  decomposes to  two equations  of a Schr\"odinger form, given for $\psi_c$ as 
$  i \hbar \pd_t \psi_{\qu}= \lf(- (\hbar^2/2 \Mcal)  \nabla^2 + V (\ub)  \rt)   \psi_{\qu }$.
Using in it (\ref{eq-DV.Fp.b-p}a) for $V_{}(\zeta)= \Vd_{\vq}^\dagsup = \frac{1}{2} \beta_\vq^\dagsup  \zeta^2 $  and  (\ref{eq-DV.Fp.b})  for $V_{ } (\xi) =V_{ \hm}(\xi)= \frac{1}{2} \beta_{\hm}  \xi^2$ for the two pure orthogonal oscillation states
along $\zeta, \xi$, one obtains the standard solutions as for a usual harmonic oscillator:  The eigen functions $\psi_\qu$ are hermit polynomials. The eigen energies are quantised, 
$$\displaylines{
\refstepcounter{equation} \label{eq-Engqns}\label{eq-Engqns-a}
\hfill
\mbox{$\Eng_{\vq 1}^\dagsup 
= \hbar \Om_{\vq}^\dagsup 
= \frac{1}{2} \beta_\vq \Acal_{\vq 1}^2  +\Ocal_\eta c^2
=M_\a^\dagsup c^2$}, 
\hfill\cr
\hfill
\mbox{$
\Eng_{\hm \Nsub }^\dagsup=\N \Eng_h +\Ocal_\eta c^2, \quad 
\Eng_h= \hbar \Om_{\hm} 
= \frac{1}{2} \beta_{\hm} \Acal_{\hm 1}^2 
= M_{ \hm} c^2,
$}
\hfill  (\ref{eq-Engqns})
}$$
where 
based on comparison with experiment \cite{jxzj-vacpot} only  the $n_\v=1$th excited state   is permissible in $\Vd_\v$, 
and $\N=0, 1,2,\ldots$ in $V_\hf$;
the terms $\frac{1}{2} \hbar  \Om_{\vq}$, $\frac{1}{2} \hbar  \Om_{h}$ are judged unphysical and dropped. 
The second of Eqs (\ref{eq-Engqns}a),(c)  identify with the classical energies but  with $ \Acal_{\vq 1}$, $\Acal_{\hm \Nsub}= \sqrt{\N} \Acal_{\hm 1}$  quantised. The last equate $\Eng_{\vq 1}^\dagsup$,$\Eng_{h \Nsub}$ with the particle rest-mass energies, 
 $M_\a^\dagsup $, $M_{\hm}$ being the rest masses, as a generalised basic assumption of the IED model. 

Assuming $q'$ is not in excited mass state in $V_\hf$,
and for the averaged $\Acal_\v$ used, (\ref{eq-Engqns-a}a)  gives also the actual 
  total  hamiltonian of $q$ oscillation in $V_\v$
(the vacuum ground mass state), and (\ref{eq-Engqns-a}b)  that of the $q$ oscillation 
- the excited mass state - in $V_\hf$. The respective particle rest masses are
$$\displaylines{
\refstepcounter{equation} \label{eq-M-af-excM-p}
\label{eq-M-af-excM}
\hfill
\mbox{$M_\a^\dagsup=M_\a+\Ocal_\eta$,} 
\ \
\hfill\cr
\hfill
M_{\hm \Nsub}^\dagsup=N M_\hm+\Ocal_\eta, \quad
M_{\hm} =M_\pi - M_{\nu_e} =\frac{\Eng_{h1}}{c^2}=\frac{2\la V_{h  }\ra}{c^2}
= \frac{9  G \Acal_{\hm 1}  }{8\pi r_1^4 c^2  }
= \Kh \Acal_{h1}.
\hfill (\ref{eq-M-af-excM})
}$$
The second of Eqs (\ref{eq-M-af-excM}c) is given by the definition for $\pim$ 
rest mass (Sec \ref{Sec-pion}) for $q=-e$ (cf Sec \ref{Sec-pi-mass-abinitio}). Indicated by experiment \cite{perkins-20} and in theory (Sec \ref{Sec-pi-mass-abinitio}), the charge 
$q$
generates a particle $\a$  either in a pure $n_\v=1$ state in $V_\v$ (i.e. $e^\m$), or a $N\ge 1$ excited state in $V_{h\Nsub}$ (e.g. $h^\m$, giving  $\pim$), and not a mixed state of both. Suppose that $q$ is in $N\ge 1$th  excited mass-state, and $q'$ is in $N'=0$ state. So $q'$ is not in the $N\ge 1 $th kinetic oscillation but shares the other half  of the potential energy of (\ref{eq-DV.Fp.b}),  of a projected average 
$\frac{D_h}{D'} \la V_{ h' \Nsub} \ra$, 
$\la V_{ h' \Nsub} \ra
=\la V_{ h \Nsub}\ra 
= \frac{1}{2}   \beta_{ \hm }     \la \xi_{q}(t)^2 \ra_{\Nsub} 
= \frac{1}{2} \Eng_{h \Nsub}$.
 $D_h/D', =\frac{1}{2}$, reflects the frictional time that $q'$, oscillating regularly in a $D'=2$ plane, spends  along the $D_h=1$ $\xi$-axis. Including the shared potential term in  (\ref{eq-Engqns-a}) applied to $q'$, the (rest) mass of $\a'$ is
$$\displaylines{
\refstepcounter{equation} \label{eq-masses-a-ap}
 \label{eq-M-af-p} \label{eq-Enqr1-b}
\hfill M_{\a'}^\dagsup
=M_{\a'} +  \frac{D_h \la V_{ h \Nsub}  \ra }{D'  c^2}+ \Ocal_\eta' 
=M_{\a'}  +\frac{1}{4}  N M_h + \Ocal_\eta', 
\quad
\mbox{$\Ocal_\eta'=  \frac{         \eta_{\v} D_{\a'}   \la  \Vd_{h\Nsub} \ra }{D_h c^2}  $.}
\hfill (\ref{eq-masses-a-ap})
}$$
Eqs (\ref{eq-M-af-excM-p}), (\ref{eq-masses-a-ap}) above, and  (\ref{eq-Mmu-rho}) later, describe a common rest mass formation scheme  due to quantised $q$ oscillation in  $\Vd_\v$, $V_h$ or $V_{h xy}$, as contrasted to relativistic mass due to  translation.   Until Sec   \ref{Sec-pi-mass-abinitio}, we can  not predict the $M_h$ value, for in (\ref{eq-M-af-excM}) $\Acal_{h1}$  is not known, also $G$, $r_1$  are not universal.

As illustrations we apply (\ref{eq-M-af-excM}a,b-c),(\ref{eq-masses-a-ap}) to two composite particles for evaluating the $\eta_\v$ or mass. First, the neutron $n(e^\m[\nubar_e],p)$\cite{jxzj-neutron} composed of    $\a,\a' = e^\m,p$  both in the $n_\v,n_\v' $ $=1,1$th states;  $D,D'=2$. In their mutual presence the $e^\m,p$ masses are each given by (\ref{eq-M-af-excM-p}a), $M_e^\dagsup = M_e+ \frac{\eta_\v D  M_p}{2D' }$, $M_p^\dagsup $ $= M_p+ \frac{\eta_\v D'  M_e}{2 D} $; their P-O current has an apparent rest mass $M_\nue\dot{=} 0$.  The $n$  mass  is  the sum
$M_n= M_e^\dagsup +M_p^\dagsup +M_\nue 
\dot{=} M_e+M_p + \Ocal_\eta$, $\Ocal_\eta = \frac{\eta_\v(M_e+M_p)}{2}$.  
Using in this the experimental $M_n, M_e,M_p $, hence $\Ocal_\eta^{exp} = 0.782$ MeV, 
gives $\eta_\v \dot{=} \frac{2 \Ocal_\eta^{exp}          }{M_e+M_p}
=1.666  \times 10^{-3}$. 
Second,  the $\Lam (h^\m[\nubar_e],p)$ hyperon (Sec \ref{Sec-pion}) composed of $\a= h^\m$ in the $N=1$th state in $V_{h} $, and $\a'=p$ in $n_\v'=1 $th state in $V_{\v'}$; $D_h=1,D'=2$. To the $h^\m,p$ masses (\ref{eq-M-af-excM}c),(\ref{eq-masses-a-ap}) apply: 
$M_h^\dagsup=M_h+\Ocal_\eta$, 
$M_h= \frac{2 \la V_{h} \ra}{c^2} = M_\pi - M_{\nu_e}$ ($M_h $ is  invariant), $M_p^\dagsup =M_p +\frac{1}{4}  M_h + \Ocal_\eta' $, 
The $\Lam$ mass is thus $M_\Lamsub= M_h^\dagsup +  M_p^\dagsup + M_\nue^\dagsup
=\frac{5}{4} M_\pi + M_p +\sum \Ocal_\eta 
\dot{=}
1113.36 \ {\rm MeV}$, 
where $\sum \Ocal_\eta = \eta_\v(\frac{D_h \la V_{\v,+e}\ra}{D' c^2} + \frac{ D' \la V_h\ra}{ D_h c^2 })= \eta_\v (\frac{1}{4} M_p + M_\pi) \dot{=}0.623$ MeV estimated using  the $\eta_\v$ value of $n$; 
 the experimental values for $M_\pi$, $M_p$, $M_\nue(\dot{=}0)$ are used. 


Finally we derive the $\mu^\m$ mass from  the projection of (\ref{eq-M-af-excM}c).
Consider a $\mu^\m$ as produced from $\pi^\m\rar \mu^\m \nubar_\mu$, of the original mass $M_\pi'=M_\pi-M_\numu=M_h + M_\nue + M_\numu$, generated by $-e$ along $\nu_{y'z}$ (Sec \ref{Sec-muon-model}.2). Its total EMR field has one component travelling on the continually reorienting $\nu_{y'z}$, which will not be readily re-absorbed by $-e$ and hence lost to the kinetic energies of $\ehxym$, $ \nubar_{e}$, $\nu_{\mu }$, and one (as two opposite but  non-cancelling standing waves) on $\nu_{xy}$, which  can be more readily re-absorbed by $-e$. So the $xy$-projection of  $M_\pi' $ gives the  $\mu^\m$ mass,  
$M_\mu (\rho) $ $= (M_\pi' )_{xy} $ 
$=[M_{h}(1+ \Ocal )]_ \xy$, 
where $\rho=r_\xy$ is  the radius of the elliptic $\nu_\xy$, $M_h=\Kh \Acal_{h1}$ [Eq  (\ref{eq-M-af-excM}c)], and $\Kh$ is a constant. 
Provided taking  $\Ocal$ as a constant of $\rho$, the $xy$-projection of the mass $M_\pi'$ reduces to a pure geometric $xy$-projection of $\Acal_{h1}$, $ \Acal_{h1 xy} (\rho) $ which  is dependent on $\rho$. At the semi-major and semi-minor axes $\rho=a, b$, $\Acal_{h1_{xy}} (a)=\Acal_{ h1} $ and $\Acal_{h1_{xy}}(b)= \Acal_{ h1} \cos \theta_\hf $. Using  their mean  
\mbox{$\overline{\Acal }_{h1_{xy} } 
= (\Acal_{h1_{xy}}  (a) \Acal_{h1_{xy}}  (b)     )^{1/2} 
= \Acal_{h1} \sqrt{ \cos \theta_\hf} 
$},
 $\cos \theta_\hf=\frac{1}{ \sqrt{3}}$ from earlier,  gives
$$\displaylines{\refstepcounter{equation} \label{eq-Mmu-rho}
\hfill
\quad
M_{\mu}
\dot{=}
\mbox{$        
M_\pi' (\Acal_{h1_{xy}}  (a) \Acal_{h1_{xy}}  (b)   )^{1/2}  /    \Acal_{h1} 
$}
= (M_\pi-M_\numu) \sqrt{ \cos \theta_\hf}
=105.860 \ {\rm MeV}.
\hfill (\ref{eq-Mmu-rho})
}$$

\section{{\it Ab initio} predictions of the muon and charged pion  masses}
\label{Sec-pi-mass-abinitio}
 \setcounter{equation}{6}

Consider first two particles $a,b$ of charges $-e,+e$ and   masses $m_a= \g_a M_a$, $m_b = \g_b M_b$, moving at relative speed $\ve$  under their Coulomb force $ F_c= - \nabla V_c = - \hbar \a c / r^2$ at a separation $r=r^0/\g$, where $V_c $ $=  - \frac{ \hbar \a c }{ r}$, $\a= \frac{e^2}{ 4\pi\ev_0 \hbar c}=1/137.036$. In the CM frame the relativistic eom is  $\mr d^2_t \rb =F_c \rhat$, where $\mr$ $=\g \mr^0 $, $\gbar=1/\sqrt{1-\ve^2/c^2}$, $\mr^0=\frac{M_a M_b}{M_a +M_b}$, $\rb =r \rhat=\rb_a-\rb_b$; $\rb_a =  \frac{m_b\rb}{m_a+m_b}$ and $\rb_b= -  \frac{m_a\rb}{m_a+m_b}$ relative to their CM  at $\Rb=0$ here. The total wavefunction $\psi_{tot}$ of the fictitious particle of  mass $\mr$ 
obeys the Klein-Gordon equation (KGE). In the extreme case either $\gbar \dot{=}1$ or $\gbar>>1$, this reduces to SQR-KGE\cite{jxzj-neutron}  $[\frac{\g  ( (p_r^2)_{op}  +(\Jcal^2)_{op} /r^2) }{(\g+1)\mr} +V_c]  \psi=\Hcal \psi $, where $p_r$ and $\Jcal$ are the radial and angular momentum operators  each of the kinetic motion of $\mr$.
For the central potential in spherical polar co-ordinates, $\psi=\Rcal(r) \Ycal(\theta,\vphi)$; the SQR-KGE separates to
$\mbox{$
[-\frac{\g \hbar^2}{(\g+1)\mr r^2}\frac{\pd }{\pd r}(r^2 \frac{\pd}{\pd r})
+ \frac{\g \lc(\lc+1) \hbar^2}{(\g+1) \mr r^2} +V_c(r)
] \Rcal(r) =\Hcal \Rcal(r)$} $, 
$\mbox{$(\Jtr^\ast{}^2)_{op} \Ycal^\ast{}(\vartheta^\ast{},\phi^\ast{})
= \Jtr^\ast{}^2  \Ycal^\ast{}(\vartheta^\ast{},\phi^\ast{})
$}$. 
The eigen functions are $\Rcal(r)= \frac{(\ka r)^\ka }{r} e^{-\ka r} \sum_{i=0,1\ldots} \frac{b_i}{(\ka r)^i }$, $\ka = \sqrt{-(\gbar+1 )\mr \Hcal /\hbar^2}$ and the spherical harmonics $\Ycal_\lcsub^{m_\lcsub} =C_{\lcsub}^{m_\lcsub} P_\lcsub^{m_\lcsub}(\cos \vartheta^\ast) e^{i m_\lcsub \phi^\ast} $. The level $n$ ($=1,2, \ldots$) eigen energies are
$\Hcal_\nc=T_\nc+ V_{c \nc}
= - \frac{\mr_\nc \ve_\nc^2}{(\gbar_\nc+1)}
= \frac{V_{c \nc}}{\gbar_\nc+1} \equiv  \frac{\hbar \a c}{(\gbar_\nc+1) r_\nc}
= -\frac{T_\nc}{\gbar_\nc}$, 
$\ve_\nc = \frac{\a c}{\nc}=\frac{\ve_1}{\nc}$, 
$r_\nc = \frac{\nc^2 \hbar}{\mr_\nc \ve_1} $, $\mr_\nc=\g_\nc \mr^0$; 
the wavelength is $\lam_n=\frac{h}{\mr \ve_n}= \frac{2\pi r_n}{n}$.
The relation $\nc^2 =\sum_{\lc=0}^{\nc-1}(2 \lc+1)$ holds.  
Hence $\Hcal_\nc = $ $ \frac{ \Hcal_\nc}{\nc^2} \sum_{\lc=0}^{\nc-1} (2 \lc+1) $ $
=  \sum_{\lc=0}^{\nc-1} \Hcal_{\nclc}$; the $\lcfoot$ state projection of   $\Hcal_n$ and
 those of $\ve_\nc,r_\nc, \lam_\nc,\Jcal_\nc, \mr_\nc$ are  
 $$\displaylines{\refstepcounter{equation} \label{eq-nc} 
\hfill
\mbox{$
\Hcal_{\nclc}
=\frac{\Hcal_\nc (2\lcfoot+1)}{\nc^2}
= - \frac{\mr_\nclc \ve_\nclc^2 }{\g_\nclc +1 }
=-  \frac{\mr_\nclc \ve_\nc^2 (2\lcfoot+1) }{(\g_\nclc +1)\nc^2 }
=\frac{V_{c \nclc}}{\g_\nclc+1} 
= - \frac{\hbar \a c}{(\g_\nclc+1) r_\nclc} 
= - \frac{ \hbar \a c (2\lcfoot+1 )}{(\g_\nclc+1)  r_\nc \nc^2}, 
$}
\hfill (\ref{eq-nc})
\cr
\refstepcounter{equation} \label{eq-lams} 
\hfill 
\mbox{$
\ve_\nclc = \frac{\ve_\nc \sqrt{2 \lcfoot+1}}{\nc}= \frac{\a c \sqrt{2\lcfoot +1}}{ \nc^2}, 
\quad 
r_\nclc=  \frac{r_\nc \nc^2 }{2\lcfoot+1} 
$}, 
\quad 
\mbox{$
\lam_\nclc
=\frac{h}{\mr_\nclc \ve_\nclc}
= \frac{ 2\pi r_\nclc      \ve_\nclc }{\a c}
= \frac{2\pi r_\nclc \sqrt{ 2\lcfoot +1}}{\nc^2} =
$} 
\hfill 
  \cr \hfill 
\mbox{$=\frac{\lam_\nc \nc}{\sqrt{2\lcfoot +1}} $,  \quad
$
\Jcal_\nclc = \rbar_\nclc \mr_\nclc \ve_\nclc 
= \rbar_\nclc  \mr_\nclc \frac{\a c \sqrt{2\lcfoot +1}}{\nc^2} =\sqrt{\lcfoot(\lcfoot+1)} \hbar
$}, \quad 
\mr_\nclc=\g_\nclc \mr^0.
\hfill (\ref{eq-lams})
}$$
For the usual positronium, $M_b=M_a=M_e$, $\ve_\nc=\frac{\a c }{\nc}$,  
$ \gbar_\nclc\dot{=} \g_\nc\dot{=}1$,   $\mr_\nclc \dot{=} \mr^0=\frac{1}{2} M_e$; 
$\lam_\nclc^0 $ $  =\frac{2h}{M_e \ve_\nclc}$ $  = \frac{ 2h \nc^2 }{M_e \a c \sqrt{2 \lcfoot+1}  }$, 
$r_\nclc^0 = \frac{\nc^2 }{   \sqrt{2 \lcfoot+1}  } \frac{ \lam_\nclc^0 }{2\pi}
=\frac{2 \hbar c \nc^4}{M_e c^2 \a (2 \lcfoot+1)} 
=\frac{ 1.058 \times 10^{-10} \nc^4 }{2\lcfoot+1} $ m, $ \Jcal_\nclc^0 = \frac{r_\nclc^0 M_e \ve_\nclc^0 }{2}=  \frac{r_\nclc^0  M_e \a c \sqrt{2\lcfoot +1}}{2\nc^2}$.

We shall show that it is possible to create  a positronium-like system, a double heavy positronium (DHP)  composed of  a relativistic electron and positron, $a,b=e^\m{}^\astp, e^\p{}^\astp$  moving at a relative speed $\ve^\astp_\nc =|\veb_a-\veb_b| =g_\nc \ve_\nc  \dot{=}c$ in their Coulomb field $V_c$, such that $e^\m{}^\astp, e^\p{}^\astp$ have each an apparent rest  mass $M_{s\nclc}^\astp= \mr_{s\nclc}^\astp+ M_e$, that is given by the eigen state $\nc, \lcfoot$ relativistic mass energy $\mr_{s\nclc}^\astp c^2$ above  the $n_\v=1$ vacuum  level for $-e$, $M_e c^2$. This implies that, for level $n$, the apparent rest mass is $ \mr_{s\nc}^\astp= \g_{s\nc}^\astp \mr^0 = \g_{s\nc}^\astp\frac{1}{2}M_e $ and has an apparent rest  wavelength $\Lam_{sn}^\astp = \frac{h}{\mr_{s\nc}^\astp \ve_\nc}= \frac{nh}{\g_{s\nc}^\astp \mr^0 \a c }$  equal to the wavelength  $\Lam_e=\frac{h}{M_e c}$ of the total  EMR of  $e^\m$. 
There thus requires $\frac{\nc}{\g_{s\nc}^\astp\frac{1}{2} \a}=1$, or 
$$\displaylines{
\refstepcounter{equation} \label{eq-g-o-a-AA} 
\hfill
\g^\astp_{s\nc} = \frac{2\nc}{\a}= 2\g_\nc^\astp, \quad 
\g_\nc^\astp= \frac{\nc}{\a}.  \quad {\rm So} \ \
\mr_{s\nc}^\astp = \g_{s\nc}^\astp \mr^0= 2 \g_\nc^\astp \mr^0 =2 \mr_\nc^\astp= \frac{\nc M_e}{\a};   
\hfill (\ref{eq-g-o-a-AA})
\cr
 {\rm and} \hfill
\cr
\refstepcounter{equation} \label{eq-mrnc}
\hfill
\mr_{s \nclc}^\astp= \mr_{s \nc}^\astp\frac{2\lc+1}{\nc^2}
= 2 \g_{ \nclc}^\astp \mr^0= 2 \mr_\nclc^\astp, 
\quad
\gbar^\astp_{\nclc}=  \frac{ \g_\nc^\astp(2\lcfoot+1)}{\nc^2}
=  \frac{ (2\lcfoot+1)}{\nc \a},
\hfill (\ref{eq-mrnc})
}$$
given using (\ref{eq-nc}); $g_n=\sqrt{\frac{\g^2-1}{\g^2}} \frac{c}{\ve_\nc} \dot{=} \frac{n}{\a}$. To achieve such dynamics  for both charges $-e,+e$, there requires in the CM frame  two fictitious particles, designating by $s=\mu,\mubar$, each having a mixed total wave  $\psi_{s } =  \psi_{\nubar} \psi_\nu e^{ i \beta_s}$, with $\psi_{\nubar},\psi_{\nu}$ satisfying SQR-KGE.
Accordingly $\mr_{s\nclc}^\astp$ is double the $\mr_{\nclc}^\astp $ of a  single relativistic positronium. The $\mr_{s\nclc}^\astp$'s are actually each moving at the speed $\ve_{\nclc}^\astp \dot{= }\ve_{\nc}^\astp \dot{=} c$; they have therefore each a relativistic wavelength
$$\displaylines{
\refstepcounter{equation} \label{eq-g-o-a-nc}
\hfill
\lam_{s \nclc}^\astp=\frac{h}{\mr^\astp_{s \nclc} c}
= \frac{ \lam^\astp_{\nclc} }{2}, 
\quad
 \lam^\astp_{\nclc}  
= \frac{h}{\mr^\astp_{\nclc} c} 
= \frac{(\a\sqrt{2\lcfoot +1})  \cdot h \cdot \nc^2}{      \nc^2  \cdot(   \gbar_{\nclc}^\astp   \frac{1}{2} M_e)\cdot 
   ( \a c \sqrt{2\lc+1})
} 
=\frac{     \a \sqrt{2\lcfoot+1}    \lam_\nclc ^\osup }{ \nc^2   \gbar^\astp_{\nclc} }. 
\hfill \cr
\hfill
(\ref{eq-g-o-a-nc})
}$$

The above system can be obtained by accelerating  $e^\m$, $e^\p$ (from rest) to $e^\m{}^\astp$, $e^\p{}^\astp$ of the CM frame linear momenta $\pb_a= - M_{s\nclc}^\astp \ve_\nclc^\astp \zhat$, $ \pb_{b } = - \pb_{a}$ in $-z,z$ directions at time $t=0$, and then subjecting them to a near head-on "quantum collision" at  $\rb_{a }$, $\rb_{b }=\frac{1}{2}\rb_{\nclc}^\astp,-\frac{1}{2} \rb_{\nclc}^\astp$, at a separation $r_{s\nclc}^\astp= \frac{1}{2}  r_{\nclc}^\astp=r_{a }=r_b$, 
with $r_{\nclc}^\astp$ corresponding through  (\ref{eq-lams}c) to $\lam_\nclc^\astp$ of (\ref{eq-g-o-a-nc}),
$$\displaylines{
\refstepcounter{equation} \label{eq-r-astp-a}
\hfill
r_{\nclc}^\astp=|\rb_{a }-\rb_{b }|
=\frac{\nc^2 }{ \sqrt{2 \lc+1} }  \frac{\lam_\nclc^\astp}{2\pi}
=\frac{r_\nclc^0 \a \sqrt{2\lcfoot+1}  }{
 \gbar^\astp_{\nclc} \nc^2}
=\frac{r_\nclc^0 \a^2  }{
 \nc \sqrt{2\lcfoot+1} }. 
\hfill (\ref{eq-r-astp-a})
}$$ 
In their central $V_c$ field $e^\m{}^\astp,e^\p{}^\astp$ are thus turned  to rotations 
at the pre-defined eigen dynamical variables  (\ref{eq-g-o-a-AA})-(\ref{eq-g-o-a-nc}). The initial time condition $\pb_{\mubar}$ $ = \pb_{b } $ $=-\pb_\mu $ $ =-\pb_{a}$ permits only states of the same $\nc$ and $m, m' $ $=-\lcfoot$,  $\lcfoot$;   $\psi_\nubar,\psi_\nu$ are thus travelling along  two orbits  $\nubar, \nu$ (two semiclassical effective circles of radii $r_{\nclc}^\astp$)  in the  $x'y'$, $x''y'' $ planes of normals $\zhat',\zhat''$  at  quantised angles $\pi-\theta_{\lcsub }, \theta_{\lcsub }$ to the $z$ axis  (Fig \ref{fig-mu-mu+fr_e-e+.eps}b); and $\beta_{\mubar}-\beta_\mu=\pi$. The angular momenta of the component  motions of $\mu$ along  $\nubar$, $\nu$ are   $\Jcalb_{\mu \nubar \nclc} ^\astp=| \rb_{\nclc}^\astp \times  \pb_{\nclc}^\astp | \zhat' 
= \Jcal^\astp_\nclc (-\cos \theta_\lcsub \zhat+ \sin \theta_\lcsub \rhat_\xy)$, 
$\Jcalb_{\mu \nu \nclc}^\astp =|\rb_{\nclc}^\astp \times  \pb_{\nclc}^\astp | \zhat'' 
= J^\astp_\nclc (\cos \theta_\lcsub \zhat+ \sin \theta_\lcsub \rhat_\xy)$,
$$\displaylines{
\refstepcounter{equation} \label{eq-Jcal-ncs}
\hfill
\mbox{$
\Jcal^\astp_\nclc  
= r_\nclc^\astp \mr_\nclc^\astp c  
=\frac{(r_\nclc^0 \a \sqrt{2\lcfoot+1}) (\g_\nclc^\astp \mr^0   ) c  }{  \gbar^\astp_{\nclc} \nc^2 } 
= \frac{r_\nclc^0 M_e \ve_\nclc }{2}
=  \frac{   r_\nclc^0  M_e         \a c \sqrt{2\lcfoot +1}}{2\nc^2}  =\sqrt{\lcfoot( \lcfoot+1) } \ \hbar=\Jcal_\nclc^0
$}
\hfill
\cr
\hfill (\ref{eq-Jcal-ncs})
}$$
for $\ve_\nclc^\astp\dot{=}c$; \mbox{$\sin \theta_\lcsub=1/\sqrt{(\lcfoot+1)}$, $\cos\theta_\lcsub= \sqrt{\lcfoot}/\sqrt{(\lcfoot+1)} $ };  $\Jcal^\astp_\nclc$ is equal to  $\Jcal_\nclc^0$ of the usual positronium. The orbits $\nubar,\nu$ undergo Thomas precession (TP) each about $z$ in the lab frame (for a usual system the TP effect gives a small energy correction through $g$ factor\cite{Jackson}); their normals $z',z''$ are thus turned to   angles $\pi -\theta_j $, $\theta_j$ to the $z$ axis  (cf Fig \ref{fig-mu_orbits.eps}a), so that  
$J^\astp_{ \mu \eta \nc j }=| \Jcalb^\astp_{\mu \eta \nclc } \pm J^\TPsup \zhat  |
= \sqrt{j(j+1)} \hbar $ for $\eta =\nubar,\nu$, $j=\lcfoot -\frac{1}{2}$. 
For the superposed total motion of $\mu (\nu.\nubar)$, $\Jb^\astp_{\mu \nc j}=\Jb^\astp_{\mu \nubar \nc j} +\Jb^\astp_{\mu \nu \nc j}= 2  \sqrt{j(j+1)} \sin \theta_j \rhat_\xy$, $J_{\mu \nc j z}= \Jcal^\astp_{ \mu \nclc z} =0$.  The  total motion of $\mu$ is 
 along the elliptic $\nu_{y'z}$ in a $y'z$ or $y''z$ plane at a random angle $\vphi$ to the $x$ axis, and is at rest on the elliptic $\nu_\xy$ in the $xy$ plane; the $xy $-components of the  $\nu, \nubar$ motions cancel out, and these also do not present for the incident $\em{}^\astp$ (and $\ep^\astp)$     but (when actually decomposed) can be produced in an internal explosive collision. For ($\nc=2$) $\lcfoot=1$, the angular motion of  $\mu$ resembles directly  that of the muon $\mu^\m$ in Sec \ref{Sec-muon-model}.2.  And similarly for $s=\mubar$ except for a phase factor  $e^{i\pi}$. 

$e^\m{}^\astp,e^\p{}^\astp$, or $\mu,\mubar $ in the CM frame, are  energetically unstable, since at $t >0$ their separation $(r_\nclc^\astp \rar) r' $ along  $\nu_{y'z}$ is variant. Under action $V_c'(<0)$, $e^\m{}^\astp,e^\p{}^\astp$ will continue to move closer,  switching to magnetic or weak interaction $V_\hf$ dominant at  $r' $ comparable to the charge size $\aav$ ($ \approx 10^{-18} \sim 10^{-17} $ m \cite{jxzj-neutron}). Here the two quanta of the total apparent rest mass energy $M_{ s\nclc}^\astp c^2 $ each are able to {\it in situ} (-- at the same positions and velocities) convert to the more stable quantum oscillation energies of the charges $-e,+e$  in their $ V_\hf$ field. Relative to the CM and  $V_\hf$, the total mass energy and (instantaneous) four momentum  (for the rotation) of $s=\mu$ (or $\mubar$) before the conversion are $\Eng' = T{}' + M_e c^2  = \g' M_e c^2 =\g_{\astp}  M_{s \nclc}^\astp c^2$, $p^i{}' =({ \Eng' /c \atop  \g_\astp  M_{s\nclc}^\astp  \ve_{\astp} })$, where  $\g_\astp =1/\sqrt{1-(\ve_\astp/c)^2}$, $\ve_{\astp}$ is the apparent velocity of $\mr_{s \nclc}^\astp$, $i=0,1,2,3$. Those after the conversion are $\Eng'' =\g''  M'' c^2$,  $p^i{}''  =({ \Eng'' /c \atop \g''  M''  \ve'' }) $, where $M''=M_{h\Nsub _{xy} }+ \sum M_\nu$, $\g'' =1/\sqrt{1-(\ve''/c)^2}$, and $\ve''$  is the velocity  of $M''$. For the {\it in situ} conversion,  $\ve_\astp=\ve''$, $\g_\astp =\g''$.  Four momentum invariance $ p_i '' p^i{}'' =p_i ' p^i {}' $ gives $M''{}^2 c^2 {\g'' }^2 (1- (\ve''/c)^2) =M_{s\nclc}^{\astp ^2} c^2 {\g_\astp}^2 (1- (\ve_\astp/c)^2)$, or $ M''=M_{s\nclc}^\astp = \mr_{s\nclc}^\astp+M_e$.

In sum, $\mu,\mubar$ can be assigned with charges $-e,+e$,  (the lab-frame) 
spins $S_{\mu z}= J_{\mu  \nc j z} + S_{\mp e} = \frac{1}{2} \hbar $ for $j =\frac{1}{2}$, and for
$n=2$,
$\lcfoot=1$  have the masses $M_{s 2,1 }^\astp $ given by (\ref{eq-Mh-qm}), which are overall identifiable to those of the muons $\mum,\mup$.   Thus when {\it in situ} converted to the $N=1$th quantum oscillations of the charges $-e,+e$ in $V_\hf$, the converted resemble in all respects the muons $\mum, \mup$ (Sec \ref{Sec-muon-model}). The un-projected $N=1$ oscillation states  of $-e,+e $  converted from the un-projected $\nc=2$ states (of mass $M_{s \nc}^\astp $ each) resemble then the pions $\pim, \pip$ in theory.  Using (\ref{eq-mrnc}) for $ \mr_{s\nclc}^\astp$, $\nc=2,\lc=1$, gives in the CM frame, and in the lab frame where the $\nu,\nubar$ precessions  cancel out, the rest mass of the muon $\mu^\m$   (or $\mup$), $M'' \equiv  M_{\mu ^\mp} $,
$$\displaylines{
\refstepcounter{equation} \label{eq-Mh-qm}
\hfill
M_\mum
=M^\astp_{s 2,1}
=\mr_{s 2,1}^\astp +M_e
= \frac{3}{4}  \mr_{s2}^\astp  +M_e 
= \frac{3}{4} (\frac{2M_e}{\a}) +M_e
=105.549  \ {\rm MeV}. 
\hfill (\ref{eq-Mh-qm})
}$$
The presence of $\mup $ does not add a friction term, so  $M_\mum^{f}=M_\mum $, since the $\mum,\mup$ rest masses defined in the $xy$ plane have no relative motion therein.
 For the pion $\pi^\m$, 
$(M'' \equiv )$ $M_\pi^\dagsup (= M_\pi+\Ocal_\eta) 
= \mr_{s2}^\astp +M_e= \frac{2M_e}{\a} +M_e=140.525 \ {\rm MeV}$ in theory;
   $\Ocal_\eta$ is a friction term in an
actual $\pi^\m, \pi^\p$ production,  not the $n=2$ states here. 
 In experiment, an $\em,\ep$ collision can directly produce $\mum,\mup$, apparently owing to the symmetric partial $\nubar, \nu$ orbits for each charge, whilst for producing $\pim,\pip$, only one $\nubar$ or $\nu$ can be attributed to each  charge. It instead takes an intermediate  bound  state  $ (\em\ep \rar)  \rho^0(770)$  to produce a $\pim,\pip$ pair.  $\rho^0$  can be  represented as consisting primarily of a  $h^\m_\Nsub,h^\p_\Nsub$ pair generated by  $-e,+e$  in  $N=2$ oscillation states in $V_\hf$,  converted  from $n=4, \lc=1$ states of the DHP.
Indicated by  its still larger experimental mass, $\rho^0$ apparently also contains two  $\em,\ep$ pairs, which are  present apparently to provide the $\nubar,\nu$ form of symmetry to each charge. Equations (\ref{eq-g-o-a-AA})-(\ref{eq-Jcal-ncs}) are general. The $\nc > 2$ levels can be expected to give rise to all the higher masses of unstable leptons ($\tau^\mp$ are the only observed ones) and composite meson particles,   typically stabilised  in presence of  secondary $\em,\ep$ pair(s).

The author expresses thanks to Professor Chairman C Burdik and the Organising Committee for the opportunity of presenting this work at the ISQS26, Tech Univ, Prague, 2019, to emeritus scientist P-I Johansson for continued moral support  and private financial support to the author's unification  research, and to Professors B Johansson and I Lindgren for giving moral support to the author's unification research, and to Dr R Dahm and Professor S Catto for useful discussions.

\vspace{2cm}


\renewcommand{\clearpage}{}

\noindent
{\large{Part D} (Abstract presented at 
QTS12, Prague, 2023)}
\vspace{0.cm}

\setcounter{section}{0}

\bibliographystyle{iopart-num}



\def\epm {\mbox{$\mathbin{e\mkern-7.7mu^{_{_{^{_{-}}}}}}$
}\hspace{-0.2cm}{}^\m }

\def\epm {\underaccent{\underline{e}}^\m} 

\def\epm{\underline{e}^\m}
\def\epmr{\epm_\rho}

\def\epmst{\cancel{e}^{\m\dastp}}

\def\rhoxy{\rho}







\title[Quantum Electromagnetic Theory of the IVBs and the Higgs, JXZJ
]{A Quantum Electromagnetic Theory of the Intermediate Vector Bosons and the Higgs}

\author{J.X. Zheng-Johansson
}
\address{
Institute of Fundamental Physics Research
} 
\address{}

\def\z{\zeta}
\def\z{\zeta}
\def\zbar{{\bar{\zeta}}}
\def\rbtd{\tilde{\rb}}
\def\rtdb{\tilde{\rb}}
\def\rbtd{\bar{\rb}}
\def\rtdb{\bar{\rb}}
\def\Ka{K}
\def\psiy{Y_{\nu\nubar}}
\def\dastpp{{}}
\def\scat{q}
\def\empt{{\mbox{\tiny${\emptyset}$}}}
\def\empty{{\mbox{\tiny${\emptyset}$}}}
\def\pse{{\mbox{\tiny{ps}}}}
\def\Q{t}
\def\Qbar{\bar{\Q}}
\def\ybar{\bar{y}}
\def\tbar{{\bar{t}}}
\def\tbar{{\bar{t}}}
\def\sbar{\bar{s}}
\def\Vmu{V_\mu}
\def\Vb{{\bf{V}}}
\def\rc{{r.c}}
\def\Zp{Z^\p{}}
\def\Zm{Z^\m}
\def\Zbar{\bar{Z}}
\def\Zr{Z_r}
\def\Zr{Z}
\def\Vbar{\bar{V}}         
\def\V{\mbox{$V\hspace{-0.17cm} \tiny{_{^\setminus}}$}}


\def\Vpm{\V^\pm}
\def\Vm{\fp^\m}
\def\Vp{\fp^\p}

\def\fpbar{Z^\m}

\def\fpbar{\Vfk^\m}
\def\fp{\Vfk^\p}
\def\fppm{\Vfk^\pm}

\def\fpsub{{\mbox{\tiny{$\fp$}}}}
\def\fpsubp{{\mbox{\tiny{$\fp{}^\p$}}}}
\def\fpbarsub{{\mbox{\tiny{$\fpbar$}}}}
\def\fppmsub{{\mbox{\tiny{$\fppm$}}}}
\def\vpsub{{\mbox{\tiny{$V^\p$}}}}

\def\app{a^\psup}

\def\abarr{{\bar{a}}}

\def\pp{{\pbar p}}
\def\pastp{p^\dastp}
\def\pbarastp{\pbar^\dastp}
\def\mup{\mu^\p}
\def\mupp{\mu^\p}

\def\mubar{{\bar{\mu}}}
\def\mumu{\mu\mu}
\def\mum{{\mu^\m}}
\def\mup{{\mu^\p}}
\def\Kh{{K_h}}
\def\Khat{\hat{K}}
\def\rhohat{\hat{\rho}}
\def\vehat{\hat{\ve}}
\def\vebhat{\hat{\veb}}

\def\pim{{\pi^\m}}
\def\pip{{\pi^\p}}
\def\pio{{\pi^0}}
\def\ein{{$e^{\m}_{in}$ }}
\def\einsub{{\mbox{$e^{\m}_{(in)}$}}}

\def\nc{n}
\def\lc{{\mbox{\small{$l$}}}}
\def\lcsub{{\mbox{\tiny{$l$}}}}
\def\lcfoot{{\mbox{\footnotesize{$l$}}}}
\def\lcfoots{{\mbox{\scriptsize{$l$}}}}
\def\lcfootss{{\mbox{\tiny{$l$}}}}

\def\nclc{{\nc,\lcsub}}



\def\ka{k}
\def\kasub{\ka}
\def\kan{\ka}
\def\kansub{\ka}
\def\kabar{{\bar{\ka}}}
\def\kbar{\bar{\ka}}
\def\kap{\mathcal{K}}

\def\Htr{\mathscr{H}}
\def\Hcal{\mathcal{H}}

\def\vebar{\bar{\ve}}

\def\jbar{\bar{j}}

\def\rbbar{\bar{\rb}}
\def\Mbar{\bar{M}}
\def\mrbar{\bar{\mr}}
\def\Rbbar{\bar{\Rb}}
\def\onebar{{\bar{1}}}
\def\tobar{{\bar{2}}}
\def\fbar{{\bar{4}}}
\def\tobarsub{{_{\bar{2}}}}
\def\onebarsub{{_{\bar{1}}}}
\def\osup{{\mbox{$\tiny{0}$}}}

\def\etaav{\bar{\eta}}
\def\etd{\tilde{e}}

\def\hm{h}
\def\tm{t}
\def\hy{t}

\def\rlar{\leftrightarrow}

\def\astt{{\mbox{{\tiny{$\wedge$}}}}}
\def\astt{{\mbox{{\tiny{$\bigtriangleup$}}}}}

\def\pn{p_n}
\def\pbn{\pb_n}
\def\En{E_n}
\def\Mn{\mathcal{M}}

\def\Ombar{{\Om\hspace{-0.25cm}^{_{\mbox{-}}} \hspace{0.13cm}}}
\def\Ombarp{{\Om\hspace{-0.2cm}^{_{\mbox{-}}} \hspace{0.13cm}}}

\def\pstru{\mathbin{p\mkern-9.2mu\mbox{\scriptsize{$-$}}}}
\def\Pstru{\mathbin{P\mkern-12.5mu^{_{\mbox{$-$}}}}}
\def\pbstru{\mathbin{\pb\mkern-9.2mu\mbox{\scriptsize{$-$}}}}
\def\pstrub{\mathbin{\pb\mkern-9.2mu\mbox{\scriptsize{$-$}}}}

\def\pmttot{\mathcal{P}}
\def\pmttotb{{\pmb{\mathcal{P}}\hspace{-0.1cm}}}
\def\pmttotbf{{\pmb{\mathcal{P}}\hspace{-0.1cm}}}



\def\pmt{p}
\def\pmtb{{\mathbf{\pmt}\hspace{-0.0cm}}}
\def\pmtbf{{\mathbf{\pmt}\hspace{-0.0cm}}}

\def\Pbstru{\mathbin{\Pb\mkern-12.5mu^{_{\mbox{$-$}}}}}
\def\Pstrub{\mathbin{\Pb\mkern-12.5mu^{_{\mbox{$-$}}}}}

\def\Pbar{{P\hspace{-0.2cm}^{{\mbox{-}}} \hspace{0.13cm}}}

\def\k{{\mbox{{\scriptsize{$\ve$}}}}}
\def\Tsub{{\mbox{{\tiny{$TP$}}}}}

\def\dagsup{\dagger}
\def\qu{c}
\def\M{M}

\def\excm{\hm^\m{}}

\def\eexcm{\hm^\m{}}

\def\eexc{{\hm}}

\def\ph{{p_{_h}}}
\def\pho{{ph}}

\def\eh{h}
\def\ehxy{{{h}_{xy}}}
\def\hxy{{{h}_{xy}}}
\def\ehxym{{h^\m_{xy}}}
\def\hxym{{h^\m_{xy}}}

\def\ehsub{{h}}

       \def\ehm{h^\m}

\def\ehmsub{{h^\m}}

\def\xysub{{{xy}}}
\def\xy{{{xy}}}
\def\em{{e^\m}}
\def\ep{{e^\p}}

\def\uq{u_q}
\def\uqb{{\bf{u}}_q}

\def\uqr{\xi_{q}}
\def\uqrb{{\pmb{\xi}}_q}
\def\uqz{\zeta_q}
\def\uqzb{{\pmb{\zeta}}_q}

\def\ubar{\bar{u}}
\def\dbar{\bar{d}}

\def\gt{>}
\def\Htr{\mathscr{H}}

\def\onebar{{\bar{1}}}
\def\tobar{{\bar{2}}}
\def\fbar{{\bar{4}}}
\def\tobarsub{{_{\bar{2}}}}
\def\onebarsub{{_{\bar{1}}}}
\def\osup{{\mbox{$\tiny{0}$}}}
\def\nc{{\bar{n}}}
\def\nc{n}

\def\lc{{\bar{l}}}
\def\etaav{\bar{\eta}}
\def\etd{\tilde{e}}

\def\hm{h}
\def\hmm{h^\m}
\def\tm{t}
\def\hy{t}

\def\astt{{\mbox{{\tiny{$\wedge$}}}}}
\def\astt{{\mbox{{\tiny{$\bigtriangleup$}}}}}

\def\pn{p_n}
\def\pbn{\pb_n}
\def\En{E_n}
\def\Mn{\mathcal{M}}

\def\Ombar{{\Om\hspace{-0.3cm}^{_{\mbox{-}}} \hspace{0.13cm}}}

\def\pstru{\mathbin{p\mkern-9.2mu\mbox{\scriptsize{$-$}}}}
\def\Pstru{\mathbin{P\mkern-12.5mu^{_{\mbox{$-$}}}}}
\def\pbstru{\mathbin{\pb\mkern-9.2mu\mbox{\scriptsize{$-$}}}}
\def\pstrub{\mathbin{\pb\mkern-9.2mu\mbox{\scriptsize{$-$}}}}

\def\Pbstru{\mathbin{\Pb\mkern-12.5mu^{_{\mbox{$-$}}}}}
\def\Pstrub{\mathbin{\Pb\mkern-12.5mu^{_{\mbox{$-$}}}}}

\def\Pbar{{P\hspace{-0.2cm}^{{\mbox{-}}} \hspace{0.13cm}}}

\def\k{{\mbox{{\scriptsize{$\ve$}}}}}
\def\Tsub{{\mbox{{\tiny{$TP$}}}}}

\def\dagsup{\dagger}
\def\qu{c}
\def\M{M}

\def\excm{\hm^\m{}}

\def\eexcm{\hm^\m{}}

\def\eexc{{\hm}}

\def\ph{{p_{_h}}}
\def\pho{{ph}}

\def\eh{h}
\def\ehxy{{{e_h}_{xy}}}
\def\ehxym{{{e_h^\m}_{xy}}}
\def\ehsub{{e_h}}
       \def\ehm{e_h^\m}
\def\ehmsub{{e_h^\m}}
\def\xysub{{{xy}}}
\def\xy{{{xy}}}
\def\em{{e^\m}}
\def\ep{{e^\p}{\hspace{-0.2cm}}}
\def\epp{{e^\p}}
\def\uq{u_q}
\def\uqb{{\bf{u}}_q}

\def\uqr{\xi_{q}}
\def\uqrb{{\pmb{\xi}}_q}
\def\uqz{\zeta_q}
\def\uqzb{{\pmb{\zeta}}_q}

\def\pim{{\pi^\m}}
\def\pio{{\pi^0}}
\def\ein{{$e^{\m}_{in}$ }}
\def\einsub{{\mbox{$e^{\m}_{(in)}$}}}

\def\self{{\rm{self}}}

\def\str{\star}
\def\astr{{a^\star}}
\def\bstr{{b^\star}}

\def\mbstr{m_b^\str}
\def\mastr{m_a^\str}
\def\MAsub{M^\Asub}
\def\mrAsub{\mr^\Asub}
\def\MBsub{M^\Bsub}
\def\mrBsub{\mr^\Bsub}

\def\Cbar{\hspace{0.02cm}C\hspace{-0.365cm}^{{  \atop  -}\hspace{0.08cm}}{}   }
\def\Cbarp{\hspace{0.02cm}C\hspace{-0.34cm}^{{  \atop  -}\hspace{0.08cm}}{}   }
\def\Ups{\mit{\Upsilon}}
\def\ebar{\bar{e}}

\def\bs{\beta}
\def\bsbar{\bar{\beta}}

\def\bp{{\beta'}}
\def\bbp{{\b\bp}}
\def\qt{{q_t}}
\def\mj{{m_j}}
\def\chat{\hat{c}}

\def\sighat{\hat{\sig}}
\def\nhat{\hat{n}}
\def\uhat{\hat{u}}
\def\rhat{\hat{r}}
\def\rbhat{\hat{\rb}}

\def\that{\hat{t}}
\def\xihat{\hat{\xi}}

\def\zetahat{\hat{\zeta}}

\def\ehat{\hat{e}}
\def\zhat{\hat{\zb}}
\def\zbhat{\hat{\zb}}
\def\yhat{\hat{\yb}}
\def\ybhat{\hat{\yb}}

\def\xhat{\hat{\xb}}
\def\xbhat{\hat{\xb}}
\def\phihat{\hat{\vphi}}
\def\vphihat{\hat{\vphi}}
\def\vphitdhat{\hat{\vphitd}}
\def\vphitd{{\tilde{\vphi}}}
\def\thetahat{\hat{\theta}}
\def\xphat{\hat{x'}}
\def\xpphat{\hat{x''}}

\def\Jcal{{\mathcal{J}}}
\def\Jcalb{{      \pmb{\mathcal{J}}   }}

\def\Jorb{{\mathcal{J}}}
\def\Jtr{{\mathcal{J}}}
\def\Jtrb{{\pmb{\mathcal{J}}}}
\def\Tsub{{\mbox{{\tiny{$TP$}}}}}

\def\Asub{{\mbox{{\tiny{$A$}}}}}
\def\Bsub{{\mbox{{\tiny{$B$}}}}}

\def\AA{\mbox{$A${\hspace{-0.25cm}}{}$^{^o}$\hspace{0.2cm}}}

\def\rhobar{\bar{\rho}}
\def\rhob{\pmb{\rho}}

\def\hfmbox{\mbox{$\frac{1}{2}$}}

\def\hfp{{\mbox{\tiny{$1/2$}}}}
\def\hfp{{\mbox{\tiny{$1\hspace{-0.06cm}/\hspace{-0.03cm}2$}}}}

\def\hf{{\mbox{\tiny{$\frac{1}{2}$}}}}
\def\hfmbox{\mbox{$\frac{1}{2}$}}
\def\hfpp{{^1\hspace{-0.06cm}\mbox{\tiny{/}}\hspace{-0.05cm}{}_2}}
\def\hfb{{}^{{\mbox{\tiny{$\frac{1}{2}$}}}}}

\def\hfp{{\mbox{\tiny{$1_{\hspace{-0.05cm}{^/\hspace{-0.05cm}2}}      $}}}}
\def\trhfp{{\mbox{\tiny{$3_{\hspace{-0.05cm}{^/\hspace{-0.05cm}2}}      $}}}}
\def\trhf{{\mbox{\tiny{$\frac{3}{2}$}}}}

\def\taum{\tau^\m}
\def\taup{\tau^\p}
\def\pbar{{\bar{p}}}
\def\nbar{\bar{n}}

\def\tr{{\mbox{\tiny{$I$}}}}
\def\tra{{\rm{tr}}}

\def\rarno{\mbox{$\rightarrow \hspace{-0.45cm}/$\hspace{0.45cm}}}

\def\ovrar{\overrightarrow}

\def\rar{\rightarrow}
\def\lar{\leftarrow}
\def\sear{\searrow}
\def\hrar{\hookrightarrow}

\def\uarsup{{\mbox{\tiny{$\uparrow\hspace{-0.09cm}$}}}}
\def\darsup{{\mbox{\tiny{$\downarrow\hspace{-0.09cm}$}}}}

\def\uar{{\mbox{$\hspace{-0.05cm}\uparrow$}}}
\def\dar{{\mbox{$\hspace{-0.05cm}\downarrow$}}}

\def\point{{\rm{point}}}
\def\j{j}
\def\jsub{{\mbox{\tiny{$j$}}}}

\def\J{J}
\def\Jscr{J}

\def\Jtot{\mbox{$\mathscr{J}$}}

\def\Jb{{\mathbf{J}}}
\def\Jscrb{{\mathbf{J}}}
\def\Jbscr{{\mathbf{J}}}

\def\Jorb{{\mathcal{J}}}
\def\Jtr{{\mathcal{J}}}
\def\Jtrb{{\pmb{\mathcal{J}}}}

\def\Tsub{{\mbox{{\tiny{$TP$}}}}}

\def\db{\pmb{\delta}}

\def\Cb{{\bf{C}}}

\def\d{\delta}
\def\GWS{{\mbox{\tiny{GWS}} }}
\def\Xsub{{\mbox{\tiny{X}} }}
\def\Isub{{\mbox{\tiny{I}} }}
\def\Rsub{{\mbox{\tiny{R}} }}
\def\Lsub{{\mbox{\tiny{L}} }}
\def\Ssub{{\mbox{\tiny{S}} }}

\def\rsub{{{_o}}}
\def\vtheta{\vartheta}
\def\pbf{{\bf{p}}}
\def\neu{{\mbox{\tiny{Neu}} }}
\def\rbav{\bar{\rb}}
\def\Fbav{\bar{\Fb}}
\def\fb{{\bf{f}}}

\def\Rsub{{\mbox{\tiny{R}} }}
\def\Lsub{{\mbox{\tiny{L}} }}
\def\HRsub{{\mbox{\tiny{H}} }}

\def\in{{in}}
\def\osup{{\mbox{\tiny{$0$}} }}
\def\inTo{{\mbox{\tiny{in$T_o$}} }}

\def\Ksub{{\mbox{\tiny{$K$}} }}
\def\Lamsub{{\mbox{\tiny{$\Lam$}} }}
\def\expm{{o}}
\def\exto{{o}}
\def\ex{{eq}}
\def\eq{{eq}}
\def\exti{{1}}
\def\ext{{ext}}

\def\sigb{{\pmb{ \sigma }}}

\def\tot{{     \mbox{\tiny{tot}} }}
\def\ef{{\rm{ef}}}
\def\Lrw{\Longrightarrow}
\def\nubar{{\bar{\nu}}}
\def\nubar{{\bar{\nu}}}
\def\nubarmu{{\bar{\nu}_{\mu}}}

\def\numu{{\nu_{\mu}}}
\def\numup{{\nu_{\mu}'}}

\def\nubarem{{\bar{\nu}_{e}}}
\def\nuepr{{\nu_e'}}
\def\nuep{{\nu_{e^+}}}
\def\nue{{\nu_{e}}}
\def\nubare{{\nubar_e}}
\def\nubarep{{\nubar_e'}}

\def\orb{{orb}}
\def\Mcm{M}
\def\mcm{M}
\def\velcm{U}
\def\Lambar{\overline{\Lam}}

\def\gbar{{\gamma\hspace{-0.125cm}{_{\mbox{-}}} \hspace{0.cm}}}

\def\Vol{\Volp}
\def\Volp{{\mbox{V\hspace{-0.3cm}$-$}}}


\def\Volp{{V\hspace{-0.25cm}^{_{\mbox{-}}} \hspace{0.01cm}}}
\def\Volp{{\mathscr{V}}}

\def\Vpm{V^\pm}
\def\Vpmsub{{\mbox{\tiny{$\Vpm$}}}}

\def\l{l}
\def\mn{{m_\lsub}}
\def\lsub{{\mbox{\tiny{$l$}}}}
\def\isub{{\mbox{\tiny{$1$}}}}

\def\Vfrak{\mathfrak{V}}
\def\Vfk{\mathfrak{V}}
\def\Vfksub{\tiny{\mbox{$\mathfrak{V}$}}}

\def\Ocal{\mathcal{O}}
\def\Rcal{\mathcal{R}}

\def\Ycal{Y}

\def\Tcal{\mathcal{T}}

\def\Rbcal{{\pmb{\mathcal{R}}}}

\def\sb{{\bf{s}}}

\def\velsp{\vel_{s_p}}
\def\velse{\vel_{s_e}}

\def\Db{{\bf{D}}}
\def\velop{\velavp}
\def\veloe{\velave}

\def\velsp{\vel^s_{p}}
\def\velse{\vel^s_{e}}
\def\velavp{\bar{\vel}^s_{p}}
\def\velsavp{\bar{\vel}^s_{p}}
        \def\velspav{\bar{\vel}^s_{p}}  
        \def\velave{\bar{\vel}^s_{e}}
\def\velsave{\bar{\vel}^s_{e}}
         \def\velseav{\bar{\vel}^s_{e}}

\def\velbavp{\bar{\velb}^s_{p}}
\def\velbave{\bar{\velb}^s_{e}}

\def\velbaveb{\bar{\velb}^s_{e}}

\def\velav{\bar{\vel}}
\def\velavb{\bar{\pmb{\vel}}}
\def\velbav{\bar{\pmb{\vel}}}

\def\aavp{\bar{a}_{p}}
\def\apav{\bar{a}_{p}}
\def\aeav{\bar{a}_{e}}
\def\aave{\bar{a}_{e}}
\def\aav{\bar{a}}
\def\aavb{\bar{{\bf{a}}}}

\def\aop{\bar{a}_{p}}
\def\aoe{\bar{a}_{e}}
\def\amp{a_{m_p}}
\def\ame{a_{m_e}}
\def\aqp{a_{p}}
\def\aqe{a_{e}}
\def\ws{\w^s}
\def\wsp{\w^s_{p}}
\def\wse{\w^s_{e}}
\def\wsa{\w^s_\a}

\def\mumu{\mu\mu}

\def\musub{\mu}
\def\musa{\mu^s_\a}

\def\velsa{\vel^s_\a}

\def\Ls{L^s}
\def\Lsp{L^s_{p}}
\def\Lse{L^s_{e}}
\def\mus{\mu^s}
\def\musp{\mu^s_{p}}
\def\muse{\mu^s_{e}}
\def\Lorb{L}
\def\worb{\w}
\def\muorb{\mu}

\def\Mu{\mathscr{M}}
\def\gb{{\bf{g}}}
\def\zuni{\hat{z}}
\def\Zuni{\hat{Z}}

\def\runi{\hat{r}}
\def\thuni{\hat{\phi}}
\def\Runi{\hat{R}}
\def\Thuni{\hat{{\mit\Phi}}}

\def\Bbext{\Bb_{ext}}
\def\Fbext{\Fb_{ext}}
\def\mrm{{\mathbin{\mu\mkern-5.6mu\mbox{-}}}}

\def\rvec{\vec{r}}
\def\Rvec{\vec{R}}

\def\suf{{}_{}}
\def\labsup{{     \mbox{\tiny{IL}} }}
\def\lab{\labsup}
\def\Lab{     {\mbox{\tiny{L}}} }
\def\Labi{     {\mbox{\tiny{       $L_1$      }}} }

\def\arssup{\star}
\def\arsup{\star}
\def\ars{\arsup}



\def\FbIL{{\bf{F}}}
\def\BbIL{{\bf{B}}}
\def\BfbIL{{\bf{B}}}
\def\MIL{M}
\def\RIL{{\sf{R}}}
\def\RbIL{{\bf{R}}}

\def\Fbicm{{\pmb{\sf{F}}}}
\def\Bbicm{{\pmb{\sf{B}}}}
\def\Bfbicm{{\pmb{\sf{B}}}}
\def\Ebicm{{\pmb{\sf{E}}}}
\def\Lbicm{{\pmb{\sf{L}}}}
\def\Sbicm{{\pmb{\sf{S}}}}
\def\Lbicm{{\pmb{\sf{L}}}}
\def\Vicm{{\sf{V}}}
\def\Hicm{{\sf{H}}}
\def\Eicm{{\sf{E}}}
\def\Ticm{{\sf{T}}}
\def\Ficm{{\sf{F}}}
\def\Licm{{\sf{L}}}

\def\Fbast{{\pmb{\sf{F}}}}
\def\Bbast{{\pmb{\sf{B}}}}
\def\Bfbast{{\pmb{\sf{B}}}}
\def\Ebast{{\pmb{\sf{E}}}}
\def\Efbast{{\pmb{\sf{E}}}}
\def\Lbast{{\pmb{\sf{L}}}}
\def\Sbast{{\pmb{\sf{S}}}}
\def\Lbast{{\pmb{\sf{L}}}}
\def\Sbast{{\pmb{\sf{S}}}}
\def\Jbast{{\pmb{\sf{J}}}}

\def\Wb{{\bf{W}}}

\def\Vast{{\sf{V}}}
\def\Hast{{\sf{H}}}
\def\East{{\sf{E}}}
\def\Tast{{\sf{T}}}
\def\Fast{{\sf{F}}}
\def\Last{{\sf{L}}}
\def\Sast{{\sf{S}}}
\def\Bast{{\sf{B}}}
\def\Mast{{\sf{M}}}
\def\Jast{{\sf{J}}}
\def\Uast{{\sf{U}}}
\def\Gast{{\sf{G}}}

\def\Vdag{{V^\dag}}
\def\Hdag{{H^\dag}}
\def\Edag{{E^\dag}}
\def\Tdag{{T^\dag}}
\def\Fdag{{F^\dag}}
\def\Ldag{{L^\dag}}
\def\Sdag{{S^\dag}}
\def\Bdag{{B^\dag}}
\def\Mdag{{M^\dag}}
\def\Jdag{{J^\dag}}
\def\Udag{{U^\dag}}

\def\Fbdag{ {{\mathbf F}^\dag} }
\def\Bbdag{{{\mathbf B}^\dag} }
\def\Bfbdag{{{\mathbf B}^\dag} }
\def\Ebdag{{{\mathbf E}^\dag} }
\def\Efbdag{{{\mathbf E}^\dag} }
\def\Lbdag{{{\mathbf L}^\dag} }
\def\Sbdag{{{\mathbf S}^\dag} }
\def\Lbdag{{{\mathbf L}^\dag} }
\def\Sbdag{{{\mathbf S}^\dag} }
\def\Jbdag{{{\mathbf J}^\dag} }

\def\Ecal{{\mathcal{E}}}

\def\astsup{{}}
\def\ast{\astsup}






\def\dastp{*}

\def\astn{{*}}
\def\astp{{*}}
\def\astpd{{**}}
\def\astpp{{*'}}

\def\ab{{\pmb{\mbox{$a$}}}}
\def\acb{{\bf{\mbox{\sf{a}}}}}
\def\ac{{\mbox{\sf{a}}}}

\def\bb{{\bf{b}}}

\def\dagsup{\dagger}


\def\mr{{\mathscr{M}}}

\def\mrb{{\pmb{\mathscr{M}}}}

\def\mrdastp{{{\mathscr{M}\hspace{-0.1cm}}^\dastp}}

\def\mrsub{{\mbox{{\tiny{$\mr$}}}}}

\def\mrrr{{\mathscr{M}^*}}
\def\mrr{{\mathscr{M}^*}}
          \def\gar{\g^*}
\def\garr{\g^*{}}
\def\velrr{\vel^\ddagger{}}
        \def\velr{\vel^\ddagger{}}
     \def\velbrr{\velb^\ddagger{}}

     \def\trr{t^*{}}
     \def\rbrr{{\bf{r}^*}}
\def\rbr{{\bf{r}^*}}

\def\rvecrr{{\vec{r}^*}}
 \def\rvecr{{\vec{r}^*}}

\def\mr{{\mathscr{M}}}
\def\mrdag{{\mathscr{M}^*\dagger}}
\def\mrlab{{\mathscr{M}}}
\def\gadag{\g^\dagger}
\def\gdag{\g^\dagger}
\def\veldag{\vel^\dagger{}}
       \def\velbdag{\velb^\dagger{}}
\def\velrr{\vel^\ddagger{}}
     \def\velbrr{\velb^\ddagger{}}

\def\rbdag{{\bf{r}^\dagger}}

\def\tdag{t^\dagger{}}
\def\rbdag{{\bf{r}^\dagger}}

\def\rlab{r}
\def\glab{\g}
\def\galab{\g}
\def\velblab{\velb}
\def\vellab{\vel}
\def\mrlab{{\mathscr{M}}}
\def\rblab{{\bf{r}}}
\def\rveclab{{\vec{r}}}
\def\rveclab{{\vec{r}}}

\def\rb{{\mathbf r}}
\def\rbtd{\bar{\rb}}
\def\tlab{t}
\def\garcmf{\g_{\lab:\cmsub}}

\def\xtd{\tilde{x}}
\def\ytd{\tilde{y}}
\def\ztd{\tilde{z}}
\def\cbtd{\tilde{\cb}}

\def\aba{{\bar{a}}}
\def\rbar{{\bar{r}}}
\def\m{{\mbox{-}}}
\def\mpr{{\mbox{-}' }}

\def\ubscr{\pmb{\mathscr{U}}}
        \def\ubscrq{{\pmb{\mathscr{U}}\hspace{-0.14cm}}_q}
         \def\ubscrqsq{{\pmb{\mathscr{U}}\hspace{-0.14cm}}_q{\hspace{-0.05cm}}}

\def\N{{\mbox{{\small{$N$}}}}}

\def\Nsub{{\mbox{{\tiny{$N$}}}}}
\def\Nsup{{\mbox{{\tiny{$N$}}}}}

\def\Vvqo{V_{\v q0}}
\def\Vvq{V_{\v q}}
\def\mub{\pmb{\mu}}
\def\taub{\pmb{\tau}}
\def\tbar{\bar{t}}
\def\thetab{\pmb{\theta}}
\def\phib{\pmb{\phi}}
\def\Phimb{\pmb{\Phim}}

\def\wb{\pmb{\w}}
\def\mb{{\mathbf{m}}}

\def\lep{l}
\def\Rbb{\mathbb{R}}
\def\Kbb{\mathbb{P}}
\def\R{r_{max}}
\def\Pscr{\mathscr{P}}
\def\Hscr{\mathscr{H}}
\def\Vscr{\mathscr{V}}
\def\Tscr{\mathscr{T}}
\def\Lscr{\mathscr{L}}

\def\Ds{\mathscr{D}}

\def\Xim{{\mit{\Xi}}}

\def\rw{\rightarrow}
\def\jm{{j\mu}}
\def\kp{{j'}}
\def\muk{{\n'}}
\def\Nssk{\Nss_1}
\def\p{{\mbox{\footnotesize{$+$}} \hspace{-0.03cm}}}
\def\p{{\mbox{\footnotesize{$+$}} \hspace{-0.03cm}}}

\def\p{\psup}

\def\psub{{\hspace{-0.05cm}\mbox{\tiny{$+$}}\hspace{0cm}}}
\def\psup{\psub}

\def\pe{\p e}
\def\pq{\p q}
\def\conm{\hspace{-0.07cm}:}

\def\H{{a_{\Sigsub} \hspace{-0.05cm}}}
\def\Hn{a_{n\Sigsub}}
\def\Ia{\mathcal{A}}
\def\hpbar{\abar}

\def\htid{\tilde{h}}

\def\hpbars{\hpbar}
\def\Zbr{\Z^\dagsup}

\def\Zbrsub{{\mbox{{\tiny{$\Zbr$}}}}}

\def\Zhat{\widehat{\Z}}
\def\Zddot{\ddot{\Z}}
\def\Zbar{\bar{\Z}}

\def\Ztd{{\tilde{\Z\hspace{0.1cm}}}}
\def\Ztdp{{\tilde{\Z}}}

\def\Vtd{{\tilde{\Vfrak\hspace{0.1cm}}}}

\def\Vtdp{\bar{V}}

\def\hpp{h^\p}
\def\hp{a}
               \def\abar{{a\hspace{-0.18cm}\mbox{{\small $^{_{_{-}}}$}}\hspace{-0.07cm}}}
                \def\abars{{\hspace{-0.03cm}a\hspace{-0.1cm}\mbox{\small{-}}\hspace{0.0cm}}}
\def\abar{{a\hspace{-0.18cm}\mbox{{\small $^{_{_{-}}}$}}\hspace{-0.07cm}}}
\def\Qcal{\mathbin{{Q}\mkern-8.5mu^{_{\mbox{\small{$\dash$}}}}\hspace{-0.04cm} }}

\def\Sa{{\mathfrak{S}}}
\def\nfrak{{\mathfrak{N}}}

\def\h{h}

\def\taubar{\bar{\tau}}
\def\taustru{\mathbin{{\tau}\mkern-10.3mu_{^{{}^{{\mbox{\tiny{$-$}}}}\hspace{-0.10cm}}} }}
\def\ho{\eta_0}
\def\hjn{\hp_{\jn}}
\def\Wst{{\mathbin{\Omega\mkern-4.1mu^{_{\mbox{\footnotesize{-}}}}}\hspace{-0.04cm}}}
\def\Wsts{{\mathbin{\Omega\mkern-4.2mu^{_{\mbox{\scriptsize{-}}}}}\hspace{-0.06cm}}}
\def\Wstsup{{\mathbin{\Omega\mkern-5.mu^{_{\mbox{\scriptsize{-}}}}}}}
\def\Wstt{{\mathbin{\Omega\mkern-3.5 mu^{_{\mbox{\scriptsize{-}}}}}}}  


\def\Nstat{{\mathcal{N}}}

\def\Nsts{{\Wsts}}
\def\Nst{{\Wst}}
\def\Nss{{\Wst}}
\def\Nstsup{{\Wstsup}}

\def\nsig{{\Sigsub\n}}
\def\nPi{{\n_{^{_\Pi}}}}

\def\pjn{p_{j\n}}
\def\jn{{jn}}
\def\xjn{{j\n}}
\def\Pcaln{\Pcal_{n}}
\def\Pcalbarn{ \overline{\Pcal_\n}}

\def\Pcalens{\Pcal_{\ens}}
\def\Pens{\Pcal_{\ens}}
\def\Pensm{\Pcal_{\ens,max}}
\def\Nstgm{\Nst_{i.g.m}}
\def\Nstam{\Nst_{i.a.m}}
\def\quadd{\ \ }
\def\qe{{\mathbin{q\mkern-11mu-}}}

\def\la{\langle}
\def\ra{\rangle}
\def\mrsub{{\mbox{{\scriptsize{$\mr$}}}}}
\def\Msub{{_M}}

\def\cmsub{{        {_{\mbox{{\tiny{CM}}}}} }}

\def\cmsub{   {\mbox{{\tiny{cm}}}} }

\def\icmsub{{\mbox{{\tiny{ICM}}}}}

\def\ncmsub{{\mbox{{\tiny{NCM}}}}}
\def\Neusub{{\mbox{{\tiny{Neut}}}}}

\def\pmsub{{\mbox{{\tiny{$\pm$}}}}}

\def\Zsub{{\mbox{{\tiny{$Z$}}}}}
\def\Zsup{{\mbox{{\tiny{$Z$}}}}}

\def\Zsubo{{\mbox{{\tiny{$Z$}}}}}
\def\Zosub{{\mbox{{\tiny{$Z^0$}}}}}

\def\Zsubp{{\mbox{{\tiny{$Z^\psup$}}}}}
\def\Zsubm{{\mbox{{\tiny{$Z^\m$}}}}}

\def\Vsubp{{\mbox{{\tiny{$\Vfk^\psup$}}}}}

\def\Hsub{{\mbox{{\tiny{H}}}}}

\def\TPsup{{\mbox{{\tiny{$TP$}}}}}
\def\TPsub{{\mbox{{\tiny{$TP$}}}}}

\def\Hsub{{\mbox{{\tiny{$H$}}}}}

\def\Fsub{{\mbox{{\tiny{$F$}}}}}
\def\Larmsub{{\mbox{{\tiny{$Larm$}}}}}

\def\Lsub{{\mbox{{\tiny{$L$}}}}}
\def\Sigsub{{\mbox{{\tiny{$\Sigma$}}}}}
\def\Wsub{{\mbox{{\tiny{$W^\pmsub$}}}}}
\def\wsub{{\mbox{\tiny{w}}}}

\def\wsubi{{\mbox{\tiny{$W_1$}}}}
\def\wpmsub{{\mbox{\tiny{$W^\pmsub$}}}}
\def\wsubpm{{\mbox{\tiny{$W^\pmsub$}}}}
\def\wpmsub{{\mbox{\tiny{$W^\pmsub$}}}}
\def\wssub{\wsub}

\def\wsubo{{\mbox{\tiny{$0$}}}}
\def\wsubthi{{\mbox{\tiny{$\theta_1$}}}}

\def\Nsub{{\mbox{{\tiny{$N$}}}}}
\def\rep{{rep}}
\def\Sig{\Sigma}
\def\bi{b^{i}}
\def\i{i}
\def\n{\nu}
\def\uscr{\mathscr{U}}
\def\uscrdotbar{\bar{\dot{\mathscr{U}}}}
\def\vac{{\rm{vac}}}
\def\DV{V_{aq}{}}
\def\DF{F_{aq}{}}
\def\DFb{\Fb_{aq}{}}

\def\Vcal{{\mathscr{V}}} 

\def\Bsub{{\mbox{\scriptsize{B}}}}

\def\Csub{{\mbox{\scriptsize{C}}}}
\def\Csubti{{\mbox{\tiny{C}}}}
\def\Dsub{{\mbox{\scriptsize{D}}}}

\def\Nsub{{{\mbox{\tiny${N}$}}}}
\def\Hsub{{{\mbox{\tiny${H}$}}}}

\def\Hbar{\bar{H}}
\def\pbar{{\bar{p}}}

\def\exc{{\rm exc}}
\def\ext{{{\rm ext}}}
\def\mini{0}

\def\Pcal{{\mathcal{P}}}
\def\Pcalb{{\pmb{{\mathcal{P}}}}}

\def\bav{{\bar{b}}}
\def\v{{\rm v}}
\def\vrm{\vel_{t}{}}
\def\vit{\vrm}
\def\vrmb{{\bf{v}}}

\def\Hbar{\bar{H}}
\def\D{\Delta}
\def\bcal{b}
\def\bbar{\mathbin{{b}\mkern-9.5mu^{{\mbox{\tiny{$-$}}}}\hspace{-0.00cm} }}
\def\bbarr{\bar{b}}
\def\bebar{{\bar{\beta}}}
\def\be{\beta}

\def\nstat{\nu}
\def\nst{\nu}
\def\engbar{\bar{\eng}}
\def\engobar{\bar{\eng}_0}
\def\psias{\psi}
\def\Phimas{\Phim}
\def\fas{f}
\def\rbb{\as}

\def\La{L}
\def\Ja{J}
\def\as{p}
\def\ioii{{\mbox{\normalsize${\frac{1}{2}}$}}}
\def\Rb{{\bf R}}

\def\xb{{\bf x}}
\def\Xb{{\bf X}}
\def\yb{{\bf y}}
\def\zb{{\bf z}}

\def\ub{{\bf{u}}}
\def\hatu{\hat{u}_q}
\def\Nsub{{{\mbox{\tiny${N}$}}}}
\def\Pisub{{{\mbox{\tiny${\mit{\Pi}}$}}}}

\def\q{\bar{q}}
\def\xdot{\dot{x}}
\def\ens{{ens}}
\def\Lcal{\mathcal{L}}
\def\Lcalb{\pmb{\mathcal{L}}}
\def\Tcal{\mathcal{T}}
\def\Kcal{{\mathcal{K}}}
\def\kcal{\kappa}
\def\kcalb{{\pmb{\kcal}}}

\def\Xcal{{\mathcal{X}}}

\def\Kcalb{  {       \pmb{   \mathcal{K}}    }                }

\def\Wvel{\Omegavel}
\def\Ncal{{\mathcal{N}}}
\def\Ncalsub{{\mbox{\tiny{$\Ncal$}}}}

\def\Dcal{{\mathcal{D}}}
\def\Dcalsub{{\mbox{\tiny{$\Dcal$}}}}

\def\Qcal{{\mathcal{Q}}}
\def\Qcalsub{{\mbox{\tiny{$\Qcal$}}}}

\def\Omegavel{\mathbin{{\mit\Omega}\mkern-13.mu^{_{\mbox{$-$}}}\hspace{-0.08cm}{}_d }}
\def\Om{{{\mit{\Omega}}}}
\def\omegavel{{\w\mbox{\hspace{-0.38cm} \vspace{0.15cm}$-$\hspace{-0.02cm}}}}
\def\wvel{\omegavel_d}

\def\Ucal{\mathcal{U}}

\def\Omegavel{\mathbin{{\mit\Omega}\mkern-13.mu^{_{\mbox{$-$}}}\hspace{-0.08cm}{}_d }}
\def\Wvel{\Omegavel}

\def\q{\mathbin{q\mkern-11mu-}}
\def\PE{\mbox{\tiny{{\rm P.E.}}}}
\def\ME{\mbox{\tiny{{\rm M.E.}}}}
\def\QM{\mbox{\tiny{{\rm QM}}}}
\def\Psub{\mbox{\tiny{{\rm P}}}}
\def\TP{{\mbox{\tiny{{\rm TP}}}}}

\def\SM{{\mbox{\tiny{{\rm SM}}}}}
\def\MT{{\mbox{\tiny{{\rm MT}}}}}

\def\ev{\epsilon}

\def\Ci{1}
\def\betamt{{\bf{b}}}
\def\kb{{\bf{k}}}
\def\qb{{\bf{q}}}
\def\kbf{{\bf{k}}}
\def\Kb{{\bf{K}}}
\def\cb{{\mathbf{c}}}

\def\orar{\overrightarrow}

\def\pbf{{\bf{p}}}
\def\pb{\pbf}

\def\Pbf{{\bf{P}}}
\def\Mbf{{\bf{M}}}
\def\pbf{{\bf{p}}}
     \def\Acal{{\cal{A}}}
\def\Bcal{{I_{{\rm{ex}}}}}
\def\Ccal{{\cal{C}}}
\def\Ccal{{\cal{C}}}

\def\psipi{\psi_{\p}(1)}
\def\psipii{\psi_{\p}(2)}
\def\psimi{\psi_{\m}(1)}
\def\psimii{\psi_{\m}(2)}

\def\ai{\alpha(1)}
\def\aii{\alpha^{'}(2)}
\def\bi{\beta^{'}(1)}
\def\bii{\beta(2)}

\def\fa{f_r}

\def\Ca{C_a}
\def\fbf{{\bf{f}}}
\def\fb{{\bf{f}}}

\def\Ocal{{\cal{O}}}
\def\psib{{\pmb{\psi}}}
\def\alphab{{\pmb{\alpha}}}
\def\sigmab{{\pmb{\sigma}}}
\def\sig{\sigma}
\def\Eb{{\bf E}}
\def\Bb{{\bf B}}
\def\ke{\kappa}
\def\nabb{{\pmb{\nabla}}}
\def\nablab{{\pmb{\nabla}}}
\def\vir{{\rm vir}}

\def\psitot{\psi}
\def\jb{{\bf{j}}}
\def\vel{upsilon}
\def\ve{\upsilon}

\def\vels{{\hspace{0.1cm}\breve{\hspace{-0.1cm}\vel}}}
\def\velsb{{\breve{\velb}}}
\def\vb{{\bf{v}}}
\def\Imtr{I}
\def\citeUnif{4?}
\def\App{}
\def\Qcal{{\mathcal{Q}}}

\def\Tcal{{\mathcal{T}}}
\def\Cross{Q}

\def\vphilim{f}
\def\ft{{\mathcal{B}}}
\def\vphibar{\mathbin{\varphi\mkern-12.5mu-}}
\def\vphi{\varphi}
\def\vphib{{\pmb{\varphi}}}

\def\med{{\med}}
\def\Mcal{{\mathfrak{M}}}
\def\Mfrak{{\mathfrak{M}}}
\def\Mca{{\mathcal{M}}}

\def\Sb{{\bf{S}}}
         \def\xia{{\mathcal{A}}}
\def\tha{\theta}

\def\xib{\pmb{\xi}}
\def\zetab{\pmb{\zeta}}
\def\nb{{\bf{n}}}
\def\zb{{\bf{z}}}
\def\nbhat{\hat{\nb}}

\def\phiv{\varphi}
\def\Lb{{\bf{L}}}
\def\velsub{_{\vel}}
\def\Jb{{\bf{J}}}
\def\Pb{{\bf{P}}}
\def\Mb{{\bf{M}}}
\def\Zo{{Z^0}}

\def\Zb{{\bf{Z}}}
\def\Zsup{{\mbox{{\tiny{$Z$}}}}}

\def\nablab{{\pmb{\nabla}}}
\def\velb{{\pmb{\vel}}}
\def\veb{{\pmb{\ve}}}

\def\Db{{\bf{D}}}

\def\Ab{{\bf{A}}_a}
\def\Abb{{\bf{A}}}

\def\vel{\upsilon}
\def\ve{\upsilon}
\def\Thm{\vartheta}
\def\Thetam{{\mit{\Theta}}}
\def\Thetamb{ \pmb{{\mit{\Theta}}}}

\def\lb{{\bf l}}
\def\ldb{{\pmb{\ld}}}
\def\ld{\ell}
\def\lp{{\ell_p}}
\def\Lcal{\mathcal{L}}
\def\ellb{{\pmb{\ell}}}
\def\vb{\velb_{t}{}}

\def\Rb{{\bf R}}
\def\pd{\partial}
\def\vphi{\varphi}

\def\psitot{\varphi}
\def\psiR{\widetilde{\psi}}
\def\psiL{\widetilde{\psi}^{{\rm vir}}}
\def\Phim{{\mit{\Phi}}}
\def\PhimR{\widetilde{ {\mit \Phi}}}
\def\PsimR{\widetilde{ {\mit \Psi}}}
\def\PsimL{{\widetilde{ {\mit \Psi}}}^{{\rm vir}}}
\def\a{\alpha}
\def\ap{{\kappa}}

\def\uav{\bar{u}}
\def\D{\Delta}
\def\th{\theta}
\def\r{{\mbox{\tiny${R}$}}}
\def\re{{\mbox{\tiny${R}$}}}
\def\Fmed{F_{{\rm a.med}}}
\def\med{{\rm med}}
\def\Lw{L_{\varphi}}
\def\Fb{{\bf{F}}}

\def\Ef{{\sf{E}}}
\def\Bf{{\sf{B}}}

\def\Efb{{ \pmb{ \Ef} }}
\def\Bfb{{ \pmb{ \Bf} }}
\def\Yb{{\bf{Y}}}

\def\Ac{ \varphi}
\def\Xsub{{\mbox{\tiny${X}$}}}
\def\Ysub{{\mbox{\tiny${Y}$}}}
\def\Zsub{{\mbox{\tiny${Z}$}}}

\def\Vsub{{\mbox{\tiny${\Vfk}$}}}

\def\MTsub{{}}

\def\Vmscr{\mathscr{V}}

\def\Nu{{\mathcal{V}}}
\def\Nubar{{\bar{{\mathcal{V}}}}}

\def\Nusub{\nu}
\def\Nubarsub{\nubar}

\def\Ksub{{\mbox{\tiny${K}$}}}
\def\W{{\mit \Omega}}
\def\Wd{\W_d{}}
\def\Nuscr{{\mathscr{V}}}
\def\Nud{\Nu_d{}}
\def\Eng{{\cal E}}
\def\eng{{\varepsilon}}
\def\vep{\varepsilon}
\def\Kmscr{{\mathscr{K}}}
\def\Lscr{{\mathscr{L}}}
           \def\engk{\Kcal}
\def\Acuni{\Ac_{{\Ksub}^\dagsup}^{\dagsup}}
\def\unduni{\Ac_{{\Ksub}^\dagger}^{\dagsup}}
\def\Acauni{\Ac_{{\Ksub}^\ddagsup}^{\ddagsup}}
\def\Acunim{{\Ac_{{\Ksub}^\dagsup}^{\dagsup *}}}
\def\undunim{{\Ac_{{\Ksub}^\dagsup}^{\dagsup}}^*}
\def\Acaunim{{\Ac_{{\Ksub}^\ddagsup}^{\ddagsup *}}}
\def\pd{\partial}
\def\Ad{ {\mit \psi}}
\def\psim{ {\mit \psi}}
\def\Kd{K_d{}}
\def\Xim{{\mit \Xi}}
\def\Sigm{{\mit \Sigma}}

\def\Lam{{{\mit \Lambda}}}

\def\lam{\lambda}
\def\dagsup{{\mbox{\tiny${\dagger}$}}}
\def\ddagsup{{\mbox{\tiny${\ddagger}$}}}

\def\psimKdK{\psim_{\Ksub,\Kdsub}}
\def\wk{1}

\def\w{\omega{}}

\def\om{\omega{}}
\def\omb{\pmb{\omega{}}}
\def\Omb{\pmb{\Omega}}
\def\ombar{\omega{}\hspace{-0.4cm}-\hspace{-0.1cm}}

\def\wrm{{\rm{w}}}
\def\wit{\w_{t}{}}
\def\witb{{\pmb{\it{w}}}}

\def\wdlow{\omega_d }
\def\g{\gamma{}} 
\def\Phimc{{\mathcal C}}
\def\Psim{{\mit \Psi}}
\def\arm{{\rm a}}
\def\brm{{\rm b}}
\def\crm{{\rm c}}
\def\drm{{\rm d}}
\def\erm{{\rm e}}
\def\frm{{\rm f}}
\def\grm{{\rm g}}
\def\hrm{{\rm h}}
\def\lf{\left}
\def\rt{\right}
\def\Kdsub{{\mbox{\tiny${K_d}$}}}
\def\psimkd{\psim_{\kdsub}}
\def\psimKd{\psim_{\Kdsub}}
\def\hquad{ \ \ } 
\def\Taum{{\mit \Gamma}}
\def\Gam{{\mit \Gamma}}

\def\dagsup{{\mbox{\tiny${\dagger}$}}}
\def\ddagsup{{\mbox{\tiny${\ddagger}$}}}

  \def\ehm{h^\m}
\def\ehmsub{{h^\m}}


\def\Ha{E}
\def\Haop{H}

\def\rtd{\bar{r}}

\def\hf{{\mbox{\tiny{$\frac{1}{2}$}}}}
\def\trhf{{\mbox{\tiny{$\frac{3}{2}$}}}}
\def\Hcal{\mathcal{H}}
\def\Hig{\Hcal}
\def\H{H}

\def\Hsub{{\mbox{{\tiny{$\H$}}}}}
\def\Zcal{\mathcal{Z}}
\def\Z{Z}

\def\Zsub{{\mbox{{\tiny{$\Z$}}}}}
\def\Zsup{{\mbox{{\tiny{$\Z$}}}}}
\def\Ztdp{{\tilde{\Z}}}
\def\Ztd{{\tilde{\Z}}}

\begin{abstract}

We extend the SQR-KGE solutions for the heavy positronium of paper \cite{jxzj-pimu} to a heavy protonium, i.e. a relativistic proton $p^*$ and antiproton $\bar{p}^*$ orbiting at speed $\upsilon_k^*=g^*_n\upsilon_n\dot{=}c$ under a formal Coulomb potential $\bar{V}_{ck}=-\alpha\hbar{}c/\bar{r}_k^*$, in this paper.
The $p^*$,$\bar{p}^*$ reduced masses in orbital (level, state) $k$ are obtained as ${\mathscr{M}}^*_{n}=\gamma_n^*(\frac{1}{2}M_p)=\frac{nM_p}{2\alpha}$, ${\mathscr{M}}^*_{n,{\mbox{\tiny{$l$}}}}=\frac{(2{\mbox{\scriptsize{$l$}}}+1)}{n^2}{\mathscr{M}}^*_{n}$, $M_p$ being the proton mass.
For $n=2$, ${\mathscr{M}}^*_{\kappa} \sim 100 $  GeV
is the striking desired mass scale of the IVB's,  
and $\bar{r}_\kappa^*\sim{}10^{-18}$ m is the scale of charge radius at which the interactions are manifestly magnetic (or weak), $\bar{V}_{m k}=-G_k/\frac{4\pi}{3}\bar{r}_{k}^{_*3}\equiv\bar{V}_{ck}$.
The precessional-orbital (P-O) angular momenta are ${\mathbf{J}}_{\eta{\mbox{\tiny{$\frac{1}{2}$}}}}=\sqrt{j(j+1)}\,\hbar\,\,\hat{\mathbf{z}}'/\hat{\mathbf{z}}''$, $J_{\pm\eta{}jz}=|{\bf{J}}_{\eta{}j}|\cos\theta_{j}=\pm j\hbar$; $n,{\mbox{\scriptsize{$l$}}}=2,1$, $j,-j=\frac{1}{2},-\frac{1}{2}$, or denoted by $(\eta=)\nu,\bar{\nu}$, are the minimal P-O states of finite ${\mathbf{J}}_{\eta{\mbox{\tiny{$\frac{1}{2}$}}}}$, $J_{\pm\eta{}\frac{1}{2}z}$. Production conditions render the invariant rotated two-obit states $\zeta^{\mbox{\tiny{$+$}}},\zeta^{\mbox{\tiny{$-$}}}$ of $p^*,\bar{p}^*$, obtained such that $\mathbf{J}_{\eta_\rho}=\pm\tilde{\bf{r}}^*_{2,1\rho}\times{\mathscr{M}}^*_{\eta 2,1\rho}{\bf{c}}=\pm J_{\eta{\mbox{\tiny{$\frac{1}{2}$}}}z}\hat{\mathbf{z}}$. The state vectors are $\psi_{\eta}^r=e^{i\theta_q}\psi_\nu,e^{-i\theta_q}\psi_{\bar{\nu}}$, as rotated by angles $-\theta_q,\theta_q=\frac{1}{2}\theta_{\mbox{\tiny{$\frac{1}{2}$}}}=27.4^o$ from $\psi_\nu,\psi_{\bar{\nu}}$ along $z',z''/\bar{z}'$. 
The coherent reduced masses described by $\psi_{\eta \rho}^r$ are ${\mathscr{M}}^*_{\eta 2,1\rho}={\mathscr{M}}^*_{\eta 2,1}\sqrt{\cos\theta_q}$ rotating on orbits $(\eta_\rho=)\nu_\rho/\bar{\nu}_\rho,\bar{\nu}_\rho$ in the $(\rho\equiv)xy$ plane. 
The $\zeta^{\mbox{\tiny{$+$}}\hspace{-0.03cm}},\zeta^{\mbox{-}}$ of aligned spins $\frac{1}{2},\frac{1}{2}$, when rotating on two-orbit $\nu_\rho\bar{\nu}_\rho$ of $n,{\mbox{\scriptsize{$l$}}}=2,1$ make up $Z$, and on $\bar{\nu}_\rho\bar{\nu}_\rho$, $\nu_{0\rho}\bar{\nu}_{0\rho}$ of $n=2$ make up $H$, which have charges zeros, spins $1,0$, and resonance masses $M_{{\mbox{{\tiny{$Z$}}}}}=\sum_\eta{\mathscr{M}}_{\eta 2,1\rho}^*={\mathscr{M}}_{2,1}^*\sqrt{\cos\theta_q}=90.9$ GeV, $M_{{\mbox{{\tiny{$H$}}}}}=M_{\mbox{{\tiny{$Z$}}}}+{\mathscr{M}}_{2,0}^*=123.0$ GeV. Upon disintegration, $Z\rightarrow{}e^{\mbox{\tiny{$+$}}}e^{\mbox{\tiny{$-$}}}, (Z'\rightarrow)$$\mu^{\mbox{\tiny{$+$}}}(h_\rho^{\mbox{\tiny{$+$}}})\mu^{\mbox{\tiny{$-$}}}(h_\rho^{\mbox{\tiny{$-$}}}), \tau^{\mbox{\tiny{$+$}}}(p_2)\tau^{\mbox{\tiny{$-$}}}(\bar{p}_2)$, or $Z\rightarrow{}Z^\pm{}A^\mp$, $Z^\pm\rightarrow{}e^\pm\nu_e/\bar{\nu}_e, \mu^\pm\nu_\mu/\bar{\nu}_\mu$. 
$\nu_e, \nubar_e$, etc designate the spin-$\frac{1}{2}$ EM radiation energy currents emitted from the $\nu$,$\nubar$- motions of $e^\pm$, etc. 
$Z'$ or $Z^\pm$ consists of $\zeta^{\mbox{\tiny{$+$}}}{}',\zeta^{\mbox{\tiny{$-$}}}{}'$ or $\zeta^{\pm\prime\prime}$ in P-O motions about orbits $(\eta'=)\nu',\bar{\nu}'$ or $\nu''/\bar{\nu}''$, such that $\mathbf{J}_{\eta'{\mbox{\tiny{$\frac{1}{2}$}}}}=J_{\eta'{\mbox{\tiny{$\frac{1}{2}$}}}}\hat{{\bf{z}}}'/\hat{{\bf{z}}}''$ are at angles $\theta_{\mbox{\tiny{$\frac{1}{2}$}}},\pi-\theta_{\mbox{\tiny{$\frac{1}{2}$}}}$ to $z$ and invariant of $\mathbf{J}_{\eta_\rho}$; and $\psi_{\eta_\rho}\rightarrow\psi_{\eta'}=e^{-i\theta_q}\psi_{\nu_\rho},e^{i\theta_q}\psi_{\bar{\nu}_\rho}$. $Z'$ or $Z^\pm$ has a coherent mass $M_{{\mbox{{\tiny{$Z$}}}}^\pm}=M_{{\mbox{{\tiny{$Z$}}}}'}=M_{{\mbox{{\tiny{$Z$}}}}}\cos\theta_q=80.7$ GeV; and $G_{2,1\rho}\rightarrow{}G_{2,1\rho}'=\frac{{\mathcal{Q}}^2\sqrt{2}\hbar^2}{M_{{\mbox{{\tiny{$Z$}}}}'}^2{}c^2}=1.70\times{}10^{-62}${}Jm$^3$, $\mathcal{Q}^2=\frac{4e^2}{3\epsilon_0}(\cos\theta_q)^{11/2}$. The $Z$, $Z^\pm$, $H$ resemble in overall respects the observational neutral and charged IVB's and Higgs. 
\end{abstract}

\section{Introduction}\label{Sec-1}
\setcounter{equation}{0}
\setcounter{figure}{0}

In the Standard Model (SM) description\cite{perkins-20,FrauenfelderHenley91,Workman.et.al2022}, the  hundreds of elementary particles experimentally observed  are divided into $2 \times 6$ leptons, hundreds of baryons and mesons made of $2 \times 6$ quarks, five or so force mediators, and in addition, the Higgs regarded as a class of its own. Regarding the weak decays subjected by all the particles except for the electron $e^\m$ and proton $p$ and  their anti particles $e^\p$, $\pbar$, or optionally also the neutrinos $\nu, \nubar$'s, the SM quark model assumes that the decay product particles, such as  $e^\m, p, \nubar_e$ from the neutron $\beta$ decay $n \rar e^\m p \nubar_e$, do not exist until $n$ decays. The emitted $e^\m p \nubar_e$ are thus instantaneously produced upon the $n$ decay. The SM has proven successful in reproducing the charges, spins, C, P and T symmetries of overall elementary  particles, has in practice guided the discoveries of many of the particles, and has, as one of the successful applications of the quantum theory, predicted their transition probabilities  - absolute or comparative. But the SM has unnatural aspects, concerning the quarks and weak decays in particular. The baryons and mesons are made of quarks instead of the particles they decay into,   which is an abrupt departure from the atomic and nuclear descriptions where the atoms or the nuclears are composed of the particles they decay into along with the electromagnetic radiation emitted. Free quarks are never observed in experiment. Furthermore, the SM weak interaction Hamiltonian $H=G/(\frac{4\pi}{3} r^3)$ is a phenomenological construction. The weak force is not predicted. The weak, strong, gravitational and electromagnetic forces are not unified.  The basic questions in common with ones to the two only permanent particles (- why?) $e^\m,p$ remain outstanding, including the origin of mass, the nature of matter waves, and the cause of gravity.

The Internally Electrodynamic (IED) particle model proposed by the author in 2000, first elaborated for $p,e^\m$ in a polarisable vacuuonic dielectric (PVD) vacuum (see \cite{jxzj-ied} for a review) and generalised to the unstable particles in weak fields in \cite{jxzj-neutron,jxzj-pimu}, is a complementary approach to  the SM, and it goes beyond the SM. The IED model was developed using the overall experimental properties of particles as input information, and was aimed from the beginning to be built on a unification scheme for the classical, quantum and relativistic mechanics and for the four fundamental forces. In terms of the IED model, a simple single charged matter particle is composed of an oscillatory charge $q =+e$ or $-e$ and the electromagnetic radiation (EMR) field generated by the charge  oscillating in its specific vacuum potential field. The vacuum is constructed based on overall relevant experimental observations to be filled of neutral but polarisable vacuuons disorderly densely packed at a mean spacing $\sim 1 \times 10^{-18}$ m, with each vacuuon made of a $p$-vaculeon of charge $+e$ at the core and a $n$-vaculeon of charge $-e$ on the enclosing envelop
\cite{jxzj-dielec-vac,jxzj-unifbook}. 
This PVD vacuum necessarily presents a deep attractive potential well to an external charge $+e$, and a shallow one to $-e$. The level $n_\v=1$ quantum excitation energies of the harmonic oscillations of the charges $+e$ and $ -e$ in their vacuum potential fields  give rise to the rest masses of the proton $p$ and electron $e^\m$  \cite{jxzj-vacpot,jxzj-mass-origin}, the former being accordingly much greater then the latter. The $p,e^\m$ are permanent as their charges are not facilitated to emit the oscillatory energy quanta except in pair annihilations.  Their antiparticles $\pbar, e^\psup $ are naturally produced in $p,\pbar$ and $e^\p,e^\m$ pair productions out of a vacuuon in the vacuum each, with identical masses within each pair  as constrained by energy-momentum invariance.

The other single charged particles, including $\pi^\pm$, $\mu^\pm$, $\tau^\pm$ and the charged IVB's, consist of the simple heavier mass states $h^\pm, h_\rhoxy^\pm$\cite{jxzj-pimu}, $p_2/\pbar_2$ and $\z''/\zbar''$ (this paper) in precessional orbital (P-O) motions in a single or combinatory of the orbitals $n,\lcfoots,j=2,1,\frac{1}{2} $, $m_j =\frac{1}{2}, -\frac{1}{2}$,  or denoted by $\nu,\nubar$. These are meta-stable in a free vacuum as the result of the energy and momentum quantisations of the $\nu$ and/or $\nubar$ orbital motions. The mass states $h^\pm, h_\rhoxy^\pm$ \cite{jxzj-pimu} (see further Sec \ref{Sec-eng-mmt-decays-n}) are generated by the charge $\pm e$ oscillations in the magnetic (or weak) potential fields $V_m=-G/$volume of other particles at a separation  $\sim 10^{-18}$ m. The displacement amplitudes of their charges and hence their masses are uniquely specified by the eigen orbital level $n$, state $n,\lcfoots=2,1$ energies of a heavy positronium (HPs) composed of a relativistic electron $e^{\m \dastp}$ and positron $e^{\p \dastp} $.
In decays, e.g. $\mu^\pm \rar e^\pm+ \nu/\nubar + \nubar'/\nu'$, the spin $s=m_j= \frac{1}{2},-\frac{1}{2}$
 currents $\nu$,$\nubar$ of pure EMR fields are emitted which represent the observational neutrinos. The remaining unstable elementary particles are composites of the simple single charged particles (or mass states), $p, \pbar$, $e^\pm$, $h^\pm, h^\pm_\rhoxy$, etc., and their P-O energy currents $\nu$ and/or $\nubar$ which they decay into. As two prototypes of these, the neutron $n$ is composed of $e^\m, p$ \cite{jxzj-neutron} (denoted as $\epm{}^\dastp, p^\dastp$ in this paper), and the $\Lam$-hyperon of  $h^\m$ (heavy electron) and $p$ \cite{jxzj-pimu}, respectively in relative relativistic P-O motions along the $\nubar_e$, $\nubar_\mu$ orbits. The quantum solutions for the HPs in \cite{jxzj-pimu} however could not have directly yielded a prediction of the $G$ for the $\pi^\pm, \mu^\pm$ productions, nor the $100$ GeV scale masses of the free or virtual intermediate particles IVB's\cite{UA1-CERN-Z-1983,UA1-CERN-1983} and the Higgs \cite{AtlasCollab2012}. At the weak interaction range $r\sim 10^{-18}$ m the magnetic (or weak) potential is a fraction of one GeV only. To produce the 100 GeV scale masses thereof would require an energy level $n>>1$ at which the $e^{\m \dastp},e^{\p \dastp}$ would have long evaporated into classical particles.

In this paper we present a relativistic quantum electromagnetic theoretic description of the IVB's and the Higgs in terms of the invariant rotated states of the heavy protonium's (HPn's). In Sec \ref{Sec-theo} we re-derive the theory of the HPs for a general HPs-like system based on both the scaled \cite{jxzj-pimu} and direct SQR-KGE solutions. In Sec \ref{Sec-IVBs-n} we apply the SQR-KGE solutions to $p, \pbar$ and obtain $p^\dastp,\pbar^\dastp$ comprising a HPn, which  minimal orbital  energies are of the striking characteristic $100$ GeV scale in question. We further seek out the invariant rotated two-orbit states, $Z(\z,\zbar)$, $H(\z,\zbar)$, of the HPn's in the minimal eigen state $n,\lcfoots =2,1$, level $n=2$ conforming to the production conditions and the spins of the observational neutral IVB and the Higgs. The $H, Z$ disintegration schemes are represented in Sec \ref{Sec-eng-mmt-decays-n}. Intermediate reverse rotated states, a neutral $Z'(\z',\zbar')$, and alternatively a charged $Z^\pm(\z''/\zbar'')$ and a $A^\mp$ at rest, are obtained for the $Z$ decay in the radial ($y$) oscillation modes of their charges. Through Secs \ref{Sec-IVBs-n}-\ref{Sec-eng-mmt-decays-n} the basic properties including charges, spins, (resonance) masses, and internal interaction potentials of $Z$, $Z^\pm$, $H$ are predicted based on the particle compositions, the spins and orbital structures of the component particles, the SQR-KGE  solutions and symmetry operations, and are compared with those of the observational neutral and charged IVB's and the Higgs. The final $Z$ decay product particles, $\mu^\pm(h_\rhoxy^\pm)$, $\tau^\pm(p_2/\pbar_2)$, due to the $y$, $z$ charge oscillation modes are compared with the observational muons and  tau particles.

\section{Relativistic quantum theory for heavy positronium-like systems}
\label{Sec-theo}
\def\fit{{\it{f\,}}}

We predicted in  \cite{jxzj-pimu} the presence of a heavy positronium (HPs), namely a Coulomb bound relativistic positron and electron,  $e^{\p \dastp}, e^{\m \dastp}$, in relative orbital motion at essentially the light speed $c$. We re-derive in this section the formal relativistic quantum theory of the HPs, generalised to a heavy positronium like system $a^\dastp, \abarr^\dastp$ here.  At the base of $a^\dastp, \abarr^\dastp$ is a usual positronium (Ps) like system, a binary $a, \abarr$ of rest masses $M_a,M_a'=M_a$ and charges $+e,-e$ at a separation $r$, that are in relative rotational motion at a velocity $\ve<<c$ under their mutual Coulomb potential $V_c=-\frac{e^2}{4\pi\ev_0 r}=-\frac{\a \hbar c }{r}$. 
Their motion in the centre of mass frame (cmf) is governed by the Schr\"odinger equation, or the SQR-KGE \cite{jxzj-pimu}, and the transformed Newtonian equation of motion, in the non-relativistic limit each.
The eigen solutions for the dynamical variables are quantised because of the wave nature of $a,\abarr$, which can be represented by the (nominal) eigen wavelength 
 $\lam_n(=\frac{\lam_n^0}{\g_n}=\frac{2\pi r_n}{n}) = \frac{h}{\mr_n \ve_n}$ of the cmf fictitious ({\it f}) particle of the (reduced) mass $\mr_n=\g_n \mr^0$ in orbital level $n=1, 2, \ldots$, where $\g_n=(1-\ve_n^2/c^2)^{-1/2} \dot{=}1$, $\mr^0 =\frac{M_a M_a'}{M_a+M_a'}=\frac{1}{2}M_a$, $\ve_n=|\veb_{a n} -\veb_{\abarr n}|=\frac{\a c}{n}$, $\lam_n^0=\frac{2\pi r_n^0}{n}$, and $r_n^0(=|\rb_{a n}^0- \rb_{\abarr n}^0|)=\frac{n^2\hbar c}{\a \mr^0 c^2} = \frac{4\pi \ev_0 n^2 \hbar^2 c^2}{e^2 \mr^0 c^2 }$.

The conception of the HPs in \cite{jxzj-pimu} is in essence the recognition, as nature has apparently long recognised, that the equation for $\lam_n$ of the usual Ps - the $a,\abarr$ here, continues to hold after dividing a factor $\g_n\equiv g_n$ through, $\frac{\lam_n}{\g_n}(=\frac{2\pi r_n}{n \g_n })=\frac{h}{\mr_n g_n \ve_n}$. Accordingly the quantisation of the (nominal) angular momentum holds, $|\frac{1}{\g_n}\rb_n \times (\mr_n g_n \veb_n)| = n \hbar$. At the limit $\ve_n =1\cdot \ve_n\rar \ve_n^\dastp=g_n^\dastp \ve_n= g_n^\dastp \frac{\a c}{n}  \dot{=} c$, hence $g_n^\dastp=\frac{n}{\a}$, and $\g_n\rar \g_n^\dastp=(1-\ve_n^\dastp{}^2/c^2)^{- 1/2}= g_n^\dastp$, we obtain a HPs like system $a^\dastp,\abarr^\dastp$ in $n$th level with the wavelength and reduced mass given in (\ref{eq-lamnst}), (\ref{eq-gna}), and the corresponding linear momentum, kinetic energy, (formal) Coulomb potential and Hamiltonian given in (\ref{eq-pn-Tn}),
 $$\displaylines{
\refstepcounter{equation} \label{eq-gn}
  \label{eq-lamnst} 
\hfill
 \mbox{$
\frac{\lam_n}{\g_n} \rar 
\frac{\lam_n^0}{\g_n^{_\dastp 2} }
 =\lam_n^\dastp =\frac{2\pi r_n^\dastp}{n}
=\frac{h}{  \mr_n^\dastp  c }
=\frac{h}{  \mr_n^\dastp \frac{1}{\g_n^\dastp} c \g_n^\dastp} 
=\frac{\Lam_n^\dastp}{\g_n^\dastp},
\quad
 r_n^\dastp=\frac{r_n^0}{\g_n^{_\dastp 2}}=\frac{1}{(n/\a)^2}n^2 r_1^0=\a^2 r_1^0,
$}
\hfill (\ref{eq-lamnst})
\cr
\refstepcounter{equation} \label{eq-gn} \label{eq-gna} 
\hfill
\mr^\astp_n= \g_n^\dastp\mr^0= \frac{n}{\a} \mr^0, 
\quad
\g_n^\dastp \equiv g_n^\dastp=\frac{c}{\ve_n}=\frac{n}{\a}, 
\quad    
\Lam_n^\astp
= \frac{h}{\mr^\astp_n \ve_n  }
=\frac{\lam_n^0}{\g_n^\dastp}
\equiv \frac{h}{\mr^0 c}, 
\hfill
 (\ref{eq-gna})
\cr
\refstepcounter{equation} \label{eq-pn-Tn}
\hfill
\mbox{
$\pmt_n=\mr_n^\dastp \ve_n^\dastp =\frac{h}{\lam_n^\dastp}$, 
\quad
$T_n=\frac{\g_n^{_\dastp} \pmt_n \ve_n^\dastp }{\g_n^\dastp+1} 
=\frac{\g_n^{_\dastp} \mr_n^\dastp \ve_n^{_\dastp 2} }{\g_n^\dastp+1} 
=\frac{\g_n^{_\dastp 4} 2 T_n^0 }{\g_n^\dastp+1}   
=-\frac{\g_n^\dastp{}^2 V_{c n}}{\g_n^\dastp+1}$, 
}
\hfill 
\cr
\hfill
\mbox{
$V_{c n}=-\frac{e^2}{ 4\pi\ev_0 r_n^\dastp}
 =\g_n^{\dastp^2} V_{c_n}^0$, 
\quad
$\Ha_n=T_n+V_{cn} = (1- \frac{\g_n^\dastp+1}{\g_n^{_\dastp 2}})T_n
= -(\frac{\g_n^{_\dastp 2}}{\g_n^\dastp +1} -1)V_{cn}$,
}
\hfill (\ref{eq-pn-Tn})
}$$
where $T_n^0=\frac{1}{2}\mr^0 \ve_n^2=-\frac{1}{2}V_{cn}^0$, $V_{cn}^0=-\frac{e^2}{4\pi \ev_0 r_n^0}$, $\Ha^0_n=T^0_n+V_{cn}^0=-T_n^0$ are the corresponding eigen dynamical variables of $a,\abarr$; (\ref{eq-gna}c) are given after (\ref{eq-lamnst}a), (\ref{eq-gna}a). By (\ref{eq-lamnst}a), $\Lam_n^\astp$ represents an apparent rest value of $\lam_n^\dastp$ before contracting by the factor $1/\g_n^\dastp $.
By the first of Eqs (\ref{eq-gna}c), $\Lam_n^\astp$ is the rest wavelength of the cmf \fit particle of an apparent rest mass $\mr_n^\dastp =\frac{M_{a^\dastp} M_{a^\dastp}'}{M_{a^\dastp} + M_{a^\dastp}'} =\frac{1}{2}M_{a^\dastp}= \g_n^\dastp \mr^0$ orbiting at the regular eigen speed $\ve_n=\frac{c}{\g_n^\dastp}=\frac{\a c}{n}$ about an orbit of radius $r_n^\dastp$ given in (\ref{eq-lamnst}b). $M_{a^\dastp}=2 \g_n^\dastp \mr^0=\g_n^\dastp M_a, M_{a^\dastp}'=M_{a^\dastp}$ are the corresponding apparent rest masses of $a^\dastp, \abarr^\dastp$.\footnote{$M_{a^\dastp}$ is relativistically augmented $\g_n^\dastp$ times from $M_a$ in the instantaneous $\veb_n$ direction that varies over time, of an average $<\veb_n>=0$. The rest masses of $a^\dastp$, $\bar{a}^\dastp$ in the vacuum are  ordinarily are $M_a,M_a$, and the total is $M=M_a+M_a'=2M_a$.} 
There exist in nature only two base particle pairs, $a,\abarr=p,\pbar$ and $e^\p, e^\m$, such that one particle in either pair, $p$ or $e^\m$, is a permanent stable particle in the vacuum. There thus exist only two so formed $a^\dastp,\abarr^\dastp$ pairs, $p^\dastp,\pbar^\dastp$ and $e^{\p \dastp},e^{\m \dastp}$. Particularly, the $p^\dastp$, $e^{\m \dastp}$ orbiting at the speed $\ve_n$ have the rotational linear momenta $M_{p^\dastp} \ve_n=2\g_n^\dastp \frac{1}{2}M_p \frac{\a}{n}c= M_p c$, $M_{e^\dastp}\ve_n=M_e c$. The $M_p c$, $M_e c$ on the right sides equal just the total linear momenta of the EMR fields comprising the $p, e^\m$ at rest in the vacuum (the IED model). This renders the rotating $p^\dastp$, $e^{\m \dastp}$ themselves as two specific {\it quasi} stationary states in the vacuum.

For $\lcfoots$th orbital state of $n$th level the linear momentum can be written $p_\nclc = \mr^\dastp_\nclc \ve_\nclc^\dastp=h/\lam_\nclc^\dastp$, where $\ve_\nclc^\dastp = g_n^\dastp \ve_\nclc\equiv \ve_n^\dastp (\dot{=}c)$, $\ve_\nclc\equiv \ve_n (=\frac{\a c}{n})$ for $\lcfoots \ge 1$, and $\ve_{n,0}=0$ according to Eq (\ref{eq-Jcal}a) below. 
For each vector $\pb_{\nclc}$ there is a $\rb_\nclc^\dastp$ (the radius of a semi-classical circular orbit) such that the angular momentum is quantised and given by the same eigenvalue solutions as for $a,\abarr$
 $$\displaylines{
\refstepcounter{equation} \label{eq-Jcal}
\hfill
\mbox{$
\Jcal_{\lsub}
= |\Jcalb_{\lsub, m}| =|\rb_{\nclc}^\dastp\times \pmtb_\nclc|
= r_{\nclc}^\dastp \mr_{\nclc}^\dastp \ve_\nclc^\dastp 
=\frac{r_\nclc^\dastp h}{\lam_\nclc^\dastp}
= \sqrt{\lcfoot(\lcfoot+1)} \hbar,
\ \ 
\Jcal_{ m z}=\Jcal_{\lsub}  \cos\theta_{m} 
= m \hbar,
$}
\hfill (\ref{eq-Jcal})
}$$
where $\lcfoots=0,1,\ldots, n-1$, $m=0, \pm 1, \ldots, \pm \lcfoots$; $\theta_{m}$ are the angles of $\Jcalb_{\lsub,m }^\dastpp$ to the $z$ axis. Each level $n$ consists of $\sum_{\lsub=0}^{n-1} (2\lcfoots+1)=n^2$ degenerate states $\lcfoots, m$
(which are the expansion terms in a series solution 
regardless of being occupied or not, e.g. by the electrons in an atom). $\Ha_n$ can be thus identically written 
$\Ha_\nc^\dastpp 
= \sum_{\lcsub=0}^{\nc-1} $ $\frac{ (2 \lcfoots+1)\Ha_\nc^\dastpp }{\nc^2}
= \sum_{\lcsub=0}^{\nc-1} \Ha_\nclc^\dastpp
$, 
where $\Ha_\nclc^\dastpp=T_\nclc+V_{c \nclc}
=\frac{ (2 \lcfoots+1)\Ha_\nc^\dastpp }{\nc^2}
$, $T_\nclc=\frac{ (2 \lcfoots+1)T_\nc^\dastpp }{\nc^2}$, and $V_{c \nclc}=\frac{ (2 \lcfoots+1)V_{c\nc}^\dastpp }{\nc^2}$ represent the partitions of $E_n,T_n,V_{cn}$
due to all of degenerate states of a specified 
$\lcfoots$. Substituting $T_n,V_{cn},E_n$  from Eqs (\ref{eq-pn-Tn}) in these gives (\ref{eq-Tps}), and constraining (\ref{eq-lamnst}), (\ref{eq-gna}) using (\ref{eq-Jcal}), (\ref{eq-Tps}) gives the corresponding $\lcfoots$th partitioned values in (\ref{eq-lam-pp}), (\ref{eq-gna-b}):
$$\displaylines{
\refstepcounter{equation} \label{eq-Tps}\label{eq-Vcdastpp}
\hfill
\mbox{$T_\nclc
=\frac{\g_n^\dastp \Jcal_\lsub^{ 2} }{(\g_n^\dastp+1) r_{n,\lsub}^{_\dastp 2}\mr^\dastp_\nclc }  
=\frac{\g_n^{_\dastp 4}2T_\nclc^0 }{\g_n^\dastp+1}
=-\frac{\g_n^{_\dastp 4}V_{c\nclc}^0 }{\g_n^\dastp+1}
=-\frac{\g_n^{_\dastp 2}V_{c\nclc} }{\g_n^\dastp+1}
$},
\quad
\mbox{$V_{c \nclc} 
= \g_n^{_\dastp 2} V_{c \nclc}^0
= - \frac{e^2}{4\pi \ev_0  b_\nclc r_\nclc^\dastp} 
=
$}\hfill
\cr
\hfill
\mbox{$
=- \frac{\mr^\dastp_{\nclc} c^2}{\g_n^\dastp}, \quad
\Ha_\nclc=T_\nclc+V_{c\nclc}
=(1-\frac{ \g_n^{_\dastp}+1  }{\g_n^{_\dastp 2}})T_\nclc
=-(\frac{\g_n^{_\dastp 2}}{ \g_n^{_\dastp}+1 }-1)V_{c\nclc},
$} 
\hfill (\ref{eq-Vcdastpp})
\cr
\refstepcounter{equation}\label{eq-lam-pp}
\hfill
\mbox{$
 \lam_{\nclc}^\dastp
=\frac{2\pi r_\nclc^\dastp}{\sqrt{\lcfoots(\lcfoots+1)} }
= \frac{h}{\mr_{\nclc}^\dastp c}  
= \frac{\Lam_\nc^\dastp}{\g_\nclc^\dastp}
= \frac{\Lam_\nclc^\dastp}{\g_\nclc^\dastp}
=\frac{\lam_n^\dastp}{\g_\nclc^0}
$},
\quad
\mbox{$
 r_\nclc^\dastp
=\frac{r_n^\dastp}{\g_\nclc^0 b_\nclc}
=\frac{r_{n,\lsub}^0}{\g_n^{_\dastp 2}}
$, 
} 
\hfill (\ref{eq-lam-pp})
\cr
\refstepcounter{equation} \label{eq-gna-b} 
\hfill
\mbox{$
\mr_{\nclc}^\astp
= \g_\nclc^\astp \mr^0= \frac{(2\lcfoots+1)\mr_n^\dastp}{n^2}= \g_n^\dastp \mr_\nclc^0
$},
\quad
\mbox{$
\g_\nclc^\astp 
=\frac{(2\lcfoots+1)\g_n^\dastp}{n^2}
=\g_\nclc^0 \g_n^\dastp
=\frac{ (2 \lcfoots +1) }{n \a },
$} 
\quad 
\Lam_\nclc^\dastp=\Lam_n^\dastp, 
\hfill (\ref{eq-gna-b})
}$$
where $T_\nclc^0=\frac{1}{2}\mr^0_\nclc \ve_n^2=-\frac{1}{2}V_{c n,\lsub}^0$, $\mr_\nclc^0=\g_\nclc^0\mr^0$,
$\g_\nclc^0 =\frac{2\lcfoots+1}{n^2}$, 
$b_\nclc=\frac{n }{\sqrt{\lcfoots(\lcfoots+1)}}$ 
for $\lcfoots\ge 1$ and $b_{n,0}=1$,
$V_{c\nclc}^0=-\frac{e^2}{4\pi \ev_0  b_\nclc r_\nclc^0}$,
$r_{n,1}^0=\frac{r_{n}^0}{\g_{n,1}^0 b_{n,1}}$. 

For either $\ka=n$ or $n,\lcfoots$, the relativistic  reduced mass energy is 
 $$\displaylines{
\refstepcounter{equation} \label{eq-mass-Eng} 
\hfill
\mbox{$
\Eng_{\ka}^\dastpp
=\frac{\g_n^{_\dastp 2} \mr_\ka^0 \ve_n^{_\dastp 2} }{
(\g_n^\dastp+1) } +  \mr^0_\ka c^2  
=\mr_\ka^\dastp c^2 = \g_n^\dastp \mr_\ka^0 c^2
=  \g_n^\dastp[a_\qe  \D V(\Acal_q) +a_r 2\pi r_\ka^\dastp  \sig_\ka  \ev_0 |\Eb_\ka|^2 ]
$}\hfill (\ref{eq-mass-Eng})
}$$
where $\mr_n^0 = \mr^0$, $\mr_{\nclc}^0=\g_\nclc^0 \mr^0$ for $\ka=n$, $\nclc$. The relation $\g_n^{_\dastp 2}\frac{\ve_n^{_\dastp 2}}{c^2}= \g_n^{_\dastp 2}-1$ is used.
The last of Eqs (\ref{eq-mass-Eng}) is a g-IED description\cite{jxzj-pimu}: The rest mass energy $\mr^0_\ka c^2$ is given by the sum of the fractions $a_\qe, a_r (=1-a_\qe)$ of the  Hamiltonians of the oscillating charge, $\frac{1}{2}\Mcal_\qe \dot{\xi}_\qe^2+ V(\xi_\qe)= \D V(\Acal_\qe)$, and of the EMR fields $\Eb_\ka(=\Eb_0 \psi_{\ka}), \Bb_\ka=\Eb_\ka/c$. Here, $\xi_\qe=\Acal_{\qe \ka} e^{-i \Om_\qe t}$ is the charge displacement about an (instantaneous) fixed position $\rb^\dastp$, $\D V(\xi_\qe)=\frac{1}{2}\beta_\qe \xi_q^2$ is the difference (magnetic) potential responsible for restoring $\qe$; and $2\pi r_\ka^\dastp\sig_\ka$ is the volume of a torus ring through which the EMR fields travel.

If $\g_n^\dastp>>1$, Eqs (\ref{eq-pn-Tn}) give $T_n^\dastpp \gg |V_{c n}^\dastpp|$, $\Ha_n\dot{=}T_n\dot{=}\Eng_n$. This is in contrast to $-V_{c n}^0=2T_n^0$ for $a,\abarr$ with $\g_n\dot{=}1$. The latter is the result of equal and opposite  binding and inertial forces, $F_{cn}^0=-\frac{\a\hbar c}{(r_n^0)^2}$ and $F_{ine,n}^0 =\frac{\mr^0 \ve_n^{ 2} }{r_n^0}$ for $\ka=n$: $\frac{\a \hbar c}{(r_n^0)^2} =\frac{\mr^0 \ve_n^2 }{r_n^0}$, or $\frac{\a \hbar c}{r_n^0} =\mr^0 \ve_n^ 2$. Multiplying both sides  by $\g_n^\dastp {}^2 =\g_n^\dastp g_n^\dastp$ gives $\frac{\a \hbar c}{r_n^0 / \g_n^\dastp{}^2} = (\g_n^\dastp \mr^0) \frac{(g_n^\dastp{}^2 \ve_n^2)}{g_n^\dastp}$. Combining further with (\ref{eq-lamnst}b), (\ref{eq-gna}b) and $g_n^\dastp \ve_n=c$ gives the corresponding relation for $a^\dastp, \abarr^\dastp$, $\frac{\a \hbar c}{r_n^\dastp } = \frac{ \mr_n^\dastp c^2 }{g_n^\dastp} $, the same as given in (\ref{eq-pn-Tn}b), or $(-F_{cn}^\dastpp=) \frac{\a \hbar c}{r_n^\dastp{}^2} =(\frac{F_{ine,n}^\dastpp}{g_n^\dastp}=) \frac{\mr^\dastp_n \ve_n^\dastp{}^2}{r_n^\dastp g_n^\dastp} $. Given $g_n^\dastp=\g_n^\dastp>>1$, there is the strong imbalance $F_{ine,n}^\dastpp >> - F_{cn}^\dastpp$; and similarly $F_{ine,n,\lsub}^\dastpp >> - F_{cn,\lsub}^\dastpp$. The $a^\dastp,\abarr^\dastp$ are hence implied to be loosely bound only, deemed unstable in the usual terms. Rather, the $a^\dastp, \abarr^\dastp$ can be meta-stablised by the momentum quantisation relations $p_{\nclc}=h/\lam_{\nclc}^\dastp$ and  (\ref{eq-Jcal}) for $\Jcal_{\lsub}, \Jcal_{m z}$ for a given $\lcfoots$,  and exist as a stand alone system. (\ref{eq-Vcdastpp})-(\ref{eq-gna-b}) thus describe the measurable dynamical variables of the system. This is in contrast to a regular stable bound atomic system $a,\abarr$, or $a,b$ in general, that is characterised by its orbital level $n$, or $n$ to $n'$ transition (and the atomic spectrum), or at most by the fine structures caused by spin- orbital ($\lcfoots$) magnetic interactions in an applied magnetic field. The partitions (\ref{eq-Vcdastpp})-(\ref{eq-gna-b}) for the latter are never manifest directly, nor usually expressed.  
 
\def\Ra{R}
The scaled quantum variables above are anticipated to be also the direct solutions to a relativistic Schr\"odinger equation, or the reduced SQR-KGE (\ref{App-SQR-KGE}), $\Haop^\dastpp_{op} \psi(\rb^\dastp)= \Ha^\dastpp \psi (\rb^\dastp)$, which we shall construct and solve in the cmf below. The total Hamiltonian operator consists of a radial kinetic term $\frac{\g^\dastp (p_r)^2}{(\g^\dastp+1) \mr^\dastp}$,  a non-inertial rotational term $T$ and a $V_c$ as given in (\ref{eq-pn-Tn}), $\Haop^\dastpp_{op}=  \frac{\g^\dastp (\pmt{\hspace{0.1cm}}^{\dastpp 2}_{r} )_{op} r^\dastp{}^2}{(\g^\dastp+1) \mr^\dastp r^\dastp{}^2} + T^\dastpp_{op} + V_c (r^\dastp)$, where $\mr^\dastp=\g^\dastp \mr^0$, $T^\dastpp_{op}= \frac{\g^\dastp(\Jcal^{\dastpp^2})_{op}}{(\g^\dastp+1) \mr^\dastp r^\dastp{}^2}$, 
$\Jcal=|\rb^\dastp \times \mathbf{ p}|$, $\pmtb= \mr^\dastp \veb^\dastp =\hbar {\pmb{\kappa}}^\dastp $, $|\rb^\dastp|=r^0/\g^\dastp{}^2$, $V_{c}(r^\dastp)=-\frac{e^2}{4\pi \ev_0 r^\dastp}=-\frac{\a \hbar c}{r^\dastp}$. In spherical polar co-ordinates, $(\pmt_r^{\dastpp 2})_{op} =
(i\hbar \frac{\pd }{\pd r^\dastp})^2 = -\frac{\hbar^2}{r^\dastp{}^2} \frac{\pd}{\pd r^\dastp}(r^\dastp{}^2 \frac{\pd }{\pd r^\dastp})$, $(\Jcal^{\dastpp^2})_{op}=\frac{\pd}{\pd \theta}(\sin \theta \frac{\pd }{\pd \theta}) + \frac{1}{\sin \theta} \frac{\pd^2}{\pd \vphi^2}$. Further for the central potential $V_{c}(r^\dastp)$, $\psi(\rb^\dastp)= \Ra(r^\dastp) Y(\theta, \vphi)$ and the SQR-KGE is decomposable into
$$\displaylines{
\refstepcounter{equation} \label{eq-gn}  \label{eq-SQR-KGE-p} 
\hfill
[ \frac{\g^\dastp( (\pmt_r^{\dastpp 2})_{op}  r^\dastp{}^2
+ \lcfoots(\lcfoots+1) \hbar^2)
  }{ (\g^\dastp+1) \mr^\dastp{} r^\dastp{}^2}
+V_c^\dastpp(r^\dastp)] \Ra =\Ha^\dastpp \Ra, 
\quad 
\mbox{$
(\Jtr^{\dastpp 2} )_{op} Y= \Jtr^{\dastpp^2}Y.
$} 
\hfill (\ref{eq-SQR-KGE-p})
}$$
(\ref{eq-SQR-KGE-p}b) is a regular angular equation of motion, its solutions are the usual spherical harmonics
$Y_{\lsub}^{m}(\theta, \vphi)=\sqrt{  \frac{(2\lcfoots+1)(\lcfoots-m)!}{4\pi (\lcfoots+m)!}} (-1)^m P_\lsub^m (\cos\theta) e^{i m \vphi}$, and $\Jcal^2=\lcfoots(\lcfoots+1)\hbar^2$. The solutions to (\ref{eq-SQR-KGE-p}a) are dependent on $\g^\dastp$. Of our interest here is  $\g^\dastp>>1$.  The preceding analysis for $\g^\dastp>>1$ yielded $F_{ine} = -\g^\dastp F_c >> -F_c$, $E= T + V_c \dot{=} T=\frac{\g^\dastp (\hbar \kappa^\dastp)^2 }{(\g^\dastp+1) \mr^\dastp } \dot{=}-\g^\dastp V_c  >> -V_c$ and $E>0$, which depict a system unstable unbound in general. Nevertheless, at a $r^\dastp $ comparable to $\lam^\dastp$, we anticipate a system stationary due to momentum quantisations, hence meta-stable, as if being stablised by an (counter balancing) binding force $\g^\dastp F_c=-F_{ine}$ and a potential $\g^\dastp V_c $. As such, we can employ the usual boundary conditions (e.g. \cite{Merzbacher}, chapter 12) to seek a normalisable eigen solution of the general form $R(r^\dastp)=\frac{u(r^\dastp)}{r^\dastp}
=\frac{1}{r^\dastp} (\rho)^{\lsub+1} e^{- \rho}\sum_{j=0,1, \ldots} a_j \rho^j $, where $\rho= \kappa^\dastp r^\dastp =\sqrt{\frac{(\g^\dastp+1) \mr^\dastp |E| }{\g^\dastp \hbar^2}} \,  r^\dastp$. Despite $\g^\dastp>>1$,  but insofar as this $\g^\dastp$ is absorbed into $\g^\dastp V_c=V'_c = \g'V_c'$ and in respect to $V'$, we have $\g'=1$ and a primed classical  kinetic energy $T'=\frac{1}{2} \mr^\dastp \ve^{_\dastp 2}=-E' =\frac{1}{2}E$. Accordingly a specific $\rho_0 \propto |\g^\dastp V_c|$ can be defined such that $\frac{(1/2)E}{|\g^\dastp V_c|}=\frac{\rho}{\rho_0}$. An integer $\frac{1}{2}\rho_0=N+\lcfoots+1=n$
is required for the convergence of the power series in $R(r^\dastp)$ at large $r^\dastp$. Hence $\frac{(1/2) E }{ |\g^\dastp V_c|}(=\frac{\rho}{\rho_0})= \sqrt{\frac{(\g^\dastp+1) \mr^\dastp E}{\g^\dastp \hbar^2}} \, \frac{r^\dastp}{2n}$. Sorting gives for integer $n$, $E_n=\frac{ \g^{_\dastp 2}\mr^\dastp e^2 |V_{cn}| \cdot r_n^\dastp }{4\pi \ev_0 n^2 \hbar^2}=\frac{\g^{_\dastp 2} \mr^\dastp e^4 }{(4\pi \ev_0)^2 n^2 \hbar^2}$. Equating the first equation  with $E_n=|\g^\dastp V_{cn}|$ gives $r_n^\dastp =\frac{(4\pi \ev_0) n^2 \hbar^2 }{ \g^\dastp e^2\mr^\dastp }= \frac{r_n^0}{\g^{_\dastp 2}}$. At the specific values $\g^\dastp\rar \g_n^\dastp=\frac{n}{\a}$, the eigen variables $\mr_n^\dastp=\g_n^\dastp \mr^0$, $r_n^\dastp$, $E_n$ here agree with the preceding scaled results.

\def\psitd{\tilde{\psi}}
\def\rbar{\bar{r}}
\def\psibar{\bar{\psi}}
\def\psitd{{\psi_{\nu\nubar}}}
\def\Hcalbar{\bar{\Hcal}}
\def\Hcaltd{\tilde{\Hcal}}
\def\Ka{K}

\section{Heavy protonium. Their rotated 
two-orbit states. The $Z,H$ particles} 
\label{Sec-IVBs-n}
\def\z{\zeta}
\setcounter{equation}{9}

Using a proton $p$ and antiproton $\pbar$ for $a,\abarr$ in the preceding representation, we thereby obtain their relativistic counterparts $p^\dastp,\pbar^\dastp$ formally Coulomb bound, termed a heavy protonium (HPn). Using the proton mass $M_p$ and hence $\mr^0=\frac{1}{2}M_p$ in (\ref{eq-lamnst})-(\ref{eq-gna}), (\ref{eq-lam-pp})-(\ref{eq-gna-b}), the $p^\dastp,\pbar^\dastp$ separations and reduced masses at the minimal (quantum excitation) level $n=2$ containing a finite $\lcfoots=1$  are given as $r_2^\dastp =r_2^0/\g_2^{_\dastp 2} =3.07 \times 10^{-18}$ m, $r_{2,1}^\dastp=r_{2,1}^0/\g_2^{_\dastp 2} 
=2.8939 \times 10^{-18}$ m,
$\mr_2^\dastp = \g_2^\dastp (\frac{1}{2}M_p) =128.58 $ GeV, $\mr_{2,1}^{\dastp}=\g_{2,1}^\dastp (\frac{1}{2}M_p)= 96.434$ GeV. These are the striking characteristic scales of the  weak interaction range and the (resonance) masses  of the observational neutral Higgs ($H$) and neutral intermediate vector boson (IVB, $Z$) 
\cite{perkins-20,FrauenfelderHenley91,Workman.et.al2022}. 
Stemming from this recognition, we conjecture that the minimal HPn's in specific dynamical configurations would  reproduce the neutral IVB and the Higgs. And these, immediately the $n,\lcfoots=2,1$ mass state, upon disintegration would reproduce the observational charged IVB ($Z^\pm$ in Sec \ref{Sec-eng-mmt-decays-n}). We shall in this section seek out the invariant rotated, combinatorial minimal orbital states of the HPn that are consistent with the production conditions and particle spins $1,0$ of the observational $Z,H$, evaluate their dynamical variables and compare these with the measured values of the $Z,H$. The neutron $n$ production which observationally involves a virtual $Z$ process is discussed in the end.

\def\Acalb{{\pmb{\Acal}}}

Consider that a $p$, $\pbar$ travel oppositely with linear momenta $\pmtb_{\pbar^\dastpp \ka} =- \pmtb_{p^\dastpp \ka} = -\mr_\ka^\dastp \cb$ up to positions $y_1, -y_1$ on the $y$ axis at time $t=0$, and  undergo a coherent near-head on (n-h.o) collision at the finite separation $y_1-(-y_1)=\bar{\rb}^\dastp_\ka$ here, with $\ka=n$ or $\nclc$. Under their binding force the $p,\pbar$ are hereof scattered into $p^\dastp, \pbar^\dastp$ rotating about their mass center at $\rb^\dastp=0$, comprising a HPn. The wave function satisfying both the SQR-KGE (Sec \ref{Sec-theo}) and the initial time conditions above is in general a superposition of the partial wave functions of two degenerate orbitals $ m$, $ m''=-m$ of common $n,\lcfoots$ in three dimensions, $\psi \hat{\rb}= (\psi_{n,\lsub,m }+\psi_{n,\lsub,m''})|_{xy}\rbhat_{xy}+ (\psi_{n,\lsub, m } +\psi_{n,\lsub, m''})|_{yz}\rbhat_{yz}$. For the $p,\pbar$ incident along $+z,-z$ (Fig \ref{fig_Z_mumu.eps}a) and 
$\lcfoots=1$, the two states are necessarily  $m,m''= 1, - 1$ (or $-1, 1$) or denoting by $\eta=\nu,\nubar $ (or $\nubar,\nu$). For the $p,\pbar$ incident along $+x,-x$ (Fig \ref{fig_Z_mumu.eps}b), the two $\lcfoots=1$ states are $m,m''=-1,-1$ (or $1,1$) or denoting by  $\eta=\nubar, \nubar$ (or $\nu,\nu$). In conformity to the decay reactions of the observational Higgs (Sec \ref{Sec-eng-mmt-decays-n}), the $H$ pertains to a (rotated) $n=2$ HPn, which contains also a $\lcfoots=0$ two-orbit of $m,m''=+0,-0$, or denoting by $\eta=\nu_0, \nubar_0$, corresponding to two opposite rotations in the $yz$ plane.
The partial orbital energies for a given $\ka$ are additive, 
$\sum_\eta \mr^\dastp_{\eta \ka } c^2 
=\sum_\eta \pmt_{\eta \ka } c
=\Eng_{\ka}  \dot{=}  T_{\ka} 
= \sum_\eta T_{\eta \ka}$; and $\sum_\eta V_{c\eta \ka} \equiv 2 \Vbar_{c\ka }
=-2 \frac{\a \hbar c}{\rtd^\dastp_{\ka} }
=-2\frac{\a \hbar c}{(2 r_{\ka}^\dastp) }
=V_{c \ka}
$, from which $\rtd_{\ka}^\dastp = 2 r_{\ka}^\dastp$.

Of our specific interest are (the occupations of) the minimal finite $\lcfoots=1$ and $m=\pm 1$ orbital states in which the HPn's are meta stablised by finite quantised angular momenta $\pm 1 \hbar$; $n=2$ is the minimal possible level and of concern in this section, hence $\ka=n,\lcfoots=2,1$. Given $\pmt_{\eta 2,1 }=\mr_{\eta 2,1}^\dastp c=\frac{h}{\lam^\dastp_{\eta 2,1}}$ on partial orbit $\eta$ of $n,\lcfoots=2,1$, there is a conjugate $\rtd_{2,1}^\dastp$ such that the angular momentum is an eigen solution to the SQR-KGE in the cmf, $\Jcalb_{\eta 1}= \rbtd_{2,1}^\dastp \times \mr_{\eta 2,1}^\dastp \cb =\Jcalb_{1}$, similarly as given by (\ref{eq-Jcal}). 
From this and $\mr_{\eta 2,1}^\dastp=\frac{1}{2}\mr_{2,1}^\dastp$, there is $\rtd_{2,1}^\dastp = 2 r_{2,1}^\dastp $ as given above. Subtracting from $ \Jcalb_{\eta 1}$ a Thomas precession (TP) term $\Jb_{\eta \TPsub}$ \cite{jxzj-neutron}, with a $z$ component $J_{\eta \TPsub z}=s_\TPsub \hbar$ and smallest finite $s_\TPsub=\frac{1}{2}$ satisfying the commutation relation (e.g. \cite{Schiff}), we obtain the $\eta$th angular momentum and its 
$z$ component in the lab frame, with $ j=\lcfoots-s_{\TPsub}=\frac{1}{2}$, $m_j=\pm j$, 
$$\displaylines{
\refstepcounter{equation} \label{eq-Jj}
\hfill 
\mbox{$
\Jb_{\eta \hf}
= \Jcalb_{\eta 1} + \Jb_{\eta \TPsub}= J_\hf \nbhat_\eta, 
\ \
J_\hf
=  \sqrt{ \frac{1}{2}(\frac{1}{2}+1)} \, \hbar
=\frac{\sqrt{3}}{2} \hbar,
\ \
J_{\pm \eta  \hf z}
=J_{\eta \hf} \cos\theta_{ \pm \hf}
=\pm \frac{1}{2} \hbar 
$}
\hfill (\ref{eq-Jj})
}$$
For the two-orbit of $\eta=\nu,\nubar$, the vectors $\Jb_{ \nu \hf }$, $\Jb_{ \nubar \hf }$ are along the orbital normals $\nb_\eta =$ $\zb'$, $\zb''$ at angles $\theta_{ \hf} 
=\arccos (J_{\nu\hf z} /J_{\nu \hf} ) 
=\arccos (1 /\sqrt{3} ) 
=54.7 ^o $,  
$\theta_{\m \hf} = \pi-\theta_\hf
=125.3^o $ 
to the $z$ axis, lying in the same $xz$ (or generally $x'z$) plane, see Fig \ref{fig_Z_mumu.eps}a and Inset 1. Their $z$ components are opposite and sum as $J_{\nu \hf z}+ J_{-\nubar \hf z}= |\rtdb^\dastp_{2,1 } \times (\mr^\dastp_{\nu 2,1 } \cb)|(\cos \theta_\hf-\cos \theta_\hf)=0$. This corresponds to a total standing wave $\psi_{\rhoxy}$ in the $xy$ plane, $\rhoxy$ being a short-hand notation of $xy$, superposed from the opposite traveling partial waves $\psi_{\eta \rhoxy}=R_{\rhoxy}Y_{1}^{\pm 1}(\frac{\pi}{2},\vphi)\exp{[\mp i(E_{\eta 2,1 \rhoxy}/\hbar)t]}$ in the $xy$ plane. For $\eta =\nubar,\nubar$, $\Jb_{\nubar\hf}, \Jb_{ \nubar\hf }$ are along the orbital normals $\nb_\eta= \bar{\zb}'$, $\zb''$ at angles $\theta_{\m \hf} = \pi-\theta_\hf=125.3^o$ each to the $z$ axis and lying in the same $xz$ (or $x'z$) plane (Fig \ref{fig_Z_mumu.eps}b). Their $z$ components are along $-z$, and sum as $J_{-\nubar \hf z}+ J_{-\nubar \hf z}= -|\rtdb^\dastp_{2,1 }\times \mr^\dastp_{\nu 2,1 } \cb|2 \cos \theta_\hf 
=-1 \hbar$. 

The Coulomb potential $\Vbar_c=- \frac{\a \hbar c}{\rtd^\dastp}$ underlining the HPn solutions here is formal. At the scale $\rtd_{\ka}^\dastp$ $\sim 10^{-18}$ m comparable with their charge radii $ a_\qe$ ($a_\qe=1.54 \times 10^{-18}$ m for the $e^\m,p$ comprising a neutron in \cite{jxzj-neutron}, being a close value to
the classical proton radius $a_p =1.53 \times 10^{-18}$ m)
the $p^\dastp,\pbar^\dastp$ inevitably interact also by a significant magnetic potential $\Vtdp_m$. Further for the extensive charges moving at essentially the speed $c$, there is a significant relativistic electric field ($|\Eb_{c}' $) profile narrowing in the directions $\bot \veb^\dastp $\cite{Jackson}, so that  $|\Eb_{c}' |=\Vbar_{c}' /(f_c e)= -\frac{f_c e}{4\pi \ev_0 \rtd_{ }^\dastp}$ is  effectively acted between two fractional charges $f_c e$'s, with $f_c <<1$\cite{jxzj-neutron}. The magnetic fields or moments by contrast are effectively produced from or at the electric current centers, and unaffected. Hence the full potential for the separation $\rtd^\dastp_\ka$ here is $\Vtdp=\Vtdp_{m }^\dastpp + \Vtdp_{c }'$ $\dot{=}$ $\Vtdp_{m}^\dastpp$. Considering $-\bar{V}_m^\dastpp \propto \frac{1}{\rtd^{_\dastp 3}}$ $ \propto \frac{1}{\bar{\Volp}^\dastpp{}}$ (see e.g. \cite{jxzj-neutron}), where $ \bar{\Volp}^\dastpp{}=\frac{4\pi}{3} \rtd^{_\dastp 3}$, and in turn demanding its equality with $\bar{V}_{c}$, this can be written for orbital $\ka(=\nclc$ or $n$),
\def\Voltd{\bar{\Vol}} 
$$\displaylines{
\refstepcounter{equation} \label{eq-VmVc} 
\hfill
\Vtdp_{m\ka}^\dastpp 
= -\frac{G_\kansub}{\Voltd_\kansub^{\dastpp}} 
\equiv \Vtdp_{c\ka}^\dastpp
=-\frac{e^2}{4\pi \ev_0 b_\ka \rtd_{\ka}^\dastp},
\ \
G_{n,1}
=\frac{e^2 \rtd_{n,1}^{_\dastp 2} }{3 \ev_0 b_{n,1}} 
=\frac{ 4 \pi \sqrt{2} \a \hbar c \rtd_{n,1}^{_\dastp 2} }{3 n} 
=\frac{ 8 \sqrt{2} e^2 \hbar^2 c^2 }{
3 n \ev_0 \mr_{n,1}^{_\dastp 2} c^4},
\hfill  (\ref{eq-VmVc}) 
}$$
and $G_n= \frac{e^2 \rtd_n^{_\dastp 2} }{3 \ev_0}
=\frac{4n^2 e^2 \hbar^2 c^2}{3 \ev_0 \mr_n^{_\dastp 2}c^4} =\frac{9}{2 \sqrt{2}n} G_{n,1} $ (for $n=2$: $G_2=1.591 G_{2,1}$),
where $b_\ka =b_{n,1}=\frac{n}{\sqrt{2}}$, $b_{n,2},\ldots$, or $b_\ka=b_n=1$;  
$\rtd_\ka^\dastp=2r_\ka^\dastp $ for both $\ka=n,1$ and $n$; (\ref{eq-lam-pp}a) for $r_{n,1}^\dastp$ are used. As to the basis for the $\Vbar_{m\ka}, \Vbar_{c\ka}$ equality:
In general, $\Vtdp_m^\dastpp$ is dependent on the $p,\pbar$ relative speed $\ve$, or $\g=(1-\ve^{ 2}/c^2)^{-1/2}$, 
and is ordinarily not uniquely defined (see \cite{jxzj-neutron}). Nevertheless, if $p,\pbar$ accommodate  their speed $\ve$ to just equal to $\ve_\ka^\dastp =g_n^\dastp\ve_n \dot{=} c$ which is a unique solution under $\Vbar_m \equiv \Vbar_{c }$, then the $p,\pbar$ can achieve the stationary eigen solutions of a HPn, in particular the rest wavelength $\Lam_n^\dastp$ of (\ref{eq-gna}c), (\ref{eq-gna-b}c), that is uniquely defined in the vacuum irrespective of the physical origin of the $\bar{V}$. The so determined  $G_\ka$ in (\ref{eq-VmVc}) is a constant for a given $\ka$. For $\ka=2,1$ of specific interest later, $G_{2,1}=\frac{4\sqrt{2}e^2 \hbar^2 c^2   }{
3  \ev_0 \mr_{2,1}^{_\dastp 2} c^4}
=2.2890 \times 10^{-62}$ J m$^3$.
 Finally, to achieve a maximal magnetic binding $|\Vbar_{m\ka}|$, the $p^\dastp, \pbar^\dastp $ (or $p,\pbar$) spins $s_{p^\dastp}, s_{\pbar^\dastp} =\frac{1}{2}, \frac{1}{2}$ are necessarily  aligned along $+z$ (or $-z$) here. 

\def\rhoxy{\rho}

Envisage that the highly energetic $p,\pbar$ collision is  instantaneous, incoherent, and not liable to form a total coherent wave $\psi \rbhat$ in three dimensions.
Instead, a desired stationary state of HPn($p^\dastp,\pbar^\dastp$) is effectively realised through its invariant rotated counterpart $A(\z,\zbar)$ of a reduced mass 
$\mr_{\ka\rhoxy}^\dastp=\frac{M_\z M_\zbar}{M_\z +M_\zbar}$ rotating along the two-orbit of $(\eta_\rhoxy=)\nu_\rhoxy, \nubar_\rhoxy$, $\ldots$, in the two dimensional $\rhoxy\equiv xy$ plane. Here, $\z,\zbar \equiv p^\dastp_{\rhoxy}, \pbar^\dastp_{\rhoxy}$, and $M_\z, M_\zbar$ are the dynamical masses of $\z,\zbar$. Designate $A$ by $Z$ if $\ka = n,\lcfoots=2,1$, $\eta_\rhoxy=\nu_\rhoxy, \nubar_\rhoxy$ (Fig \ref{fig_Z_mumu.eps}c), and by $H$ if $\ka =n=2$, $\eta_\rhoxy=\nubar_\rhoxy, \nubar_\rhoxy$ (Fig \ref{fig_Z_mumu.eps}d) and $\nu_0,\nubar_0$.
 Specifically, the rotated state $Z$ or $H$ is achieved such that its partial P-O angular momenta $\Jb_{\eta_\rhoxy}$, being along the $\nu_\rhoxy,\nubar_\rhoxy$ or $\nubar_\rhoxy,\nubar_\rhoxy$ orbital normals $z,-z$, or $-z,-z$, are obtained through the invariant rotations of $\Jb_{\eta \hf}$ along $z',z''$ or $\bar{z}',z''$ given in (\ref{eq-Jj}a), about the $y$ axis   by angles $-\theta_\hf,+\theta_\hf$ each as,
$$\displaylines{
\refstepcounter{equation} \label{eq-Jbp1}
\hfill
\mbox{$
\Jb_{\eta_\rhoxy} = \pm J_{\eta_\rhoxy} \zbhat,
\
J_{\eta_\rhoxy}
=\rtd^\dastp_{2,1 \rhoxy}  \mr^\dastp_{\eta 2,1 \rhoxy} c  
\equiv J_{ \eta \hf  z}(= J_{\eta \hf} \cos \theta_\hf)
= \rtd^\dastp_{2,1} \mr^\dastp_{\eta 2,1}c \cdot \frac{1}{2}\cos\theta_1
=  \frac{1}{2} \hbar 
$},
\hfill (\ref{eq-Jbp1})
}$$
where $J_{\eta \hf z}=\Jcal_{\eta 1 z}-J_{\eta \TPsub z}$, $\Jcal_{\eta 1 z}=|\rtdb_{2,1}^{\dastp} \times \mr_{\eta 2,1}^{\dastp} \cb| \cos \theta_1$ as given in (\ref{eq-Jcal}), and $J_{\eta \TPsub z} =\frac{1}{2}\Jcal_{\eta 1 z}$ are used. As the specific dynamical states of $p,\pbar$, as $p^\dastp,\pbar^\dastp$ are, the $\z,\zbar$ spins are the same as the $p,\pbar$ spins, $s_{\z}, s_{\zbar}=\frac{1}{2}, \frac{1}{2}$, and are similarly aligned along $z$ here for a maximal magnetic binding. The vector sums of the $\z,\zbar$ spins and their P-O angular momenta, given using (\ref{eq-Jbp1}) for the two-orbits of $\eta_{\rhoxy}=\nu_\rhoxy,\nubar_\rhoxy$, and $\nubar_\rhoxy,\nubar_\rhoxy$ give the total angular momenta (or spins) of $A=Z, H$,
 $$\displaylines{
\refstepcounter{equation} \label{eq-S_Z}
\hfill
\mbox{$S_\Asub=s_\Asub \hbar
= (s_{\z}+ s_{\zbar})\hbar \pm J_{\nu_\rhoxy}+J_{\nubar_\rhoxy} 
= (\frac{1}{2}+\frac{1}{2}) \hbar \pm ( \frac{1}{2}\mp \frac{1}{2})\hbar $}
= \lf\{ {1\hbar, \atop 0,}\rt.  
\quad {s_\Zsub=1, \atop s_\Hsub=0.}   
\hfill  (\ref{eq-S_Z})
}$$
The integers $s_\Zsub,s_\Hsub$ define $Z, H$ as bosons each. The $\z,\zbar$ charges $+e,-e$ sum to the $Z$ or $H$ charge, $q_\Hsub=q_\Zsub= +e-e=0$.

Conforming to the $\Jb_{\eta \hf} \rar $ $\Jb_{\eta_\rhoxy}$ invariance in (\ref{eq-Jbp1}) for $Z$ say, the state vectors of the HPn, $\psi_\eta\equiv \psi_{2,\lsub, \pm m}$'s which are in the Hilbert space along the $\nu,\nubar$ orbital normals $z',z''$ (Fig \ref{fig_Z_mumu.eps}a) 
, are rotated about the $y$ axis (by angles $-\theta_q,+\theta_q$) to $\psi_{\eta}^r = \tilde{U}_\Rsub \psi_\nu, U_\Rsub \psi_\nubar$ along $n',n''$ in Fig \ref{fig_Z_mumu.eps}c. Here $\tilde{U}_\Rsub=U_\Rsub^{-1}=U_\Rsub^\dagger$ is unitary. Similarly, those of the HPn along $\bar{z}',z''$ in Fig \ref{fig_Z_mumu.eps}b are rotated to $\nbar', n''$ of $H$ in Fig \ref{fig_Z_mumu.eps}d. For $\eta=\nu$, $\tilde{U}_\Rsub= \exp\lf( -\frac{i}{\hbar} (-\theta_\hf) \yb \cdot \Jb_{\nu \hf} \rt)=\exp\lf( i   \frac{1}{2} \theta_\hf  \yb \cdot \sigb_y  \rt)= I \cos(\frac{1}{2} \theta_\hf) + i\sig_y \sin (\frac{1}{2} \theta_\hf)$, where $\Jb_{\nu \hf}= j \sigb \hbar|_{j=1/2}=\frac{1}{2} \sigb \hbar$, $\sigb =\sig \zbhat' =\sig_x \xbhat+ \sig_y \ybhat +\sig_z \zbhat $ is a hermitian unit-vector operator in $\zb'$ direction satisfying desired commutation relations, and $\ybhat \cdot \Jb_{\nu \hf} =\frac{1}{2} \sig_y \hbar$ here. Using the Pauli matrices, 
 $\sig_y=\lf({ 0 \quad -i \atop i \quad 0}\rt)$ here, we have  
\def\qb{{\bf{q}}}
$$\displaylines{
\refstepcounter{equation} \label{eq-U-Ry-a-n}
\hfill
\tilde{U}_\Rsub
=\lf({ \ \ \cos \theta_\scat \quad \sin\theta_\scat  
   \atop 
          -\sin\theta_\scat  \quad \cos \theta_\scat                        }\rt)
=e^{i \theta_\scat}, 
\ \ U_\Rsub
= e^{-i \theta_q},
\ \  \theta_\scat=\mbox{$\frac{1}{2}$} \theta_\hf
= \mbox{$\frac{1}{2}$}  
\,{\rm acos}(\frac{1}{\sqrt{3}})
= 27.368 ^o. 
\hfill (\ref{eq-U-Ry-a-n})
}$$
The subscript $q$ of $\theta_q$ refers to the momentum transfer illustrated in Sec \ref{Sec-eng-mmt-decays-n}. 
Hence $\psi_{\nu}^r= e^{i\theta_\scat}\psi_{\nu}$ along $n'$. And similarly $\psi_{\nubar}^r= e^{-i\theta_\scat}\psi_{\nubar}$ along $n''$ or $\bar{n}'$ as rotated about the $y$ axis from $z''$ or $\bar{z}'$. The transition matrices are $M_{\eta \eta^r}= \int \psi_{\eta }^{r*} H_I \psi_{\eta}^r d^3 x= \int \psi_{\eta}^{r*} H_I' \psi_{\eta}^r d^3 x$, where e.g for $\eta=\nu$, $\psi_{\nu}^{r*}=(\tilde{U}_\Rsub \psi_\nu)^* = \psi_{\nu}^* \tilde{U}_\Rsub^\dagsup= \psi_{\nu}^* U_\Rsub$, $H_I'=\tilde{U}_\Rsub^\dagsup H_\Isub= U_\Rsub H_\Isub$. The $\fit$ particles described by $\psi_{\eta}^r$ rotate about the orbits $\nu^r,\nubar^r$ of normals $n',n''$ and a radius $\rtd_{2,1}^{\dastp r}$ each. Their vector masses are defined to be in the positive $n', \nbar''$ directions, $\mrb_{\eta 2,1}^{\dastp r}= \mr_{\nu 2,1}^{\dastp} \nbhat', \mr_{\nubar 2,1}^{\dastp} \hat{{\bf{\nbar}}}''$, so as to be consistent with the energy addition. In terms of the g-IED description (\ref{eq-mass-Eng}), the rest masses $\mr_{\eta2,1}^{0 }$ here are generated by the (reduced) charge $\qe$ partial oscillations about (instantaneous) fixed positions $\rtdb_{2,1}^{\dastp}$, of displacement amplitudes $\Acal_{\qe \eta}^r$ along radial $\rb^{\dastp r}$ directions in the $\nu^r$, $\nubar^r$ planes in the the magnetic $\Vbar_{m2,1}(r^{\dastp r})$ field (\ref{eq-VmVc}) (setting simply $a_\qe=1, a_r=0$ here), $\mr_{\eta2,1}^{0} c^2 =\D \Vbar_{m2,1}(\Acal_{\qe\eta}^r)=\frac{1}{2}[\Vbar_{m2,1}(\rtd^\dastp_{2,1} +\Acal_{\qe\eta}^r) -\Vbar_{m2,1}(\rtd^\dastp_{2,1})]= \frac{9G_{2,1}\Acal_{\qe\eta}^r}{ 8 \pi \rtd_{2,1}^{_\dastp 4}}$\cite{jxzj-pimu}, from which and (\ref{eq-Jcal}), $\Acal_{\qe \eta}^r=\frac{1}{2}\Acal_{\qe}=\frac{4}{9}\rtd_{2,1}^\dastp$. And $\mr_{\eta 2,1}^{\dastp r} 
= \g_{2}^\dastp \mr_{\eta 2,1}^{ 0} $. Or in vector forms, 
$\mrb_{\eta 2,1}^{\dastp r} = \g_{2}^\dastp \frac{9G_{2,1}\Acalb_{\qe\eta}^r \times \veb_2^\dastp}{8 \pi \rtd_{2,1}^{_\dastp 4} \ve_2^\dastp c^2}$, $\Acalb_{\qe \eta}^r= \Acal_{\qe \eta}^r \widehat{\rb^{\dastp r}}$.

The rotated vector $\Jb_{\eta_\rhoxy}=\pm J_{\eta \hf z} \zbhat$ quanta along $\pm \zb$ in (\ref{eq-Jbp1}) on the other hand are the direct results of the coherent wave $\psi_{\eta_\rhoxy}$ motions of the component masses $\mr_{\eta \ka \rhoxy}^\dastp$ along orbits $\nu_\rhoxy,\nubar_\rhoxy$ in the $xy$ plane, as projected from the $\psi_{\eta}^r$ wave motions of the masses $\mr_{\eta \ka }^{\dastp r}$. For $\ka=n,\lcfoots=2,1$, $\nu_\rhoxy,\nubar_\rhoxy$ are elliptics each having a semi-major and semi-minor axes $a=\rtd_{2,1}^{\dastp r}$, $b=\rtd_{2,1}^{\dastp r} \cos \theta_q$, as $xy$ projected from the (semi-classically circular) $\nu^r,\nubar^r$. In terms of the g-IED description, from the perspective of the original HPn of the specified $G_{2,1}$ and $\bar{\Volp}_{2,1}=\frac{4\pi}{3}\rtd_{2,1}^{_\dastp 3}$, the component masses $\mr_{\eta 2,1\rhoxy}^\dastp$ are the direct results of the geometrical projections of the $\Acalb_{\qe \eta}^r$ vectors on to the $xy$ plane, being $\Acal_{\qe\eta}$, $\Acal_{\qe\eta} \cos \theta_q$ at $a, b$. Their geometric means $\Acal_{\qe\eta\rhoxy}=\sqrt{\Acal_{\qe\eta} \cdot \Acal_{\qe\eta} \cos \theta_q}$ aree apparently faithful representations of the original HPn. Summing over $\eta$ gives for the two-orbit $\Acal_{\qe\rhoxy}=\sum_\eta \Acal_{\qe\eta \rhoxy}=\Acal_{\qe} \sqrt{\cos \theta_q}$, and accordingly $\mr_{ 2,1\rhoxy}^\dastp=\g_2^\dastp \frac{9G_{2,1}\Acal_{\qe\rhoxy} }{8 \pi \rtd_{2,1}^{_\dastp 4}}=\g_2^\dastp \frac{9G_{2,1}\Acal_{\qe }\sqrt{\cos\theta_q} }{8 \pi \rtd_{2,1}^{_\dastp 4}}
=\mr_{ 2,1}^\dastp \sqrt{\cos\theta_q}$. For $\ka=n=2$, the $\lcfoots=0$ waves $\psi_{2,0}$, accordingly $\Acal_{\qe 2,0}$, $\mr_{2,0}^\dastp$ are ordinarily spherical isotropic (but lying momentarily in the $yz$ plane at the time of production), hence $\mr_{2,0\rhoxy}^\dastp= \mr_{2,0}^\dastp$, and $\mr_{2 \rhoxy}^\dastp= \mr_{2,1 \rhoxy}^\dastp+\mr_{2,0}^\dastp $.

The $\fit$ particle masses, or the reduced masses $\mr_{\ka \rhoxy}^\dastp$ of $\zeta,\zbar$ characterise directly  the resonance energies of the $Z, H$ productions from $p,\pbar$ collisions, and in turn the de-excitation energies upon the $Z, H$ disintegrations (Sec \ref{Sec-eng-mmt-decays-n}). Call these crudely the  masses ($M_\Zsub,M_\Hsub$) of $Z, H$.
Using (\ref{eq-gna}a), (\ref{eq-gna-b}a) for $\mr^\dastp_{\ka}$, $\mr^0=\frac{1}{2}M_p$ and (\ref{eq-U-Ry-a-n}c) for $\theta_q$, the $Z,H$ (resonance) masses and their vector forms, defined to be in the positive normals $+z$ directions, are given as
\def\zbhat{\hat{\zb}}
 $$\displaylines{
\refstepcounter{equation} \label{eq-M_Z} 
\hfill
M_\Zsub
= \mr^{\dastp }_{2,1\rhoxy}
=\sum_\eta \mr^{\dastp }_{\eta 2,1\rhoxy}
=\frac{3M_p}{4\a}\sqrt{\cos{\theta_q}}
=90.876 \, {\rm GeV}, 
\quad \Mb_\Zsub= M_\Zsub \zbhat,
\hfill (\ref{eq-M_Z}) 
\cr
\refstepcounter{equation} \label{eq-M_Z_H}
\hfill
M_\Hsub= \mr_{2 \rhoxy}^\dastp
=\mr^{\dastp }_{2,1\rhoxy}+ \mr_{2,0}^\dastp
=\frac{3M_p}{4\a}(\sqrt{\cos{\theta_q}}+ \frac{1}{3})
=123.02\, {\rm GeV}, 
\quad \Mb_\Hsub= M_\Hsub \zbhat.
\hfill (\ref{eq-M_Z_H})
}$$
In their mutual presence, the vacuum potentials $V_{\v}'$ to their charges, and accordingly the masses of their base particles $p,\pbar$ of $\zeta,\bar{\zeta}$ are in general augmented by a frictional term $\Ocal_{\eta p}
= \frac{\eta_\v D |V_\v|'}{D' c^2}$\cite{jxzj-pimu} each, and $\mr_{\eta\ka}^\dastp$ by $\Ocal^\dastp_\eta =\frac{1}{2}\g_{2,1}^\dastp\Ocal_{\eta p}\sqrt{\cos \theta_q}$. The \fit particles of $Z$, $H$ are respectively at rest and rotating in the $xy$ plane, hence are of a zero and finite $\sum \Ocal_{\eta }^\dastp$, and $M^\dagsup_\Zsub=M_\Zsub$, $M_\Hsub^\dagsup=M_\Hsub+\sum \Ocal_{\eta}^\dastp$. Although, for the latter $\sum \Ocal_{\eta}^\dastp$ is negligibly small if using the $\eta_\v$ value for neutron\cite{jxzj-pimu}.
In general, in accordance with the original HPn's, the $\z,\zbar$ in general have each also a relative rotation in the $yz$ plane, $p_{\ka yz} = \g' M_\Asub c = \mr_{\ka}^\dastp c \sqrt{\sin \theta_{q}}=\mr_{\ka}^\dastp c$, $\g'= \sqrt{\sin \theta_q}/\sqrt{\cos \theta_q}$ for $Z$.

Using (\ref{eq-M_Z}) for $\mr_{\eta 2,1 \rhoxy}^\dastp$  in (\ref{eq-Jbp1}) gives $\rtd_{2,1\rhoxy}^\dastp(=\frac{\hbar}{2 \mr_{\eta 2,1 \rhoxy}^\dastp c })=\rtd_{2,1}^\dastp \frac{1}{2} \cos \theta_1/\sqrt{\cos \theta_q} (=\sqrt{ab}= \rtd_{2,1}^{\dastp r}\sqrt{\cos \theta_q}   )=2.1714 \times 10^{-18}$ m.
In respect to the final $\zeta,\zbar$ rotated from the original HPn in the specified volume $\bar{\Volp}_{2,1}$, the interaction constant pertains to a rotated $G_{2,1\rhoxy}$ from $G_{2,1}$ of (\ref{eq-VmVc}). In turn the formal Coulomb potential can be directly written for $\zeta,\zbar$ at their separation $\rtd_{2,1\rhoxy}^\dastp$,
thus $\Vbar_{m2,1\rhoxy}=-\frac{G_{2,1\rhoxy} }{ \bar{\Volp}_{2,1}} =-\frac{e^2}{ 4\pi \ev_0 b_{2,1} \rtd_{2,1 \rhoxy}^\dastp}
=-\frac{\sqrt{2} \mr^\dastp_{2,1 \rhoxy} c^2}{  \g_2^{\dastp }}=-0.46892$ GeV.
Apparently $\rtd_{2,0\rhoxy}^\dastp = \rtd_{2,0}^\dastp $, 
$\Vbar_{2,0\rhoxy}=\Vbar_{2,0}$, and $\Vbar_{m2 \rhoxy}=\Vbar_{m 2,1\rhoxy}+\Vbar_{m 2,0}$.

In sum, the rotated states $Z,H$ of the HPn's have 
spins $1,0$ and charges $0,0$ given by and after (\ref{eq-S_Z}), which are identical to those of the observational neutral IVB ($Z$) and Higgs ($H$). Their 
(reduced) masses $M_\Zsub,M_\Hsub$ given in (\ref{eq-M_Z}),(\ref{eq-M_Z_H}) can be compared with the experimental  (resonance) masses of the observational $Z,H$, $M_\Zsup^{exp},M_\Hsub^{exp} = 91.187(6)$, $125.25$ GeV  \cite{Workman.et.al2022}, and their decay reactions (Sec \ref{Sec-eng-mmt-decays-n}) with the experimental decay reactions of the $Z,H$. 
%

\begin{figure}[t] 
\vspace{-0.1cm}
\begin{flushleft}
\includegraphics[width=1.13\textwidth]{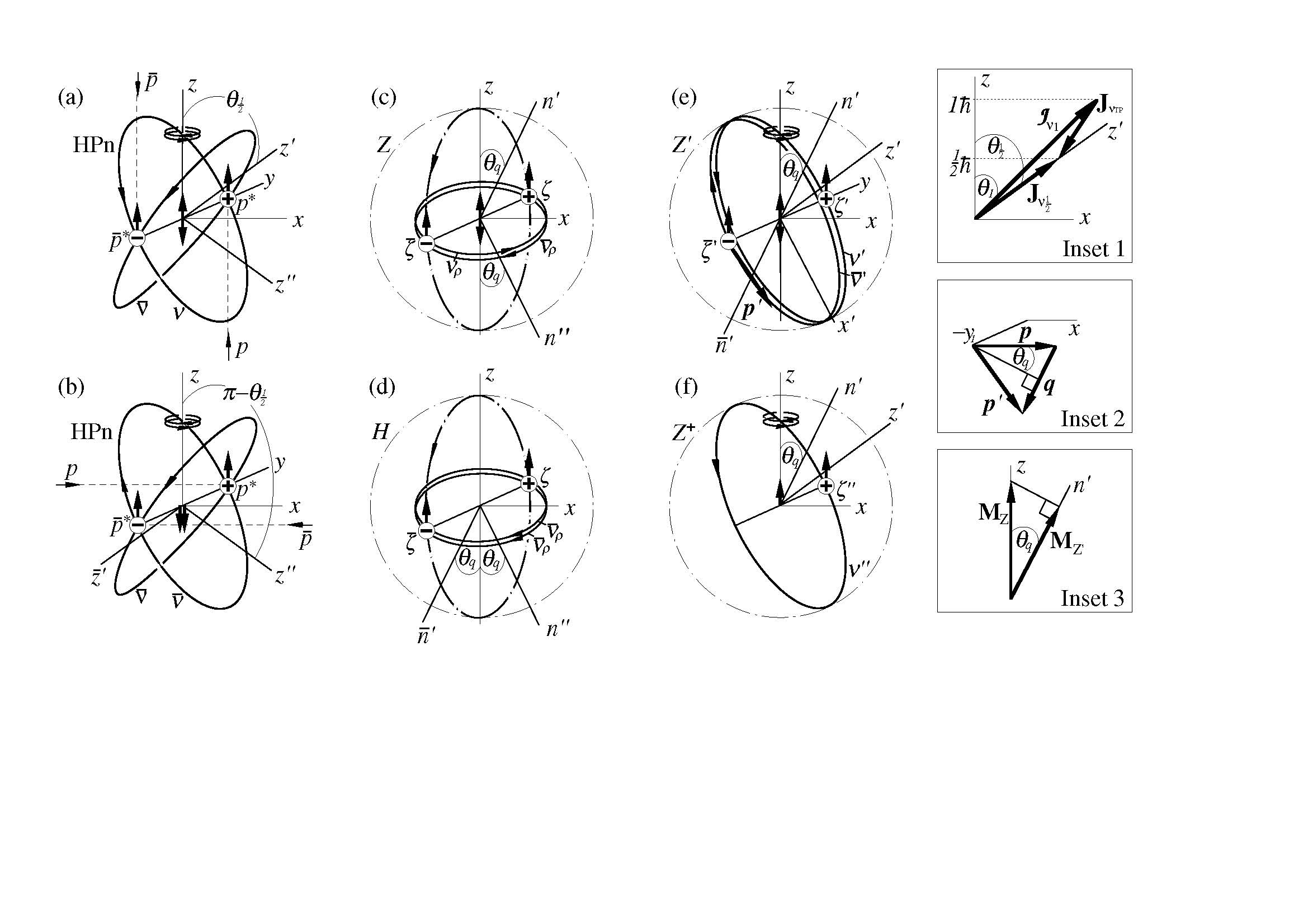} 
\end{flushleft}
\vspace{-4.2cm}
\caption{
(a,b) Schematic structures of HPn's composed each of $p^\dastp, \pbar^\dastp$ in relative P-O motions along a two-orbit of $(\eta=)\nu,\nubar$ in (a), and of $\nubar,\nubar$ in (b). The $p^\dastp, \pbar^\dastp$ are produced by a n-h.o collision of $p,\pbar$ (at $+y_1,-y_1$) on the $y$ axis, that are incident in $+z,-z$ directions in (a), and along $+x,-x$ in (b). The $\nu,\nubar,\nubar$ orbital normals $z',z'',\bar{z}'$ are at quantised angles $\theta_\hf, \theta_{-\hf}(=\pi- \theta_\hf), \theta_{-\hf}$ to $z$. (c,d) The invariant rotated HPn's, $Z(\z,\zbar)$, $H(\z,\zbar)$, with $\z,\zbar$ rotating on a two-orbit of $(\eta_\rhoxy=)\nu_\rhoxy,\nubar_\rhoxy$ in (c) and of $\nubar_\rhoxy,\nubar_\rhoxy$ in (d) in the $xy$ plane. The state vectors $\psi_\eta^r= e^{i\theta_q}\psi_\nu,e^{-i\theta_q}\psi_\nubar$ along $n'/\bar{n}', n''$ in (c)/(d) are rotated about $y$ by angles $-\theta_q/\theta_q, \theta_q=\frac{1}{2}\theta_\hf$ from $\psi_\eta$ along $z'/\bar{z}',z''$ in (a)/(b). (e) The reverse invariant rotated state $Z'(\z',\zbar')$ of $Z$, with $\z',\zbar'$ in P-O motions along the two-orbit of $(\eta'=)\nu',\nubar'$ of normals $z',\bar{z}'$, and of state vectors $\psi_{\eta'}=U_\Rsub \psi_{\eta_\rhoxy^r}$ along $n',\bar{n}'$ as rotated from the coherence components $\psi_{\eta_\rhoxy^r}$ along $z,-z$ under $U_\Rsub=e^{-i \theta_q}$. (f) The positive charged state of $Z'$, $Z^\p(\zeta'')$, with $\zeta''$ in P-O motion along orbit $\nu'$. Inset 1: Vector relations between the $\nu$ orbital angular momenta $\Jb_{\nu \hf}$, $\Jcalb_{\nu 1}$ in the lab, cm frames, and a precession term $\Jb_{\nu\TPsub}$. Inset 2: The rotation $U_\Rsub$ of $\psi_{\nu_\rho}$ to $\psi_{\nu'}$ is illustrated as an elastic scattering of the $\fit$ particle of $Z$, incident with a linear momentum $\pb= M_\Zsub c \, \xbhat $ parallel to the $x$ axis, into that of $Z'$ with $\pb'=|\pb| \, \xbhat'$ parallel to $x'$ at angle $\theta_\hf$ from $x$; $\pb'= p\cos \theta_q \hat{\qb}\times \ybhat + \frac{1}{2}\qb $, and $|\qb|=|\pb'-\pb|=2p \sin \theta_q$ is the momentum transfer. Inset 3: The $Z'$ coherent mass vector $\Mb_{\Zsub '}$ in $\nbhat$ direction as cosine projected from the $Z$ mass vector $\Mb_\Zsub$.
}\label{fig_Z_mumu.eps}
\label{fig_Z_mu++mu-_cm_axr.eps} 
\vspace{-0.3cm}
\end{figure}
%

Finally, instead of the $Z$ production by the high energy $p,\pbar$ collision at the separation $\rtd_{2,1\rhoxy}^\dastp \sim 10^{-18}$ m, consider a low energy  $e^\m$ orbiting about $p$ (at rest) in the $xy$ plane say in orbital state $n,\lcfoots=1, 0$ at a separation $\sim 1 \cdot 10^{-10}$ m at time $t=0$; choose the  $p$ spin $s_p \sigb=\frac{1}{2} \sigb$ vector $\sigb$ as the $z$ direction. Let the $e^\m$  be then distablised, and firstly accelerated under the $e^\m$-$p$ Coulomb force inward, first to a distance $r_{2,1(p)}^0
= 2 \sqrt{2} r_{2(e)}^0/3 \g_{ep}
=2.17 \times 10^{-13}$ m,  where $\g_{ep}= M_p/M_e$, 
and $e^\m, p$ has a kinetic energy $T_{e}^0 \dot{=}\g_{ep}M_e c^2 =M_p c^2$. Suppose that at about $r \lesssim r_{2,1(p)}^0$, $e^\m$ experiences firstly a short range repulsion $F_r$ and then a short range attraction $F_a$ (the strong force origin proposition\cite{jxzj-neutron}), 
which form an energy barrier $\D V \sim -T_{e}^0$ that $e^\m$ is just able to hop over to the low $r$ side. This $e^\m$,  denoting as $\epm$, is thereof  further accelerated (by $F_a$ and in turn by $-\pd V_{m}/\pd r^\dastp$) in the $xy$ plane down to $\rtd_{2,1\rhoxy}^\dastp \sim 10^{-18}$ m, so swiftly that its initial energy $T_{e}^0(=M_p c^2)=M_{\epm} c^2$ manifests as a rest mass energy in one oscillation cycle, so that $p, \epm$ become $ \z, {\epm}^\dastp$ of a reduced mass $\mr_{2,1\rhoxy}^\dastp = \g_{2,1}^\dastp(\frac{1}{2} M_p)\sqrt{\cos \theta_q }$, on orbit $\nubar_\rhoxy$ in the $xy$ plane, mimicking the partial $\nubar$ state of $Z(\z,\zbar)$. The latter further transition to $\z', {\epm}^{\dastp\prime}$ 
of a coherent mass $M_{\Zsub'}=\mr_{2,1\rhoxy}^{\dastp\prime}$ (Sec \ref{Sec-eng-mmt-decays-n}), in P-O motion about a single orbit $m_j'=-j'=-\frac{1}{2}$, or $\nubar'$, mimicking the partial $\nubar'$ state of $Z'(\z',\zbar')$. The total angular momentum of the final state is $J_{f z}= (s_e+s_p -j')\hbar =\frac{1}{2} \hbar =J_{iz}$, being invariant of  the initial $J_{iz}=s_p\hbar$.
To the vacuum, ${\epm}^{\dastp\prime}$ remains to be $e^\m$ of the rest mass $M_e$. The total reduced mass energy can be written $\Ecal_{\nubar' 2,1 \rhoxy}'  = \hbar \Kcal_{2,1\rhoxy}' c=\frac{\g_2^\dastp \mr_{2,1 \rhoxy}^{\dastp\prime} \ve_2^{\dastp 2} }{\g_2^\dastp+1}+ \frac{\g_{ep} \frac{1}{2}\g_{ep} M_e \ve_{ep}^2}{\g_{ep}+1}+ M_e c^2$. The total space function thus writes $\psi_{\nubar' 2,1 \rhoxy}^r =e^{- i \Kcal_{2,1\rhoxy}' r^\dastp }\dot{=}\Psim_{\nubar'2,1\rhoxy}^r \Psim_{ep} \Psim_e $, which is the convolution of two fast oscillating $\Psim_{\nubar' 2,1 \rhoxy}= e^{-i k_{\nubar 2,1\rhoxy}r^\dastp}$, $\Psim_{ep}=e^{-i\frac{\g_{ep}M_e c}{2\hbar} r^\dastp}$ and a slow varying or essentially constant $\Psim_e\dot{=} e^{-i \frac{M_e c}{\hbar} r^\dastp}$. The process $e^\m+p (\rar \epm +p) \rar {\epm}^{\dastp} + \z +\nubar_\rhoxy \rar {\epm}^{\dastp \prime}+ \z' +\nubar'$ represents the neutron $n$ production in terms of the $e^\m,p$ neutron model\cite{jxzj-neutron}.

\section{$Z$, $H$ disintegrations. The intermediate mass states $Z', Z^\pm$}
\label{Sec-eng-mmt-decays-n}
 \label{subsec-Sec-IVBs-1-iii-n}

\setcounter{equation}{16}

\def\rhobhat{\hat{\mathbf{\rho}}}
\def\ybhat{\hat{\yb}} 
\def\xbhat{\hat{\xb}} 
\def\qbhat{\hat{{\bf q}}}
\def\zpp{\z^{\p \prime}}
\def\zmp{\z^{\m \prime}}

If supplied with a threshold energy $H_\Isub=-\bar{V}_m$  at time $t'(=t-\tau)=0$, the bound $\z,\zbar$ of $\Z$ or $\H$ tend to be destabilised. These tend to separate apart 
by means of {\it in situ} down conversions to lighter particles so as to gain a sufficient requisite kinetic energy. $H$ has a similar structure to $Z$, and would on energy basis first down convert to $\Z$, or the $Z'$ later, plus a mass state (denoting as $Z^\dag$) associated with its extra energy $E_{2,0}$. Our main subject henceforth is to formulate a formal down conversion, hence disintegration scheme of $\Z$ based on the dynamics of the charges and energy-momentum conditions of the component particles. Throughout the discussions we shall assume no external torque is applied. 

In general, $\z,\zbar$ of charges $+e,-e$ would naturally first {\it in situ} down convert to two lighter mass states $\be',\bebar'$ generated by the same two charges $(\qe=)+e,-e$ at the initial separation $\rtd_{2,1\rhoxy}^\dastp$, or to a lighter $\be'/\bebar'$ and a $A^\pm$ at rest, depending on the activation symmetries. In the process the $\z,\zbar$ charges, of initial oscillation displacements $\xib_\qe =\xi_\qe \rbhat_{xy}$ each in level $N$ (or $n_\v=1$ in $V_{\v +e}$ here) and in $\ka$th orbital motions, can be generally regarded as first instantly retarded by ceasing the rotational oscillations. EMR fields are being emitted and re-absorbed, thereby re-driving the charges into oscillations of primed displacements $\xib_{\qe}'$ each in $N'$th level and in $\ka'$th orbital motions,
thereby forming $\be',\bebar'$. At least one of these 
ought to be a stationary state (the principles of least action and maximum entropy \cite{jxzj-Planck}). 
For $\rtd_{2,1\rhoxy}^\dastp$ is comparable with the vacuum spacing $\sim 1 \times 10^{-18}$ m, the charges are in effect situated in an isotropic potential field due either to $\Vbar_m$ or $V_{\v\qe}$. Their displacements can therefore have  three equal probable orthogonal orientations, $\xib_{\qe}' = \xi_{\qe}' \xbhat, \xi_{\qe}' \ybhat, \xi_\qe'\zbhat$, thereby possibly generating three distinct pairs of $\be', \bebar'$, or of $\be'/\bebar',A^\mp$, equal probably.

Consider first the symmetric activation, and that the re-driven oscillations of $+e,-e$ are along the radial directions along $y$ in Fig \ref{fig_Z_mumu.eps}e, $\xib_{\qe}'= \xi_{\qe}' \ybhat$, as are of the $+e,-e$ of $Z(\z,\zbar)$. The final $\beta', \bar{\beta}'$ are thus implied to possess HPn kind of internal motions, or be of a "weak" kind. As such, and for these to be separately stationary, $\z, \zbar$ are stemmed to first adiabatic transform to two intermediate mass states $\z',\zbar'$, comprising $Z'$ of a reduced mass $\mr^{\dastp\prime}_{2,1\rhoxy}=\frac{1}{2}M_{\zeta'}$, which orbital motions in the cmf are separately composed of the direct eigen states 
given by the SQR-KGE solutions (Sec \ref{Sec-IVBs-n}). 
Moreover, their P-O angular momenta $J_{\eta'j }= \sqrt{j'(j'+1)}\hbar$'s, need be invariant of the initial $\Jb_{\eta_\rhoxy}$, and similarly their vector sums. The minimals of such are the two P-O states $n',\lcfoots',j'= 2,1, \frac{1}{2}$,  $m_j'= +\frac{1}{2},-\frac{1}{2}$, or denoting by $(\eta'=)\nu',\nubar'$, which orbits remain coinciding and having opposite aligned normals $z',\bar{z}'$. Accordingly $\Jb_{\eta'\hf} =\rtdb^\dastp{}'_{2,1\rhoxy} \times (\mr^{\dastp\prime}_{2,1\rhoxy}\cb') +\Jb_{\eta' \TPsub}=J_{\eta'\hf } \zbhat'/\hat{\bar{\zb}}'$, which are also desired invariant rotations, by angles $\theta_\hf$ about the $y$ axis from the initial $\Jb_{\eta_\rhoxy}$ along $+z,-z$ to $z',\bar{z}'$ directions, such that their $z$ components are equal, 
 $$\displaylines{
\refstepcounter{equation} \label{eq-Jetaphfz}
\hfill
\mbox{$
J_{\pm \eta' \hf z}  (=\pm J_{\nu' \hf} \cos \theta_\hf)
=\Jcal_{\pm \eta' 1 z}\mp J_{\nu' \TPsub z}
=\pm |\rtd^{\dastp \prime}_{2,1 \rhoxy} \times 
\mr_{\eta' 2,1\rhoxy}^{\dastp \prime} \cb| \frac{1}{2} \cos \theta_1 
\equiv J_{\eta_\rhoxy}
=\pm  \frac{1}{2} \hbar 
$}
\hfill    (\ref{eq-Jetaphfz})
}$$ 

\def\xb{{\mathbf{x}}}
The state vectors satisfying (\ref{eq-Jetaphfz}) are rotated under $U_\Rsub$ of (\ref{eq-U-Ry-a-n}) to
$\psi_{\eta'}^r= U_\Rsub \psi_{\eta_\rhoxy} =e^{-i \theta_q}\psi_{\eta_\rhoxy}$ along $ n',\bar{n}'$ at angles $\theta_q, \pi-\theta_q$ to $z$. The masses described by $\psi_{\eta'}^r$ are $\mr_{\eta' 2,1\rhoxy}^{\dastp r}
= e^{-i \theta_q} \mr_{\eta 2,1\rhoxy}^{\dastp}
= \mr_{\eta 2,1\rhoxy}^{\dastp}(\cos \theta_q -i \sin \theta_q)$. The sine components, $\mr^{\dastp i.m}_{\eta 2,1\rhoxy}=-i \mr^{\dastp }_{\eta 2,1\rhoxy} \sin \theta_q$, are imaginary and associated with the rotations $U_\Rsub$ of the orbital normals from $z,-z$ to $n', \bar{n}'$. In the (elastic) scattering description (Fig \ref{fig_Z_mumu.eps}, Inset 2), the $\mr^{\dastp i.m}_{\eta 2,1\rhoxy}$ are associated each with one half the momentum transfer $\qb= \mr^{\dastp}_{\eta 2,1\rhoxy} c \sin \theta_q$ in the direction  $n'$ normal to the (instantaneous) rotation plane, which vector products with $\rbtd^{\dastp \prime}_{2,1 \rhoxy}$ averaged over time give zero contribution to $J_{\pm \eta' \hf z}$. The cosine projections $\mr_{\eta' 2,1\rhoxy}^{\dastp \prime }=\mr_{\eta 2,1\rhoxy}^{\dastp}\cos \theta_q$ [cf Eq (\ref{eq-MZpGp}a) and Fig \ref{fig_Z_mumu.eps}, Inset 3] are real and are coherent components rotating in the plane  normal to $n'$, and the mass vectors are defined in the positive $n'$ direction. Their sum over $\eta'=\nu',\nubar'$ gives the total (resonance) mass of $Z'$ in (\ref{eq-MZpGp}a), this combined with (\ref{eq-MZpGp}a), (\ref{eq-Jetaphfz}), (\ref{eq-Jbp1}) gives the $\eta'$ orbital radii $\rtd^{\dastp \prime}_{2,1\rhoxy}=\frac{\rtd^{\dastp}_{2,1} }{\sqrt{\cos \theta_q } \cos \theta_q}
=6.9158 \times10^{-18}$ m,
 and in turn the real potential and interaction constant projected from (the planes of normals) along $z',z''$ to $n'$ in (\ref{eq-MZpGp}b,c) below, with $M_\Zsub$, $\theta_q$ given by (\ref{eq-M_Z}), (\ref{eq-U-Ry-a-n}c), 
$$\displaylines{
\hfill 
M_{\Zsub'}
=\sum{}_{\eta'} \mr_{\eta' 2,1 \rhoxy}^{\dastp\prime}
=\mr_{ 2,1 \rhoxy}^{\dastp} \cos \theta_q
=M_\Zsub \cos \theta_q= 80.704 \, {\rm GeV},
\ \
\Vbar_{m 2,1\rhoxy }'=-\frac{e^2 \cos \theta_q}{
4\pi \ev_0 b_{2,1} \rtd^{\dastp  \prime}_{2,1 \rhoxy}  }=
 \hfill 
\cr
\refstepcounter{equation} \label{eq-MZpGp} 
\hfill
 = - \frac{G_{2,1\rhoxy}' }{ \bar{\Vol}_{2,1} }, 
\ \  
G_{2,1\rhoxy}'
=G_{2,1} \sqrt{\cos \theta_q} \cos^2 \theta_q
=\frac{4\sqrt{2}e^2 \hbar^2 c^2 
(\cos \theta_q)^{\frac{11}{2}}
}{
3 \ev_0 M_{\Zsub'}^2 c^4} 
=1.70 \times 10^{-62} \, {\rm Jm}^3 
 \hfill (\ref{eq-MZpGp})
}$$

At essentially the same time $t'=+0$ when their charges remain at $-y_1,y_1$, the $\z',\zbar'$ can proceed to {\it in situ} down transition to two final lower mass states $h^\p_\rhoxy,\h^\m_\rhoxy$, called a $(\rhoxy\equiv) xy$ projected\footnote{Refers to the geometric $xy$ projection from the $\pi$ mass in \cite{jxzj-pimu}.} heavy positron and electron in \cite{jxzj-pimu}; their spins remain aligned along $z$. In terms of the g-IED model, the $h^\p_\rhoxy,\h^\m_\rhoxy$ are generated by the charges $+e,-e$ of oscillations $\xi'_{\qe \Nsub'(e)}=N'\Acal_{\qe(e)}e^{-i\Om_{\qe(e)} t}$ in radial ($y$) directions in the $xy$ plane, in $N'=1$th level in one half the difference potential field for each charge, $\D \Vbar_{m2,1\rhoxy}'(\xi'_{\qe 1(e)}) =\frac{1}{2}[V_{m2,1 \rhoxy}'(\rtd^{\dastp \prime}_{2,1\rhoxy}+ \xi'_{\qe 1 (e)})-V_{m2,1 \rhoxy}'(\rtd^{\dastp\prime}_{2,1\rhoxy})]
=\frac{9 G_{2,1\rhoxy}' \xi'_{\qe 1(e)} }{
8 \pi \rtd^{_\dastp 4}_{2,1}}\approx \frac{1}{2}\beta_\qe \xi_{\qe 1(e)}^{\prime 2}$. Their rest masses ($M_{h_\rhoxy}$'s) are determined  by the $N'=1$ total mechanical energy $\frac{1}{2}\Mfrak_\qe \dot{\xi}^{\prime 2}_{\qe 1(e)}+ \frac{1}{2} \beta_\qe \xi^{\prime 2}_{\qe 1(e)}=\D \Vbar_{m2,1\rhoxy}'(\Acal_{\qe(e)})
=\frac{9 G_{2,1\rhoxy}' \Acal_{\qe(e)} }{8 \pi \rtd^{_\dastp 4}_{2,1}}=M_{h_\rhoxy}c^2$, where $\Mfrak_\qe=\beta_\qe/\Om_\qe^2$. $\Acal_{\qe(e)}$ and hence $M_{h_\rhoxy}$ are free parameters to be constrained such that the $h_\rho^-,h_\rho^+$  (separately) are energetically equivalent to certain lower stationary states in both the particle field and free vacuum in the sense discussed after (\ref{eq-pn-Tn}). Such states are uniquely provided by a stationary HPs($e^{\p \dastp}, e^{\m \dastp}$) in the minimal cmf orbital state $n',\lcfoots'=2,1$ \cite{jxzj-pimu} of a rest reduced mass given using (\ref{eq-gna-b}) for $a,\bar{a}=e^\p,e^\m$ here (see further \ref{AppErrota}), $\frac{1}{2} M_{h_\rhoxy}=\mr_{2,1 (e)}^\dastp =\g_{2,1}^\dastp \frac{1}{2}M_e$, or $M_{h_\rhoxy}=\g_{2,1}^\dastp M_e=\frac{3M_e}{2 \a } =105.038$ MeV, and equilibrium orbital radius $\rtd_{2,1(e)}^\dastp =2r_{2,1(e)}^\dastp$, $r_{2,1(e)}^\dastp =r_{2,1(e)}^0/\g_2^{\dastp 2} (=5.3248 \times 10^{-15} $ m) given using (\ref{eq-lam-pp}).
At the initial separation $\rtd_{2,1\rhoxy}^{\dastp \prime}$ of $\zeta',\zbar'$ here, $h_\rhoxy^\p,h^\m_\rhoxy$  maintain opposite P-O motions at speeds  $\ve'= \sqrt{(\g'{}^2-1)/\g'{}^2}\, c \dot{=}c$ along the two coinciding partial orbits of states $n',\lcfoots',j'=2,1, \frac{1}{2}, m_j'=\frac{1}{2},-\frac{1}{2}$, or denoting by $(\eta_e=)\nu_e, \nubar_e$ here. The momenta are conserved,  
$\g' \mr_{\eta_e 2,1(e)}^\dastp \ve' = \mr_{\eta' 2,1 \rhoxy}^{\dastp \prime} c= \frac{M_p}{M_e} \mr_{2,1(e)}^\dastp \cos^{3/2} \theta_q$; this gives $\g'= \frac{M_p}{M_e} (\cos \theta_q)^{3/2}$.  

The spatially oriented orbits $\nu_e, \nubar_e$ are yet incompatible with the charge oscillations in the $xy$ plane. An explosive collision is prompted to produce two coinciding opposite energy currents in P-O states $n'',\lcfoots'', j'' =2,1, \frac{1}{2}$,  $m_j'' =-\frac{1}{2}, +\frac{1}{2}$, or denoted by $\nubar_\mu,\nu_\mu$, with the $\nubar_\mu,\nu_\mu$ orbital energies provided by halving those of $\nu_e, \nubar_e$. The $\nubar_\mu,\nu_\mu$ orbital normals $z'',\bar{z}''$ are opposite aligned at angles $\pi-\theta_\hf,\theta_\hf$ to $z$, and are mirror reflections of $z', \bar{z}'$ about the $x$ axis. The final separately meta-stable ($h_\rhoxy^\p,\nu_e, \nubar_\mu)$, $(h_\rhoxy^\m,\nubar_e, \nu_\mu)$, denoting by $\mu^\p,\mu^\m$, have charges $+e,-e$, spins $\frac{1}{2},\frac{1}{2}$, and  a rest mass $M_\mu=M_{h_\rhoxy}+M_{\nu_\mu}+M_{\nu_e}\dot{=}105.29 $ MeV each ($M_{\nu_\mu}\dot{=}0.25$ MeV, $M_{\nu_e} \dot{=}0$ are used), which are identified in \cite{jxzj-pimu} as the observational muons of the measured masses 105.66 MeV each. Hence the $Z$ disintegration or decay equation $Z \rar Z' \rar \mu^\p + \mu^\m$. For the $\mu^\p, \mu^\m$ are {\it in situ}  produced in the magnetic (or weak) potential $\Vbar_{m 2,1\rhoxy}'=-G_{2,1\rhoxy}'/\bar{\Volp}_{2,1}$ [Eq (\ref{eq-MZpGp}b,c)] field, the constant  $G_{2,1\rhoxy}'$ can compare with the Fermi constant ($G^{exp}$) of a measured value $G^{exp}=1.4358 \times 10^{-62}$ Jm$^3$, with the predicted value in (\ref{eq-MZpGp}c) being appreciably larger. If the present model construction for $\mu^-,\mu^+$ is to be judged realistic, the $G_{2,1\rhoxy}'$ value in (\ref{eq-MZpGp}c) is an {\it ab initio} calculation. The $G^{exp}$ value on the other hand is determined based on the (accurate) experimental life times of such particles as $\mu^\m$, and on the theoretical assumptions of complete wave function overlap and a form factor 1 for point particle which both tend to yield a smaller $G^{exp}$.

Consider secondly $\xib_\qe'= \xi_\qe' \xbhat$, and accordingly the resulting EMR $\Eb$ fields are along $x$, which are transverse to the charge and EMR-field travel directions $+z,-z$ at positions $y_1,-y_1$. The $\zeta,\bar{\zeta}$ can naturally {\it in situ} down convert a (relativistic) positron $e^\p$ and electron $e^\m$ generated by the charge $+e,-e$ oscillations $\xib_{\qe(e)}'$, such that $e^\m$ is a level $n_\n=1$ stationary mass state in the vacuum potential $V_{\v,-e }$ field to the charge $-e$ (the IED model), and of a rest mass $M_e$. 
$e^\p$ acquires the same rest mass and relativistic linear momentum as result of the pair production.
Given the $H_\Isub$ supplied, these are in effect un-bound here and able to travel off their orbit in the opposite tangential directions parallel to $z,-z$. Hence the reaction $Z \rar e^\p+e^\m$. 

Consider thirdly $\xib_\qe' =\xi_\qe' \zbhat$ and accordingly the resulting EMR $\Eb$ fields are along $z$. These are not radial, nor transverse to their rotational velocities at $y_1,-y_1$. But the $\xib_\qe' $'s become radial when the charges have traveled (about orbits $\nu_{yz}, \nubar_{yz}$) to $z_1,-z_1$ on the $z$ axis. Here the down conversion can proceed to the final rotated states of $p_2$ (the level $n_\v=2$ mass state of charge $+e$ oscillation in the vacuum, of a rest mass $M_{p_2}=2 M_p$) and $\pbar_2$, denoted by $\z_2^0, \zbar_2^0$ and called a rotated hyper-proton and hyper-antiproton here, of the rest masses $M_{\z^0_2}= 2M_p \sqrt{\cos\theta_q} = 1768.5 $ MeV each. At the {\it in situ} separation $\rtd_{2,1(p)}^\dastp$, $\z_2^0,\zbar_2^0$ maintain spins $\frac{1}{2},\frac{1}{2}$ aligned along $z$, and in opposite P-O motions along two (rotated) coinciding orbits of states $n',\lcfoots',j'=4,1, \frac{1}{2}$, $m_j'=\frac{1}{2}, -\frac{1}{2}$, or denoting by $\nu_{p_2}, \nubar_{p_2}$. The $n\rar n'$ transition is equi-separation, $\rtd_{4,1(p_2)}^\dastp= 4 \rtd_{1,1(p_2)}^\dastp=2 \rtd_{1,1(p)}^\dastp=\rtd_{2,1(p)}^\dastp$, and equi-energy and equi-momentum, with the $xy$-projected reduced mass $\mr_{4,1 \rhoxy(p_2)}^\dastp 
=   \g_{4,1}^\dastp \times (\frac{1}{2} M_{\z^0_2})
=\frac{3}{4 \a } \times (\frac{1}{2} M_{p_2} \sqrt{\cos\theta_q})=\frac{3}{2 \a } \times (\frac{1}{2}M_p) \sqrt{\cos\theta_q}= \mr_{2,1 \rhoxy(p)}^\dastp $.
Similarly as the $Z\rar \mu^\p,\mu^\m$ process, two coinciding opposite energy currents $\nubar'', \nu''$ are produced by an explosive collision. The final two separable {\it quasi} stationary components $\tau^\p(p_2,\nu_{p_2}, \nubar'')$, $\tau^\m(\pbar_2,\nubar_{p_2}, \nubar)$, have charges $+e,-e$, spins $s_{\tau^\p} (=s_{\z^0_2}+j-j),s_{\tau^\m}=\frac{1}{2}, \frac{1}{2}$, and rest masses $M_\tau = M_{\z^0_2}+M_{\nu_{p_2}}$ each.
These properties and the $\tau^\p, \tau^\m$ decay reactions (internal work) can be compared with those of the observational tau particles of a measured mass $1784$ MeV each. Hence the decay reaction $Z\rar \tau^\p +\tau^\m$.

Consider finally the asymmetric activation of $Z$ such that $\zbar$ (or $\z$) is scattered into $A^\m$ (or $A^\p$) at rest, having effectively an infinite mass $M_{\Asub} =\infty$, and its spin is in effect de-polarised. $\z$ (or $\zbar$) alone assumes the total rotation of the cmf mass $\mr_{2,1\rhoxy}^\dastp $ momentarily along a single orbit $\nu_\rhoxy$ of radius $r_{\zeta }=r_{2,1\rhoxy}^\dastp$ about the $A^\m$ (or $A^\p$) at $r_A=0$. Its spin remains along $z$. Similarly as the $Z\rar Z'$ process, $(\z,\nu_\rhoxy)$ subsequently first transforms to an intermediate rotated state $Z^\p(\z^{ \prime\prime}, \nu'')$ [or $Z^\m(\zbar^{ \prime\prime}, \nu'')$]; see Fig \ref{fig_Z_mumu.eps}f. $\z^{\prime\prime} /\bar{\z}''$ is re-generated by the charge $\pm e$ oscillation $\xib_\qe'= \xi_\qe' \ybhat$, and is in P-O motion about $A^\mp$ each in the minimal quantum state $n'',\lcfoots'', j'', m_j''=2,1,\frac{1}{2},\frac{1}{2}$, or denoted by $\nu''$, with the orbital normal $z'$ at angle $\theta_\hf$ to $z$. The P-O angular momentum, $\Jb_{\nu''\hf}= \rb_{2,1\rho}^{\dastp \prime} \times \mr_{2,1\rho}^{\dastp \prime}\cb''= J_{\nu'' \hf} \zbhat'$, is invariant of $\Jb_{\nu_\rhoxy}$, such that $J_{\nu''\hf z}=J_{\nu_\rhoxy z}= \frac{1}{2}\hbar$. The $\z^{\prime\prime} /\bar{\z}''$ spin maintains polarised along $z$, $\Sb_{\zeta''/\bar{\zeta}''} =s_{\zeta''/\bar{\zeta}''} \hbar \zbhat$, $s_{\zeta''/\bar{\zeta}''}=\frac{1}{2}$. Hence the spin of $Z^\pm$ is $S_{\Zsub^\pm}=s_{\Zsub^\pm}\hbar= s_{\zeta''/\zbar''}\hbar + \la\Jb_{\nu''\hf} \ra = (\frac{1}{2}+ \frac{1}{2}) \hbar =1 \hbar$, being invarint of the $Z$ spin. The state vector is rotated under $U_\Rsub$ of (\ref{eq-U-Ry-a-n}) to the final $\psi_{\nu''}=e^{-i \theta_q} \psi_{\nu_\rhoxy}$. The reduced mass of $Z^\pm, A^\mp$ is $\mr_{2,1 \rhoxy}^{\dastp r}= e^{-i \theta_q} \mr_{2,1 \rhoxy}^{\dastp}$. Its cosine term gives the coherent reduced mass similarly as in (\ref{eq-MZpGp}a), with $M_{\Zsub^\pm} = M_{\zeta''}+ M_{\nu''}\dot{=} M_{\zeta''}$, 
 $$\displaylines{
\refstepcounter{equation} \label{eq-MZpm}
\hfill
\frac{M_{\zeta'' }M_\Asub   }{M_{\zeta''  }+M_\Asub }
\dot{=}M_{\Zsub^\pm }= \mr_{2,1\rhoxy}^\dastp \cos \theta_q
= M_{\Zsub'}=80.704 \ {\rm GeV};
\hfill (\ref{eq-MZpm})
}$$
and the same $\bar{V}_{m2,1\rhoxy}', G_{2,1\rhoxy}'$ for $Z^\pm$ as given in (\ref{eq-MZpGp}b,c). Similarly, the final decay reactions are $Z^\pm \rar \mu^\pm(h_\rhoxy^\pm, \nu_e/\bar{\bar{\nu}}_e, \nubar_\mu/\nu_\mu)+ \nu_\mu/\nubar_\mu $, $e^\pm + \nu_e/\nubar_e$. In the last reactions, $\nu_e,\nubar_e$ are each  the same  spin up  energy currents $\nu''$, but travel off in the $-z,+z$ directions (opposite to the $e^+, e^-$ travel directions $+z,-z$), being untiparallel and parallel to their $\nu''$- spin vectors and thus having helicities $-1$, $+1$, hence no bar over the first $\nu_e$ and a bar over the second $\nu_e$. Exceptionally, the $\tau^\pm$ decay product channel is absent, given that the charge $\pm e$ on the single orbit $\nu''$ can never access $\pm z_1$ on the $z$ axis for a down conversion to $p_2/\pbar_2$ thereof. In sum, the theoretical $Z^\p, Z^\m$ of charges $+e,-e$, spins $s_{\Zsub^\p },s_{\Zsub^\m}=1,1$, and the coherent reduced masses given by (\ref{eq-MZpm}) each, can be compared with the observational charged IVB's of measured (resonance) masses $80.377$ GeV\cite{Workman.et.al2022} each.

\begin{appendix}

\section[\qquad\qquad \quad The SQR-KGE - The relativistic Schr\"ordinger equation]{The SQR-KGE - The relativistic Schr\"ordinger equation}\label{App-SQR-KGE}

Given a particle of rest mass $M$ traveling at linear velocity $\ve$ in a potential field $V$, the Klein-Goldon equation (KGE) is $[(E_{tot}-V)^2 -M^{ 2} c^4]_{op} \psi_{tot} = \pmt^2 c^2 \psi_{tot}$, 
where $E_{tot}=m c^2 +V=E+Mc^2 $ is the total Hamiltonian, $E=T+V=\frac{\g m \ve^2}{(\g+1)}+V$, 
$m=\g M$, $\g=[1-(\ve/c)^2]^{-1/2}$ or $\g^2-1 $ $ =\g^2(\ve/c)^2$, and $\pmt^2= m^2 c^2 -M^{2} c^2 =(\g^2-1)M^{2} c^2= \g^2 (\frac{\ve^2}{c^2})^2 M^2 c^2= m^2 \ve^2$.
Adding $M^2 c^4 \psi_{tot}$ on both sides, the KGE becomes $[(E_{tot}-V)^2]_{op} \psi_{tot}= m^2 c^4 \psi_{tot}$. Or in the square root (SQR) form, $(E_{tot}-V)_{op} \psi= m c^2 \psi$, where $\psi=\sqrt{\psi_{tot}}$. In turn adding $(-Mc^2 +V)\psi$ on both sides, with $(E_{tot}-Mc^2)_{op}=E_{op}$, $T_{op} =E_{op}-V= (\frac{\g m\ve^2}{\g+1} )_{op} = \frac{\g (p^2)_{op} r^2 }{ (\g+1) m r^2} $, and $m c^2-Mc^2 +V= T+V=E$, we obtain the (reduced) SQR-KGE - or the relativistic Schr\"odiger equation, 
$$\displaylines{
\refstepcounter{equation} \label{sqr-kge}
\hfill
 [\frac{\g (p^2)_{op} r^2 }{(\g+1) m r^2} +V]\psi=E \psi
\hfill (\ref{sqr-kge})
}$$
For rotation, reserving $(p^2)_{op}$ for the non-inertial rotational motion, and using $(p_r^2)_{op}$ for an inertial radial motion, (\ref{sqr-kge}) becomes $[ \frac{ \g \lf( (\pmt_{r}^2)_{op} + (\pmt^2)_{op}  \rt) r^2 }{(\g+1) m r^2}+ V]\psi = E \psi$, the (reduced) SQR-KGE or relativistic Schr\"odinger equation used in the main text. This reduces in the limit $\ve/c<<1$,  $\g \dot{=}1$, $m\dot{=}M$ to the non relativistic form $[\frac{ 1}{2M }\lf((p^2)_{op} + (p_r^2)_{op} \rt) +V] \psi =E \psi$.

\section[\qquad \qquad \quad Errorta]{Errata for paper \cite{jxzj-pimu} }  \label{AppErrota}
 
(i)
In \cite{jxzj-pimu}, the mass of $h^\pm_\rho$,  
$2 \times \mr^\dastp_{2,1}$, (and similarly $2 \times \mr^\dastp_{2} $  of $h^\pm$), was inappropriately  argued to result from a double energy level of the HPs (hence DHPs) that would be physical only if the two energy quanta that are both formed and observed in the cmf.
(ii)
The exact total  mass energy of the $\fit$ particle of $e^{\p\dastp},e^{\m \dastp}$ is $\mr_{2,1(e)}^\dastp c^2$ as given using (\ref{eq-mass-Eng}), hence $M_{h_\rho}= 2\mr_{2,1(e)}^\dastp$. The $M_{h_\rho}$ mass ($2\mr_{2,1(e)}^\dastp +M_e$) given in \cite{jxzj-pimu} incorrectly contains a term $M_e$.

\end{appendix}


\eject

\end{document}